\documentclass[%
 reprint, twocolumn,
 superscriptaddress,
 nofootinbib,
 amsmath,amssymb,
 aps, prd,
]{revtex4-2} 

\usepackage[table,svgnames,dvipsnames]{xcolor}
\usepackage[modulo,mathlines]{lineno}
\usepackage[normalem]{ulem}
\usepackage{aas_macros}
\usepackage{subfigure}
\usepackage{graphicx} 
\usepackage{graphics}
\usepackage{dcolumn} 
\usepackage{bm} 
\usepackage{cases} 
\usepackage{booktabs}
\usepackage{comment}
\usepackage{multirow}
\usepackage{makecell}
\usepackage{siunitx}
\usepackage{tabularx}
\usepackage{xspace}
\usepackage{soul} 
\graphicspath{{figures/}}
\usepackage{fontawesome}
\usepackage{hyperref} 
\hypersetup{colorlinks=true,citecolor=blue,filecolor=blue,urlcolor=blue,linkcolor=blue}
\usepackage{adjustbox}
\usepackage{orcidlink} 
\usepackage{hanging} 

\renewcommand{\arraystretch}{1.4}
\def\be{\begin{equation}}
\def\ee{\end{equation}}

\def\ba#1\ea{\begin{align*}#1\end{align*}}

\renewcommand{\emph}[1]{\textit{#1}}

\usepackage[nameinlink,noabbrev]{cleveref}
\crefname{equation}{Eq.}{Eqs.}
\crefname{section}{Section}{Sections}
\crefname{figure}{Figure}{Figures}
\crefname{table}{Table}{Tables}
\crefname{appendix}{Appendix}{Appendices}
\Crefname{figure}{Figure}{Figures}
\Crefname{equation}{Equation}{Equations}
\Crefname{table}{Table}{Tables}
\crefformat{equation}{\eqA eq.~\eqB #2#1#3)}
\newcommand{\eqA}{}
\newcommand{\eqB}{(}

\newcommand{\mksym}[1]{\ifmmode {\rm #1}\else #1\fi}

\newcommand{\Om}{\Omega_\mathrm{m}}
\newcommand{\Ocdm}{\Omega_\mathrm{c}}
\newcommand{\Ob}{\Omega_\mathrm{b}}

\newcommand{\lcdm}{$\Lambda$CDM} 
 
\newcommand{\wowacdm}{$w_0\wa$CDM}

\newcommand{\Hrd}{H_0r_\mathrm{d}}
\newcommand{\rd}{r_\mathrm{d}}

\newcommand{\sumnu}{\sum m_\nu}

\newcommand{\wa}{w_a}
\newcommand{\dchisq}{\Delta\chi^2_\mathrm{MAP}}

\definecolor{otherblue}{rgb}{0.20, 0.73, 0.92}

\begin{document}
\title{Cosmological implications of DESI DR2 BAO measurements in light of the latest ACT DR6 CMB data
}


\author{\orcidlink{0000-0003-1481-4294}C.~Garcia-Quintero}
\thanks{NASA Einstein Fellow \\E-Mail: \url{cgarciaquintero@cfa.harvard.edu}}
\affiliation{Center for Astrophysics $|$ Harvard \& Smithsonian, 60 Garden Street, Cambridge, MA 02138, USA}

\author{\orcidlink{0000-0002-3397-3998}H.~E.~Noriega}
\affiliation{Instituto de Ciencias F\'{\i}sicas, Universidad Nacional Aut\'onoma de M\'exico, Av. Universidad s/n, Cuernavaca, Morelos, C.~P.~62210, M\'exico}
\affiliation{Instituto de F\'{\i}sica, Universidad Nacional Aut\'{o}noma de M\'{e}xico,  Circuito de la Investigaci\'{o}n Cient\'{\i}fica, Ciudad Universitaria, Cd. de M\'{e}xico  C.~P.~04510,  M\'{e}xico}
\author{\orcidlink{0000-0003-0920-2947}A.~de~Mattia}
\affiliation{IRFU, CEA, Universit\'{e} Paris-Saclay, F-91191 Gif-sur-Yvette, France}
\author{\orcidlink{0000-0001-5998-3986}A.~Aviles}
\affiliation{Instituto de Ciencias F\'{\i}sicas, Universidad Nacional Aut\'onoma de M\'exico, Av. Universidad s/n, Cuernavaca, Morelos, C.~P.~62210, M\'exico}
\author{\orcidlink{0009-0004-2558-5655}K.~Lodha}
\affiliation{Korea Astronomy and Space Science Institute, 776, Daedeokdae-ro, Yuseong-gu, Daejeon 34055, Republic of Korea}
\affiliation{University of Science and Technology, 217 Gajeong-ro, Yuseong-gu, Daejeon 34113, Republic of Korea}
\author{\orcidlink{0009-0006-7300-6616}D.~Chebat}
\affiliation{IRFU, CEA, Universit\'{e} Paris-Saclay, F-91191 Gif-sur-Yvette, France}
\author{\orcidlink{0000-0001-6423-9799}J.~Rohlf}
\affiliation{Physics Dept., Boston University, 590 Commonwealth Avenue, Boston, MA 02215, USA}
\author{\orcidlink{0000-0001-9070-3102}S.~Nadathur}
\affiliation{Institute of Cosmology and Gravitation, University of Portsmouth, Dennis Sciama Building, Portsmouth, PO1 3FX, UK}
\author{\orcidlink{0000-0002-2207-6108}W.~Elbers}
\affiliation{Institute for Computational Cosmology, Department of Physics, Durham University, South Road, Durham DH1 3LE, UK}

\author{J.~Aguilar}
\affiliation{Lawrence Berkeley National Laboratory, 1 Cyclotron Road, Berkeley, CA 94720, USA}

\author{\orcidlink{0000-0001-6098-7247}S.~Ahlen}
\affiliation{Department of Physics, Boston University, 590 Commonwealth Avenue, Boston, MA 02215 USA}

\author{O.~Alves}
\affiliation{University of Michigan, 500 S. State Street, Ann Arbor, MI 48109, USA}

\author{\orcidlink{0000-0002-4118-8236}U.~Andrade}
\affiliation{Leinweber Center for Theoretical Physics, University of Michigan, 450 Church Street, Ann Arbor, Michigan 48109-1040, USA}
\affiliation{University of Michigan, 500 S. State Street, Ann Arbor, MI 48109, USA}

\author{\orcidlink{0000-0003-4162-6619}S.~Bailey}
\affiliation{Lawrence Berkeley National Laboratory, 1 Cyclotron Road, Berkeley, CA 94720, USA}

\author{\orcidlink{0000-0001-5537-4710}S.~BenZvi}
\affiliation{Department of Physics \& Astronomy, University of Rochester, 206 Bausch and Lomb Hall, P.O. Box 270171, Rochester, NY 14627-0171, USA}

\author{\orcidlink{0000-0001-9712-0006}D.~Bianchi}
\affiliation{Dipartimento di Fisica ``Aldo Pontremoli'', Universit\`a degli Studi di Milano, Via Celoria 16, I-20133 Milano, Italy}
\affiliation{INAF-Osservatorio Astronomico di Brera, Via Brera 28, 20122 Milano, Italy}

\author{D.~Brooks}
\affiliation{Department of Physics \& Astronomy, University College London, Gower Street, London, WC1E 6BT, UK}

\author{E.~Burtin}
\affiliation{IRFU, CEA, Universit\'{e} Paris-Saclay, F-91191 Gif-sur-Yvette, France}

\author{\orcidlink{0000-0002-8215-7292}R.~Calderon}
\affiliation{CEICO, Institute of Physics of the Czech Academy of Sciences, Na Slovance 1999/2, 182 21, Prague, Czech Republic.}

\author{\orcidlink{0000-0003-3044-5150}A.~Carnero Rosell}
\affiliation{Departamento de Astrof\'{\i}sica, Universidad de La Laguna (ULL), E-38206, La Laguna, Tenerife, Spain}
\affiliation{Instituto de Astrof\'{\i}sica de Canarias, C/ V\'{\i}a L\'{a}ctea, s/n, E-38205 La Laguna, Tenerife, Spain}

\author{P.~Carrilho}
\affiliation{Institute for Astronomy, University of Edinburgh, Royal Observatory, Blackford Hill, Edinburgh EH9 3HJ, UK}

\author{\orcidlink{0000-0001-7316-4573}F.~J.~Castander}
\affiliation{Institut d'Estudis Espacials de Catalunya (IEEC), c/ Esteve Terradas 1, Edifici RDIT, Campus PMT-UPC, 08860 Castelldefels, Spain}
\affiliation{Institute of Space Sciences, ICE-CSIC, Campus UAB, Carrer de Can Magrans s/n, 08913 Bellaterra, Barcelona, Spain}

\author{\orcidlink{0000-0001-8996-4874}E.~Chaussidon}
\affiliation{Lawrence Berkeley National Laboratory, 1 Cyclotron Road, Berkeley, CA 94720, USA}

\author{T.~Claybaugh}
\affiliation{Lawrence Berkeley National Laboratory, 1 Cyclotron Road, Berkeley, CA 94720, USA}

\author{\orcidlink{0000-0002-5954-7903}S.~Cole}
\affiliation{Institute for Computational Cosmology, Department of Physics, Durham University, South Road, Durham DH1 3LE, UK}

\author{\orcidlink{0000-0002-2169-0595}A.~Cuceu}
\thanks{NASA Einstein Fellow \\E-Mail: \url{cgarciaquintero@cfa.harvard.edu}}
\affiliation{Lawrence Berkeley National Laboratory, 1 Cyclotron Road, Berkeley, CA 94720, USA}

\author{\orcidlink{0000-0003-3660-4028}R.~de Belsunce}
\affiliation{Lawrence Berkeley National Laboratory, 1 Cyclotron Road, Berkeley, CA 94720, USA}

\author{\orcidlink{0000-0002-1769-1640}A.~de la Macorra}
\affiliation{Instituto de F\'{\i}sica, Universidad Nacional Aut\'{o}noma de M\'{e}xico,  Circuito de la Investigaci\'{o}n Cient\'{\i}fica, Ciudad Universitaria, Cd. de M\'{e}xico  C.~P.~04510,  M\'{e}xico}

\author{\orcidlink{0000-0002-7311-4506}N.~Deiosso}
\affiliation{CIEMAT, Avenida Complutense 40, E-28040 Madrid, Spain}

\author{\orcidlink{0000-0003-0928-2000}J.~Della~Costa}
\affiliation{Department of Astronomy, San Diego State University, 5500 Campanile Drive, San Diego, CA 92182, USA}
\affiliation{NSF NOIRLab, 950 N. Cherry Ave., Tucson, AZ 85719, USA}

\author{\orcidlink{0000-0002-4928-4003}Arjun~Dey}
\affiliation{NSF NOIRLab, 950 N. Cherry Ave., Tucson, AZ 85719, USA}

\author{\orcidlink{0000-0002-5665-7912}Biprateep~Dey}
\affiliation{Department of Astronomy \& Astrophysics, University of Toronto, Toronto, ON M5S 3H4, Canada}
\affiliation{Department of Physics \& Astronomy and Pittsburgh Particle Physics, Astrophysics, and Cosmology Center (PITT PACC), University of Pittsburgh, 3941 O'Hara Street, Pittsburgh, PA 15260, USA}

\author{\orcidlink{0000-0002-3369-3718}Z.~Ding}
\affiliation{University of Chinese Academy of Sciences, Nanjing 211135, People's Republic of China.}

\author{P.~Doel}
\affiliation{Department of Physics \& Astronomy, University College London, Gower Street, London, WC1E 6BT, UK}

\author{\orcidlink{0000-0002-3033-7312}A.~Font-Ribera}
\affiliation{Institut de F\'{i}sica d’Altes Energies (IFAE), The Barcelona Institute of Science and Technology, Edifici Cn, Campus UAB, 08193, Bellaterra (Barcelona), Spain}

\author{\orcidlink{0000-0002-2890-3725}J.~E.~Forero-Romero}
\affiliation{Departamento de F\'isica, Universidad de los Andes, Cra. 1 No. 18A-10, Edificio Ip, CP 111711, Bogot\'a, Colombia}
\affiliation{Observatorio Astron\'omico, Universidad de los Andes, Cra. 1 No. 18A-10, Edificio H, CP 111711 Bogot\'a, Colombia}

\author{E.~Gaztañaga}
\affiliation{Institut d'Estudis Espacials de Catalunya (IEEC), c/ Esteve Terradas 1, Edifici RDIT, Campus PMT-UPC, 08860 Castelldefels, Spain}
\affiliation{Institute of Cosmology and Gravitation, University of Portsmouth, Dennis Sciama Building, Portsmouth, PO1 3FX, UK}
\affiliation{Institute of Space Sciences, ICE-CSIC, Campus UAB, Carrer de Can Magrans s/n, 08913 Bellaterra, Barcelona, Spain}

\author{\orcidlink{0000-0003-0265-6217}H.~Gil-Mar\'in}
\affiliation{Departament de F\'{\i}sica Qu\`{a}ntica i Astrof\'{\i}sica, Universitat de Barcelona, Mart\'{\i} i Franqu\`{e}s 1, E08028 Barcelona, Spain}
\affiliation{Institut d'Estudis Espacials de Catalunya (IEEC), c/ Esteve Terradas 1, Edifici RDIT, Campus PMT-UPC, 08860 Castelldefels, Spain}
\affiliation{Institut de Ci\`encies del Cosmos (ICCUB), Universitat de Barcelona (UB), c. Mart\'i i Franqu\`es, 1, 08028 Barcelona, Spain.}

\author{\orcidlink{0000-0003-3142-233X}S.~Gontcho A Gontcho}
\affiliation{Lawrence Berkeley National Laboratory, 1 Cyclotron Road, Berkeley, CA 94720, USA}

\author{G.~Gutierrez}
\affiliation{Fermi National Accelerator Laboratory, PO Box 500, Batavia, IL 60510, USA}

\author{\orcidlink{0000-0001-9822-6793}J.~Guy}
\affiliation{Lawrence Berkeley National Laboratory, 1 Cyclotron Road, Berkeley, CA 94720, USA}

\author{\orcidlink{0000-0003-1197-0902}C.~Hahn}
\affiliation{Steward Observatory, University of Arizona, 933 N. Cherry Avenue, Tucson, AZ 85721, USA}
\affiliation{Steward Observatory, University of Arizona, 933 N. Cherry Avenue, Tucson, AZ 85721, USA}

\author{\orcidlink{0000-0002-9136-9609}H.~K.~Herrera-Alcantar}
\affiliation{Institut d'Astrophysique de Paris. 98 bis boulevard Arago. 75014 Paris, France}
\affiliation{IRFU, CEA, Universit\'{e} Paris-Saclay, F-91191 Gif-sur-Yvette, France}

\author{\orcidlink{0000-0002-6550-2023}K.~Honscheid}
\affiliation{Center for Cosmology and AstroParticle Physics, The Ohio State University, 191 West Woodruff Avenue, Columbus, OH 43210, USA}
\affiliation{Department of Physics, The Ohio State University, 191 West Woodruff Avenue, Columbus, OH 43210, USA}
\affiliation{The Ohio State University, Columbus, 43210 OH, USA}

\author{\orcidlink{0000-0002-1081-9410}C.~Howlett}
\affiliation{School of Mathematics and Physics, University of Queensland, Brisbane, QLD 4072, Australia}

\author{\orcidlink{0000-0001-6558-0112}D.~Huterer}
\affiliation{Department of Physics, University of Michigan, 450 Church Street, Ann Arbor, MI 48109, USA}
\affiliation{University of Michigan, 500 S. State Street, Ann Arbor, MI 48109, USA}

\author{\orcidlink{0000-0002-6024-466X}M.~Ishak}
\affiliation{Department of Physics, The University of Texas at Dallas, 800 W. Campbell Rd., Richardson, TX 75080, USA}

\author{\orcidlink{0000-0002-0000-2394}S.~Juneau}
\affiliation{NSF NOIRLab, 950 N. Cherry Ave., Tucson, AZ 85719, USA}

\author{R.~Kehoe}
\affiliation{Department of Physics, Southern Methodist University, 3215 Daniel Avenue, Dallas, TX 75275, USA}

\author{\orcidlink{0000-0002-8828-5463}D.~Kirkby}
\affiliation{Department of Physics and Astronomy, University of California, Irvine, 92697, USA}

\author{\orcidlink{0000-0001-6356-7424}A.~Kremin}
\affiliation{Lawrence Berkeley National Laboratory, 1 Cyclotron Road, Berkeley, CA 94720, USA}

\author{O.~Lahav}
\affiliation{Department of Physics \& Astronomy, University College London, Gower Street, London, WC1E 6BT, UK}

\author{\orcidlink{0000-0002-6731-9329}C.~Lamman}
\affiliation{Center for Astrophysics $|$ Harvard \& Smithsonian, 60 Garden Street, Cambridge, MA 02138, USA}

\author{\orcidlink{0000-0003-1838-8528}M.~Landriau}
\affiliation{Lawrence Berkeley National Laboratory, 1 Cyclotron Road, Berkeley, CA 94720, USA}

\author{\orcidlink{0000-0001-7178-8868}L.~Le~Guillou}
\affiliation{Sorbonne Universit\'{e}, CNRS/IN2P3, Laboratoire de Physique Nucl\'{e}aire et de Hautes Energies (LPNHE), FR-75005 Paris, France}

\author{\orcidlink{0000-0002-3677-3617}A.~Leauthaud}
\affiliation{Department of Astronomy and Astrophysics, UCO/Lick Observatory, University of California, 1156 High Street, Santa Cruz, CA 95064, USA}
\affiliation{Department of Astronomy and Astrophysics, University of California, Santa Cruz, 1156 High Street, Santa Cruz, CA 95065, USA}

\author{\orcidlink{0000-0003-1887-1018}M.~E.~Levi}
\affiliation{Lawrence Berkeley National Laboratory, 1 Cyclotron Road, Berkeley, CA 94720, USA}

\author{\orcidlink{0000-0003-3616-6486}Q.~Li}
\affiliation{Department of Physics and Astronomy, The University of Utah, 115 South 1400 East, Salt Lake City, UT 84112, USA}

\author{\orcidlink{0000-0003-4962-8934}M.~Manera}
\affiliation{Departament de F\'{i}sica, Serra H\'{u}nter, Universitat Aut\`{o}noma de Barcelona, 08193 Bellaterra (Barcelona), Spain}
\affiliation{Institut de F\'{i}sica d’Altes Energies (IFAE), The Barcelona Institute of Science and Technology, Edifici Cn, Campus UAB, 08193, Bellaterra (Barcelona), Spain}

\author{\orcidlink{0000-0002-4279-4182}P.~Martini}
\affiliation{Center for Cosmology and AstroParticle Physics, The Ohio State University, 191 West Woodruff Avenue, Columbus, OH 43210, USA}
\affiliation{Department of Astronomy, The Ohio State University, 4055 McPherson Laboratory, 140 W 18th Avenue, Columbus, OH 43210, USA}
\affiliation{The Ohio State University, Columbus, 43210 OH, USA}

\author{\orcidlink{0000-0001-6957-772X}W.~L.~Matthewson}
\affiliation{Korea Astronomy and Space Science Institute, 776, Daedeokdae-ro, Yuseong-gu, Daejeon 34055, Republic of Korea}

\author{\orcidlink{0000-0002-1125-7384}A.~Meisner}
\affiliation{NSF NOIRLab, 950 N. Cherry Ave., Tucson, AZ 85719, USA}

\author{\orcidlink{0000-0001-9497-7266}J.~Mena-Fern\'andez}
\affiliation{Laboratoire de Physique Subatomique et de Cosmologie, 53 Avenue des Martyrs, 38000 Grenoble, France}

\author{R.~Miquel}
\affiliation{Instituci\'{o} Catalana de Recerca i Estudis Avan\c{c}ats, Passeig de Llu\'{\i}s Companys, 23, 08010 Barcelona, Spain}
\affiliation{Institut de F\'{i}sica d’Altes Energies (IFAE), The Barcelona Institute of Science and Technology, Edifici Cn, Campus UAB, 08193, Bellaterra (Barcelona), Spain}

\author{\orcidlink{0000-0002-2733-4559}J.~Moustakas}
\affiliation{Department of Physics and Astronomy, Siena College, 515 Loudon Road, Loudonville, NY 12211, USA}

\author{A.~Muñoz-Gutiérrez}
\affiliation{Instituto de F\'{\i}sica, Universidad Nacional Aut\'{o}noma de M\'{e}xico,  Circuito de la Investigaci\'{o}n Cient\'{\i}fica, Ciudad Universitaria, Cd. de M\'{e}xico  C.~P.~04510,  M\'{e}xico}

\author{\orcidlink{0000-0001-8684-2222}J.~ A.~Newman}
\affiliation{Department of Physics \& Astronomy and Pittsburgh Particle Physics, Astrophysics, and Cosmology Center (PITT PACC), University of Pittsburgh, 3941 O'Hara Street, Pittsburgh, PA 15260, USA}

\author{\orcidlink{0000-0002-1544-8946}G.~Niz}
\affiliation{Departamento de F\'{\i}sica, DCI-Campus Le\'{o}n, Universidad de Guanajuato, Loma del Bosque 103, Le\'{o}n, Guanajuato C.~P.~37150, M\'{e}xico}
\affiliation{Instituto Avanzado de Cosmolog\'{\i}a A.~C., San Marcos 11 - Atenas 202. Magdalena Contreras. Ciudad de M\'{e}xico C.~P.~10720, M\'{e}xico}

\author{\orcidlink{0000-0002-4637-2868}E.~Paillas}
\affiliation{Steward Observatory, University of Arizona, 933 N. Cherry Avenue, Tucson, AZ 85721, USA}

\author{\orcidlink{0000-0003-3188-784X}N.~Palanque-Delabrouille}
\affiliation{IRFU, CEA, Universit\'{e} Paris-Saclay, F-91191 Gif-sur-Yvette, France}
\affiliation{Lawrence Berkeley National Laboratory, 1 Cyclotron Road, Berkeley, CA 94720, USA}

\author{\orcidlink{0000-0001-9685-5756}J.~Pan}
\affiliation{University of Michigan, 500 S. State Street, Ann Arbor, MI 48109, USA}

\author{\orcidlink{0000-0002-0644-5727}W.~J.~Percival}
\affiliation{Department of Physics and Astronomy, University of Waterloo, 200 University Ave W, Waterloo, ON N2L 3G1, Canada}
\affiliation{Perimeter Institute for Theoretical Physics, 31 Caroline St. North, Waterloo, ON N2L 2Y5, Canada}
\affiliation{Waterloo Centre for Astrophysics, University of Waterloo, 200 University Ave W, Waterloo, ON N2L 3G1, Canada}

\author{\orcidlink{0000-0001-7145-8674}F.~Prada}
\affiliation{Instituto de Astrof\'{i}sica de Andaluc\'{i}a (CSIC), Glorieta de la Astronom\'{i}a, s/n, E-18008 Granada, Spain}

\author{\orcidlink{0000-0001-6979-0125}I.~P\'erez-R\`afols}
\affiliation{Departament de F\'isica, EEBE, Universitat Polit\`ecnica de Catalunya, c/Eduard Maristany 10, 08930 Barcelona, Spain}

\author{\orcidlink{0000-0001-7144-2349}M.~Rashkovetskyi}
\affiliation{Center for Astrophysics $|$ Harvard \& Smithsonian, 60 Garden Street, Cambridge, MA 02138, USA}

\author{\orcidlink{0000-0002-3500-6635}C.~Ravoux}
\affiliation{Universit\'{e} Clermont-Auvergne, CNRS, LPCA, 63000 Clermont-Ferrand, France}

\author{\orcidlink{0000-0002-7522-9083}A.~J.~Ross}
\affiliation{Center for Cosmology and AstroParticle Physics, The Ohio State University, 191 West Woodruff Avenue, Columbus, OH 43210, USA}
\affiliation{Department of Astronomy, The Ohio State University, 4055 McPherson Laboratory, 140 W 18th Avenue, Columbus, OH 43210, USA}
\affiliation{The Ohio State University, Columbus, 43210 OH, USA}

\author{G.~Rossi}
\affiliation{Department of Physics and Astronomy, Sejong University, 209 Neungdong-ro, Gwangjin-gu, Seoul 05006, Republic of Korea}

\author{\orcidlink{0000-0002-9646-8198}E.~Sanchez}
\affiliation{CIEMAT, Avenida Complutense 40, E-28040 Madrid, Spain}

\author{D.~Schlegel}
\affiliation{Lawrence Berkeley National Laboratory, 1 Cyclotron Road, Berkeley, CA 94720, USA}

\author{M.~Schubnell}
\affiliation{Department of Physics, University of Michigan, 450 Church Street, Ann Arbor, MI 48109, USA}
\affiliation{University of Michigan, 500 S. State Street, Ann Arbor, MI 48109, USA}

\author{\orcidlink{0000-0002-6588-3508}H.~Seo}
\affiliation{Department of Physics \& Astronomy, Ohio University, 139 University Terrace, Athens, OH 45701, USA}

\author{\orcidlink{0000-0001-6815-0337}A.~Shafieloo}
\affiliation{Korea Astronomy and Space Science Institute, 776, Daedeokdae-ro, Yuseong-gu, Daejeon 34055, Republic of Korea}
\affiliation{University of Science and Technology, 217 Gajeong-ro, Yuseong-gu, Daejeon 34113, Republic of Korea}

\author{\orcidlink{0000-0002-3461-0320}J.~Silber}
\affiliation{Lawrence Berkeley National Laboratory, 1 Cyclotron Road, Berkeley, CA 94720, USA}

\author{D.~Sprayberry}
\affiliation{NSF NOIRLab, 950 N. Cherry Ave., Tucson, AZ 85719, USA}

\author{\orcidlink{0000-0003-1704-0781}G.~Tarl\'{e}}
\affiliation{University of Michigan, 500 S. State Street, Ann Arbor, MI 48109, USA}

\author{P.~Taylor}
\affiliation{The Ohio State University, Columbus, 43210 OH, USA}

\author{\orcidlink{0000-0003-3841-1836}M.~Vargas-Maga\~na}
\affiliation{Instituto de F\'{\i}sica, Universidad Nacional Aut\'{o}noma de M\'{e}xico,  Circuito de la Investigaci\'{o}n Cient\'{\i}fica, Ciudad Universitaria, Cd. de M\'{e}xico  C.~P.~04510,  M\'{e}xico}

\author{\orcidlink{0000-0002-1748-3745}M.~Walther}
\affiliation{Excellence Cluster ORIGINS, Boltzmannstrasse 2, D-85748 Garching, Germany}
\affiliation{University Observatory, Faculty of Physics, Ludwig-Maximilians-Universit\"{a}t, Scheinerstr. 1, 81677 M\"{u}nchen, Germany}

\author{B.~A.~Weaver}
\affiliation{NSF NOIRLab, 950 N. Cherry Ave., Tucson, AZ 85719, USA}

\author{\orcidlink{0000-0001-5146-8533}C.~Yèche}
\affiliation{IRFU, CEA, Universit\'{e} Paris-Saclay, F-91191 Gif-sur-Yvette, France}

\author{\orcidlink{0000-0002-7305-9578}P.~Zarrouk}
\affiliation{Sorbonne Universit\'{e}, CNRS/IN2P3, Laboratoire de Physique Nucl\'{e}aire et de Hautes Energies (LPNHE), FR-75005 Paris, France}

\author{Z.~Zhai}
\affiliation{Department of Astronomy, School of Physics and Astronomy, Shanghai Jiao Tong University, Shanghai 200240, China}

\author{\orcidlink{0000-0002-1991-7295}C.~Zhao}
\affiliation{Department of Astronomy, Tsinghua University, 30 Shuangqing Road, Haidian District, Beijing, China, 100190}

\author{\orcidlink{0000-0001-5381-4372}R.~Zhou}
\affiliation{Lawrence Berkeley National Laboratory, 1 Cyclotron Road, Berkeley, CA 94720, USA}

\collaboration{The DESI collaboration}

\begin{abstract}
We report cosmological results from the Dark Energy Spectroscopic Instrument (DESI) measurements of baryon acoustic oscillations (BAO) when combined with recent data from the Atacama Cosmology Telescope (ACT). By jointly analyzing ACT and {\it Planck} data and applying conservative cuts to overlapping multipole ranges, we assess how different {\it Planck}+ACT dataset combinations affect consistency with DESI. While ACT alone exhibits a tension with DESI exceeding 3$\sigma$ within the $\Lambda$CDM model, this discrepancy is reduced when ACT is analyzed in combination with {\it Planck}. For our baseline DESI DR2 BAO+{\it Planck} PR4+ACT likelihood combination, the preference for evolving dark energy over a cosmological constant is about 3$\sigma$, increasing to over 4$\sigma$ with the inclusion of Type Ia supernova data. While the dark energy results remain quite consistent across various combinations of {\it Planck} and ACT likelihoods with those obtained by the DESI collaboration, the constraints on neutrino mass are more sensitive, ranging from $\sumnu < 0.061$ eV in our baseline analysis, to $\sumnu < 0.077$ eV (95\% confidence level) in the CMB likelihood combination chosen by ACT when imposing the physical prior $\sumnu>0$ eV.

\end{abstract}

\maketitle

\section{Introduction}

The current understanding of the expansion of the Universe has been built upon rapidly evolving observations from a variety of cosmological probes. Each cosmological dataset provides information about the Universe at different epochs.
The Cosmic Microwave Background (CMB) data provide a description of the Universe at redshift $z\approx 1090$, a mere 400,000 years after the Big Bang. 
Baryon acoustic oscillations (BAO) from galaxy surveys constrain the expansion history of the Universe at $0.1<z<4.2$, exploring the matter dominated era and the recent era of cosmic acceleration. Type Ia Supernovae (SNe Ia) data were used to discover dark energy \cite{1998AJ....116.1009R,1999ApJ...517..565P} and constrain the expansion history at low redshifts. The combination of these cosmological probes across the evolving Universe provides quantitative information on the temporal evolution of the dark energy density.

The most precise BAO measurements to date come from the Dark Energy Spectroscopic Instrument (DESI). DESI is installed on the Mayall Telescope at the Kitt Peak National Observatory \cite{Snowmass2013.Levi,DESI2016a.Science,DESI2016b.Instr,DESI2022.KP1.Instr}, and uses robotically controlled fiber-optic cables \cite{FocalPlane.Silber.2023,FiberSystem.Poppett.2024,Corrector.Miller.2023,Spectro.Pipeline.Guy.2023,SurveyOps.Schlafly.2023} to simultaneously capture light from up to 5,000 pre-selected galaxies and quasars across an eight-square-degree field of view. By analyzing the spectra of collected light, DESI has obtained redshift information on over 30 million extragalactic objects during the first three years of observations \cite{DESI2023a.KP1.SV,DESI2023b.KP1.EDR,DESI2024.I.DR1}. BAO measurements from more than 14 million galaxies and quasars \cite{Y3.clust-s1.Andrade.2025,DESI.DR2.BAO.cosmo}, as well as from 820,000 Lyman-$\alpha$ forest spectra \cite{Y3.lya-s1.Casas.2025,Y3.lya-s2.Brodzeller.2025,DESI.DR2.BAO.lya}
have significantly contributed to current knowledge of cosmological parameters. 

Recently, the DESI collaboration presented a cosmological analysis based on the latest BAO measurements from its Data Release 2 (DR2) \cite{DESI.DR2.BAO.cosmo}. This analysis points to discrepancies between datasets becoming more relevant within the $\Lambda$CDM model, highlighting evolving dark energy as a possible solution (or alternatively, unrecognized systematics in one or more datasets, or a rare statistical fluctuation), and also reporting updated constraints on the neutrino mass (with an extended analysis allowing an effective neutrino mass parameter that allows negative values \cite{Y3.cpe-s2.Elbers.2025}). In particular, while the DESI DR2 BAO results are broadly consistent with the picture of the  $\Lambda$CDM cosmological model, they exhibit a 2.3$\sigma$ tension \cite{DESI.DR2.BAO.cosmo} with {\it Planck} CMB data (including external CMB lensing data from \cite{Qu:2023}) \cite{DESI2024.I.DR1,DESI2024.II.KP3,DESI2024.III.KP4,DESI2024.IV.KP6,DESI2024.VI.KP7A,DESI2024.VII.KP7B}. On the other hand, a time evolving dark energy component is favored by the joint analysis of these datasets, a scenario in which the datasets are consistent. This behavior can be deduced through either a simple CPL parameterization ($w_0w_a \rm CDM$) \cite{Chevallier:2001,Linder2003}, or other dark energy reconstruction methods \cite{Y3.cpe-s1.Lodha.2025}. Using the $w_0w_a$CDM parameterization, DESI BAO DR2 combined with CMB temperature and polarization anisotropies, as well as CMB lensing, shows evolving dark energy is preferred at $3.1\sigma$, increasing up to $4.2\sigma$ when including SNe data \cite{DESI.DR2.BAO.cosmo} (this preference is also supported by the Dark Energy Survey BAO and SN combined analysis \cite{DES:2025bxy}). 

Around the same time as the DESI DR2 results were published, the Atacama Cosmology Telescope (ACT) collaboration also presented their final results on the CMB power spectrum measurements and cosmological implications \cite{ACT:2025fju,2025arXiv250314451N,ACT:2025tim}, based on their Data Release 6 (DR6) with five years of observations. The ACT results provide measurements on the small-scale region of the CMB spectrum, adding extra information to the damping tail of the CMB not covered by the {\it Planck} space-based mission. Conversely, {\it Planck} is able to resolve anisotropies in the temperature and polarization of the CMB maps at large scales not accessible to ACT ($\ell<600$), making the two CMB datasets complementary to each other.

Given the mild discrepancy between DESI BAO and {\it Planck} CMB (including ACT DR6 CMB lensing), it becomes important to evaluate the effect of the final ACT DR6 dataset in the context of DESI BAO, whether analyzed individually or combined with {\it Planck}. A conservative combination of ACT and {\it Planck} involves considering non-overlapping multipole ranges of each CMB spectrum.  While in \cite{ACT:2025fju} the complete ACT dataset was complemented with CMB measurements from {\it Planck} up to $\ell=1000$ to provide a complete CMB picture, other combinations with different multipole cuts are of interest, notably combinations between ACT and {\it Planck} based on the multipole ranges where each CMB experiment is more precise. Different combinations may impact parameters such as the physical cold-dark-matter density $\Omega_\mathrm{c} h^2$, which can affect the discrepancy between DESI and CMB, and impact the evidence in favor of $w_0w_a$CDM. While ACT alone reports a higher $\Omega_\mathrm{c} h^2$ value compared to {\it Planck} \cite{ACT:2025fju} (a $\sim$1.5$\sigma$ shift), the P-ACT combination used by the ACT collaboration measures a slightly lower $\Omega_\mathrm{c} h^2$ (where $h$ is the hubble constant normalized to 100 km s$^{-1}$ Mpc$^{-1}$) with respect to {\it Planck}.

In addition, BAO helps break geometric degeneracies in CMB constraints on neutrino mass. DESI reported a 95\% upper limit of $\sum m_\nu < 0.064$ eV for the sum of neutrino masses in the $\Lambda$CDM model when combined with {\it Planck} CMB data, that  changes to $\sum m_\nu<0.078$ eV with an alternative CMB likelihood \cite{DESI.DR2.BAO.cosmo}. The constraints on the neutrino mass is relaxed to $\sum m_\nu<0.16$ eV in a $w_0w_a$CDM model. These differences in the neutrino mass constraints further justify a joint DESI+ACT analysis that covers the neutrino sector.

The purpose of this work is to explore the cosmological implications of including the latest ACT power spectrum data and likelihoods within the BAO+CMB+SNe combination already analyzed by the DESI Collaboration, and assess the robustness of the conclusions made by DESI in light of the latest ACT data. We extend the analysis to different combinations of the ACT and {\it Planck} datasets based on other multipole cuts that can be more precise at measuring $\Omega_c h^2$ and other parameters.

In \cref{sec:method_data}, we describe the datasets, likelihoods, and methodology.
In \cref{sec:consistency}, we test the consistency of DESI and CMB data within the $\Lambda$CDM model. In \cref{sec:dark_energy}, we present the results of dark energy for different datasets, including ACT DR6 power spectra. In \cref{sec:neutrinos}, we present constraints on the neutrino masses. Finally, we present our conclusions in \cref{sec:conclusions}.

\section{Methodology and data}
\label{sec:method_data} 

\begin{table*}
\centering
\resizebox{\textwidth}{!}{
\renewcommand{\arraystretch}{1.3}
\begin{tabular}{ p{3.5cm} | p{18cm} }
\hline \hline
Datasets & Description  \\ \hline
\multicolumn{2}{l}{\textbf{1. BAO data}}  \\ \hline
DESI DR2 & BAO measurements from  DESI DR2 in the range $0.1<z<4.2$ \cite{DESI.DR2.BAO.cosmo, DESI.DR2.BAO.lya}. \\ \hline 
\multicolumn{2}{l}{\textbf{2. SNe Ia data}} \\ \hline
Pantheon+ & A compilation of 1550 spectroscopically-confirmed SNe Ia in the range $0.001<z<2.26$~\cite{Scolnic:2021amr}. \\
Union3 & A compilation of 2087 SNe-Ia (among which 1363 SNe Ia are common to Pantheon+) that were analyzed through an updated Bayesian framework~\cite{Rubin:2023}. \\
DESY5 & A compilation of 1635 SNe Ia in the redshift range $0.10<z<1.13$ complemented by an external sample consisting of 194 SNe Ia common to Pantheon+ in the range $0.025<z<0.10$~\cite{DES:2024tys}. \\ \hline
\multicolumn{2}{l}{\textbf{3. CMB standalone likelihoods (including CMB lensing)}} \\ \hline 
low-$\ell$ TT & {\it Planck} 2018 PR3 low-$\ell$ \texttt{Commander} likelihood for TT in the range $2\leq \ell<30$ \cite{Planck-2018-likelihoods, Planck-2018-cosmology}. \\
low-$\ell$ EE \texttt{SimAll} & {\it Planck} 2018 PR3 low-$\ell$ \texttt{SimAll}  likelihood for EE in the range $2\leq \ell<30$ \cite{Planck-2018-likelihoods, Planck-2018-cosmology}. \\
low-$\ell$ EE \texttt{SRoll2} & Alternative low-$\ell$ likelihood for EE based on the \texttt{SRoll2} code in the range $2\leq \ell<30$ \cite{2020_Sroll2_Pagano}. \\
high-$\ell$ PR3 & {\it Planck} PR3 \texttt{Plik\_lite} likelihood for the high-$\ell$ CMB TT, TE, EE spectra from $\ell=30$ up to $\ell=2500$ \cite{Planck-2018-likelihoods, Planck-2018-cosmology}. \\
high-$\ell$ PR4 & {\it Planck} PR4 high-$\ell$ temperature and polarization likelihood using \texttt{NPIPE} maps. The high-$\ell$ TT, TE, EE spectra from {\it Planck} extends from $\ell=30$ up to $\ell=2500$ \cite{Efstathiou:2021, Rosenberg:2022}. \\
ACT DR6 & Power spectra from the anisotropies in the temperature and polarization CMB maps from the 6th data release of the Atacama Cosmology Telescope. The CMB power spectra extends from $\ell=600$ up to $\ell=8500$ \cite{ACT:2025fju}. \\
CMB lensing & Combination of the CMB lensing measurements from the reconstruction of the CMB lensing potential using {\it Planck} PR4 \texttt{NPIPE} maps \cite{Carron:2022}, and the CMB lensing measurements from the ACT Data Release 6 (DR6), which consists of five seasons of CMB temperature and polarization observations, with 67\% of sky fraction overlap with {\it Planck} \cite{Madhavacheril:ACT-DR6,Qu:2023}. \\ \hline
\multicolumn{2}{l}{\textbf{4. Main CMB combinations}} \\ \hline
ACT & low-$\ell$ EE \texttt{SRoll2} + ACT DR6 + CMB lensing \\
P-ACT & low-$\ell$ TT + low-$\ell$ EE \texttt{SRoll2} + high-$\ell$ PR3 ($\ell<1000$ TT, $\ell<600$ TE, EE) + ACT DR6 + CMB lensing \\
PR4+ACT & low-$\ell$ TT + low-$\ell$ EE \texttt{SimAll} + high-$\ell$ PR4 ($\ell<2000$ TT, $\ell<1000$ TE, EE) + ACT DR6 ($\ell \geq 2000$ TT, $\ell \geq 1000$ TE, EE) + CMB lensing \\ \hline
\multicolumn{2}{l}{\textbf{5. Additional CMB combinations studied}} \\ \hline
ACT (no CMB lensing) & low-$\ell$ EE \texttt{SRoll2} + ACT DR6 (same as ACT base in \cite{ACT:2025fju}) \\
ACT (low-$\ell$ TT, EE) & low-$\ell$ TT + low-$\ell$ EE \texttt{SimAll} + ACT DR6 + CMB lensing \\
PR4 & low-$\ell$ TT + low-$\ell$ EE \texttt{SimAll} + high-$\ell$ PR4 + CMB lensing (same as baseline CMB in \cite{DESI.DR2.BAO.cosmo}) \\
PR4$_{(1000,600)}$+ACT & low-$\ell$ TT + low-$\ell$ EE \texttt{SimAll} + high-$\ell$ PR4 ($\ell<1000$ TT, $\ell<600$ TE, EE) + ACT DR6 + CMB lensing \\
PR4$_{(\texttt{SRoll2})}$+ACT & low-$\ell$ TT + low-$\ell$ EE \texttt{SRoll2} + high-$\ell$ PR4 ($\ell<2000$ TT, $\ell<1000$ TE, EE) + ACT DR6 ($\ell \geq 2000$ TT, $\ell \geq 1000$ TE, EE) + CMB lensing \\
\hline\hline
\end{tabular}
 }
\caption{Summary of the primary data sets (1-3) and CMB combinations (4-5) used in this work. For the CMB data, we also indicate the individual likelihood packages that were used in the fits.}
\label{tab:datasets}
\end{table*}

The purpose of this work is to explore the cosmological implications of incorporating the latest ACT power spectrum data \cite{ACT:2025fju} into a combined analysis with BAO, CMB, and SNe observations, as presented in \cite{DESI.DR2.BAO.cosmo}. We also consider the CMB data from {\it Planck} \cite{Planck-2018-overview} and assess how the inclusion of ACT data affects the cosmological parameters. As reported in \cite{ACT:2025fju}, ACT measures a higher $\Ocdm h^2$ value compared to {\it Planck} (a 1.5$\sigma$ shift). Also, it was observed that the P-ACT combination defined in \cite{ACT:2025fju} favors a lower value of $\Ocdm h^2$ compared to both {\it Planck} (0.4$\sigma$ lower) and ACT (1.9$\sigma$ lower). Since the CMB measurement of $\Ocdm h^2$ is key for the DESI results as it can affect the discrepancy between DESI and CMB, the preference for $w_0w_a$CDM and the neutrino mass constraints, the exploration of these CMB datasets and how to combine Planck with ACT becomes relevant. Additionally, the results presented in \cite{DESI.DR2.BAO.cosmo} and \cite{ACT:2025fju} use a different version for the low-$\ell$ EE likelihood, which can have an impact on the neutrino mass constraints and deserves exploration.

The CMB high-$\ell$ power spectra from {\it Planck} and ACT are combined in a conservative way, without modeling a covariance between the surveys, but rather by applying simple data cuts.  ACT  resolves temperature anisotropies at smaller angular scales compared to {\it Planck}: its CMB spectra start at $\ell=600$ and extend up to $\ell=8500$. On the other hand, {\it Planck} can resolve the anisotropies well on large scales, even in the low-$\ell$ regime ($\ell<30$) where the distribution of the CMB spectra is non-Gaussian, while its high-$\ell$ spectra are measured up to $\ell=2500$. In our analysis, we consider the official {\it Planck} Release 3 (PR3) 
\cite{Planck-2018-likelihoods,2020A&A...641A...8P,Planck-2018-cosmology}, as well as {\it Planck} Release 4 (PR4) \cite{Planck-2020-NPIPE}, which is a reanalysis using the NPIPE processing pipeline and CMB lensing reconstruction of NPIPE maps \cite{Carron:2022}. Throughout this work, we focus on the PR4 \texttt{CamSpec} high-$\ell$ likelihood, which provides $\sim$8--10\% tighter constraints than PR3 \texttt{Plik} \cite{Rosenberg:2022} on relevant parameters for this work such as $\Ocdm h^2$ and $\Om$, when analyzing CMB alone.

We summarize the datasets used in this work in \cref{tab:datasets} and define the variations under which we combine CMB datasets. 
Motivated by the ACT baseline data as defined in \cite{ACT:2025fju}, we use the ACT DR6 high-$\ell$ spectra \texttt{MFLike} in combination with the low-$\ell$ EE {\it Planck} data analyzed through the \texttt{SRoll2} likelihood code \cite{2020_Sroll2_Pagano}\footnote{The inclusion of the low-$\ell$ EE data helps to break degeneracies between the optical depth parameter $\tau$ and the amplitude of primordial scalar fluctuations $A_{\rm s}$.} (see also \cite{Delouis:2019bub,deBelsunce:2021mec} for works related to the \texttt{SRoll2} maps) but with the inclusion of CMB lensing from the combination of {\it Planck} NPIPE and ACT DR6 as described in \cite{Madhavacheril:ACT-DR6}. We label this combination simply as ACT \footnote{Note that this would be equivalent to ACT-L, in the nomenclature used in \cite{ACT:2025fju}.}. Since the ACT data cannot fully constrain the first two CMB acoustic peaks, it is useful to complement it with data from other CMB surveys.  We define an analogous combination to P-ACT as used in \cite{ACT:2025fju}, based on a combination of ACT DR6 and PR3 (\texttt{plik\_lite})\footnote{The absolute calibration parameter $A_{Planck}$ is shared between the two CMB likelihoods.}, but with the addition of CMB lensing. We consider fits to the PR4 data only, matching the baseline CMB combination presented in \cite{DESI.DR2.BAO.cosmo}. Since {\it Planck} and ACT probe a common part of the CMB spectra, we test different combinations between the two by means of $\ell$ cuts. We consider combinations between PR4 and ACT DR6 with a baseline low-$\ell$ TT \texttt{Commander} and low-$\ell$ EE \texttt{SimAll} data from {\it Planck} ($\ell<30$) \cite{Planck-2018-likelihoods}, CMB lensing, and  mixed high-$\ell$ information from PR4 and ACT DR6 split by a multipole cut $\ell_{\text{TT}}$ and a common cut $\ell_{\text{TE,EE}}$. This cut uses PR4 TT information for $\ell<\ell_\text{TT}$ and then ACT DR6 information for higher values of $\ell$ up to $\ell=8500$. Similarly, the common TE and EE cut uses PR4 data for $\ell<\ell_\text{TE,EE}$ and ACT DR6 for the rest of the $\ell$ range up to $\ell=8500$. We define the explicit cuts we use below.

We construct a CMB combination that can be regarded as maximal in the information included from ACT DR6, using its full CMB spectra. However, since the ACT DR6 data are highly impacted by atmospheric noise in the range $600<\ell<1000$ \cite{2025arXiv250314451N}, we increase the Planck TT coverage and use $\ell_\mathrm{TT}=1000$ and $\ell_{\text{TE,EE}}=600$, as done for P-ACT. This combination is analogous to P-ACT in this work but uses PR4 \texttt{CamSpec}. We label this combination as  PR4$_{(1000,600)}$+ACT. We also test a combination with $\ell_\text{TT}=2000$ and $\ell_\text{TE,EE}=1000$ motivated by Figure 12 in \cite{ACT:2025fju}. This combination corresponds approximately to the multipole range where the precision of ACT DR6 in measuring the TT spectrum across frequency channels begins to become comparable to, or slightly exceed, that of {\it Planck}, around $\ell \sim$ 2000. Similarly, $\ell=1000$ is a point where ACT DR6 is roughly more precise in measuring TE compared to {\it Planck}, while $\ell=1000$ is also the scale at which approximately the white noise transition occurs for the ACT DR6 EE polarization data. We label this combination simply as PR4+ACT and consider it as our baseline CMB combination. We note that this choice of $\ell$ cuts may not correspond to the configuration that minimizes cosmological parameter uncertainties, but rather should provide a fair approximation. For the latter set of cuts, we also test the effect of the \texttt{SRoll2} likelihood by defining an analogous variation where we replace the low-$\ell$ EE \texttt{SimAll} likelihood with the low-$\ell$ EE \texttt{SRoll2} likelihood and label this PR4$_\text{(SRoll2)}$+ACT. \cref{tab:datasets} summarizes the CMB variations used in this paper. We discuss the effect on the parameter space for some of these combinations with respect to ACT DR6 and PR4 in \cref{appendix}.

We use DESI DR2 BAO data (see \cite{DESI.DR2.BAO.cosmo}  for definitions of co-moving distances $D_\mathrm{M}$, $D_\mathrm{H}$, and $D_\mathrm{V}$) containing measurements of $D_{\rm V}/r_{\rm d}$ at redshifts $0.1 < z < 0.4$ for the BGS tracer, and measurements of $D_{\rm H}/r_{\rm d}$ and $D_{\rm M}/r_{\rm d}$ for the rest of the tracers, LRGs at $0.4<z<0.6$ and $0.6<z<0.8$, a combined tracer LRG+ELG at $0.8<z<1.1$, the ELG tracer at $1.1<z<1.6$, QSO at $0.8<z<2.1$ plus the Ly$\alpha$ forest and the correlation with QSO positions, at $1.8<z<4.2$. Here, $\rd$ represents the sound horizon at the drag epoch, when acoustic waves stall in the primordial plasma as baryons cease to feel the `drag' from the photons.

We also used SNe Ia data from Pantheon+ that consists of 1550 spectroscopically classified type Ia SNe \cite{Brout:2022}. Similarly, we use the Union3 sample that consists of 2087 SNe Ia and uses an alternative analysis framework based on Bayesian hierarchical modeling using Unity 1.5 \cite{Rubin:2023}. We also include SNe Ia data from the DESY5 sample with 1635 photometrically classified SNe \cite{DES:2024tys}.

We test deviations from $\Lambda$CDM, corresponding to dynamical dark energy, by means of the $w_0w_a \rm CDM$ parameterization based on an equation of state of dark energy given by $w(a)=w_0+w_a(1-a)$, with $w_0$ and $w_a$ as free parameters. Finally, when performing constraints on the sum of neutrino masses, we adopt a physical prior $\sum m_\nu>0$ and assume 3 degenerate states.
Throughout this work, we adopt the priors and tension metrics as described in \cite{DESI.DR2.BAO.cosmo}.

\section{Consistency of DESI DR2 BAO and CMB within $\Lambda$CDM}
\label{sec:consistency}

\begin{figure*}
    \centering
    \includegraphics[width=0.53\textwidth]{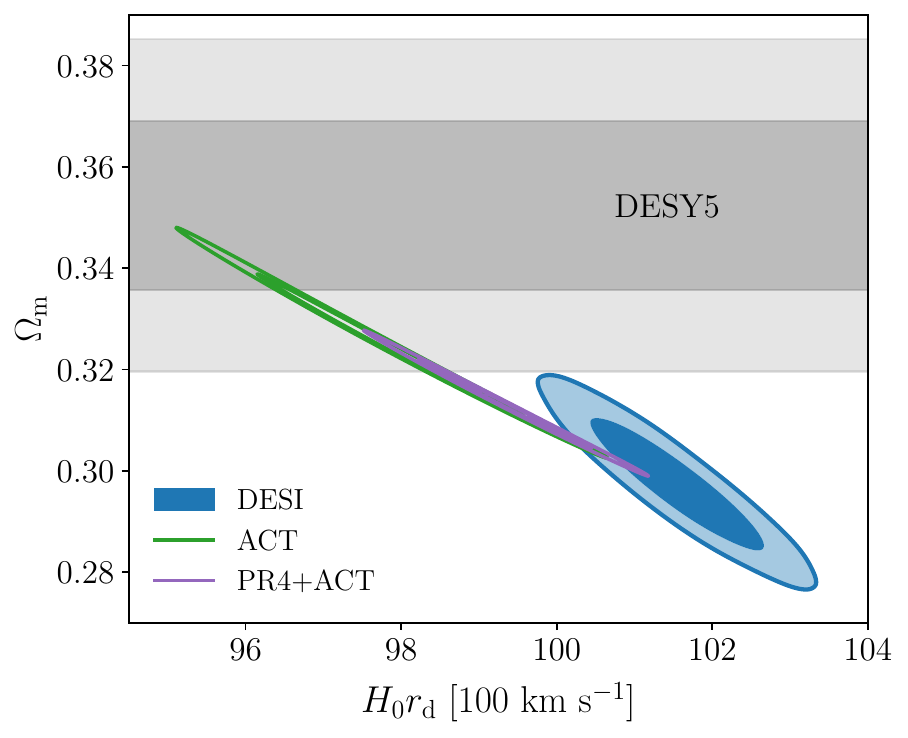}
    \includegraphics[width=0.43\textwidth]{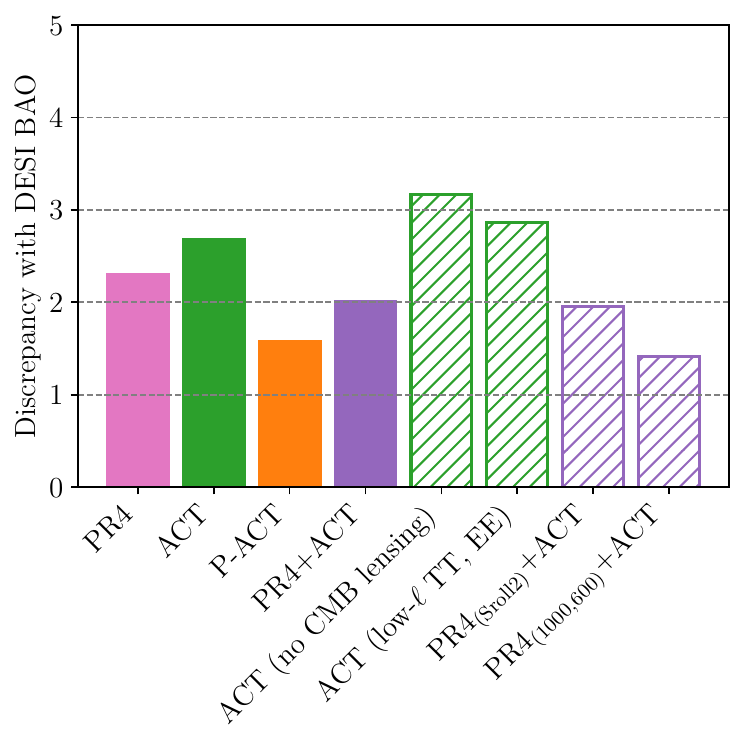}

    \caption{
    \textit{Left:} The 68\% and 95\% constraints on the $H_0\rd$-$\Om$ 2D parameter space for DESI DR2 BAO, ACT, PR4+ACT, and DESY5 SNe Ia, under the $\Lambda$CDM model. The combination based on ACT alone with \texttt{SRoll2} is shown in green, while our baseline combination PR4+ACT is shown in purple. The ACT dataset shows a 2.7$\sigma$ tension with DESI (3.2$\sigma$ if CMB lensing is excluded), while the combination of PR4+ACT shows a 2.0$\sigma$ tension with DESI, once CMB lensing and low-$\ell$ TT data have been also included.
    \textit{Right:} Tension between DESI DR2 BAO and different CMB variations (expressed in n$\sigma$ units), in the $\Lambda$CDM model. The tension is calculated given the 2D posterior distributions of $\Om$ and $\Hrd$. The first bars shown are PR4 (pink), ACT (green), P-ACT (orange) and PR4+ACT (purple). The hatched bars of the corresponding color represent variations of the CMB dataset.
    }
    \label{fig:lcdm_constraints}
\end{figure*}

In this section, we explore the consistency between DESI BAO data and CMB across the CMB combinations described in \cref{sec:method_data}, assuming the $\Lambda$CDM model. In \cite{DESI.DR2.BAO.cosmo} it was pointed out that DESI DR2 BAO data show a discrepancy of 2.3$\sigma$ with PR4. It is interesting to assess how this discrepancy stands with the latest CMB spectra from ACT, which provide precision measurements of the small-scale CMB anisotropies in temperature and polarization, as well as tight constraints on the CMB damping tail. At the cosmological parameter level, the quantitative comparison between BAO and CMB data is performed by analyzing the 2D posteriors on the combination $H_0\rd$-$\Om$, which are the two cosmological parameters constrained by BAO data. The left panel of \cref{fig:lcdm_constraints} shows the comparison of two CMB combinations (ACT in green, and PR4+ACT in purple) with DESI BAO and DESY5 SNe data. We observe that the discrepancy between DESI and CMB occurs along the degeneracy direction of constant $\Om h^3=\text{const.}$ \cite{Percival2002:astro-ph/0206256}, roughly preserving the location of the first acoustic peak. This ensures that DESI is consistent with the acoustic angular scale $\theta_*$ predicted from CMB. In the following, we describe how the consistency between CMB and DESI changes across the various CMB combinations.

\begin{figure*}[t]
    \centering
    \includegraphics[width=0.992\columnwidth]{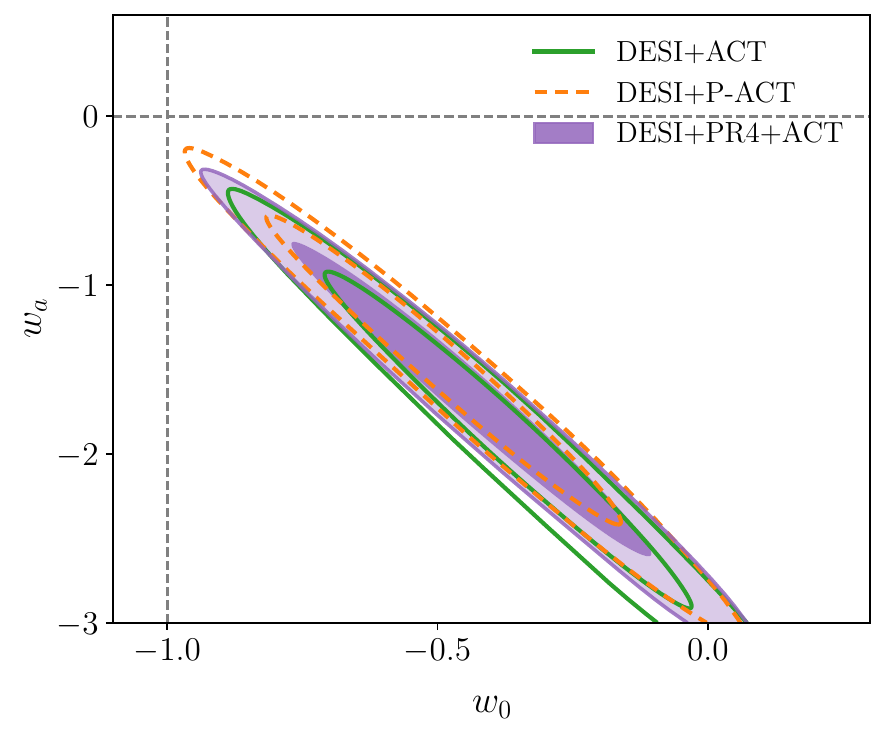}
    \includegraphics[width=\columnwidth]{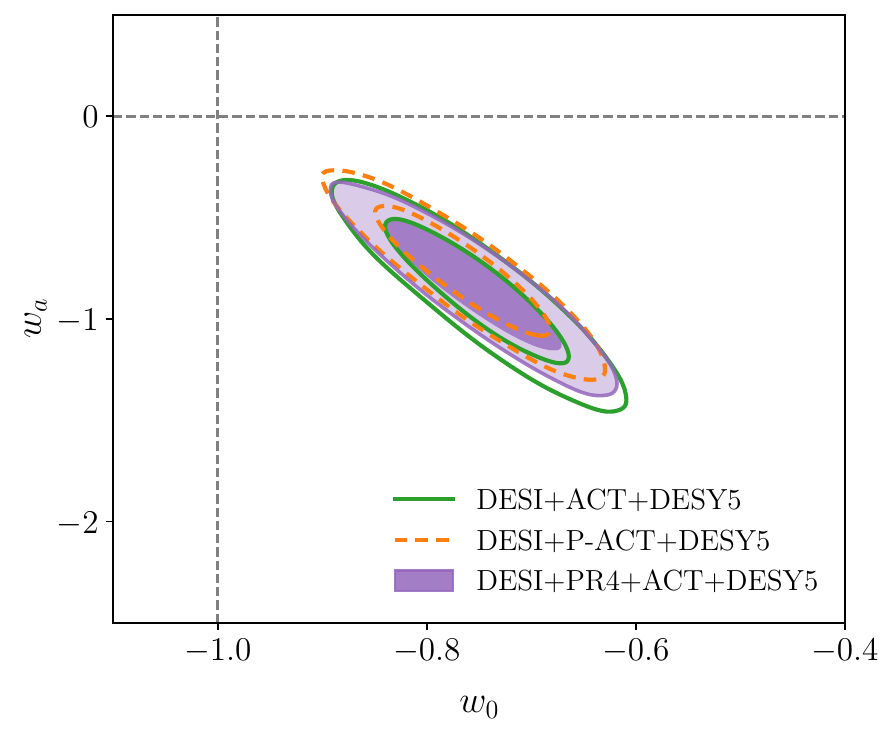}
    \caption{\textit{Left:} The 68\% and 95\% confidence contours in the $w_0$-$w_a$ plane using DESI DR2 BAO data in combination with CMB data. The blue contour describes the results from the combination DESI+ACT, while the unfilled orange contour shows DESI in combination with the P-ACT dataset described in \cite{ACT:2025fju}. The combination between DESI and our baseline CMB dataset with mixed multipole cuts in both {\it Planck} PR4 and ACT is shown in the unfilled green contour. The intersection of the two straight dashed gray lines represents the $\Lambda$CDM model.
    \textit{Right:} Similar to the left panel but now including DESY5 data but using a different ranges for $w_0$ and $w_a$.} 
    \label{fig:w0wacd_desi_cmb_constraints}
\end{figure*}

The base ACT combination  (without CMB lensing) shown in \cite{ACT:2025fju} reports high values of $\Ocdm h^2$ and $\Ob h^2$ and a lower $H_0$ with respect to {\it Planck}. This further pushes the ACT (no CMB lensing) data away from DESI, leading to a moderate 3.2$\sigma$ tension. The inclusion of CMB lensing reduces this tension to 2.7$\sigma$. Interestingly, this is the opposite effect observed in \cite{DESI.DR2.BAO.cosmo}, where excluding CMB lensing elongates the CMB contours reducing the discrepancy between DESI and PR4 data. The discrepancy persists even when the low-$\ell$ \texttt{SRoll2} likelihood is replaced with low-$\ell$ \texttt{SimAll} and low-$\ell$ TT information is included. However, the combination of P-ACT with the CMB lensing included alleviates the discrepancy leading to a 1.6$\sigma$ difference with DESI. This pull of the CMB contours towards the DESI best-fit value is related to the higher values of $\Ob h^2$ and the spectral index $n_{\rm s}$ measured by ACT with respect to {\it Planck}, as well as to parameter correlations, as explained in \cref{appendix}.

We further explore how DESI compares with a combination of PR4 and ACT. While ACT is more consistent with PR3 as compared to PR4 as pointed out in \cite{ACT:2025fju}, here we assess the consistency between ACT and PR4 under the assumed $\ell$ cuts before combining them and verify that the overall discrepancy is 2.0$\sigma$ or less in a six-parameter space under $\Lambda$CDM, and that individual parameters show discrepancies below this threshold. Our baseline CMB dataset labeled as PR4+ACT shows a 14\% and 23\% precision improvement in $n_{\rm s}$ and $\Ob h^2$ parameters, respectively, with respect to PR4. Also, it shows only a mild discrepancy of 2.0$\sigma$ with DESI (or 1.9$\sigma$ for PR4$_{(\texttt{SRoll2})}$+ACT). Applying a cut analogous to the one assumed in the P-ACT combination but using PR4, leads to a 1.4$\sigma$ discrepancy with DESI, for our PR4$_{(1000,600)}$+ACT combination. We also tested that removing the $600<\ell<1000$ TT data from {\it Planck} to avoid any overlaps in $\ell$ while still using a dataset that is ``maximal'' in ACT increases the discrepancy with DESI to 2.0$\sigma$.

We summarize the results in the right-hand-side panel of \cref{fig:lcdm_constraints} highlighting the three base CMB combinations for the rest of this work: ACT (the new CMB dataset), P-ACT (the official combination using PR3) and PR4+ACT (our baseline combination using PR4).

Finally, the left panel of \cref{fig:lcdm_constraints} also shows the SNe Ia data from DESY5 (for a more comprehensive comparison of all cosmological probes), which favors a high value of $\Om$ with respect to DESI and CMB. Other SNe Ia datasets (not shown in \cref{fig:lcdm_constraints}), Pantheon+ and Union3, show a slightly better consistency with DESI. If we combine DESI with P-ACT, the $\Om$ tension between DESI+CMB and SNe Ia ranges from 1.6$\sigma$ for Pantheon+, to 2.0$\sigma$ for Union3, and to 3.0$\sigma$ for DESY5. We do not observe a significant change in these numbers using different CMB datasets. Overall, the combination between DESI and any CMB combination of {\it Planck} and ACT shows a mild to moderate discrepancy with the $\Om$ measurement from SNe Ia, under $\Lambda$CDM.

\section{Dark energy constraints}
\label{sec:dark_energy}

\begin{figure*}[t]
    \centering
    \includegraphics[width=1.0\textwidth]{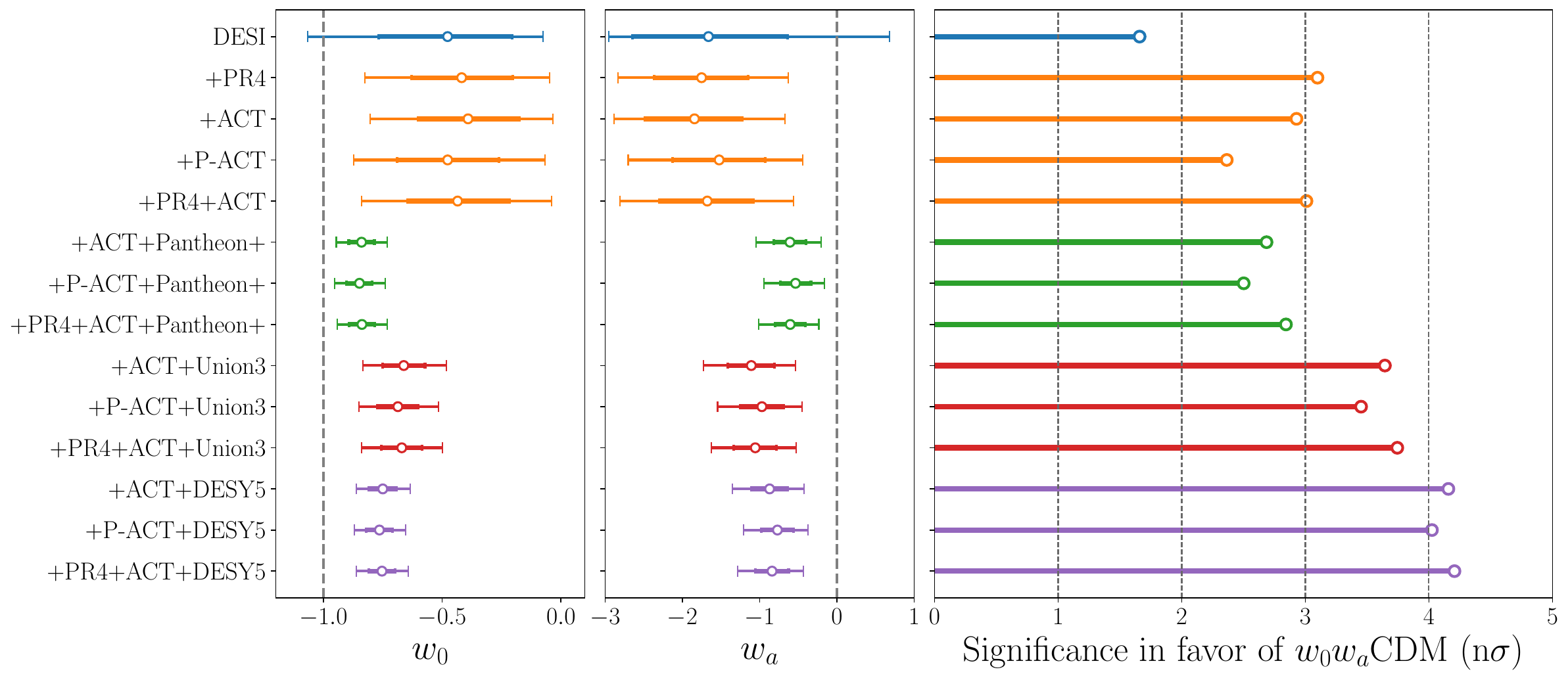}
    \caption{
    1D constraints on $w_0$ (left panel) and $w_a$ (middle panel), highlighting the robustness of the results against variations in the CMB dataset chosen for the analysis. In the left and central panels, the thick bars represent the 68\% errors while the thin bars correspond to the 95\% errors. The vertical black dashed lines represent the $\Lambda$CDM value for $w_0$ and $w_a$. The right panel shows the corresponding significance in favor of the $w_0w_a$CDM model.
    }
    \label{fig:w0wa_constraints}
\end{figure*}

We proceed to explore how the constraints on dark energy as presented in \cite{DESI.DR2.BAO.cosmo} depend on the ACT results. We parameterize deviations from the cosmological constant using the $w_0w_a \rm CDM$ \cite{Chevallier:2001,Linder2003}
dark energy equation of state given by $w(a)=w_0+(1-a)w_a$.
As demonstrated in \cite{DESI.DR2.BAO.cosmo}, $w_0w_a$CDM provides enough flexibility to jointly fit current BAO, CMB, and SNe Ia data while preserving consistency between the datasets; thus, we combine all three datasets into our analysis for $w_0w_a$CDM constraints.

The combination of DESI DR2 BAO with ACT, P-ACT and PR4+ACT are shown in the $w_0,w_a$ plane in the left-hand-side panel of \cref{fig:w0wacd_desi_cmb_constraints}. We observe that DESI+ACT  pulls the constraints away from the $(w_0,w_a)=(-1,0)$ limit, leading to a 2.9$\sigma$ preference in favor of $w_0w_a$CDM. Interestingly, ACT provides a CMB dataset that is well fit by $A_\text{Lens}\approx 1$, where the parameter $A_\text{Lens}$ can lead to an artificial smoothing of the CMB peaks \cite{Calabrese08}. Thus, it does not suffer from the so-called lensing anomaly. For DESI+P-ACT, the preference for a departure from $\Lambda$CDM diminishes to 2.4$\sigma$. On the other hand, the combination DESI+PR4+ACT shows a 3.0$\sigma$ preference in favor of $w_0w_a$CDM. For this combination, the errors on $w_0$ and $w_a$ remain unchanged compared to P-ACT. Since the PR4+ACT combination roughly chooses a sweet spot between {\it Planck} and ACT that provides more precise error bars on CMB parameters, we include more data from {\it Planck} compared to the P-ACT combination. 

\begin{table}
\centering
\resizebox{\columnwidth}{!}{
    \begin{tabular}{lccc}
    \toprule
    Datasets & $\dchisq$ & Significance & $\Delta$(DIC) \\
    \midrule
    DESI+ACT & $-11.4$ & $2.9\sigma$  & $-7.8$ \\
    DESI+ACT+Pantheon+ & $-9.9$ & $2.7\sigma$  & $-6.3$ \\
    DESI+ACT+Union3 & $-16.4$ & $3.6\sigma$  & $-13.0$ \\
    DESI+ACT+DESY5 & $-20.7$ & $4.2\sigma$  & $-16.8$ \\
    \hline
    DESI+P-ACT & $-8.0$ & $2.4\sigma$  & $-5.4$ \\
    DESI+P-ACT+Pantheon+ & $-8.8$ & $2.5\sigma$  & $-4.5$ \\
    DESI+P-ACT+Union3 & $-15.0$ & $3.5\sigma$  & $-10.9$ \\
    DESI+P-ACT+DESY5 & $-19.5$ & $4.0\sigma$  & $-15.2$ \\
    \hline
    DESI+PR4+ACT & $-11.9$ & $3.0\sigma$  & $-7.8$ \\
    DESI+PR4+ACT+Pantheon+ & $-10.8$ & $2.8\sigma$  & $-6.5$ \\
    DESI+PR4+ACT+Union3 & $-17.2$ & $3.7\sigma$  & $-13.8$ \\
    DESI+PR4+ACT+DESY5 & $-21.1$ & $4.2\sigma$  & $-17.2$ \\

    \bottomrule
    \end{tabular}
}
\caption{Table summarizing the results on the difference in the effective $\chi^2_{\rm MAP}$ value for the best-fit \wowacdm\ model relative to the best \lcdm\ model with  $w_0=-1$, $\wa=0$, and its corresponding significance level in a frequentist representation. The last column shows the results for the deviance information criteria, $\Delta(\mathrm{DIC})=\mathrm{DIC}_{w_0\wa\mathrm{CDM}}-\mathrm{DIC}_{\Lambda\mathrm{CDM}}$. 
\label{tab:dchisq}
}
\end{table}

The difference between the evidence in favor of $w_0w_a$CDM when DESI is combined with ACT versus that when DESI is combined with P-ACT or PR4+ACT can be linked to the difference in the measured $\Omega_\mathrm{c} h^2$ in the $\Lambda$CDM model by these respective CMB datasets. For example, a high CMB measurement of $\Omega_\mathrm{c} h^2$ in $\Lambda$CDM would lead to a high $\Om$ and a low $H_0\rd$ with respect to DESI, pulling the contours away from $w_0=-1$ and $w_a=0$ in a $w_0w_a$CDM model. In contrast, a lower measurement of $\Omega_\mathrm{c} h^2$ with respect to, e.g. PR4, would lead to more consistent results with DESI. This is also related to the importance of jointly matching physical density parameters like $\Omega_{c}h^2$ from the CMB with BAO, given the consistency of BAO with the CMB-predicted acoustic scale $\theta_*$ and that most of the significance in favor of $w_0w_a$CDM comes from CMB priors on ($\theta_*$, $\Omega_{b}h^2$, $\Omega_{bc}h^2$), as discussed in \cite{DESI.DR2.BAO.cosmo}. As shown in \cref{appendix}, ACT alone shows a high value of $\Omega_\mathrm{c} h^2$ with respect to PR4, although with larger error bars. On the other hand, P-ACT shows the opposite behavior measuring a lower $\Omega_\mathrm{c} h^2$ compared to PR4, while in PR4+ACT the $\Omega_\mathrm{c} h^2$ measurement is still lower than the PR4 prediction but the constraint is 19\% tighter than P-ACT. In terms of the ACT data, it was reported that there is a strong dependency on the inferred values of $\Ocdm h^2$ and $H_0$ given the polarization efficiency calibration choices \cite{ACT:2025fju}, which may impact the determination of these parameters.

The right-hand-side panel of \cref{fig:w0wacd_desi_cmb_constraints} shows the constraints for DESI+DESY5 in combination with the three CMB baseline datasets. We can see that once the background cosmology is set by BAO and SNe Ia data, the effect of assuming a different CMB variation leads to only minor changes. As summarized in \cref{fig:w0wa_constraints}, we find that the significance in favor of $w_0w_a$CDM ranges from 2.5$\sigma$ to 4.2$\sigma$, depending on the SNe Ia dataset used, while the assumed CMB variation has little impact on it. \cref{tab:dchisq} summarizes the results on the effective difference on the $\chi^2_\text{MAP}$, the corresponding significance in the frequentist representation, and the results for the deviance information criterion analysis, as presented in \cite{Y3.cpe-s1.Lodha.2025}. Once SNe Ia is included, the combination providing the mildest tension with $\Lambda$CDM is DESI+P-ACT+Pantheon+, with a 2.5$\sigma$ significance in favor of $w_0w_a$CDM. However, replacing Pantheon+ with Union3 and DESY5 increases the significance in favor of $w_0w_a$CDM to 3.5$\sigma$ and 4.0$\sigma$, respectively. The results from our baseline CMB variations provide compatible results, with tensions ranging between 2.8$\sigma$ (using Pantheon+) up to 4.2$\sigma$ (using DESY5). We also test the effect of using an alternative low-$\ell$ EE likelihood, replacing \texttt{SRoll2} with \texttt{SimAll} in the ACT combination, and find that changing the low-$\ell$ polarization likelihood has little impact in the significance in favor of $w_0w_a$CDM. This highlights the robustness of the results presented in \cite{DESI.DR2.BAO.cosmo}, after including newer CMB datasets based on ACT or combinations of it with {\it Planck}.

Finally, we summarize the results from the parameter constraints in \cref{tab:constraints} for $\Lambda$CDM and $w_0w_a$CDM\footnote{While this paper was under final review, \cite{Mirpoorian:2025rfp} presented constraints on $w_0w_a$CDM from DESI+P-ACT and DESI+P-ACT+Pantheon+. We find consistent results with theirs.}, as well as models with varying neutrino mass as discussed in the following section.

\begin{table*}
    \centering
    \resizebox{\linewidth}{!}{
\begin{tabular}{lccccc}
\toprule
\toprule
Model/Dataset & $\Omega_{\rm m}$ & $H_0$ [km s$^{-1}$ Mpc$^{-1}$] & $\sum m_\nu$ [eV] & $w_0$ & $w_a$ \\
\midrule
$\bm{\Lambda}$\textbf{CDM} &  &  &  &  &  \\
DESI+ACT & $0.3003\pm 0.0039$ & $68.48\pm 0.29$ & --- & --- & --- \\
DESI+P-ACT  & $0.3003\pm 0.0035$ & $68.43\pm 0.27$ & --- & --- & --- \\
DESI+PR4+ACT & $0.3019\pm 0.0035$ & $68.28\pm 0.26$ & --- & --- & --- \\
\hline
$\bm{w_0w_a}$\textbf{CDM} &  &  &  &  &  \\
DESI+ACT & $0.355^{+0.022}_{-0.020}$ & $63.6^{+1.6}_{-2.0}$ & --- & $-0.39^{+0.23}_{-0.19}$ & $-1.84\pm 0.59$ \\
DESI+ACT+Pantheon+ & $0.3108\pm 0.0057$ & $67.72\pm 0.60$ & --- & $-0.839\pm 0.055$ & $-0.61^{+0.22}_{-0.20}$ \\
DESI+ACT+Union3 &  $0.3274\pm 0.0088$ & $66.09\pm 0.85$ & --- & $-0.661\pm 0.089$ & $-1.11^{+0.32}_{-0.28}$ \\
DESI+ACT+DESY5 &  $0.3188\pm 0.0058$ & $66.94\pm 0.57$ & --- & $-0.750\pm 0.058$ & $-0.87^{+0.25}_{-0.22}$ \\
[-1.0ex]
\multicolumn{6}{c}{\makebox[\linewidth]{\dotfill}} \\
%
DESI+P-ACT  & $0.347^{+0.020}_{-0.023}$ & $64.1\pm 1.9$ & --- & $-0.48\pm 0.21$ & $-1.52^{+0.64}_{-0.56}$ \\
DESI+P-ACT+Pantheon+  & $0.3098\pm 0.0056$ & $67.62\pm 0.60$ & --- & $-0.848\pm 0.054$ & $-0.54^{+0.21}_{-0.18}$ \\
DESI+P-ACT+Union3  &  $0.3251\pm 0.0085$ & $66.08\pm 0.84$ & --- & $-0.686\pm 0.086$ & $-0.97^{+0.30}_{-0.26}$ \\
DESI+P-ACT+DESY5 &  $0.3175\pm 0.0055$ & $66.85\pm 0.56$ & --- & $-0.764\pm 0.056$ & $-0.77^{+0.22}_{-0.20}$ \\
[-1.0ex]
\multicolumn{6}{c}{\makebox[\linewidth]{\dotfill}} \\
%
DESI+PR4+ACT &  $0.350\pm 0.021$ & $63.8^{+1.8}_{-2.0}$ & --- & $-0.43\pm 0.21$ & $-1.68\pm 0.58$ \\
DESI+PR4+ACT+Pantheon+ &  $0.3107\pm 0.0056$ & $67.59\pm 0.59$ & --- & $-0.837\pm 0.054$ & $-0.60^{+0.21}_{-0.19}$ \\
DESI+PR4+ACT+Union3 &  $0.3265\pm 0.0085$ & $66.00\pm 0.84$ & --- & $-0.670\pm 0.086$ & $-1.06^{+0.29}_{-0.26}$ \\
DESI+PR4+ACT+DESY5&  $0.3182\pm 0.0055$ & $66.83\pm 0.56$ & --- & $-0.753\pm 0.056$ & $-0.84^{+0.23}_{-0.20}$ \\
\hline
$\bm{\Lambda}$\textbf{CDM+}$\bm{\sum m_\nu}$ &  &  &  &  &  \\
DESI+ACT &   $0.2992\pm 0.0039$ & $68.63\pm 0.31$ & $<0.0733$ & --- & --- \\
DESI+P-ACT  & $0.2987\pm 0.0037$ & $68.61\pm 0.29$ & $<0.0768$ & --- & --- \\
DESI+PR4+ACT & $0.2999\pm 0.0036$ & $68.50\pm 0.28$ & $<0.0606$ & --- & ---\\
\hline
$\bm{w_0w_a}$\textbf{CDM+}$\bm{\sum m_\nu}$ &  &  &  &  &  \\
DESI+ACT &  $0.355^{+0.024}_{-0.020}$ & $63.7^{+1.6}_{-2.2}$ & $<0.170$ & $-0.39^{+0.25}_{-0.19}$ & $-1.85^{+0.61}_{-0.75}$ \\
DESI+ACT+Pantheon+ &  $0.3105\pm 0.0058$ & $67.72\pm 0.60$ & $<0.124$ & $-0.843\pm 0.056$ & $-0.57^{+0.24}_{-0.21}$ \\
DESI+ACT+Union3 &  $0.3273\pm 0.0090$ & $66.10\pm 0.85$ & $<0.147$ & $-0.665\pm 0.092$ & $-1.09^{+0.36}_{-0.30}$ \\
DESI+ACT+DESY5 &  $0.3186\pm 0.0058$ & $66.95\pm 0.57$ & $<0.136$ & $-0.753\pm 0.059$ & $-0.85^{+0.26}_{-0.23}$ \\
[-1.0ex]
\multicolumn{6}{c}{\makebox[\linewidth]{\dotfill}} \\
%
DESI+P-ACT  &  $0.349\pm 0.022$ & $63.9^{+1.8}_{-2.1}$ & $<0.186$ & $-0.45\pm 0.22$ & $-1.62^{+0.73}_{-0.65}$ \\
DESI+P-ACT+Pantheon+  &  $0.3095\pm 0.0057$ & $67.64\pm 0.60$ & $<0.131$ & $-0.852\pm 0.055$ & $-0.51^{+0.23}_{-0.19}$ \\
DESI+P-ACT+Union3  &  $0.3253\pm 0.0089$ & $66.08\pm 0.85$ & $<0.155$ & $-0.687\pm 0.090$ & $-0.97^{+0.34}_{-0.28}$ \\
DESI+P-ACT+DESY5 &  $0.3173\pm 0.0058$ & $66.87\pm 0.56$ & $<0.149$ & $-0.766\pm 0.058$ & $-0.76^{+0.26}_{-0.21}$ \\
[-1.0ex]
\multicolumn{6}{c}{\makebox[\linewidth]{\dotfill}} \\
%
DESI+PR4+ACT &  $0.350\pm 0.022$ & $63.9^{+1.8}_{-2.1}$ & $<0.152$ & $-0.44\pm 0.21$ & $-1.66\pm 0.62$ \\
DESI+PR4+ACT+Pantheon+ &  $0.3099\pm 0.0057$ & $67.64\pm 0.60$ & $<0.108$ & $-0.846\pm 0.054$ & $-0.55^{+0.22}_{-0.19}$ \\
DESI+PR4+ACT+Union3 & $0.3259\pm 0.0087$ & $66.05\pm 0.84$ & $<0.128$ & $-0.678\pm 0.088$ & $-1.01^{+0.32}_{-0.27}$ \\
DESI+PR4+ACT+DESY5& $0.3177\pm 0.0057$ & $66.86\pm 0.57$ & $<0.122$ & $-0.761\pm 0.057$ & $-0.79^{+0.25}_{-0.21}$ \\
\bottomrule
\bottomrule
\end{tabular}}
    \caption{Summary table of key cosmological parameter constraints from DESI DR2 BAO (labeled simply as DESI) in combination with external datasets for the \lcdm\, and extended models. We report the mean value and the 68\% confidence for all parameters, except for the total neutrino mass, for which the 95\% upper limit is quoted.
    }
    \label{tab:constraints}
\end{table*}

\section{Neutrino mass constraints}
\label{sec:neutrinos}

Despite the fact that BAO data alone cannot constrain the total neutrino mass, $\sumnu$, determining $\Om$ and $H_0 \rd$ from low redshifts significantly enhances the CMB sensitivity to $\sumnu$ via the neutrino-induced shift in the angular diameter distance to last scattering. CMB photons are sensitive to the neutrino mass through lensing, since neutrinos suppress the growth of structures below their free-streaming scale. Low-$\ell$ E-mode polarization of the CMB also plays an important indirect role in constraining neutrino masses by breaking the degeneracy between the optical depth to reionization and the amplitude of primordial scalar fluctuations measured from high-$\ell$ multipoles. This in turn gives a more accurate estimation of the lensing potential, which is suppressed on small scales by massive neutrinos. Therefore, the upper limits on $\sumnu$ obtained from the combination of DESI and CMB depend on both the BAO constraining power and the choice of the CMB likelihood, as different likelihoods slightly vary in the amount of lensing power they infer from the lensed TT, TE, and EE spectra at both low- and high-$\ell$.

The inferred constraints on $\sumnu$ from BAO+CMB depend on the underlying cosmological model. In the following, we focus on the $\Lambda$CDM and $w_0w_a$CDM models with free total neutrino mass, assuming three degenerate mass eigenstates. We also adopt a minimal physical prior, $\sumnu > 0$ eV, noting that scenarios allowing for a negative effective neutrino mass have recently gained interest \cite{Green24,Craig24,Elbers24, Y3.cpe-s2.Elbers.2025}.

Recently, the ACT collaboration set an upper bound on the total neutrino mass of $\sum m_\nu < 0.082\,\text{eV}$ at 95\% confidence level (c.l.) under $\Lambda$CDM \cite{ACT:2025tim}, based on the combination of DESI DR1 BAO measurements and what we refer to as P-ACT. Using the updated DESI DR2 BAO data, this limit tightens to $\sum m_\nu < 0.077\,\text{eV}$ at 95\% c.l., representing about a 6\% reduction in the upper bound due to the improved BAO measurements in DESI DR2 with respect to DR1. 
This constraint is, however, about 20\% weaker than the baseline bound recently reported by the DESI collaboration, $\sumnu < 0.064$ eV at 95\%  c.l. \cite{DESI.DR2.BAO.cosmo}, which combines DESI DR2 BAO with {\it Planck} PR4. The difference between the P-ACT and PR4 combinations primarily arises from the treatment of low-$\ell$ EE and high-$\ell$ TT likelihoods, including specific choices of cuts.

\begin{figure}
    \centering
    \includegraphics[width=0.48\textwidth]{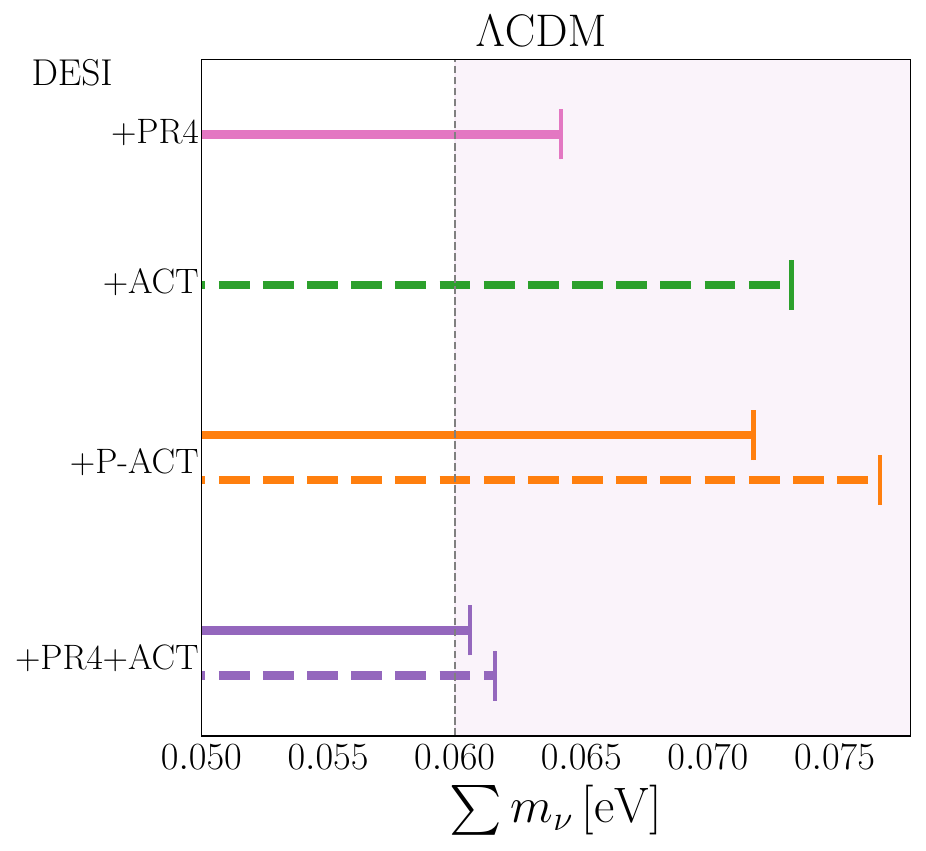}
    \caption{Whisker plots showing the 95\% confidence constraints on $\sumnu$ from the combination of DESI DR2 BAO with various CMB likelihoods under the \lcdm\ model. Dashed lines correspond to constraints obtained using the low-$\ell$ EE \texttt{SRoll2} likelihood, while solid lines use low-$\ell$ EE \texttt{SimAll} likelihood in each corresponding combination. The vertical dashed line and shaded region indicate the minimal sum of neutrino masses for the normal ($\sum m_\nu > 0.06$ eV) mass ordering. 
    }
    \label{fig:1d_neutrinos}
\end{figure}

\begin{figure}
    \centering
    \includegraphics[width=0.48\textwidth]{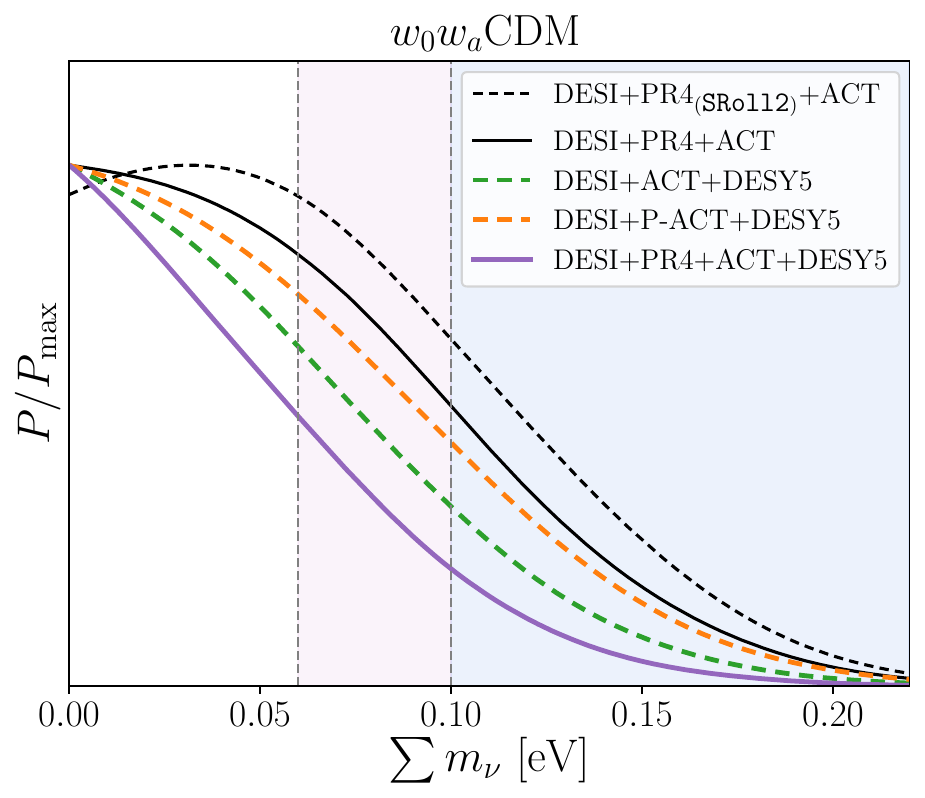}
    \caption{1D marginalized posterior constraints on $\sumnu$ from various dataset combinations within the $w_0w_a$CDM model. 
    The black curves show results from combining DESI DR2 BAO with PR4+ACT. We also present posteriors obtained using DESI DR2 BAO with the main CMB combinations, all combined with DESY5. Dashed curves correspond to datasets that include the low-$\ell$ EE \texttt{SRoll2} likelihood, while solid curves instead use the low-$\ell$ EE \texttt{SimAll} likelihood. The vertical dashed lines and shaded regions indicate the minimum sum of neutrino masses allowed for the normal ($\sum m_\nu > 0.06$ eV) and inverted ($\sum m_\nu > 0.10$ eV) mass orderings.
    }
    \label{fig:1d_neutrinos_w_wa}
\end{figure}

However, we stress that these results, along with those presented below, are influenced by the prior $\sumnu > 0$ eV. This dependence has been discussed in \cite{DESI.DR2.BAO.cosmo}, and more thoroughly in \cite{Y3.cpe-s2.Elbers.2025}, that employed a profile likelihood analysis to quantify the impact of the prior and further investigated the implications of effective neutrino negative masses \cite{Elbers24}. A similar profile likelihood analysis is presented in \cref{appendix_B}.

\cref{fig:1d_neutrinos} illustrates how the constraints on the neutrino mass are influenced by the main CMB combinations — ACT, P-ACT, and PR4+ACT — when combined with DESI DR2 BAO. For comparison, we include the baseline results from \cite{DESI.DR2.BAO.cosmo} in magenta. For the PR4+ACT combination, we observe that by cutting the PR4 likelihood at the multipoles  $\ell = 2000$ in TT and $\ell = 1000$ in TEEE, and merging with ACT data starting from these multipole cuts, tightens the upper bound to $\sumnu < 0.061$ eV at 95\% c.l. (solid purple line) for \lcdm. This results in a reduction of approximately 5\% compared to the baseline results from \cite{DESI.DR2.BAO.cosmo}, providing the tightest constraints on the total neutrino mass to date, derived exclusively from BAO and CMB datasets.

The difference in constraining power between PR4+ACT and the baseline results from \cite{DESI.DR2.BAO.cosmo} stems not only from the inclusion of ACT data, but also from the specific choice of the $\ell$-range where \textit{Planck} is cut and ACT is added.  In the PR4+ACT combination, we select the region of the spectrum where the ACT signal roughly exhibits lower uncertainty compared to \textit{Planck} across most frequency spectra (see Figure 12 of \cite{ACT:2025fju}), leading to tighter neutrino mass constraints.

Additionally, \cref{fig:1d_neutrinos} includes the constraints from the joint DESI+ACT analysis, which, despite covering a smaller portion of the CMB power spectrum, yields an upper bound of $\sumnu < 0.073$ eV (95\% c.l.), similar to those obtained with other CMB likelihood combinations.

In the $w_0w_a$CDM scenario, the combination of DESI DR2 BAO and the main CMB datasets (without SNe) yields marginalized 1D posterior distributions that peak at positive values of $\sumnu$, consistent with the findings of \cite{DESI.DR2.BAO.cosmo}. This behavior is found across all main CMB dataset combinations. To avoid overcrowding, \cref{fig:1d_neutrinos_w_wa} shows only the posteriors for ${\rm PR4}_{(\texttt{SRoll2})}$+ACT (dashed black) and PR4+ACT (solid black), the latter being the only DESI+CMB combination whose posterior peaks at zero. This case also provides the most stringent constraint on the neutrino mass within $w_0w_a$CDM, with $\sum m_\nu < 0.15$ eV (95\% c.l.).

Although the DESI+CMB results are largely consistent with positive neutrino masses, they also exhibit a preference for higher values of $\Omega_m$ (see \cref{tab:constraints}), which are ``stabilized'' when including information from SNe Ia datasets. In that case, all posteriors peak at $\sum m_\nu = 0$, and they would reach a maximum at negative $\sumnu$ values if extrapolated.\footnote{This feature has motivated interest in exploring the implications of negative effective neutrino masses \cite{Elbers24,Craig24,Green24,Y3.cpe-s2.Elbers.2025}.} This behavior is illustrated in \cref{fig:1d_neutrinos_w_wa} for DESY5 and also holds when considering either Pantheon+ or Union3. These findings are consistent with those reported in \cite{DESI.DR2.BAO.cosmo,DESI2024.VII.KP7B}, and we explore them in more detail using the profile likelihood analysis in \cref{appendix_B}.

Finally, we highlight the impact of the $\ell<30$ multipoles polarization data. In particular, we find that across all analyses, replacing the baseline low-$\ell$ EE likelihood with \texttt{SRoll2} shifts the posterior peak toward higher values, as illustrated in \cref{fig:1d_neutrinos_w_wa}. This shift apparently loosens the neutrino mass constraints by up to 7\%, in both the $\Lambda$CDM and $w_0w_a$CDM models.

\section{Conclusions}
\label{sec:conclusions}

In this paper, we reanalyze the cosmological results on the evidence for evolving dark energy and neutrino mass constraints of the official DESI DR2 BAO analysis presented in \cite{DESI.DR2.BAO.cosmo} by incorporating the latest ACT DR6 CMB data. The ACT DR6 data predict higher values for the physical densities of both baryons $\Ob h^2$ and cold dark matter $\Ocdm h^2$, compared to {\it Planck}. Within the $\Lambda$CDM model, this results in a discrepancy with DESI at a level exceeding $3\sigma$, which is larger than the 2.0$\sigma$ discrepancy observed with {\it Planck} PR4 (without CMB lensing). Since the ACT DR6 data cover a wide range of multipoles, partially sharing angular scales with {\it Planck} (in the range $600 \leq \ell \leq 2500$), we combine these two CMB datasets using multipole cuts to avoid overlap. Along with ACT, we focus on two other CMB data combinations, namely, P-ACT (based on PR3, matching the combination presented in \cite{ACT:2025fju}), and PR4+ACT (our baseline CMB dataset). The PR4+ACT combination, based on PR4, uses cuts in the common multipole range between PR4 and ACT that lead to tighter constraints, showing a precision improvement of 14\% in $n_{\rm s}$ and 23\% in $\Ob h^2$ with respect to PR4, and a 5\% precision improvement over other combinations with a different $\ell$-cut scheme such as PR4$_{(1000,600)}$+ACT.

We explore the evidence for evolving dark energy under these three CMB datasets using ACT DR6 and find that DESI+ACT shows a 2.9$\sigma$ evidence in favor of the $w_0w_a$CDM model. This evidence is reduced to 2.4$\sigma$ when using DESI+P-ACT. This is due to the pull from P-ACT towards lower values of $\Ocdm h^2$, compared to those from ACT and {\it Planck} individually, in the $\Lambda$CDM model. The combination DESI+PR4+ACT, which provides tighter constraints on cosmological parameters, leads to a 3.0$\sigma$ significance in favor of $w_0w_a$CDM. We also test the inclusion of the three SNe Ia datasets, namely Pantheon+, Union3 and DESY5, and find that variations in the CMB dataset leads to at most 0.3$\sigma$ differences and that the evidence for evolving dark energy can go up to 4.0$\sigma$. Therefore, we conclude that the results presented in \cite{DESI.DR2.BAO.cosmo} are robust in light of the new ACT CMB data.

We also present updated constraints in the neutrino mass bounds from cosmology and find that, in $\Lambda$CDM and assuming a physical prior $\sum m_\nu > 0$ eV, DESI+ACT yields a neutrino mass constraint of $\sum m_\nu<0.073$ eV (95\% c.l.). Our baseline CMB dataset, PR4+ACT, imposes an upper bound of $\sum m_\nu<0.061$ eV (95\% c.l.) when combined with DESI. This represents a 5\% reduction in the error compared to the neutrino mass constraints in \cite{DESI.DR2.BAO.cosmo}, for the DESI+CMB baseline combination. Extending the background to an evolving dark energy component parameterized by $w_0$ and $w_a$ gives an upper mass limit of $\sum m_\nu<0.17$ eV and $\sum m_\nu<0.15$ eV at 95\% c.l., for DESI+ACT and DESI+PR4+ACT, respectively. Consistent with \cite{DESI.DR2.BAO.cosmo}, we find that combining DESI BAO with CMB data yields a preference for positive neutrino masses. However, this preference vanishes when SNe data are included in the analysis. Finally, we find that the use of the low-$\ell$ EE \texttt{SRoll2} likelihood can relax the constraints on neutrino mass compared to the low-$\ell$ EE \texttt{SimAll} likelihood by up to 7\%, mostly due to a shift of the best-fit $\sum m_\nu$ towards larger values.

Overall, we find the results presented in \cite{DESI.DR2.BAO.cosmo} to be robust under the inclusion of the ACT data for the CMB combinations tested in this work. A joint treatment of {\it Planck} and ACT DR6 covariance could eventually coalesce to a consolidated CMB dataset. 

\section{Data Availability}
The data used in this analysis will be made public with Data Release 2 (details in \url{https://data.desi.lbl.gov/doc/releases/}).
The data corresponding to the figures in this paper will be available in a Zenodo repository.

\acknowledgments

We thank Martin White for useful comments on the manuscript while serving as internal reviewer.

CGQ acknowledges the support provided by NASA through the NASA Hubble Fellowship grant HST-HF2-51554.001-A awarded by the Space Telescope Science Institute, which is operated by the Association of Universities for Research in Astronomy, Inc., for NASA, under contract NAS5-26555. HN and AA acknowledge support by SECIHTI grant CBF2023-2024-162 and PAPIIT IA101825. HN acknowledges support by PAPIIT- IN101124. JR acknowledges funding from US Department of Energy grant DE-SC0016021. SN acknowledges support from an STFC Ernest Rutherford Fellowship, grant reference ST/T005009/2. WE acknowledges STFC Consolidated Grant ST/X001075/1 and support from the European Research Council (ERC) Advanced Investigator grant DMIDAS (GA 786910).

This material is based upon work supported by the U.S.\ Department of Energy (DOE), Office of Science, Office of High-Energy Physics, under Contract No.\ DE–AC02–05CH11231, and by the National Energy Research Scientific Computing Center, a DOE Office of Science User Facility under the same contract. Additional support for DESI was provided by the U.S. National Science Foundation (NSF), Division of Astronomical Sciences under Contract No.\ AST-0950945 to the NSF National Optical-Infrared Astronomy Research Laboratory; the Science and Technology Facilities Council of the United Kingdom; the Gordon and Betty Moore Foundation; the Heising-Simons Foundation; the French Alternative Energies and Atomic Energy Commission (CEA); the National Council of Humanities, Science and Technology of Mexico (CONAHCYT); the Ministry of Science and Innovation of Spain (MICINN), and by the DESI Member Institutions: \url{https://www.desi. lbl.gov/collaborating-institutions}. 

The authors are honored to be permitted to conduct scientific research on I’oligam Du’ag (Kitt Peak), a mountain with particular significance to the Tohono O’odham Nation.



\bibliographystyle{mod-apsrev4-2} 
\bibliography{Y1KP7a_references,references, DESI_supporting_papers}

\begin{thebibliography}{59}%
\makeatletter
\providecommand \@ifxundefined [1]{%
 \@ifx{#1\undefined}
}%
\providecommand \@ifnum [1]{%
 \ifnum #1\expandafter \@firstoftwo
 \else \expandafter \@secondoftwo
 \fi
}%
\providecommand \@ifx [1]{%
 \ifx #1\expandafter \@firstoftwo
 \else \expandafter \@secondoftwo
 \fi
}%
\providecommand \natexlab [1]{#1}%
\providecommand \enquote  [1]{``#1''}%
\providecommand \bibnamefont  [1]{#1}%
\providecommand \bibfnamefont [1]{#1}%
\providecommand \citenamefont [1]{#1}%
\providecommand \href@noop [0]{\@secondoftwo}%
\providecommand \href [0]{\begingroup \@sanitize@url \@href}%
\providecommand \@href[1]{\@@startlink{#1}\@@href}%
\providecommand \@@href[1]{\endgroup#1\@@endlink}%
\providecommand \@sanitize@url [0]{\catcode `\\12\catcode `\$12\catcode `\&12\catcode `\#12\catcode `\^12\catcode `\_12\catcode `\%12\relax}%
\providecommand \@@startlink[1]{}%
\providecommand \@@endlink[0]{}%
\providecommand \url  [0]{\begingroup\@sanitize@url \@url }%
\providecommand \@url [1]{\endgroup\@href {#1}{\urlprefix }}%
\providecommand \urlprefix  [0]{URL }%
\providecommand \Eprint [0]{\href }%
\providecommand \doibase [0]{https://doi.org/}%
\providecommand \selectlanguage [0]{\@gobble}%
\providecommand \bibinfo  [0]{\@secondoftwo}%
\providecommand \bibfield  [0]{\@secondoftwo}%
\providecommand \translation [1]{[#1]}%
\providecommand \BibitemOpen [0]{}%
\providecommand \bibitemStop [0]{}%
\providecommand \bibitemNoStop [0]{.\EOS\space}%
\providecommand \EOS [0]{\spacefactor3000\relax}%
\providecommand \BibitemShut  [1]{\csname bibitem#1\endcsname}%
\let\auto@bib@innerbib\@empty
\bibitem [{\citenamefont {{Riess}}\ \emph {et~al.}(1998)\citenamefont {{Riess}}, \citenamefont {{Filippenko}}, \citenamefont {{Challis}}, \citenamefont {{Clocchiatti}}, \citenamefont {{Diercks}}, \citenamefont {{Garnavich}}, \citenamefont {{Gilliland}}, \citenamefont {{Hogan}}, \citenamefont {{Jha}}, \citenamefont {{Kirshner}}, \citenamefont {{Leibundgut}}, \citenamefont {{Phillips}}, \citenamefont {{Reiss}}, \citenamefont {{Schmidt}}, \citenamefont {{Schommer}}, \citenamefont {{Smith}}, \citenamefont {{Spyromilio}}, \citenamefont {{Stubbs}}, \citenamefont {{Suntzeff}},\ and\ \citenamefont {{Tonry}}}]{1998AJ....116.1009R}%
  \BibitemOpen
  \bibfield  {author} {\bibinfo {author} {\bibfnamefont {A.~G.}\ \bibnamefont {{Riess}}}, \bibinfo {author} {\bibfnamefont {A.~V.}\ \bibnamefont {{Filippenko}}}, \bibinfo {author} {\bibfnamefont {P.}~\bibnamefont {{Challis}}}, \bibinfo {author} {\bibfnamefont {A.}~\bibnamefont {{Clocchiatti}}}, \bibinfo {author} {\bibfnamefont {A.}~\bibnamefont {{Diercks}}}, \bibinfo {author} {\bibfnamefont {P.~M.}\ \bibnamefont {{Garnavich}}}, \bibinfo {author} {\bibfnamefont {R.~L.}\ \bibnamefont {{Gilliland}}}, \bibinfo {author} {\bibfnamefont {C.~J.}\ \bibnamefont {{Hogan}}}, \bibinfo {author} {\bibfnamefont {S.}~\bibnamefont {{Jha}}}, \bibinfo {author} {\bibfnamefont {R.~P.}\ \bibnamefont {{Kirshner}}}, \bibinfo {author} {\bibfnamefont {B.}~\bibnamefont {{Leibundgut}}}, \bibinfo {author} {\bibfnamefont {M.~M.}\ \bibnamefont {{Phillips}}}, \bibinfo {author} {\bibfnamefont {D.}~\bibnamefont {{Reiss}}}, \bibinfo {author} {\bibfnamefont {B.~P.}\ \bibnamefont {{Schmidt}}}, \bibinfo {author} {\bibfnamefont {R.~A.}\
  \bibnamefont {{Schommer}}}, \bibnamefont {and~others},\ }\href {https://doi.org/10.1086/300499} {\bibfield  {journal} {\bibinfo  {journal} {\aj}\ }\textbf {\bibinfo {volume} {116}},\ \bibinfo {pages} {1009} (\bibinfo {year} {1998})},\ \Eprint {https://arxiv.org/abs/astro-ph/9805201} {arXiv:astro-ph/9805201 [astro-ph]} \BibitemShut {NoStop}%
\bibitem [{\citenamefont {{Perlmutter}}\ \emph {et~al.}(1999)\citenamefont {{Perlmutter}}, \citenamefont {{Aldering}}, \citenamefont {{Goldhaber}}, \citenamefont {{Knop}}, \citenamefont {{Nugent}}, \citenamefont {{Castro}}, \citenamefont {{Deustua}}, \citenamefont {{Fabbro}}, \citenamefont {{Goobar}}, \citenamefont {{Groom}}, \citenamefont {{Hook}}, \citenamefont {{Kim}}, \citenamefont {{Kim}}, \citenamefont {{Lee}}, \citenamefont {{Nunes}}, \citenamefont {{Pain}}, \citenamefont {{Pennypacker}}, \citenamefont {{Quimby}}, \citenamefont {{Lidman}}, \citenamefont {{Ellis}}, \citenamefont {{Irwin}}, \citenamefont {{McMahon}}, \citenamefont {{Ruiz-Lapuente}}, \citenamefont {{Walton}}, \citenamefont {{Schaefer}}, \citenamefont {{Boyle}}, \citenamefont {{Filippenko}}, \citenamefont {{Matheson}}, \citenamefont {{Fruchter}}, \citenamefont {{Panagia}}, \citenamefont {{Newberg}}, \citenamefont {{Couch}},\ and\ \citenamefont {{Project}}}]{1999ApJ...517..565P}%
  \BibitemOpen
  \bibfield  {author} {\bibinfo {author} {\bibfnamefont {S.}~\bibnamefont {{Perlmutter}}}, \bibinfo {author} {\bibfnamefont {G.}~\bibnamefont {{Aldering}}}, \bibinfo {author} {\bibfnamefont {G.}~\bibnamefont {{Goldhaber}}}, \bibinfo {author} {\bibfnamefont {R.~A.}\ \bibnamefont {{Knop}}}, \bibinfo {author} {\bibfnamefont {P.}~\bibnamefont {{Nugent}}}, \bibinfo {author} {\bibfnamefont {P.~G.}\ \bibnamefont {{Castro}}}, \bibinfo {author} {\bibfnamefont {S.}~\bibnamefont {{Deustua}}}, \bibinfo {author} {\bibfnamefont {S.}~\bibnamefont {{Fabbro}}}, \bibinfo {author} {\bibfnamefont {A.}~\bibnamefont {{Goobar}}}, \bibinfo {author} {\bibfnamefont {D.~E.}\ \bibnamefont {{Groom}}}, \bibinfo {author} {\bibfnamefont {I.~M.}\ \bibnamefont {{Hook}}}, \bibinfo {author} {\bibfnamefont {A.~G.}\ \bibnamefont {{Kim}}}, \bibinfo {author} {\bibfnamefont {M.~Y.}\ \bibnamefont {{Kim}}}, \bibinfo {author} {\bibfnamefont {J.~C.}\ \bibnamefont {{Lee}}}, \bibinfo {author} {\bibfnamefont {N.~J.}\ \bibnamefont {{Nunes}}}, \bibnamefont
  {and~others},\ }\href {https://doi.org/10.1086/307221} {\bibfield  {journal} {\bibinfo  {journal} {\apj}\ }\textbf {\bibinfo {volume} {517}},\ \bibinfo {pages} {565} (\bibinfo {year} {1999})},\ \Eprint {https://arxiv.org/abs/astro-ph/9812133} {arXiv:astro-ph/9812133 [astro-ph]} \BibitemShut {NoStop}%
\bibitem [{\citenamefont {{Levi}}\ \emph {et~al.}(2013)\citenamefont {{Levi}}, \citenamefont {{Bebek}}, \citenamefont {{Beers}}, \citenamefont {{Blum}}, \citenamefont {{Cahn}}, \citenamefont {{Eisenstein}}, \citenamefont {{Flaugher}}, \citenamefont {{Honscheid}}, \citenamefont {{Kron}}, \citenamefont {{Lahav}}, \citenamefont {{McDonald}}, \citenamefont {{Roe}}, \citenamefont {{Schlegel}},\ and\ \citenamefont {{representing the DESI collaboration}}}]{Snowmass2013.Levi}%
  \BibitemOpen
  \bibfield  {author} {\bibinfo {author} {\bibfnamefont {M.}~\bibnamefont {{Levi}}}, \bibinfo {author} {\bibfnamefont {C.}~\bibnamefont {{Bebek}}}, \bibinfo {author} {\bibfnamefont {T.}~\bibnamefont {{Beers}}}, \bibinfo {author} {\bibfnamefont {R.}~\bibnamefont {{Blum}}}, \bibinfo {author} {\bibfnamefont {R.}~\bibnamefont {{Cahn}}}, \bibinfo {author} {\bibfnamefont {D.}~\bibnamefont {{Eisenstein}}}, \bibinfo {author} {\bibfnamefont {B.}~\bibnamefont {{Flaugher}}}, \bibinfo {author} {\bibfnamefont {K.}~\bibnamefont {{Honscheid}}}, \bibinfo {author} {\bibfnamefont {R.}~\bibnamefont {{Kron}}}, \bibinfo {author} {\bibfnamefont {O.}~\bibnamefont {{Lahav}}}, \bibinfo {author} {\bibfnamefont {P.}~\bibnamefont {{McDonald}}}, \bibinfo {author} {\bibfnamefont {N.}~\bibnamefont {{Roe}}}, \bibinfo {author} {\bibfnamefont {D.}~\bibnamefont {{Schlegel}}},\ \bibnamefont {and}\ \bibinfo {author} {\bibnamefont {{representing the DESI collaboration}}},\ }\href@noop {} {\bibfield  {journal} {\bibinfo  {journal} {arXiv
  e-prints}\ ,\ \bibinfo {eid} {arXiv:1308.0847}} (\bibinfo {year} {2013})},\ \Eprint {https://arxiv.org/abs/1308.0847} {arXiv:1308.0847 [astro-ph.CO]} \BibitemShut {NoStop}%
\bibitem [{\citenamefont {{DESI Collaboration}}\ \emph {et~al.}(2016{\natexlab{a}})\citenamefont {{DESI Collaboration}}, \citenamefont {{Aghamousa}}, \citenamefont {{Aguilar}}, \citenamefont {{Ahlen}}, \citenamefont {{Alam}}, \citenamefont {{Allen}}, \citenamefont {{Allende Prieto}}, \citenamefont {{Annis}}, \citenamefont {{Bailey}}, \citenamefont {{Balland}}, \citenamefont {{Ballester}}, \citenamefont {{Baltay}}, \citenamefont {{Beaufore}}, \citenamefont {{Bebek}}, \citenamefont {{Beers}}, \citenamefont {{Bell}}, \citenamefont {{Bernal}}, \citenamefont {{Besuner}}, \citenamefont {{Beutler}}, \citenamefont {{Blake}}, \citenamefont {{Bleuler}}, \citenamefont {{Blomqvist}}, \citenamefont {{Blum}}, \citenamefont {{Bolton}}, \citenamefont {{Briceno}}, \citenamefont {{Brooks}}, \citenamefont {{Brownstein}}, \citenamefont {{Buckley-Geer}}, \citenamefont {{Burden}}, \citenamefont {{Burtin}}, \citenamefont {{Busca}}, \citenamefont {{Cahn}}, \citenamefont {{Cai}}, \citenamefont {{Cardiel-Sas}}, \citenamefont
  {{Carlberg}}, \citenamefont {{Carton}}, \citenamefont {{Casas}}, \citenamefont {{Castander}}, \citenamefont {{Cervantes-Cota}}, \citenamefont {{Claybaugh}}, \citenamefont {{Close}}, \citenamefont {{Coker}}, \citenamefont {{Cole}}, \citenamefont {{Comparat}}, \citenamefont {{Cooper}}, \citenamefont {{Cousinou}}, \citenamefont {{Crocce}}, \citenamefont {{Cuby}}, \citenamefont {{Cunningham}}, \citenamefont {{Davis}}, \citenamefont {{Dawson}}, \citenamefont {{de la Macorra}}, \citenamefont {{De Vicente}}, \citenamefont {{Delubac}}, \citenamefont {{Derwent}}, \citenamefont {{Dey}}, \citenamefont {{Dhungana}}, \citenamefont {{Ding}}, \citenamefont {{Doel}}, \citenamefont {{Duan}}, \citenamefont {{Ealet}}, \citenamefont {{Edelstein}}, \citenamefont {{Eftekharzadeh}}, \citenamefont {{Eisenstein}}, \citenamefont {{Elliott}}, \citenamefont {{Escoffier}}, \citenamefont {{Evatt}}, \citenamefont {{Fagrelius}}, \citenamefont {{Fan}}, \citenamefont {{Fanning}}, \citenamefont {{Farahi}}, \citenamefont {{Farihi}},
  \citenamefont {{Favole}}, \citenamefont {{Feng}}, \citenamefont {{Fernandez}}, \citenamefont {{Findlay}}, \citenamefont {{Finkbeiner}}, \citenamefont {{Fitzpatrick}}, \citenamefont {{Flaugher}}, \citenamefont {{Flender}}, \citenamefont {{Font-Ribera}}, \citenamefont {{Forero-Romero}}, \citenamefont {{Fosalba}}, \citenamefont {{Frenk}}, \citenamefont {{Fumagalli}}, \citenamefont {{Gaensicke}}, \citenamefont {{Gallo}}, \citenamefont {{Garcia-Bellido}}, \citenamefont {{Gaztanaga}}, \citenamefont {{Pietro Gentile Fusillo}}, \citenamefont {{Gerard}}, \citenamefont {{Gershkovich}}, \citenamefont {{Giannantonio}}, \citenamefont {{Gillet}}, \citenamefont {{Gonzalez-de-Rivera}}, \citenamefont {{Gonzalez-Perez}}, \citenamefont {{Gott}}, \citenamefont {{Graur}}, \citenamefont {{Gutierrez}}, \citenamefont {{Guy}}, \citenamefont {{Habib}}, \citenamefont {{Heetderks}}, \citenamefont {{Heetderks}}, \citenamefont {{Heitmann}}, \citenamefont {{Hellwing}}, \citenamefont {{Herrera}}, \citenamefont {{Ho}}, \citenamefont
  {{Holland}}, \citenamefont {{Honscheid}}, \citenamefont {{Huff}}, \citenamefont {{Hutchinson}}, \citenamefont {{Huterer}}, \citenamefont {{Hwang}}, \citenamefont {{Illa Laguna}}, \citenamefont {{Ishikawa}}, \citenamefont {{Jacobs}}, \citenamefont {{Jeffrey}}, \citenamefont {{Jelinsky}}, \citenamefont {{Jennings}}, \citenamefont {{Jiang}}, \citenamefont {{Jimenez}}, \citenamefont {{Johnson}}, \citenamefont {{Joyce}}, \citenamefont {{Jullo}}, \citenamefont {{Juneau}}, \citenamefont {{Kama}}, \citenamefont {{Karcher}}, \citenamefont {{Karkar}}, \citenamefont {{Kehoe}}, \citenamefont {{Kennamer}}, \citenamefont {{Kent}}, \citenamefont {{Kilbinger}}, \citenamefont {{Kim}}, \citenamefont {{Kirkby}}, \citenamefont {{Kisner}}, \citenamefont {{Kitanidis}}, \citenamefont {{Kneib}}, \citenamefont {{Koposov}}, \citenamefont {{Kovacs}}, \citenamefont {{Koyama}}, \citenamefont {{Kremin}}, \citenamefont {{Kron}}, \citenamefont {{Kronig}}, \citenamefont {{Kueter-Young}}, \citenamefont {{Lacey}}, \citenamefont {{Lafever}},
  \citenamefont {{Lahav}}, \citenamefont {{Lambert}}, \citenamefont {{Lampton}}, \citenamefont {{Landriau}}, \citenamefont {{Lang}}, \citenamefont {{Lauer}}, \citenamefont {{Le Goff}}, \citenamefont {{Le Guillou}}, \citenamefont {{Le Van Suu}}, \citenamefont {{Lee}}, \citenamefont {{Lee}}, \citenamefont {{Leitner}}, \citenamefont {{Lesser}}, \citenamefont {{Levi}}, \citenamefont {{L'Huillier}}, \citenamefont {{Li}}, \citenamefont {{Liang}}, \citenamefont {{Lin}}, \citenamefont {{Linder}}, \citenamefont {{Loebman}}, \citenamefont {{Luki{\'c}}}, \citenamefont {{Ma}}, \citenamefont {{MacCrann}}, \citenamefont {{Magneville}}, \citenamefont {{Makarem}}, \citenamefont {{Manera}}, \citenamefont {{Manser}}, \citenamefont {{Marshall}}, \citenamefont {{Martini}}, \citenamefont {{Massey}}, \citenamefont {{Matheson}}, \citenamefont {{McCauley}}, \citenamefont {{McDonald}}, \citenamefont {{McGreer}}, \citenamefont {{Meisner}}, \citenamefont {{Metcalfe}}, \citenamefont {{Miller}}, \citenamefont {{Miquel}}, \citenamefont
  {{Moustakas}}, \citenamefont {{Myers}}, \citenamefont {{Naik}}, \citenamefont {{Newman}}, \citenamefont {{Nichol}}, \citenamefont {{Nicola}}, \citenamefont {{Nicolati da Costa}}, \citenamefont {{Nie}}, \citenamefont {{Niz}}, \citenamefont {{Norberg}}, \citenamefont {{Nord}}, \citenamefont {{Norman}}, \citenamefont {{Nugent}}, \citenamefont {{O'Brien}}, \citenamefont {{Oh}}, \citenamefont {{Olsen}}, \citenamefont {{Padilla}}, \citenamefont {{Padmanabhan}}, \citenamefont {{Padmanabhan}}, \citenamefont {{Palanque-Delabrouille}}, \citenamefont {{Palmese}}, \citenamefont {{Pappalardo}}, \citenamefont {{P{\^a}ris}}, \citenamefont {{Park}}, \citenamefont {{Patej}}, \citenamefont {{Peacock}}, \citenamefont {{Peiris}}, \citenamefont {{Peng}}, \citenamefont {{Percival}}, \citenamefont {{Perruchot}}, \citenamefont {{Pieri}}, \citenamefont {{Pogge}}, \citenamefont {{Pollack}}, \citenamefont {{Poppett}}, \citenamefont {{Prada}}, \citenamefont {{Prakash}}, \citenamefont {{Probst}}, \citenamefont {{Rabinowitz}},
  \citenamefont {{Raichoor}}, \citenamefont {{Ree}}, \citenamefont {{Refregier}}, \citenamefont {{Regal}}, \citenamefont {{Reid}}, \citenamefont {{Reil}}, \citenamefont {{Rezaie}}, \citenamefont {{Rockosi}}, \citenamefont {{Roe}}, \citenamefont {{Ronayette}}, \citenamefont {{Roodman}}, \citenamefont {{Ross}}, \citenamefont {{Ross}}, \citenamefont {{Rossi}}, \citenamefont {{Rozo}}, \citenamefont {{Ruhlmann-Kleider}}, \citenamefont {{Rykoff}}, \citenamefont {{Sabiu}}, \citenamefont {{Samushia}}, \citenamefont {{Sanchez}}, \citenamefont {{Sanchez}}, \citenamefont {{Schlegel}}, \citenamefont {{Schneider}}, \citenamefont {{Schubnell}}, \citenamefont {{Secroun}}, \citenamefont {{Seljak}}, \citenamefont {{Seo}}, \citenamefont {{Serrano}}, \citenamefont {{Shafieloo}}, \citenamefont {{Shan}}, \citenamefont {{Sharples}}, \citenamefont {{Sholl}}, \citenamefont {{Shourt}}, \citenamefont {{Silber}}, \citenamefont {{Silva}}, \citenamefont {{Sirk}}, \citenamefont {{Slosar}}, \citenamefont {{Smith}}, \citenamefont {{Smoot}},
  \citenamefont {{Som}}, \citenamefont {{Song}}, \citenamefont {{Sprayberry}}, \citenamefont {{Staten}}, \citenamefont {{Stefanik}}, \citenamefont {{Tarle}}, \citenamefont {{Sien Tie}}, \citenamefont {{Tinker}}, \citenamefont {{Tojeiro}}, \citenamefont {{Valdes}}, \citenamefont {{Valenzuela}}, \citenamefont {{Valluri}}, \citenamefont {{Vargas-Magana}}, \citenamefont {{Verde}}, \citenamefont {{Walker}}, \citenamefont {{Wang}}, \citenamefont {{Wang}}, \citenamefont {{Weaver}}, \citenamefont {{Weaverdyck}}, \citenamefont {{Wechsler}}, \citenamefont {{Weinberg}}, \citenamefont {{White}}, \citenamefont {{Yang}}, \citenamefont {{Yeche}}, \citenamefont {{Zhang}}, \citenamefont {{Zhao}}, \citenamefont {{Zheng}}, \citenamefont {{Zhou}}, \citenamefont {{Zhou}}, \citenamefont {{Zhu}}, \citenamefont {{Zou}},\ and\ \citenamefont {{Zu}}}]{DESI2016a.Science}%
  \BibitemOpen
  \bibfield  {author} {\bibinfo {author} {\bibnamefont {{DESI Collaboration}}}, \bibinfo {author} {\bibfnamefont {A.}~\bibnamefont {{Aghamousa}}}, \bibinfo {author} {\bibfnamefont {J.}~\bibnamefont {{Aguilar}}}, \bibinfo {author} {\bibfnamefont {S.}~\bibnamefont {{Ahlen}}}, \bibinfo {author} {\bibfnamefont {S.}~\bibnamefont {{Alam}}}, \bibinfo {author} {\bibfnamefont {L.~E.}\ \bibnamefont {{Allen}}}, \bibinfo {author} {\bibfnamefont {C.}~\bibnamefont {{Allende Prieto}}}, \bibinfo {author} {\bibfnamefont {J.}~\bibnamefont {{Annis}}}, \bibinfo {author} {\bibfnamefont {S.}~\bibnamefont {{Bailey}}}, \bibinfo {author} {\bibfnamefont {C.}~\bibnamefont {{Balland}}}, \bibinfo {author} {\bibfnamefont {O.}~\bibnamefont {{Ballester}}}, \bibinfo {author} {\bibfnamefont {C.}~\bibnamefont {{Baltay}}}, \bibinfo {author} {\bibfnamefont {L.}~\bibnamefont {{Beaufore}}}, \bibinfo {author} {\bibfnamefont {C.}~\bibnamefont {{Bebek}}}, \bibinfo {author} {\bibfnamefont {T.~C.}\ \bibnamefont {{Beers}}}, \bibnamefont {and~others},\
  }\href@noop {} {\bibfield  {journal} {\bibinfo  {journal} {arXiv e-prints}\ ,\ \bibinfo {eid} {arXiv:1611.00036}} (\bibinfo {year} {2016}{\natexlab{a}})},\ \Eprint {https://arxiv.org/abs/1611.00036} {arXiv:1611.00036 [astro-ph.IM]} \BibitemShut {NoStop}%
\bibitem [{\citenamefont {{DESI Collaboration}}\ \emph {et~al.}(2016{\natexlab{b}})\citenamefont {{DESI Collaboration}}, \citenamefont {{Aghamousa}}, \citenamefont {{Aguilar}}, \citenamefont {{Ahlen}}, \citenamefont {{Alam}}, \citenamefont {{Allen}}, \citenamefont {{Allende Prieto}}, \citenamefont {{Annis}}, \citenamefont {{Bailey}}, \citenamefont {{Balland}}, \citenamefont {{Ballester}}, \citenamefont {{Baltay}}, \citenamefont {{Beaufore}}, \citenamefont {{Bebek}}, \citenamefont {{Beers}}, \citenamefont {{Bell}}, \citenamefont {{Bernal}}, \citenamefont {{Besuner}}, \citenamefont {{Beutler}}, \citenamefont {{Blake}}, \citenamefont {{Bleuler}}, \citenamefont {{Blomqvist}}, \citenamefont {{Blum}}, \citenamefont {{Bolton}}, \citenamefont {{Briceno}}, \citenamefont {{Brooks}}, \citenamefont {{Brownstein}}, \citenamefont {{Buckley-Geer}}, \citenamefont {{Burden}}, \citenamefont {{Burtin}}, \citenamefont {{Busca}}, \citenamefont {{Cahn}}, \citenamefont {{Cai}}, \citenamefont {{Cardiel-Sas}}, \citenamefont
  {{Carlberg}}, \citenamefont {{Carton}}, \citenamefont {{Casas}}, \citenamefont {{Castander}}, \citenamefont {{Cervantes-Cota}}, \citenamefont {{Claybaugh}}, \citenamefont {{Close}}, \citenamefont {{Coker}}, \citenamefont {{Cole}}, \citenamefont {{Comparat}}, \citenamefont {{Cooper}}, \citenamefont {{Cousinou}}, \citenamefont {{Crocce}}, \citenamefont {{Cuby}}, \citenamefont {{Cunningham}}, \citenamefont {{Davis}}, \citenamefont {{Dawson}}, \citenamefont {{de la Macorra}}, \citenamefont {{De Vicente}}, \citenamefont {{Delubac}}, \citenamefont {{Derwent}}, \citenamefont {{Dey}}, \citenamefont {{Dhungana}}, \citenamefont {{Ding}}, \citenamefont {{Doel}}, \citenamefont {{Duan}}, \citenamefont {{Ealet}}, \citenamefont {{Edelstein}}, \citenamefont {{Eftekharzadeh}}, \citenamefont {{Eisenstein}}, \citenamefont {{Elliott}}, \citenamefont {{Escoffier}}, \citenamefont {{Evatt}}, \citenamefont {{Fagrelius}}, \citenamefont {{Fan}}, \citenamefont {{Fanning}}, \citenamefont {{Farahi}}, \citenamefont {{Farihi}},
  \citenamefont {{Favole}}, \citenamefont {{Feng}}, \citenamefont {{Fernandez}}, \citenamefont {{Findlay}}, \citenamefont {{Finkbeiner}}, \citenamefont {{Fitzpatrick}}, \citenamefont {{Flaugher}}, \citenamefont {{Flender}}, \citenamefont {{Font-Ribera}}, \citenamefont {{Forero-Romero}}, \citenamefont {{Fosalba}}, \citenamefont {{Frenk}}, \citenamefont {{Fumagalli}}, \citenamefont {{Gaensicke}}, \citenamefont {{Gallo}}, \citenamefont {{Garcia-Bellido}}, \citenamefont {{Gaztanaga}}, \citenamefont {{Pietro Gentile Fusillo}}, \citenamefont {{Gerard}}, \citenamefont {{Gershkovich}}, \citenamefont {{Giannantonio}}, \citenamefont {{Gillet}}, \citenamefont {{Gonzalez-de-Rivera}}, \citenamefont {{Gonzalez-Perez}}, \citenamefont {{Gott}}, \citenamefont {{Graur}}, \citenamefont {{Gutierrez}}, \citenamefont {{Guy}}, \citenamefont {{Habib}}, \citenamefont {{Heetderks}}, \citenamefont {{Heetderks}}, \citenamefont {{Heitmann}}, \citenamefont {{Hellwing}}, \citenamefont {{Herrera}}, \citenamefont {{Ho}}, \citenamefont
  {{Holland}}, \citenamefont {{Honscheid}}, \citenamefont {{Huff}}, \citenamefont {{Hutchinson}}, \citenamefont {{Huterer}}, \citenamefont {{Hwang}}, \citenamefont {{Illa Laguna}}, \citenamefont {{Ishikawa}}, \citenamefont {{Jacobs}}, \citenamefont {{Jeffrey}}, \citenamefont {{Jelinsky}}, \citenamefont {{Jennings}}, \citenamefont {{Jiang}}, \citenamefont {{Jimenez}}, \citenamefont {{Johnson}}, \citenamefont {{Joyce}}, \citenamefont {{Jullo}}, \citenamefont {{Juneau}}, \citenamefont {{Kama}}, \citenamefont {{Karcher}}, \citenamefont {{Karkar}}, \citenamefont {{Kehoe}}, \citenamefont {{Kennamer}}, \citenamefont {{Kent}}, \citenamefont {{Kilbinger}}, \citenamefont {{Kim}}, \citenamefont {{Kirkby}}, \citenamefont {{Kisner}}, \citenamefont {{Kitanidis}}, \citenamefont {{Kneib}}, \citenamefont {{Koposov}}, \citenamefont {{Kovacs}}, \citenamefont {{Koyama}}, \citenamefont {{Kremin}}, \citenamefont {{Kron}}, \citenamefont {{Kronig}}, \citenamefont {{Kueter-Young}}, \citenamefont {{Lacey}}, \citenamefont {{Lafever}},
  \citenamefont {{Lahav}}, \citenamefont {{Lambert}}, \citenamefont {{Lampton}}, \citenamefont {{Landriau}}, \citenamefont {{Lang}}, \citenamefont {{Lauer}}, \citenamefont {{Le Goff}}, \citenamefont {{Le Guillou}}, \citenamefont {{Le Van Suu}}, \citenamefont {{Lee}}, \citenamefont {{Lee}}, \citenamefont {{Leitner}}, \citenamefont {{Lesser}}, \citenamefont {{Levi}}, \citenamefont {{L'Huillier}}, \citenamefont {{Li}}, \citenamefont {{Liang}}, \citenamefont {{Lin}}, \citenamefont {{Linder}}, \citenamefont {{Loebman}}, \citenamefont {{Luki{\'c}}}, \citenamefont {{Ma}}, \citenamefont {{MacCrann}}, \citenamefont {{Magneville}}, \citenamefont {{Makarem}}, \citenamefont {{Manera}}, \citenamefont {{Manser}}, \citenamefont {{Marshall}}, \citenamefont {{Martini}}, \citenamefont {{Massey}}, \citenamefont {{Matheson}}, \citenamefont {{McCauley}}, \citenamefont {{McDonald}}, \citenamefont {{McGreer}}, \citenamefont {{Meisner}}, \citenamefont {{Metcalfe}}, \citenamefont {{Miller}}, \citenamefont {{Miquel}}, \citenamefont
  {{Moustakas}}, \citenamefont {{Myers}}, \citenamefont {{Naik}}, \citenamefont {{Newman}}, \citenamefont {{Nichol}}, \citenamefont {{Nicola}}, \citenamefont {{Nicolati da Costa}}, \citenamefont {{Nie}}, \citenamefont {{Niz}}, \citenamefont {{Norberg}}, \citenamefont {{Nord}}, \citenamefont {{Norman}}, \citenamefont {{Nugent}}, \citenamefont {{O'Brien}}, \citenamefont {{Oh}}, \citenamefont {{Olsen}}, \citenamefont {{Padilla}}, \citenamefont {{Padmanabhan}}, \citenamefont {{Padmanabhan}}, \citenamefont {{Palanque-Delabrouille}}, \citenamefont {{Palmese}}, \citenamefont {{Pappalardo}}, \citenamefont {{P{\^a}ris}}, \citenamefont {{Park}}, \citenamefont {{Patej}}, \citenamefont {{Peacock}}, \citenamefont {{Peiris}}, \citenamefont {{Peng}}, \citenamefont {{Percival}}, \citenamefont {{Perruchot}}, \citenamefont {{Pieri}}, \citenamefont {{Pogge}}, \citenamefont {{Pollack}}, \citenamefont {{Poppett}}, \citenamefont {{Prada}}, \citenamefont {{Prakash}}, \citenamefont {{Probst}}, \citenamefont {{Rabinowitz}},
  \citenamefont {{Raichoor}}, \citenamefont {{Ree}}, \citenamefont {{Refregier}}, \citenamefont {{Regal}}, \citenamefont {{Reid}}, \citenamefont {{Reil}}, \citenamefont {{Rezaie}}, \citenamefont {{Rockosi}}, \citenamefont {{Roe}}, \citenamefont {{Ronayette}}, \citenamefont {{Roodman}}, \citenamefont {{Ross}}, \citenamefont {{Ross}}, \citenamefont {{Rossi}}, \citenamefont {{Rozo}}, \citenamefont {{Ruhlmann-Kleider}}, \citenamefont {{Rykoff}}, \citenamefont {{Sabiu}}, \citenamefont {{Samushia}}, \citenamefont {{Sanchez}}, \citenamefont {{Sanchez}}, \citenamefont {{Schlegel}}, \citenamefont {{Schneider}}, \citenamefont {{Schubnell}}, \citenamefont {{Secroun}}, \citenamefont {{Seljak}}, \citenamefont {{Seo}}, \citenamefont {{Serrano}}, \citenamefont {{Shafieloo}}, \citenamefont {{Shan}}, \citenamefont {{Sharples}}, \citenamefont {{Sholl}}, \citenamefont {{Shourt}}, \citenamefont {{Silber}}, \citenamefont {{Silva}}, \citenamefont {{Sirk}}, \citenamefont {{Slosar}}, \citenamefont {{Smith}}, \citenamefont {{Smoot}},
  \citenamefont {{Som}}, \citenamefont {{Song}}, \citenamefont {{Sprayberry}}, \citenamefont {{Staten}}, \citenamefont {{Stefanik}}, \citenamefont {{Tarle}}, \citenamefont {{Sien Tie}}, \citenamefont {{Tinker}}, \citenamefont {{Tojeiro}}, \citenamefont {{Valdes}}, \citenamefont {{Valenzuela}}, \citenamefont {{Valluri}}, \citenamefont {{Vargas-Magana}}, \citenamefont {{Verde}}, \citenamefont {{Walker}}, \citenamefont {{Wang}}, \citenamefont {{Wang}}, \citenamefont {{Weaver}}, \citenamefont {{Weaverdyck}}, \citenamefont {{Wechsler}}, \citenamefont {{Weinberg}}, \citenamefont {{White}}, \citenamefont {{Yang}}, \citenamefont {{Yeche}}, \citenamefont {{Zhang}}, \citenamefont {{Zhao}}, \citenamefont {{Zheng}}, \citenamefont {{Zhou}}, \citenamefont {{Zhou}}, \citenamefont {{Zhu}}, \citenamefont {{Zou}},\ and\ \citenamefont {{Zu}}}]{DESI2016b.Instr}%
  \BibitemOpen
  \bibfield  {author} {\bibinfo {author} {\bibnamefont {{DESI Collaboration}}}, \bibinfo {author} {\bibfnamefont {A.}~\bibnamefont {{Aghamousa}}}, \bibinfo {author} {\bibfnamefont {J.}~\bibnamefont {{Aguilar}}}, \bibinfo {author} {\bibfnamefont {S.}~\bibnamefont {{Ahlen}}}, \bibinfo {author} {\bibfnamefont {S.}~\bibnamefont {{Alam}}}, \bibinfo {author} {\bibfnamefont {L.~E.}\ \bibnamefont {{Allen}}}, \bibinfo {author} {\bibfnamefont {C.}~\bibnamefont {{Allende Prieto}}}, \bibinfo {author} {\bibfnamefont {J.}~\bibnamefont {{Annis}}}, \bibinfo {author} {\bibfnamefont {S.}~\bibnamefont {{Bailey}}}, \bibinfo {author} {\bibfnamefont {C.}~\bibnamefont {{Balland}}}, \bibinfo {author} {\bibfnamefont {O.}~\bibnamefont {{Ballester}}}, \bibinfo {author} {\bibfnamefont {C.}~\bibnamefont {{Baltay}}}, \bibinfo {author} {\bibfnamefont {L.}~\bibnamefont {{Beaufore}}}, \bibinfo {author} {\bibfnamefont {C.}~\bibnamefont {{Bebek}}}, \bibinfo {author} {\bibfnamefont {T.~C.}\ \bibnamefont {{Beers}}}, \bibnamefont {and~others},\
  }\href@noop {} {\bibfield  {journal} {\bibinfo  {journal} {arXiv e-prints}\ ,\ \bibinfo {eid} {arXiv:1611.00037}} (\bibinfo {year} {2016}{\natexlab{b}})},\ \Eprint {https://arxiv.org/abs/1611.00037} {arXiv:1611.00037 [astro-ph.IM]} \BibitemShut {NoStop}%
\bibitem [{\citenamefont {{DESI Collaboration}}\ \emph {et~al.}(2022)\citenamefont {{DESI Collaboration}}, \citenamefont {{Abareshi}}, \citenamefont {{Aguilar}}, \citenamefont {{Ahlen}}, \citenamefont {{Alam}}, \citenamefont {{Alexander}}, \citenamefont {{Alfarsy}}, \citenamefont {{Allen}}, \citenamefont {{Allende Prieto}}, \citenamefont {{Alves}}, \citenamefont {{Ameel}}, \citenamefont {{Armengaud}}, \citenamefont {{Asorey}}, \citenamefont {{Aviles}}, \citenamefont {{Bailey}}, \citenamefont {{Balaguera-Antol{\'\i}nez}}, \citenamefont {{Ballester}}, \citenamefont {{Baltay}}, \citenamefont {{Bault}}, \citenamefont {{Beltran}}, \citenamefont {{Benavides}}, \citenamefont {{BenZvi}}, \citenamefont {{Berti}}, \citenamefont {{Besuner}}, \citenamefont {{Beutler}}, \citenamefont {{Bianchi}}, \citenamefont {{Blake}}, \citenamefont {{Blanc}}, \citenamefont {{Blum}}, \citenamefont {{Bolton}}, \citenamefont {{Bose}}, \citenamefont {{Bramall}}, \citenamefont {{Brieden}}, \citenamefont {{Brodzeller}}, \citenamefont
  {{Brooks}}, \citenamefont {{Brownewell}}, \citenamefont {{Buckley-Geer}}, \citenamefont {{Cahn}}, \citenamefont {{Cai}}, \citenamefont {{Canning}}, \citenamefont {{Capasso}}, \citenamefont {{Carnero Rosell}}, \citenamefont {{Carton}}, \citenamefont {{Casas}}, \citenamefont {{Castander}}, \citenamefont {{Cervantes-Cota}}, \citenamefont {{Chabanier}}, \citenamefont {{Chaussidon}}, \citenamefont {{Chuang}}, \citenamefont {{Circosta}}, \citenamefont {{Cole}}, \citenamefont {{Cooper}}, \citenamefont {{da Costa}}, \citenamefont {{Cousinou}}, \citenamefont {{Cuceu}}, \citenamefont {{Davis}}, \citenamefont {{Dawson}}, \citenamefont {{de la Cruz-Noriega}}, \citenamefont {{de la Macorra}}, \citenamefont {{de Mattia}}, \citenamefont {{Della Costa}}, \citenamefont {{Demmer}}, \citenamefont {{Derwent}}, \citenamefont {{Dey}}, \citenamefont {{Dey}}, \citenamefont {{Dhungana}}, \citenamefont {{Ding}}, \citenamefont {{Dobson}}, \citenamefont {{Doel}}, \citenamefont {{Donald-McCann}}, \citenamefont {{Donaldson}},
  \citenamefont {{Douglass}}, \citenamefont {{Duan}}, \citenamefont {{Dunlop}}, \citenamefont {{Edelstein}}, \citenamefont {{Eftekharzadeh}}, \citenamefont {{Eisenstein}}, \citenamefont {{Enriquez-Vargas}}, \citenamefont {{Escoffier}}, \citenamefont {{Evatt}}, \citenamefont {{Fagrelius}}, \citenamefont {{Fan}}, \citenamefont {{Fanning}}, \citenamefont {{Fawcett}}, \citenamefont {{Ferraro}}, \citenamefont {{Ereza}}, \citenamefont {{Flaugher}}, \citenamefont {{Font-Ribera}}, \citenamefont {{Forero-Romero}}, \citenamefont {{Frenk}}, \citenamefont {{Fromenteau}}, \citenamefont {{G{\"a}nsicke}}, \citenamefont {{Garcia-Quintero}}, \citenamefont {{Garrison}}, \citenamefont {{Gazta{\~n}aga}}, \citenamefont {{Gerardi}}, \citenamefont {{Gil-Mar{\'\i}n}}, \citenamefont {{Gontcho a Gontcho}}, \citenamefont {{Gonzalez-Morales}}, \citenamefont {{Gonzalez-de-Rivera}}, \citenamefont {{Gonzalez-Perez}}, \citenamefont {{Gordon}}, \citenamefont {{Graur}}, \citenamefont {{Green}}, \citenamefont {{Grove}}, \citenamefont
  {{Gruen}}, \citenamefont {{Gutierrez}}, \citenamefont {{Guy}}, \citenamefont {{Hahn}}, \citenamefont {{Harris}}, \citenamefont {{Herrera}}, \citenamefont {{Herrera-Alcantar}}, \citenamefont {{Honscheid}}, \citenamefont {{Howlett}}, \citenamefont {{Huterer}}, \citenamefont {{Ir{\v{s}}i{\v{c}}}}, \citenamefont {{Ishak}}, \citenamefont {{Jelinsky}}, \citenamefont {{Jiang}}, \citenamefont {{Jimenez}}, \citenamefont {{Jing}}, \citenamefont {{Joyce}}, \citenamefont {{Jullo}}, \citenamefont {{Juneau}}, \citenamefont {{Kara{\c{c}}ayl{\i}}}, \citenamefont {{Karamanis}}, \citenamefont {{Karcher}}, \citenamefont {{Karim}}, \citenamefont {{Kehoe}}, \citenamefont {{Kent}}, \citenamefont {{Kirkby}}, \citenamefont {{Kisner}}, \citenamefont {{Kitaura}}, \citenamefont {{Koposov}}, \citenamefont {{Kov{\'a}cs}}, \citenamefont {{Kremin}}, \citenamefont {{Krolewski}}, \citenamefont {{L'Huillier}}, \citenamefont {{Lahav}}, \citenamefont {{Lambert}}, \citenamefont {{Lamman}}, \citenamefont {{Lan}}, \citenamefont {{Landriau}},
  \citenamefont {{Lane}}, \citenamefont {{Lang}}, \citenamefont {{Lange}}, \citenamefont {{Lasker}}, \citenamefont {{Le Guillou}}, \citenamefont {{Leauthaud}}, \citenamefont {{Le Van Suu}}, \citenamefont {{Levi}}, \citenamefont {{Li}}, \citenamefont {{Magneville}}, \citenamefont {{Manera}}, \citenamefont {{Manser}}, \citenamefont {{Marshall}}, \citenamefont {{Martini}}, \citenamefont {{McCollam}}, \citenamefont {{McDonald}}, \citenamefont {{Meisner}}, \citenamefont {{Mena-Fern{\'a}ndez}}, \citenamefont {{Meneses-Rizo}}, \citenamefont {{Mezcua}}, \citenamefont {{Miller}}, \citenamefont {{Miquel}}, \citenamefont {{Montero-Camacho}}, \citenamefont {{Moon}}, \citenamefont {{Moustakas}}, \citenamefont {{Mueller}}, \citenamefont {{Mu{\~n}oz-Guti{\'e}rrez}}, \citenamefont {{Myers}}, \citenamefont {{Nadathur}}, \citenamefont {{Najita}}, \citenamefont {{Napolitano}}, \citenamefont {{Neilsen}}, \citenamefont {{Newman}}, \citenamefont {{Nie}}, \citenamefont {{Ning}}, \citenamefont {{Niz}}, \citenamefont {{Norberg}},
  \citenamefont {{Noriega}}, \citenamefont {{O'Brien}}, \citenamefont {{Obuljen}}, \citenamefont {{Palanque-Delabrouille}}, \citenamefont {{Palmese}}, \citenamefont {{Zhiwei}}, \citenamefont {{Pappalardo}}, \citenamefont {{PENG}}, \citenamefont {{Percival}}, \citenamefont {{Perruchot}}, \citenamefont {{Pogge}}, \citenamefont {{Poppett}}, \citenamefont {{Porredon}}, \citenamefont {{Prada}}, \citenamefont {{Prochaska}}, \citenamefont {{Pucha}}, \citenamefont {{P{\'e}rez-Fern{\'a}ndez}}, \citenamefont {{P{\'e}rez-R{\`a}fols}}, \citenamefont {{Rabinowitz}}, \citenamefont {{Raichoor}}, \citenamefont {{Ramirez-Solano}}, \citenamefont {{Ram{\'\i}rez-P{\'e}rez}}, \citenamefont {{Ravoux}}, \citenamefont {{Reil}}, \citenamefont {{Rezaie}}, \citenamefont {{Rocher}}, \citenamefont {{Rockosi}}, \citenamefont {{Roe}}, \citenamefont {{Roodman}}, \citenamefont {{Ross}}, \citenamefont {{Rossi}}, \citenamefont {{Ruggeri}}, \citenamefont {{Ruhlmann-Kleider}}, \citenamefont {{Sabiu}}, \citenamefont {{Safonova}}, \citenamefont
  {{Said}}, \citenamefont {{Saintonge}}, \citenamefont {{Salas Catonga}}, \citenamefont {{Samushia}}, \citenamefont {{Sanchez}}, \citenamefont {{Saulder}}, \citenamefont {{Schaan}}, \citenamefont {{Schlafly}}, \citenamefont {{Schlegel}}, \citenamefont {{Schmoll}}, \citenamefont {{Scholte}}, \citenamefont {{Schubnell}}, \citenamefont {{Secroun}}, \citenamefont {{Seo}}, \citenamefont {{Serrano}}, \citenamefont {{Sharples}}, \citenamefont {{Sholl}}, \citenamefont {{Silber}}, \citenamefont {{Silva}}, \citenamefont {{Sirk}}, \citenamefont {{Siudek}}, \citenamefont {{Smith}}, \citenamefont {{Sprayberry}}, \citenamefont {{Staten}}, \citenamefont {{Stupak}}, \citenamefont {{Tan}}, \citenamefont {{Tarl{\'e}}}, \citenamefont {{Tie}}, \citenamefont {{Tojeiro}}, \citenamefont {{Ure{\~n}a-L{\'o}pez}}, \citenamefont {{Valdes}}, \citenamefont {{Valenzuela}}, \citenamefont {{Valluri}}, \citenamefont {{Vargas-Maga{\~n}a}}, \citenamefont {{Verde}}, \citenamefont {{Walther}}, \citenamefont {{Wang}}, \citenamefont {{Wang}},
  \citenamefont {{Weaver}}, \citenamefont {{Weaverdyck}}, \citenamefont {{Wechsler}}, \citenamefont {{Wilson}}, \citenamefont {{Yang}}, \citenamefont {{Yu}}, \citenamefont {{Yuan}}, \citenamefont {{Y{\`e}che}}, \citenamefont {{Zhang}}, \citenamefont {{Zhang}}, \citenamefont {{Zhao}}, \citenamefont {{Zhou}}, \citenamefont {{Zhou}}, \citenamefont {{Zou}}, \citenamefont {{Zou}}, \citenamefont {{Zou}}, \citenamefont {{Zu}},\ and\ \citenamefont {{DESI Collaboration}}}]{DESI2022.KP1.Instr}%
  \BibitemOpen
  \bibfield  {author} {\bibinfo {author} {\bibnamefont {{DESI Collaboration}}}, \bibinfo {author} {\bibfnamefont {B.}~\bibnamefont {{Abareshi}}}, \bibinfo {author} {\bibfnamefont {J.}~\bibnamefont {{Aguilar}}}, \bibinfo {author} {\bibfnamefont {S.}~\bibnamefont {{Ahlen}}}, \bibinfo {author} {\bibfnamefont {S.}~\bibnamefont {{Alam}}}, \bibinfo {author} {\bibfnamefont {D.~M.}\ \bibnamefont {{Alexander}}}, \bibinfo {author} {\bibfnamefont {R.}~\bibnamefont {{Alfarsy}}}, \bibinfo {author} {\bibfnamefont {L.}~\bibnamefont {{Allen}}}, \bibinfo {author} {\bibfnamefont {C.}~\bibnamefont {{Allende Prieto}}}, \bibinfo {author} {\bibfnamefont {O.}~\bibnamefont {{Alves}}}, \bibinfo {author} {\bibfnamefont {J.}~\bibnamefont {{Ameel}}}, \bibinfo {author} {\bibfnamefont {E.}~\bibnamefont {{Armengaud}}}, \bibinfo {author} {\bibfnamefont {J.}~\bibnamefont {{Asorey}}}, \bibinfo {author} {\bibfnamefont {A.}~\bibnamefont {{Aviles}}}, \bibinfo {author} {\bibfnamefont {S.}~\bibnamefont {{Bailey}}}, \bibnamefont {and~others},\
  }\href {https://doi.org/10.3847/1538-3881/ac882b} {\bibfield  {journal} {\bibinfo  {journal} {\aj}\ }\textbf {\bibinfo {volume} {164}},\ \bibinfo {eid} {207} (\bibinfo {year} {2022})},\ \Eprint {https://arxiv.org/abs/2205.10939} {arXiv:2205.10939 [astro-ph.IM]} \BibitemShut {NoStop}%
\bibitem [{\citenamefont {{Silber}}\ \emph {et~al.}(2023)\citenamefont {{Silber}}, \citenamefont {{Fagrelius}}, \citenamefont {{Fanning}}, \citenamefont {{Schubnell}}, \citenamefont {{Aguilar}}, \citenamefont {{Ahlen}}, \citenamefont {{Ameel}}, \citenamefont {{Ballester}}, \citenamefont {{Baltay}}, \citenamefont {{Bebek}}, \citenamefont {{Benton Beard}}, \citenamefont {{Besuner}}, \citenamefont {{Cardiel-Sas}}, \citenamefont {{Casas}}, \citenamefont {{Castander}}, \citenamefont {{Claybaugh}}, \citenamefont {{Dobson}}, \citenamefont {{Duan}}, \citenamefont {{Dunlop}}, \citenamefont {{Edelstein}}, \citenamefont {{Emmet}}, \citenamefont {{Elliott}}, \citenamefont {{Evatt}}, \citenamefont {{Gershkovich}}, \citenamefont {{Guy}}, \citenamefont {{Harris}}, \citenamefont {{Heetderks}}, \citenamefont {{Heetderks}}, \citenamefont {{Honscheid}}, \citenamefont {{Illa}}, \citenamefont {{Jelinsky}}, \citenamefont {{Jelinsky}}, \citenamefont {{Jimenez}}, \citenamefont {{Karcher}}, \citenamefont {{Kent}}, \citenamefont
  {{Kirkby}}, \citenamefont {{Kneib}}, \citenamefont {{Lambert}}, \citenamefont {{Lampton}}, \citenamefont {{Leitner}}, \citenamefont {{Levi}}, \citenamefont {{McCauley}}, \citenamefont {{Meisner}}, \citenamefont {{Miller}}, \citenamefont {{Miquel}}, \citenamefont {{Mundet}}, \citenamefont {{Poppett}}, \citenamefont {{Rabinowitz}}, \citenamefont {{Reil}}, \citenamefont {{Roman}}, \citenamefont {{Schlegel}}, \citenamefont {{Serrano}}, \citenamefont {{Van Shourt}}, \citenamefont {{Sprayberry}}, \citenamefont {{Tarl{\'e}}}, \citenamefont {{Tie}}, \citenamefont {{Weaverdyck}}, \citenamefont {{Zhang}}, \citenamefont {{Azzaro}}, \citenamefont {{Bailey}}, \citenamefont {{Becerril}}, \citenamefont {{Blackwell}}, \citenamefont {{Bouri}}, \citenamefont {{Brooks}}, \citenamefont {{Buckley-Geer}}, \citenamefont {{Castro}}, \citenamefont {{Derwent}}, \citenamefont {{Dey}}, \citenamefont {{Dhungana}}, \citenamefont {{Doel}}, \citenamefont {{Eisenstein}}, \citenamefont {{Fahim}}, \citenamefont {{Garcia-Bellido}},
  \citenamefont {{Gazta{\~n}aga}}, \citenamefont {{A Gontcho}}, \citenamefont {{Gutierrez}}, \citenamefont {{H{\"o}rler}}, \citenamefont {{Kehoe}}, \citenamefont {{Kisner}}, \citenamefont {{Kremin}}, \citenamefont {{Kronig}}, \citenamefont {{Landriau}}, \citenamefont {{Le Guillou}}, \citenamefont {{Martini}}, \citenamefont {{Moustakas}}, \citenamefont {{Palanque-Delabrouille}}, \citenamefont {{Peng}}, \citenamefont {{Percival}}, \citenamefont {{Prada}}, \citenamefont {{Allende Prieto}}, \citenamefont {{de Rivera}}, \citenamefont {{Sanchez}}, \citenamefont {{Sanchez}}, \citenamefont {{Sharples}}, \citenamefont {{Soares-Santos}}, \citenamefont {{Schlafly}}, \citenamefont {{Weaver}}, \citenamefont {{Zhou}}, \citenamefont {{Zhu}}, \citenamefont {{Zou}},\ and\ \citenamefont {{DESI Collaboration}}}]{FocalPlane.Silber.2023}%
  \BibitemOpen
  \bibfield  {author} {\bibinfo {author} {\bibfnamefont {J.~H.}\ \bibnamefont {{Silber}}}, \bibinfo {author} {\bibfnamefont {P.}~\bibnamefont {{Fagrelius}}}, \bibinfo {author} {\bibfnamefont {K.}~\bibnamefont {{Fanning}}}, \bibinfo {author} {\bibfnamefont {M.}~\bibnamefont {{Schubnell}}}, \bibinfo {author} {\bibfnamefont {J.~N.}\ \bibnamefont {{Aguilar}}}, \bibinfo {author} {\bibfnamefont {S.}~\bibnamefont {{Ahlen}}}, \bibinfo {author} {\bibfnamefont {J.}~\bibnamefont {{Ameel}}}, \bibinfo {author} {\bibfnamefont {O.}~\bibnamefont {{Ballester}}}, \bibinfo {author} {\bibfnamefont {C.}~\bibnamefont {{Baltay}}}, \bibinfo {author} {\bibfnamefont {C.}~\bibnamefont {{Bebek}}}, \bibinfo {author} {\bibfnamefont {D.}~\bibnamefont {{Benton Beard}}}, \bibinfo {author} {\bibfnamefont {R.}~\bibnamefont {{Besuner}}}, \bibinfo {author} {\bibfnamefont {L.}~\bibnamefont {{Cardiel-Sas}}}, \bibinfo {author} {\bibfnamefont {R.}~\bibnamefont {{Casas}}}, \bibinfo {author} {\bibfnamefont {F.~J.}\ \bibnamefont {{Castander}}},
  \bibnamefont {and~others},\ }\href {https://doi.org/10.3847/1538-3881/ac9ab1} {\bibfield  {journal} {\bibinfo  {journal} {\aj}\ }\textbf {\bibinfo {volume} {165}},\ \bibinfo {eid} {9} (\bibinfo {year} {2023})},\ \Eprint {https://arxiv.org/abs/2205.09014} {arXiv:2205.09014 [astro-ph.IM]} \BibitemShut {NoStop}%
\bibitem [{\citenamefont {{Poppett}}\ \emph {et~al.}(2024)\citenamefont {{Poppett}}, \citenamefont {{Tyas}}, \citenamefont {{Aguilar}}, \citenamefont {{Bebek}}, \citenamefont {{Bramall}}, \citenamefont {{Claybaugh}}, \citenamefont {{Edelstein}}, \citenamefont {{Fagrelius}}, \citenamefont {{Heetderks}}, \citenamefont {{Jelinsky}}, \citenamefont {{Jelinsky}}, \citenamefont {{Lafever}}, \citenamefont {{Lambert}}, \citenamefont {{Lampton}}, \citenamefont {{Levi}}, \citenamefont {{Martini}}, \citenamefont {{Rockosi}}, \citenamefont {{Schmoll}}, \citenamefont {{Sharples}}, \citenamefont {{Sirk}}, \citenamefont {{Wishnow}}, \citenamefont {{Yu}}, \citenamefont {{Ahlen}}, \citenamefont {{Bault}}, \citenamefont {{BenZvi}}, \citenamefont {{Brooks}}, \citenamefont {{Cole}}, \citenamefont {{de la Macorra}}, \citenamefont {{Dey}}, \citenamefont {{Doel}}, \citenamefont {{Fanning}}, \citenamefont {{Font-Ribera}}, \citenamefont {{Forero-Romero}}, \citenamefont {{Gazta{\~n}aga}}, \citenamefont {{Gontcho A Gontcho}},
  \citenamefont {{Gonzalez-Morales}}, \citenamefont {{Hahn}}, \citenamefont {{Honscheid}}, \citenamefont {{Jimenez}}, \citenamefont {{Juneau}}, \citenamefont {{Kirkby}}, \citenamefont {{Kremin}}, \citenamefont {{Landriau}}, \citenamefont {{Le Guillou}}, \citenamefont {{Manera}}, \citenamefont {{Meisner}}, \citenamefont {{Miquel}}, \citenamefont {{Moustakas}}, \citenamefont {{Mueller}}, \citenamefont {{Mu{\~n}oz-Guti{\'e}rrez}}, \citenamefont {{Myers}}, \citenamefont {{Nie}}, \citenamefont {{Niz}}, \citenamefont {{Palanque-Delabrouille}}, \citenamefont {{Percival}}, \citenamefont {{Prada}}, \citenamefont {{Rabinowitz}}, \citenamefont {{Rezaie}}, \citenamefont {{Rossi}}, \citenamefont {{Sanchez}}, \citenamefont {{Schlafly}}, \citenamefont {{Schlegel}}, \citenamefont {{Schubnell}}, \citenamefont {{Seo}}, \citenamefont {{Sprayberry}}, \citenamefont {{Tarl{\'e}}}, \citenamefont {{Vargas-Maga{\~n}a}}, \citenamefont {{Weaver}},\ and\ \citenamefont {{Zhou}}}]{FiberSystem.Poppett.2024}%
  \BibitemOpen
  \bibfield  {author} {\bibinfo {author} {\bibfnamefont {C.}~\bibnamefont {{Poppett}}}, \bibinfo {author} {\bibfnamefont {L.}~\bibnamefont {{Tyas}}}, \bibinfo {author} {\bibfnamefont {J.}~\bibnamefont {{Aguilar}}}, \bibinfo {author} {\bibfnamefont {C.}~\bibnamefont {{Bebek}}}, \bibinfo {author} {\bibfnamefont {D.}~\bibnamefont {{Bramall}}}, \bibinfo {author} {\bibfnamefont {T.}~\bibnamefont {{Claybaugh}}}, \bibinfo {author} {\bibfnamefont {J.}~\bibnamefont {{Edelstein}}}, \bibinfo {author} {\bibfnamefont {P.}~\bibnamefont {{Fagrelius}}}, \bibinfo {author} {\bibfnamefont {H.}~\bibnamefont {{Heetderks}}}, \bibinfo {author} {\bibfnamefont {P.}~\bibnamefont {{Jelinsky}}}, \bibinfo {author} {\bibfnamefont {S.}~\bibnamefont {{Jelinsky}}}, \bibinfo {author} {\bibfnamefont {R.}~\bibnamefont {{Lafever}}}, \bibinfo {author} {\bibfnamefont {A.}~\bibnamefont {{Lambert}}}, \bibinfo {author} {\bibfnamefont {M.}~\bibnamefont {{Lampton}}}, \bibinfo {author} {\bibfnamefont {M.~E.}\ \bibnamefont {{Levi}}}, \bibnamefont
  {and~others},\ }\href {https://doi.org/10.3847/1538-3881/ad76a4} {\bibfield  {journal} {\bibinfo  {journal} {\aj}\ }\textbf {\bibinfo {volume} {168}},\ \bibinfo {eid} {245} (\bibinfo {year} {2024})}\BibitemShut {NoStop}%
\bibitem [{\citenamefont {{Miller}}\ \emph {et~al.}(2024)\citenamefont {{Miller}}, \citenamefont {{Doel}}, \citenamefont {{Gutierrez}}, \citenamefont {{Besuner}}, \citenamefont {{Brooks}}, \citenamefont {{Gallo}}, \citenamefont {{Heetderks}}, \citenamefont {{Jelinsky}}, \citenamefont {{Kent}}, \citenamefont {{Lampton}}, \citenamefont {{Levi}}, \citenamefont {{Liang}}, \citenamefont {{Meisner}}, \citenamefont {{Sholl}}, \citenamefont {{Silber}}, \citenamefont {{Sprayberry}}, \citenamefont {{Aguilar}}, \citenamefont {{de la Macorra}}, \citenamefont {{Eisenstein}}, \citenamefont {{Fanning}}, \citenamefont {{Font-Ribera}}, \citenamefont {{Gazta{\~n}aga}}, \citenamefont {{Gontcho A Gontcho}}, \citenamefont {{Honscheid}}, \citenamefont {{Jimenez}}, \citenamefont {{Joyce}}, \citenamefont {{Kehoe}}, \citenamefont {{Kisner}}, \citenamefont {{Kremin}}, \citenamefont {{Landriau}}, \citenamefont {{Le Guillou}}, \citenamefont {{Magneville}}, \citenamefont {{Martini}}, \citenamefont {{Miquel}}, \citenamefont {{Moustakas}},
  \citenamefont {{Nie}}, \citenamefont {{Percival}}, \citenamefont {{Poppett}}, \citenamefont {{Prada}}, \citenamefont {{Rossi}}, \citenamefont {{Schlegel}}, \citenamefont {{Schubnell}}, \citenamefont {{Seo}}, \citenamefont {{Sharples}}, \citenamefont {{Tarl{\'e}}}, \citenamefont {{Vargas-Maga{\~n}a}}, \citenamefont {{Zhou}},\ and\ \citenamefont {{the DESI Collaboration}}}]{Corrector.Miller.2023}%
  \BibitemOpen
  \bibfield  {author} {\bibinfo {author} {\bibfnamefont {T.~N.}\ \bibnamefont {{Miller}}}, \bibinfo {author} {\bibfnamefont {P.}~\bibnamefont {{Doel}}}, \bibinfo {author} {\bibfnamefont {G.}~\bibnamefont {{Gutierrez}}}, \bibinfo {author} {\bibfnamefont {R.}~\bibnamefont {{Besuner}}}, \bibinfo {author} {\bibfnamefont {D.}~\bibnamefont {{Brooks}}}, \bibinfo {author} {\bibfnamefont {G.}~\bibnamefont {{Gallo}}}, \bibinfo {author} {\bibfnamefont {H.}~\bibnamefont {{Heetderks}}}, \bibinfo {author} {\bibfnamefont {P.}~\bibnamefont {{Jelinsky}}}, \bibinfo {author} {\bibfnamefont {S.~M.}\ \bibnamefont {{Kent}}}, \bibinfo {author} {\bibfnamefont {M.}~\bibnamefont {{Lampton}}}, \bibinfo {author} {\bibfnamefont {M.~E.}\ \bibnamefont {{Levi}}}, \bibinfo {author} {\bibfnamefont {M.}~\bibnamefont {{Liang}}}, \bibinfo {author} {\bibfnamefont {A.}~\bibnamefont {{Meisner}}}, \bibinfo {author} {\bibfnamefont {M.~J.}\ \bibnamefont {{Sholl}}}, \bibinfo {author} {\bibfnamefont {J.~H.}\ \bibnamefont {{Silber}}}, \bibnamefont
  {and~others},\ }\href {https://doi.org/10.3847/1538-3881/ad45fe} {\bibfield  {journal} {\bibinfo  {journal} {\aj}\ }\textbf {\bibinfo {volume} {168}},\ \bibinfo {eid} {95} (\bibinfo {year} {2024})},\ \Eprint {https://arxiv.org/abs/2306.06310} {arXiv:2306.06310 [astro-ph.IM]} \BibitemShut {NoStop}%
\bibitem [{\citenamefont {{Guy}}\ \emph {et~al.}(2023)\citenamefont {{Guy}}, \citenamefont {{Bailey}}, \citenamefont {{Kremin}}, \citenamefont {{Alam}}, \citenamefont {{Alexander}}, \citenamefont {{Allende Prieto}}, \citenamefont {{BenZvi}}, \citenamefont {{Bolton}}, \citenamefont {{Brooks}}, \citenamefont {{Chaussidon}}, \citenamefont {{Cooper}}, \citenamefont {{Dawson}}, \citenamefont {{de la Macorra}}, \citenamefont {{Dey}}, \citenamefont {{Dey}}, \citenamefont {{Dhungana}}, \citenamefont {{Eisenstein}}, \citenamefont {{Font-Ribera}}, \citenamefont {{Forero-Romero}}, \citenamefont {{Gazta{\~n}aga}}, \citenamefont {{Gontcho A Gontcho}}, \citenamefont {{Green}}, \citenamefont {{Honscheid}}, \citenamefont {{Ishak}}, \citenamefont {{Kehoe}}, \citenamefont {{Kirkby}}, \citenamefont {{Kisner}}, \citenamefont {{Koposov}}, \citenamefont {{Lan}}, \citenamefont {{Landriau}}, \citenamefont {{Le Guillou}}, \citenamefont {{Levi}}, \citenamefont {{Magneville}}, \citenamefont {{Manser}}, \citenamefont {{Martini}},
  \citenamefont {{Meisner}}, \citenamefont {{Miquel}}, \citenamefont {{Moustakas}}, \citenamefont {{Myers}}, \citenamefont {{Newman}}, \citenamefont {{Nie}}, \citenamefont {{Palanque-Delabrouille}}, \citenamefont {{Percival}}, \citenamefont {{Poppett}}, \citenamefont {{Prada}}, \citenamefont {{Raichoor}}, \citenamefont {{Ravoux}}, \citenamefont {{Ross}}, \citenamefont {{Schlafly}}, \citenamefont {{Schlegel}}, \citenamefont {{Schubnell}}, \citenamefont {{Sharples}}, \citenamefont {{Tarl{\'e}}}, \citenamefont {{Weaver}}, \citenamefont {{Y{\'e}che}}, \citenamefont {{Zhou}}, \citenamefont {{Zhou}},\ and\ \citenamefont {{Zou}}}]{Spectro.Pipeline.Guy.2023}%
  \BibitemOpen
  \bibfield  {author} {\bibinfo {author} {\bibfnamefont {J.}~\bibnamefont {{Guy}}}, \bibinfo {author} {\bibfnamefont {S.}~\bibnamefont {{Bailey}}}, \bibinfo {author} {\bibfnamefont {A.}~\bibnamefont {{Kremin}}}, \bibinfo {author} {\bibfnamefont {S.}~\bibnamefont {{Alam}}}, \bibinfo {author} {\bibfnamefont {D.~M.}\ \bibnamefont {{Alexander}}}, \bibinfo {author} {\bibfnamefont {C.}~\bibnamefont {{Allende Prieto}}}, \bibinfo {author} {\bibfnamefont {S.}~\bibnamefont {{BenZvi}}}, \bibinfo {author} {\bibfnamefont {A.~S.}\ \bibnamefont {{Bolton}}}, \bibinfo {author} {\bibfnamefont {D.}~\bibnamefont {{Brooks}}}, \bibinfo {author} {\bibfnamefont {E.}~\bibnamefont {{Chaussidon}}}, \bibinfo {author} {\bibfnamefont {A.~P.}\ \bibnamefont {{Cooper}}}, \bibinfo {author} {\bibfnamefont {K.}~\bibnamefont {{Dawson}}}, \bibinfo {author} {\bibfnamefont {A.}~\bibnamefont {{de la Macorra}}}, \bibinfo {author} {\bibfnamefont {A.}~\bibnamefont {{Dey}}}, \bibinfo {author} {\bibfnamefont {B.}~\bibnamefont {{Dey}}}, \bibnamefont
  {and~others},\ }\href {https://doi.org/10.3847/1538-3881/acb212} {\bibfield  {journal} {\bibinfo  {journal} {\aj}\ }\textbf {\bibinfo {volume} {165}},\ \bibinfo {eid} {144} (\bibinfo {year} {2023})},\ \Eprint {https://arxiv.org/abs/2209.14482} {arXiv:2209.14482 [astro-ph.IM]} \BibitemShut {NoStop}%
\bibitem [{\citenamefont {{Schlafly}}\ \emph {et~al.}(2023)\citenamefont {{Schlafly}}, \citenamefont {{Kirkby}}, \citenamefont {{Schlegel}}, \citenamefont {{Myers}}, \citenamefont {{Raichoor}}, \citenamefont {{Dawson}}, \citenamefont {{Aguilar}}, \citenamefont {{Allende Prieto}}, \citenamefont {{Bailey}}, \citenamefont {{BenZvi}}, \citenamefont {{Bermejo-Climent}}, \citenamefont {{Brooks}}, \citenamefont {{de la Macorra}}, \citenamefont {{Dey}}, \citenamefont {{Doel}}, \citenamefont {{Fanning}}, \citenamefont {{Font-Ribera}}, \citenamefont {{Forero-Romero}}, \citenamefont {{Garc{\'\i}a-Bellido}}, \citenamefont {{Gontcho A Gontcho}}, \citenamefont {{Guy}}, \citenamefont {{Hahn}}, \citenamefont {{Honscheid}}, \citenamefont {{Ishak}}, \citenamefont {{Juneau}}, \citenamefont {{Kehoe}}, \citenamefont {{Kisner}}, \citenamefont {{Kremin}}, \citenamefont {{Landriau}}, \citenamefont {{Lang}}, \citenamefont {{Lasker}}, \citenamefont {{Levi}}, \citenamefont {{Magneville}}, \citenamefont {{Manser}}, \citenamefont
  {{Martini}}, \citenamefont {{Meisner}}, \citenamefont {{Miquel}}, \citenamefont {{Moustakas}}, \citenamefont {{Newman}}, \citenamefont {{Nie}}, \citenamefont {{Palanque-Delabrouille}}, \citenamefont {{Percival}}, \citenamefont {{Poppett}}, \citenamefont {{Rockosi}}, \citenamefont {{Ross}}, \citenamefont {{Rossi}}, \citenamefont {{Tarl{\'e}}}, \citenamefont {{Weaver}}, \citenamefont {{Y{\`e}che}}, \citenamefont {{Zhou}},\ and\ \citenamefont {{DESI Collaboration}}}]{SurveyOps.Schlafly.2023}%
  \BibitemOpen
  \bibfield  {author} {\bibinfo {author} {\bibfnamefont {E.~F.}\ \bibnamefont {{Schlafly}}}, \bibinfo {author} {\bibfnamefont {D.}~\bibnamefont {{Kirkby}}}, \bibinfo {author} {\bibfnamefont {D.~J.}\ \bibnamefont {{Schlegel}}}, \bibinfo {author} {\bibfnamefont {A.~D.}\ \bibnamefont {{Myers}}}, \bibinfo {author} {\bibfnamefont {A.}~\bibnamefont {{Raichoor}}}, \bibinfo {author} {\bibfnamefont {K.}~\bibnamefont {{Dawson}}}, \bibinfo {author} {\bibfnamefont {J.}~\bibnamefont {{Aguilar}}}, \bibinfo {author} {\bibfnamefont {C.}~\bibnamefont {{Allende Prieto}}}, \bibinfo {author} {\bibfnamefont {S.}~\bibnamefont {{Bailey}}}, \bibinfo {author} {\bibfnamefont {S.}~\bibnamefont {{BenZvi}}}, \bibinfo {author} {\bibfnamefont {J.}~\bibnamefont {{Bermejo-Climent}}}, \bibinfo {author} {\bibfnamefont {D.}~\bibnamefont {{Brooks}}}, \bibinfo {author} {\bibfnamefont {A.}~\bibnamefont {{de la Macorra}}}, \bibinfo {author} {\bibfnamefont {A.}~\bibnamefont {{Dey}}}, \bibinfo {author} {\bibfnamefont {P.}~\bibnamefont {{Doel}}},
  \bibnamefont {and~others},\ }\href {https://doi.org/10.3847/1538-3881/ad0832} {\bibfield  {journal} {\bibinfo  {journal} {\aj}\ }\textbf {\bibinfo {volume} {166}},\ \bibinfo {eid} {259} (\bibinfo {year} {2023})},\ \Eprint {https://arxiv.org/abs/2306.06309} {arXiv:2306.06309 [astro-ph.CO]} \BibitemShut {NoStop}%
\bibitem [{\citenamefont {{DESI Collaboration}}\ \emph {et~al.}(2024{\natexlab{a}})\citenamefont {{DESI Collaboration}}, \citenamefont {{Adame}}, \citenamefont {{Aguilar}}, \citenamefont {{Ahlen}}, \citenamefont {{Alam}}, \citenamefont {{Aldering}}, \citenamefont {{Alexander}}, \citenamefont {{Alfarsy}}, \citenamefont {{Allende Prieto}}, \citenamefont {{Alvarez}}, \citenamefont {{Alves}}, \citenamefont {{Anand}}, \citenamefont {{Andrade-Oliveira}}, \citenamefont {{Armengaud}}, \citenamefont {{Asorey}}, \citenamefont {{Avila}}, \citenamefont {{Aviles}}, \citenamefont {{Bailey}}, \citenamefont {{Balaguera-Antol{\'\i}nez}}, \citenamefont {{Ballester}}, \citenamefont {{Baltay}}, \citenamefont {{Bault}}, \citenamefont {{Bautista}}, \citenamefont {{Behera}}, \citenamefont {{Beltran}}, \citenamefont {{BenZvi}}, \citenamefont {{Beraldo e Silva}}, \citenamefont {{Bermejo-Climent}}, \citenamefont {{Berti}}, \citenamefont {{Besuner}}, \citenamefont {{Beutler}}, \citenamefont {{Bianchi}}, \citenamefont {{Blake}},
  \citenamefont {{Blum}}, \citenamefont {{Bolton}}, \citenamefont {{Brieden}}, \citenamefont {{Brodzeller}}, \citenamefont {{Brooks}}, \citenamefont {{Brown}}, \citenamefont {{Buckley-Geer}}, \citenamefont {{Burtin}}, \citenamefont {{Cabayol-Garcia}}, \citenamefont {{Cai}}, \citenamefont {{Canning}}, \citenamefont {{Cardiel-Sas}}, \citenamefont {{Carnero Rosell}}, \citenamefont {{Castander}}, \citenamefont {{Cervantes-Cota}}, \citenamefont {{Chabanier}}, \citenamefont {{Chaussidon}}, \citenamefont {{Chaves-Montero}}, \citenamefont {{Chen}}, \citenamefont {{Chen}}, \citenamefont {{Chuang}}, \citenamefont {{Claybaugh}}, \citenamefont {{Cole}}, \citenamefont {{Cooper}}, \citenamefont {{Cuceu}}, \citenamefont {{Davis}}, \citenamefont {{Dawson}}, \citenamefont {{de Belsunce}}, \citenamefont {{de la Cruz}}, \citenamefont {{de la Macorra}}, \citenamefont {{de Mattia}}, \citenamefont {{Demina}}, \citenamefont {{Demirbozan}}, \citenamefont {{DeRose}}, \citenamefont {{Dey}}, \citenamefont {{Dey}}, \citenamefont
  {{Dhungana}}, \citenamefont {{Ding}}, \citenamefont {{Ding}}, \citenamefont {{Doel}}, \citenamefont {{Doshi}}, \citenamefont {{Douglass}}, \citenamefont {{Edge}}, \citenamefont {{Eftekharzadeh}}, \citenamefont {{Eisenstein}}, \citenamefont {{Elliott}}, \citenamefont {{Escoffier}}, \citenamefont {{Fagrelius}}, \citenamefont {{Fan}}, \citenamefont {{Fanning}}, \citenamefont {{Fawcett}}, \citenamefont {{Ferraro}}, \citenamefont {{Ereza}}, \citenamefont {{Flaugher}}, \citenamefont {{Font-Ribera}}, \citenamefont {{Forero-S{\'a}nchez}}, \citenamefont {{Forero-Romero}}, \citenamefont {{Frenk}}, \citenamefont {{G{\"a}nsicke}}, \citenamefont {{Garc{\'\i}a}}, \citenamefont {{Garc{\'\i}a-Bellido}}, \citenamefont {{Garcia-Quintero}}, \citenamefont {{Garrison}}, \citenamefont {{Gil-Mar{\'\i}n}}, \citenamefont {{Golden-Marx}}, \citenamefont {{Gontcho A Gontcho}}, \citenamefont {{Gonzalez-Morales}}, \citenamefont {{Gonzalez-Perez}}, \citenamefont {{Gordon}}, \citenamefont {{Graur}}, \citenamefont {{Green}}, \citenamefont
  {{Gruen}}, \citenamefont {{Guy}}, \citenamefont {{Hadzhiyska}}, \citenamefont {{Hahn}}, \citenamefont {{Han}}, \citenamefont {{Hanif}}, \citenamefont {{Herrera-Alcantar}}, \citenamefont {{Honscheid}}, \citenamefont {{Hou}}, \citenamefont {{Howlett}}, \citenamefont {{Huterer}}, \citenamefont {{Ir{\v{s}}i{\v{c}}}}, \citenamefont {{Ishak}}, \citenamefont {{Jana}}, \citenamefont {{Jiang}}, \citenamefont {{Jimenez}}, \citenamefont {{Jing}}, \citenamefont {{Joudaki}}, \citenamefont {{Jullo}}, \citenamefont {{Joyce}}, \citenamefont {{Juneau}}, \citenamefont {{Kizhuprakkat}}, \citenamefont {{Kara{\c{c}}ayl{\i}}}, \citenamefont {{Karim}}, \citenamefont {{Kehoe}}, \citenamefont {{Kent}}, \citenamefont {{Khederlarian}}, \citenamefont {{Kim}}, \citenamefont {{Kirkby}}, \citenamefont {{Kisner}}, \citenamefont {{Kitaura}}, \citenamefont {{Kneib}}, \citenamefont {{Koposov}}, \citenamefont {{Kov{\'a}cs}}, \citenamefont {{Kremin}}, \citenamefont {{Krolewski}}, \citenamefont {{L'Huillier}}, \citenamefont {{Lahav}},
  \citenamefont {{Lambert}}, \citenamefont {{Lamman}}, \citenamefont {{Lan}}, \citenamefont {{Landriau}}, \citenamefont {{Lang}}, \citenamefont {{Lange}}, \citenamefont {{Lasker}}, \citenamefont {{Le Guillou}}, \citenamefont {{Leauthaud}}, \citenamefont {{Levi}}, \citenamefont {{Li}}, \citenamefont {{Linder}}, \citenamefont {{Lyons}}, \citenamefont {{Magneville}}, \citenamefont {{Manera}}, \citenamefont {{Manser}}, \citenamefont {{Margala}}, \citenamefont {{Martini}}, \citenamefont {{McDonald}}, \citenamefont {{Medina}}, \citenamefont {{Medina-Varela}}, \citenamefont {{Meisner}}, \citenamefont {{Mena-Fern{\'a}ndez}}, \citenamefont {{Meneses-Rizo}}, \citenamefont {{Mezcua}}, \citenamefont {{Miquel}}, \citenamefont {{Montero-Camacho}}, \citenamefont {{Moon}}, \citenamefont {{Moore}}, \citenamefont {{Moustakas}}, \citenamefont {{Mueller}}, \citenamefont {{Mundet}}, \citenamefont {{Mu{\~n}oz-Guti{\'e}rrez}}, \citenamefont {{Myers}}, \citenamefont {{Nadathur}}, \citenamefont {{Napolitano}}, \citenamefont
  {{Neveux}}, \citenamefont {{Newman}}, \citenamefont {{Nie}}, \citenamefont {{Niz}}, \citenamefont {{Norberg}}, \citenamefont {{Noriega}}, \citenamefont {{Paillas}}, \citenamefont {{Palanque-Delabrouille}}, \citenamefont {{Palmese}}, \citenamefont {{Zhiwei}}, \citenamefont {{Parkinson}}, \citenamefont {{Penmetsa}}, \citenamefont {{Percival}}, \citenamefont {{P{\'e}rez-Fern{\'a}ndez}}, \citenamefont {{P{\'e}rez-R{\`a}fols}}, \citenamefont {{Pieri}}, \citenamefont {{Poppett}}, \citenamefont {{Porredon}}, \citenamefont {{Prada}}, \citenamefont {{Pucha}}, \citenamefont {{Raichoor}}, \citenamefont {{Ram{\'\i}rez-P{\'e}rez}}, \citenamefont {{Ramirez-Solano}}, \citenamefont {{Rashkovetskyi}}, \citenamefont {{Ravoux}}, \citenamefont {{Rocher}}, \citenamefont {{Rockosi}}, \citenamefont {{Ross}}, \citenamefont {{Rossi}}, \citenamefont {{Ruggeri}}, \citenamefont {{Ruhlmann-Kleider}}, \citenamefont {{Sabiu}}, \citenamefont {{Said}}, \citenamefont {{Saintonge}}, \citenamefont {{Samushia}}, \citenamefont {{Sanchez}},
  \citenamefont {{Saulder}}, \citenamefont {{Schaan}}, \citenamefont {{Schlafly}}, \citenamefont {{Schlegel}}, \citenamefont {{Scholte}}, \citenamefont {{Schubnell}}, \citenamefont {{Seo}}, \citenamefont {{Shafieloo}}, \citenamefont {{Sharples}}, \citenamefont {{Sheu}}, \citenamefont {{Silber}}, \citenamefont {{Sinigaglia}}, \citenamefont {{Siudek}}, \citenamefont {{Slepian}}, \citenamefont {{Smith}}, \citenamefont {{Sprayberry}}, \citenamefont {{Stephey}}, \citenamefont {{Su{\'a}rez-P{\'e}rez}}, \citenamefont {{Sun}}, \citenamefont {{Tan}}, \citenamefont {{Tarl{\'e}}}, \citenamefont {{Tojeiro}}, \citenamefont {{Ure{\~n}a-L{\'o}pez}}, \citenamefont {{Vaisakh}}, \citenamefont {{Valcin}}, \citenamefont {{Valdes}}, \citenamefont {{Valluri}}, \citenamefont {{Vargas-Maga{\~n}a}}, \citenamefont {{Variu}}, \citenamefont {{Verde}}, \citenamefont {{Walther}}, \citenamefont {{Wang}}, \citenamefont {{Wang}}, \citenamefont {{Weaver}}, \citenamefont {{Weaverdyck}}, \citenamefont {{Wechsler}}, \citenamefont {{White}},
  \citenamefont {{Xie}}, \citenamefont {{Yang}}, \citenamefont {{Y{\`e}che}}, \citenamefont {{Yu}}, \citenamefont {{Yuan}}, \citenamefont {{Zhang}}, \citenamefont {{Zhang}}, \citenamefont {{Zhao}}, \citenamefont {{Zheng}}, \citenamefont {{Zhou}}, \citenamefont {{Zhou}}, \citenamefont {{Zou}}, \citenamefont {{Zou}}, \citenamefont {{Zu}},\ and\ \citenamefont {{DESI Collaboration}}}]{DESI2023a.KP1.SV}%
  \BibitemOpen
  \bibfield  {author} {\bibinfo {author} {\bibnamefont {{DESI Collaboration}}}, \bibinfo {author} {\bibfnamefont {A.~G.}\ \bibnamefont {{Adame}}}, \bibinfo {author} {\bibfnamefont {J.}~\bibnamefont {{Aguilar}}}, \bibinfo {author} {\bibfnamefont {S.}~\bibnamefont {{Ahlen}}}, \bibinfo {author} {\bibfnamefont {S.}~\bibnamefont {{Alam}}}, \bibinfo {author} {\bibfnamefont {G.}~\bibnamefont {{Aldering}}}, \bibinfo {author} {\bibfnamefont {D.~M.}\ \bibnamefont {{Alexander}}}, \bibinfo {author} {\bibfnamefont {R.}~\bibnamefont {{Alfarsy}}}, \bibinfo {author} {\bibfnamefont {C.}~\bibnamefont {{Allende Prieto}}}, \bibinfo {author} {\bibfnamefont {M.}~\bibnamefont {{Alvarez}}}, \bibinfo {author} {\bibfnamefont {O.}~\bibnamefont {{Alves}}}, \bibinfo {author} {\bibfnamefont {A.}~\bibnamefont {{Anand}}}, \bibinfo {author} {\bibfnamefont {F.}~\bibnamefont {{Andrade-Oliveira}}}, \bibinfo {author} {\bibfnamefont {E.}~\bibnamefont {{Armengaud}}}, \bibinfo {author} {\bibfnamefont {J.}~\bibnamefont {{Asorey}}}, \bibnamefont
  {and~others},\ }\href {https://doi.org/10.3847/1538-3881/ad0b08} {\bibfield  {journal} {\bibinfo  {journal} {\aj}\ }\textbf {\bibinfo {volume} {167}},\ \bibinfo {eid} {62} (\bibinfo {year} {2024}{\natexlab{a}})},\ \Eprint {https://arxiv.org/abs/2306.06307} {arXiv:2306.06307 [astro-ph.CO]} \BibitemShut {NoStop}%
\bibitem [{\citenamefont {{DESI Collaboration}}\ \emph {et~al.}(2024{\natexlab{b}})\citenamefont {{DESI Collaboration}}, \citenamefont {{Adame}}, \citenamefont {{Aguilar}}, \citenamefont {{Ahlen}}, \citenamefont {{Alam}}, \citenamefont {{Aldering}}, \citenamefont {{Alexander}}, \citenamefont {{Alfarsy}}, \citenamefont {{Allende Prieto}}, \citenamefont {{Alvarez}}, \citenamefont {{Alves}}, \citenamefont {{Anand}}, \citenamefont {{Andrade-Oliveira}}, \citenamefont {{Armengaud}}, \citenamefont {{Asorey}}, \citenamefont {{Avila}}, \citenamefont {{Aviles}}, \citenamefont {{Bailey}}, \citenamefont {{Balaguera-Antol{\'\i}nez}}, \citenamefont {{Ballester}}, \citenamefont {{Baltay}}, \citenamefont {{Bault}}, \citenamefont {{Bautista}}, \citenamefont {{Behera}}, \citenamefont {{Beltran}}, \citenamefont {{BenZvi}}, \citenamefont {{Beraldo e Silva}}, \citenamefont {{Bermejo-Climent}}, \citenamefont {{Berti}}, \citenamefont {{Besuner}}, \citenamefont {{Beutler}}, \citenamefont {{Bianchi}}, \citenamefont {{Blake}},
  \citenamefont {{Blum}}, \citenamefont {{Bolton}}, \citenamefont {{Brieden}}, \citenamefont {{Brodzeller}}, \citenamefont {{Brooks}}, \citenamefont {{Brown}}, \citenamefont {{Buckley-Geer}}, \citenamefont {{Burtin}}, \citenamefont {{Cabayol-Garcia}}, \citenamefont {{Cai}}, \citenamefont {{Canning}}, \citenamefont {{Cardiel-Sas}}, \citenamefont {{Carnero Rosell}}, \citenamefont {{Castander}}, \citenamefont {{Cervantes-Cota}}, \citenamefont {{Chabanier}}, \citenamefont {{Chaussidon}}, \citenamefont {{Chaves-Montero}}, \citenamefont {{Chen}}, \citenamefont {{Chen}}, \citenamefont {{Chuang}}, \citenamefont {{Claybaugh}}, \citenamefont {{Cole}}, \citenamefont {{Cooper}}, \citenamefont {{Cuceu}}, \citenamefont {{Davis}}, \citenamefont {{Dawson}}, \citenamefont {{de Belsunce}}, \citenamefont {{de la Cruz}}, \citenamefont {{de la Macorra}}, \citenamefont {{Della Costa}}, \citenamefont {{de Mattia}}, \citenamefont {{Demina}}, \citenamefont {{Demirbozan}}, \citenamefont {{DeRose}}, \citenamefont {{Dey}}, \citenamefont
  {{Dey}}, \citenamefont {{Dhungana}}, \citenamefont {{Ding}}, \citenamefont {{Ding}}, \citenamefont {{Doel}}, \citenamefont {{Doshi}}, \citenamefont {{Douglass}}, \citenamefont {{Edge}}, \citenamefont {{Eftekharzadeh}}, \citenamefont {{Eisenstein}}, \citenamefont {{Elliott}}, \citenamefont {{Ereza}}, \citenamefont {{Escoffier}}, \citenamefont {{Fagrelius}}, \citenamefont {{Fan}}, \citenamefont {{Fanning}}, \citenamefont {{Fawcett}}, \citenamefont {{Ferraro}}, \citenamefont {{Flaugher}}, \citenamefont {{Font-Ribera}}, \citenamefont {{Forero-Romero}}, \citenamefont {{Forero-S{\'a}nchez}}, \citenamefont {{Frenk}}, \citenamefont {{G{\"a}nsicke}}, \citenamefont {{Garc{\'\i}a}}, \citenamefont {{Garc{\'\i}a-Bellido}}, \citenamefont {{Garcia-Quintero}}, \citenamefont {{Garrison}}, \citenamefont {{Gil-Mar{\'\i}n}}, \citenamefont {{Golden-Marx}}, \citenamefont {{Gontcho A Gontcho}}, \citenamefont {{Gonzalez-Morales}}, \citenamefont {{Gonzalez-Perez}}, \citenamefont {{Gordon}}, \citenamefont {{Graur}}, \citenamefont
  {{Green}}, \citenamefont {{Gruen}}, \citenamefont {{Guy}}, \citenamefont {{Hadzhiyska}}, \citenamefont {{Hahn}}, \citenamefont {{Han}}, \citenamefont {{Hanif}}, \citenamefont {{Herrera-Alcantar}}, \citenamefont {{Honscheid}}, \citenamefont {{Hou}}, \citenamefont {{Howlett}}, \citenamefont {{Huterer}}, \citenamefont {{Ir{\v{s}}i{\v{c}}}}, \citenamefont {{Ishak}}, \citenamefont {{Jacques}}, \citenamefont {{Jana}}, \citenamefont {{Jiang}}, \citenamefont {{Jimenez}}, \citenamefont {{Jing}}, \citenamefont {{Joudaki}}, \citenamefont {{Joyce}}, \citenamefont {{Jullo}}, \citenamefont {{Juneau}}, \citenamefont {{Kara{\c{c}}ayl{\i}}}, \citenamefont {{Karim}}, \citenamefont {{Kehoe}}, \citenamefont {{Kent}}, \citenamefont {{Khederlarian}}, \citenamefont {{Kim}}, \citenamefont {{Kirkby}}, \citenamefont {{Kisner}}, \citenamefont {{Kitaura}}, \citenamefont {{Kizhuprakkat}}, \citenamefont {{Kneib}}, \citenamefont {{Koposov}}, \citenamefont {{Kov{\'a}cs}}, \citenamefont {{Kremin}}, \citenamefont {{Krolewski}},
  \citenamefont {{L'Huillier}}, \citenamefont {{Lahav}}, \citenamefont {{Lambert}}, \citenamefont {{Lamman}}, \citenamefont {{Lan}}, \citenamefont {{Landriau}}, \citenamefont {{Lang}}, \citenamefont {{Lange}}, \citenamefont {{Lasker}}, \citenamefont {{Leauthaud}}, \citenamefont {{Le Guillou}}, \citenamefont {{Levi}}, \citenamefont {{Li}}, \citenamefont {{Linder}}, \citenamefont {{Lyons}}, \citenamefont {{Magneville}}, \citenamefont {{Manera}}, \citenamefont {{Manser}}, \citenamefont {{Margala}}, \citenamefont {{Martini}}, \citenamefont {{McDonald}}, \citenamefont {{Medina}}, \citenamefont {{Medina-Varela}}, \citenamefont {{Meisner}}, \citenamefont {{Mena-Fern{\'a}ndez}}, \citenamefont {{Meneses-Rizo}}, \citenamefont {{Mezcua}}, \citenamefont {{Miquel}}, \citenamefont {{Montero-Camacho}}, \citenamefont {{Moon}}, \citenamefont {{Moore}}, \citenamefont {{Moustakas}}, \citenamefont {{Mueller}}, \citenamefont {{Mundet}}, \citenamefont {{Mu{\~n}oz-Guti{\'e}rrez}}, \citenamefont {{Myers}}, \citenamefont
  {{Nadathur}}, \citenamefont {{Napolitano}}, \citenamefont {{Neveux}}, \citenamefont {{Newman}}, \citenamefont {{Nie}}, \citenamefont {{Nikutta}}, \citenamefont {{Niz}}, \citenamefont {{Norberg}}, \citenamefont {{Noriega}}, \citenamefont {{Paillas}}, \citenamefont {{Palanque-Delabrouille}}, \citenamefont {{Palmese}}, \citenamefont {{Pan}}, \citenamefont {{Parkinson}}, \citenamefont {{Penmetsa}}, \citenamefont {{Percival}}, \citenamefont {{P{\'e}rez-Fern{\'a}ndez}}, \citenamefont {{P{\'e}rez-R{\`a}fols}}, \citenamefont {{Pieri}}, \citenamefont {{Poppett}}, \citenamefont {{Porredon}}, \citenamefont {{Pothier}}, \citenamefont {{Prada}}, \citenamefont {{Pucha}}, \citenamefont {{Raichoor}}, \citenamefont {{Ram{\'\i}rez-P{\'e}rez}}, \citenamefont {{Ramirez-Solano}}, \citenamefont {{Rashkovetskyi}}, \citenamefont {{Ravoux}}, \citenamefont {{Rocher}}, \citenamefont {{Rockosi}}, \citenamefont {{Ross}}, \citenamefont {{Rossi}}, \citenamefont {{Ruggeri}}, \citenamefont {{Ruhlmann-Kleider}}, \citenamefont {{Sabiu}},
  \citenamefont {{Said}}, \citenamefont {{Saintonge}}, \citenamefont {{Samushia}}, \citenamefont {{Sanchez}}, \citenamefont {{Saulder}}, \citenamefont {{Schaan}}, \citenamefont {{Schlafly}}, \citenamefont {{Schlegel}}, \citenamefont {{Scholte}}, \citenamefont {{Schubnell}}, \citenamefont {{Seo}}, \citenamefont {{Shafieloo}}, \citenamefont {{Sharples}}, \citenamefont {{Sheu}}, \citenamefont {{Silber}}, \citenamefont {{Sinigaglia}}, \citenamefont {{Siudek}}, \citenamefont {{Slepian}}, \citenamefont {{Smith}}, \citenamefont {{Soumagnac}}, \citenamefont {{Sprayberry}}, \citenamefont {{Stephey}}, \citenamefont {{Su{\'a}rez-P{\'e}rez}}, \citenamefont {{Sun}}, \citenamefont {{Tan}}, \citenamefont {{Tarl{\'e}}}, \citenamefont {{Tojeiro}}, \citenamefont {{Ure{\~n}a-L{\'o}pez}}, \citenamefont {{Vaisakh}}, \citenamefont {{Valcin}}, \citenamefont {{Valdes}}, \citenamefont {{Valluri}}, \citenamefont {{Vargas-Maga{\~n}a}}, \citenamefont {{Variu}}, \citenamefont {{Verde}}, \citenamefont {{Walther}}, \citenamefont {{Wang}},
  \citenamefont {{Wang}}, \citenamefont {{Weaver}}, \citenamefont {{Weaverdyck}}, \citenamefont {{Wechsler}}, \citenamefont {{White}}, \citenamefont {{Xie}}, \citenamefont {{Yang}}, \citenamefont {{Y{\`e}che}}, \citenamefont {{Yu}}, \citenamefont {{Yuan}}, \citenamefont {{Zhang}}, \citenamefont {{Zhang}}, \citenamefont {{Zhao}}, \citenamefont {{Zheng}}, \citenamefont {{Zhou}}, \citenamefont {{Zhou}}, \citenamefont {{Zou}}, \citenamefont {{Zou}},\ and\ \citenamefont {{Zu}}}]{DESI2023b.KP1.EDR}%
  \BibitemOpen
  \bibfield  {author} {\bibinfo {author} {\bibnamefont {{DESI Collaboration}}}, \bibinfo {author} {\bibfnamefont {A.~G.}\ \bibnamefont {{Adame}}}, \bibinfo {author} {\bibfnamefont {J.}~\bibnamefont {{Aguilar}}}, \bibinfo {author} {\bibfnamefont {S.}~\bibnamefont {{Ahlen}}}, \bibinfo {author} {\bibfnamefont {S.}~\bibnamefont {{Alam}}}, \bibinfo {author} {\bibfnamefont {G.}~\bibnamefont {{Aldering}}}, \bibinfo {author} {\bibfnamefont {D.~M.}\ \bibnamefont {{Alexander}}}, \bibinfo {author} {\bibfnamefont {R.}~\bibnamefont {{Alfarsy}}}, \bibinfo {author} {\bibfnamefont {C.}~\bibnamefont {{Allende Prieto}}}, \bibinfo {author} {\bibfnamefont {M.}~\bibnamefont {{Alvarez}}}, \bibinfo {author} {\bibfnamefont {O.}~\bibnamefont {{Alves}}}, \bibinfo {author} {\bibfnamefont {A.}~\bibnamefont {{Anand}}}, \bibinfo {author} {\bibfnamefont {F.}~\bibnamefont {{Andrade-Oliveira}}}, \bibinfo {author} {\bibfnamefont {E.}~\bibnamefont {{Armengaud}}}, \bibinfo {author} {\bibfnamefont {J.}~\bibnamefont {{Asorey}}}, \bibnamefont
  {and~others},\ }\href {https://doi.org/10.3847/1538-3881/ad3217} {\bibfield  {journal} {\bibinfo  {journal} {\aj}\ }\textbf {\bibinfo {volume} {168}},\ \bibinfo {eid} {58} (\bibinfo {year} {2024}{\natexlab{b}})},\ \Eprint {https://arxiv.org/abs/2306.06308} {arXiv:2306.06308 [astro-ph.CO]} \BibitemShut {NoStop}%
\bibitem [{\citenamefont {{DESI Collaboration}}\ \emph {et~al.}(2025{\natexlab{a}})\citenamefont {{DESI Collaboration}}, \citenamefont {{Abdul-Karim}}, \citenamefont {{Adame}}, \citenamefont {{Aguado}}, \citenamefont {{Aguilar}}, \citenamefont {{Ahlen}}, \citenamefont {{Alam}}, \citenamefont {{Aldering}}, \citenamefont {{Alexander}}, \citenamefont {{Alfarsy}}, \citenamefont {{Allen}}, \citenamefont {{Allende Prieto}}, \citenamefont {{Alves}}, \citenamefont {{Anand}}, \citenamefont {{Andrade}}, \citenamefont {{Armengaud}}, \citenamefont {{Avila}}, \citenamefont {{Aviles}}, \citenamefont {{Awan}}, \citenamefont {{Bailey}}, \citenamefont {{Baleato Lizancos}}, \citenamefont {{Ballester}}, \citenamefont {{Bault}}, \citenamefont {{Bautista}}, \citenamefont {{BenZvi}}, \citenamefont {{Beraldo e Silva}}, \citenamefont {{Bermejo-Climent}}, \citenamefont {{Beutler}}, \citenamefont {{Bianchi}}, \citenamefont {{Blake}}, \citenamefont {{Blum}}, \citenamefont {{Bolton}}, \citenamefont {{Bonici}}, \citenamefont {{Brieden}},
  \citenamefont {{Brodzeller}}, \citenamefont {{Brooks}}, \citenamefont {{Buckley-Geer}}, \citenamefont {{Burtin}}, \citenamefont {{Canning}}, \citenamefont {{Carnero Rosell}}, \citenamefont {{Carr}}, \citenamefont {{Carrilho}}, \citenamefont {{Casas}}, \citenamefont {{Castander}}, \citenamefont {{Cereskaite}}, \citenamefont {{Cervantes-Cota}}, \citenamefont {{Chaussidon}}, \citenamefont {{Chaves-Montero}}, \citenamefont {{Chen}}, \citenamefont {{Chen}}, \citenamefont {{Claybaugh}}, \citenamefont {{Cole}}, \citenamefont {{Cooper}}, \citenamefont {{Cousinou}}, \citenamefont {{Cuceu}}, \citenamefont {{Davis}}, \citenamefont {{Dawson}}, \citenamefont {{de Belsunce}}, \citenamefont {{de la Cruz}}, \citenamefont {{de la Macorra}}, \citenamefont {{de Mattia}}, \citenamefont {{Deiosso}}, \citenamefont {{Della Costa}}, \citenamefont {{Demina}}, \citenamefont {{Demirbozan}}, \citenamefont {{DeRose}}, \citenamefont {{Dey}}, \citenamefont {{Dey}}, \citenamefont {{Ding}}, \citenamefont {{Ding}}, \citenamefont {{Doel}},
  \citenamefont {{Douglass}}, \citenamefont {{Dowicz}}, \citenamefont {{Ebina}}, \citenamefont {{Edelstein}}, \citenamefont {{Eisenstein}}, \citenamefont {{Elbers}}, \citenamefont {{Emas}}, \citenamefont {{Escoffier}}, \citenamefont {{Fagrelius}}, \citenamefont {{Fan}}, \citenamefont {{Fanning}}, \citenamefont {{Fawcett}}, \citenamefont {{Fern{\'a}ndez-Garc{\'\i}a}}, \citenamefont {{Ferraro}}, \citenamefont {{Findlay}}, \citenamefont {{Font-Ribera}}, \citenamefont {{Forero-Romero}}, \citenamefont {{Forero-S{\'a}nchez}}, \citenamefont {{Frenk}}, \citenamefont {{G{\"a}nsicke}}, \citenamefont {{Galbany}}, \citenamefont {{Garc{\'\i}a-Bellido}}, \citenamefont {{Garcia-Quintero}}, \citenamefont {{Garrison}}, \citenamefont {{Gazta{\~n}aga}}, \citenamefont {{Gil-Mar{\'\i}n}}, \citenamefont {{Gnedin}}, \citenamefont {{Gontcho}}, \citenamefont {{Gonzalez-Morales}}, \citenamefont {{Gonzalez-Perez}}, \citenamefont {{Gordon}}, \citenamefont {{Graur}}, \citenamefont {{Green}}, \citenamefont {{Gruen}}, \citenamefont
  {{Gsponer}}, \citenamefont {{Guandalin}}, \citenamefont {{Gutierrez}}, \citenamefont {{Guy}}, \citenamefont {{Hahn}}, \citenamefont {{Han}}, \citenamefont {{Han}}, \citenamefont {{He}}, \citenamefont {{Herrera-Alcantar}}, \citenamefont {{Honscheid}}, \citenamefont {{Hou}}, \citenamefont {{Howlett}}, \citenamefont {{Huterer}}, \citenamefont {{Ir{\v{s}}i{\v{c}}}}, \citenamefont {{Ishak}}, \citenamefont {{Jacques}}, \citenamefont {{Jimenez}}, \citenamefont {{Jing}}, \citenamefont {{Joachimi}}, \citenamefont {{Joudaki}}, \citenamefont {{Joyce}}, \citenamefont {{Jullo}}, \citenamefont {{Juneau}}, \citenamefont {{Kara{\c{c}}ayl{\i}}}, \citenamefont {{Karim}}, \citenamefont {{Kehoe}}, \citenamefont {{Kent}}, \citenamefont {{Khederlarian}}, \citenamefont {{Kirkby}}, \citenamefont {{Kisner}}, \citenamefont {{Kitaura}}, \citenamefont {{Kizhuprakkat}}, \citenamefont {{Kong}}, \citenamefont {{Koposov}}, \citenamefont {{Kremin}}, \citenamefont {{Krolewski}}, \citenamefont {{Lahav}}, \citenamefont {{Lai}}, \citenamefont
  {{Lamman}}, \citenamefont {{Lan}}, \citenamefont {{Landriau}}, \citenamefont {{Lang}}, \citenamefont {{Lange}}, \citenamefont {{Lasker}}, \citenamefont {{Le Goff}}, \citenamefont {{Le Guillou}}, \citenamefont {{Leauthaud}}, \citenamefont {{Levi}}, \citenamefont {{Li}}, \citenamefont {{Li}}, \citenamefont {{Lodha}}, \citenamefont {{Lokken}}, \citenamefont {{Luo}}, \citenamefont {{Magneville}}, \citenamefont {{Manera}}, \citenamefont {{Manser}}, \citenamefont {{Margala}}, \citenamefont {{Martini}}, \citenamefont {{Maus}}, \citenamefont {{McCullough}}, \citenamefont {{McDonald}}, \citenamefont {{Medina}}, \citenamefont {{Medina-Varela}}, \citenamefont {{Meisner}}, \citenamefont {{Mena-Fern{\'a}ndez}}, \citenamefont {{Menegas}}, \citenamefont {{Mezcua}}, \citenamefont {{Miquel}}, \citenamefont {{Montero-Camacho}}, \citenamefont {{Moon}}, \citenamefont {{Moustakas}}, \citenamefont {{Mu{\~n}oz-Guti{\'e}rrez}}, \citenamefont {{Mu{\~n}oz-Santos}}, \citenamefont {{Myers}}, \citenamefont {{Myles}}, \citenamefont
  {{Nadathur}}, \citenamefont {{Najita}}, \citenamefont {{Napolitano}}, \citenamefont {{Newman}}, \citenamefont {{Nikakhtar}}, \citenamefont {{Nikutta}}, \citenamefont {{Niz}}, \citenamefont {{Noriega}}, \citenamefont {{Padmanabhan}}, \citenamefont {{Paillas}}, \citenamefont {{Palanque-Delabrouille}}, \citenamefont {{Palmese}}, \citenamefont {{Pan}}, \citenamefont {{Pan}}, \citenamefont {{Parkinson}}, \citenamefont {{Peacock}}, \citenamefont {{Percival}}, \citenamefont {{P{\'e}rez-Fern{\'a}ndez}}, \citenamefont {{P{\'e}rez-R{\`a}fols}},\ and\ \citenamefont {{Peterson}}}]{DESI2024.I.DR1}%
  \BibitemOpen
  \bibfield  {author} {\bibinfo {author} {\bibnamefont {{DESI Collaboration}}}, \bibinfo {author} {\bibfnamefont {M.}~\bibnamefont {{Abdul-Karim}}}, \bibinfo {author} {\bibfnamefont {A.~G.}\ \bibnamefont {{Adame}}}, \bibinfo {author} {\bibfnamefont {D.}~\bibnamefont {{Aguado}}}, \bibinfo {author} {\bibfnamefont {J.}~\bibnamefont {{Aguilar}}}, \bibinfo {author} {\bibfnamefont {S.}~\bibnamefont {{Ahlen}}}, \bibinfo {author} {\bibfnamefont {S.}~\bibnamefont {{Alam}}}, \bibinfo {author} {\bibfnamefont {G.}~\bibnamefont {{Aldering}}}, \bibinfo {author} {\bibfnamefont {D.~M.}\ \bibnamefont {{Alexander}}}, \bibinfo {author} {\bibfnamefont {R.}~\bibnamefont {{Alfarsy}}}, \bibinfo {author} {\bibfnamefont {L.}~\bibnamefont {{Allen}}}, \bibinfo {author} {\bibfnamefont {C.}~\bibnamefont {{Allende Prieto}}}, \bibinfo {author} {\bibfnamefont {O.}~\bibnamefont {{Alves}}}, \bibinfo {author} {\bibfnamefont {A.}~\bibnamefont {{Anand}}}, \bibinfo {author} {\bibfnamefont {U.}~\bibnamefont {{Andrade}}}, \bibnamefont
  {and~others},\ }\href {https://doi.org/10.48550/arXiv.2503.14745} {\bibfield  {journal} {\bibinfo  {journal} {arXiv e-prints}\ ,\ \bibinfo {eid} {arXiv:2503.14745}} (\bibinfo {year} {2025}{\natexlab{a}})},\ \Eprint {https://arxiv.org/abs/2503.14745} {arXiv:2503.14745 [astro-ph.CO]} \BibitemShut {NoStop}%
\bibitem [{\citenamefont {{Andrade}}\ \emph {et~al.}(2025)\citenamefont {{Andrade}}, \citenamefont {{Paillas}}, \citenamefont {{Mena-Fern{\'a}ndez}}, \citenamefont {{Li}}, \citenamefont {{Ross}}, \citenamefont {{Nadathur}}, \citenamefont {{Rashkovetskyi}}, \citenamefont {{P{\'e}rez-Fern{\'a}ndez}}, \citenamefont {{Seo}}, \citenamefont {{Sanders}}, \citenamefont {{Alves}}, \citenamefont {{Chen}}, \citenamefont {{Deiosso}}, \citenamefont {{Abdul-Karim}}, \citenamefont {{Ahlen}}, \citenamefont {{Armengaud}}, \citenamefont {{Aviles}}, \citenamefont {{Bianchi}}, \citenamefont {{Brieden}}, \citenamefont {{Brodzeller}}, \citenamefont {{Brooks}}, \citenamefont {{Burtin}}, \citenamefont {{Calderon}}, \citenamefont {{Canning}}, \citenamefont {{Carnero Rosell}}, \citenamefont {{Casas}}, \citenamefont {{Castander}}, \citenamefont {{Charles}}, \citenamefont {{Chaussidon}}, \citenamefont {{Chaves-Montero}}, \citenamefont {{Claybaugh}}, \citenamefont {{Cole}}, \citenamefont {{Cuceu}}, \citenamefont {{Dawson}}, \citenamefont
  {{de la Macorra}}, \citenamefont {{de Mattia}}, \citenamefont {{Della Costa}}, \citenamefont {{Dey}}, \citenamefont {{Dey}}, \citenamefont {{Ding}}, \citenamefont {{Doel}}, \citenamefont {{Eisenstein}}, \citenamefont {{Elbers}}, \citenamefont {{Fern{\'a}ndez-Garc{\'\i}a}}, \citenamefont {{Ferraro}}, \citenamefont {{Font-Ribera}}, \citenamefont {{Forero-Romero}}, \citenamefont {{Garcia-Quintero}}, \citenamefont {{Garrison}}, \citenamefont {{Gazta{\~n}aga}}, \citenamefont {{Gil-Mar{\'\i}n}}, \citenamefont {{Gontcho}}, \citenamefont {{Gonzalez-Morales}}, \citenamefont {{Gordon}}, \citenamefont {{Gutierrez}}, \citenamefont {{Guy}}, \citenamefont {{Hahn}}, \citenamefont {{He}}, \citenamefont {{Herrera-Alcantar}}, \citenamefont {{Honscheid}}, \citenamefont {{Howlett}}, \citenamefont {{Huterer}}, \citenamefont {{Ishak}}, \citenamefont {{Juneau}}, \citenamefont {{Kehoe}}, \citenamefont {{Kirkby}}, \citenamefont {{Kisner}}, \citenamefont {{Kremin}}, \citenamefont {{Lahav}}, \citenamefont {{Lamman}}, \citenamefont
  {{Landriau}}, \citenamefont {{Le Guillou}}, \citenamefont {{Leauthaud}}, \citenamefont {{Levi}}, \citenamefont {{Magneville}}, \citenamefont {{Manera}}, \citenamefont {{Martini}}, \citenamefont {{Matthewson}}, \citenamefont {{Meisner}}, \citenamefont {{Miquel}}, \citenamefont {{Moustakas}}, \citenamefont {{Mu{\~n}oz-Guti{\'e}rrez}}, \citenamefont {{Mu{\~n}oz-Santos}}, \citenamefont {{Myers}}, \citenamefont {{Napolitano}}, \citenamefont {{Newman}}, \citenamefont {{Noriega}}, \citenamefont {{Palanque-Delabrouille}}, \citenamefont {{Pan}}, \citenamefont {{Percival}}, \citenamefont {{P{\'e}rez-R{\`a}fols}}, \citenamefont {{Poppett}}, \citenamefont {{Prada}}, \citenamefont {{Raichoor}}, \citenamefont {{Ram{\'\i}rez-P{\'e}rez}}, \citenamefont {{Ravoux}}, \citenamefont {{Rossi}}, \citenamefont {{Ruggeri}}, \citenamefont {{Samushia}}, \citenamefont {{Sanchez}}, \citenamefont {{Schlegel}}, \citenamefont {{Schubnell}}, \citenamefont {{Sinigaglia}}, \citenamefont {{Sprayberry}}, \citenamefont {{Tan}}, \citenamefont
  {{Tarl{\'e}}}, \citenamefont {{Taylor}}, \citenamefont {{Turner}}, \citenamefont {{Vaisakh}}, \citenamefont {{Vargas-Maga{\~n}a}}, \citenamefont {{Walther}}, \citenamefont {{Weaver}}, \citenamefont {{White}}, \citenamefont {{Wolfson}}, \citenamefont {{Yu}}, \citenamefont {{Y{\`e}che}}, \citenamefont {{Zarrouk}}, \citenamefont {{Zhou}},\ and\ \citenamefont {{Zou}}}]{Y3.clust-s1.Andrade.2025}%
  \BibitemOpen
  \bibfield  {author} {\bibinfo {author} {\bibfnamefont {U.}~\bibnamefont {{Andrade}}}, \bibinfo {author} {\bibfnamefont {E.}~\bibnamefont {{Paillas}}}, \bibinfo {author} {\bibfnamefont {J.}~\bibnamefont {{Mena-Fern{\'a}ndez}}}, \bibinfo {author} {\bibfnamefont {Q.}~\bibnamefont {{Li}}}, \bibinfo {author} {\bibfnamefont {A.~J.}\ \bibnamefont {{Ross}}}, \bibinfo {author} {\bibfnamefont {S.}~\bibnamefont {{Nadathur}}}, \bibinfo {author} {\bibfnamefont {M.}~\bibnamefont {{Rashkovetskyi}}}, \bibinfo {author} {\bibfnamefont {A.}~\bibnamefont {{P{\'e}rez-Fern{\'a}ndez}}}, \bibinfo {author} {\bibfnamefont {H.}~\bibnamefont {{Seo}}}, \bibinfo {author} {\bibfnamefont {N.}~\bibnamefont {{Sanders}}}, \bibinfo {author} {\bibfnamefont {O.}~\bibnamefont {{Alves}}}, \bibinfo {author} {\bibfnamefont {X.}~\bibnamefont {{Chen}}}, \bibinfo {author} {\bibfnamefont {N.}~\bibnamefont {{Deiosso}}}, \bibinfo {author} {\bibfnamefont {M.}~\bibnamefont {{Abdul-Karim}}}, \bibinfo {author} {\bibfnamefont {S.}~\bibnamefont {{Ahlen}}},
  \bibnamefont {and~others},\ }\href {https://doi.org/10.48550/arXiv.2503.14742} {\bibfield  {journal} {\bibinfo  {journal} {arXiv e-prints}\ ,\ \bibinfo {eid} {arXiv:2503.14742}} (\bibinfo {year} {2025})},\ \Eprint {https://arxiv.org/abs/2503.14742} {arXiv:2503.14742 [astro-ph.CO]} \BibitemShut {NoStop}%
\bibitem [{\citenamefont {{DESI Collaboration}}\ \emph {et~al.}(2025{\natexlab{b}})\citenamefont {{DESI Collaboration}}, \citenamefont {{Abdul-Karim}}, \citenamefont {{Aguilar}}, \citenamefont {{Ahlen}}, \citenamefont {{Alam}}, \citenamefont {{Allen}}, \citenamefont {{Allende Prieto}}, \citenamefont {{Alves}}, \citenamefont {{Anand}}, \citenamefont {{Andrade}}, \citenamefont {{Armengaud}}, \citenamefont {{Aviles}}, \citenamefont {{Bailey}}, \citenamefont {{Baltay}}, \citenamefont {{Bansal}}, \citenamefont {{Bault}}, \citenamefont {{Behera}}, \citenamefont {{BenZvi}}, \citenamefont {{Bianchi}}, \citenamefont {{Blake}}, \citenamefont {{Brieden}}, \citenamefont {{Brodzeller}}, \citenamefont {{Brooks}}, \citenamefont {{Buckley-Geer}}, \citenamefont {{Burtin}}, \citenamefont {{Calderon}}, \citenamefont {{Canning}}, \citenamefont {{Carnero Rosell}}, \citenamefont {{Carrilho}}, \citenamefont {{Casas}}, \citenamefont {{Castander}}, \citenamefont {{Cereskaite}}, \citenamefont {{Charles}}, \citenamefont {{Chaussidon}},
  \citenamefont {{Chaves-Montero}}, \citenamefont {{Chebat}}, \citenamefont {{Chen}}, \citenamefont {{Claybaugh}}, \citenamefont {{Cole}}, \citenamefont {{Cooper}}, \citenamefont {{Cuceu}}, \citenamefont {{Dawson}}, \citenamefont {{de la Macorra}}, \citenamefont {{de Mattia}}, \citenamefont {{Deiosso}}, \citenamefont {{Della Costa}}, \citenamefont {{Demina}}, \citenamefont {{Dey}}, \citenamefont {{Dey}}, \citenamefont {{Ding}}, \citenamefont {{Doel}}, \citenamefont {{Edelstein}}, \citenamefont {{Eisenstein}}, \citenamefont {{Elbers}}, \citenamefont {{Fagrelius}}, \citenamefont {{Fanning}}, \citenamefont {{Fern{\'a}ndez-Garc{\'\i}a}}, \citenamefont {{Ferraro}}, \citenamefont {{Font-Ribera}}, \citenamefont {{Forero-Romero}}, \citenamefont {{Frenk}}, \citenamefont {{Garcia-Quintero}}, \citenamefont {{Garrison}}, \citenamefont {{Gazta{\~n}aga}}, \citenamefont {{Gil-Mar{\'\i}n}}, \citenamefont {{Gontcho}}, \citenamefont {{Gonzalez}}, \citenamefont {{Gonzalez-Morales}}, \citenamefont {{Gordon}}, \citenamefont
  {{Green}}, \citenamefont {{Gutierrez}}, \citenamefont {{Guy}}, \citenamefont {{Hadzhiyska}}, \citenamefont {{Hahn}}, \citenamefont {{He}}, \citenamefont {{Herbold}}, \citenamefont {{Herrera-Alcantar}}, \citenamefont {{Ho}}, \citenamefont {{Honscheid}}, \citenamefont {{Howlett}}, \citenamefont {{Huterer}}, \citenamefont {{Ishak}}, \citenamefont {{Juneau}}, \citenamefont {{Kamble}}, \citenamefont {{Kara{\c{c}}ayl{\i}}}, \citenamefont {{Kehoe}}, \citenamefont {{Kent}}, \citenamefont {{Kim}}, \citenamefont {{Kirkby}}, \citenamefont {{Kisner}}, \citenamefont {{Koposov}}, \citenamefont {{Kremin}}, \citenamefont {{Krolewski}}, \citenamefont {{Lahav}}, \citenamefont {{Lamman}}, \citenamefont {{Landriau}}, \citenamefont {{Lang}}, \citenamefont {{Lasker}}, \citenamefont {{Le Goff}}, \citenamefont {{Le Guillou}}, \citenamefont {{Leauthaud}}, \citenamefont {{Levi}}, \citenamefont {{Li}}, \citenamefont {{Li}}, \citenamefont {{Lodha}}, \citenamefont {{Lokken}}, \citenamefont {{Lozano-Rodr{\'\i}guez}}, \citenamefont
  {{Magneville}}, \citenamefont {{Manera}}, \citenamefont {{Martini}}, \citenamefont {{Matthewson}}, \citenamefont {{Meisner}}, \citenamefont {{Mena-Fern{\'a}ndez}}, \citenamefont {{Menegas}}, \citenamefont {{Mergulh{\~a}o}}, \citenamefont {{Miquel}}, \citenamefont {{Moustakas}}, \citenamefont {{Mu{\~n}oz-Guti{\'e}rrez}}, \citenamefont {{Mu{\~n}oz-Santos}}, \citenamefont {{Myers}}, \citenamefont {{Nadathur}}, \citenamefont {{Naidoo}}, \citenamefont {{Napolitano}}, \citenamefont {{Newman}}, \citenamefont {{Niz}}, \citenamefont {{Noriega}}, \citenamefont {{Paillas}}, \citenamefont {{Palanque-Delabrouille}}, \citenamefont {{Pan}}, \citenamefont {{Peacock}}, \citenamefont {{Pellejero Ibanez}}, \citenamefont {{Percival}}, \citenamefont {{P{\'e}rez-Fern{\'a}ndez}}, \citenamefont {{P{\'e}rez-R{\`a}fols}}, \citenamefont {{Pieri}}, \citenamefont {{Poppett}}, \citenamefont {{Prada}}, \citenamefont {{Rabinowitz}}, \citenamefont {{Raichoor}}, \citenamefont {{Ram{\'\i}rez-P{\'e}rez}}, \citenamefont {{Rashkovetskyi}},
  \citenamefont {{Ravoux}}, \citenamefont {{Rich}}, \citenamefont {{Rocher}}, \citenamefont {{Rockosi}}, \citenamefont {{Rohlf}}, \citenamefont {{Rom{\'a}n-Herrera}}, \citenamefont {{Ross}}, \citenamefont {{Rossi}}, \citenamefont {{Ruggeri}}, \citenamefont {{Ruhlmann-Kleider}}, \citenamefont {{Samushia}}, \citenamefont {{Sanchez}}, \citenamefont {{Sanders}}, \citenamefont {{Schlegel}}, \citenamefont {{Schubnell}}, \citenamefont {{Seo}}, \citenamefont {{Shafieloo}}, \citenamefont {{Sharples}}, \citenamefont {{Silber}}, \citenamefont {{Sinigaglia}}, \citenamefont {{Sprayberry}}, \citenamefont {{Tan}}, \citenamefont {{Tarl{\'e}}}, \citenamefont {{Taylor}}, \citenamefont {{Turner}}, \citenamefont {{Ure{\~n}a-L{\'o}pez}}, \citenamefont {{Vaisakh}}, \citenamefont {{Valdes}}, \citenamefont {{Valogiannis}}, \citenamefont {{Vargas-Maga{\~n}a}}, \citenamefont {{Verde}}, \citenamefont {{Walther}}, \citenamefont {{Weaver}}, \citenamefont {{Weinberg}}, \citenamefont {{White}}, \citenamefont {{Wolfson}}, \citenamefont
  {{Y{\`e}che}}, \citenamefont {{Yu}}, \citenamefont {{Zaborowski}}, \citenamefont {{Zarrouk}}, \citenamefont {{Zhai}}, \citenamefont {{Zhang}}, \citenamefont {{Zhao}}, \citenamefont {{Zhao}}, \citenamefont {{Zhou}},\ and\ \citenamefont {{Zou}}}]{DESI.DR2.BAO.cosmo}%
  \BibitemOpen
  \bibfield  {author} {\bibinfo {author} {\bibnamefont {{DESI Collaboration}}}, \bibinfo {author} {\bibfnamefont {M.}~\bibnamefont {{Abdul-Karim}}}, \bibinfo {author} {\bibfnamefont {J.}~\bibnamefont {{Aguilar}}}, \bibinfo {author} {\bibfnamefont {S.}~\bibnamefont {{Ahlen}}}, \bibinfo {author} {\bibfnamefont {S.}~\bibnamefont {{Alam}}}, \bibinfo {author} {\bibfnamefont {L.}~\bibnamefont {{Allen}}}, \bibinfo {author} {\bibfnamefont {C.}~\bibnamefont {{Allende Prieto}}}, \bibinfo {author} {\bibfnamefont {O.}~\bibnamefont {{Alves}}}, \bibinfo {author} {\bibfnamefont {A.}~\bibnamefont {{Anand}}}, \bibinfo {author} {\bibfnamefont {U.}~\bibnamefont {{Andrade}}}, \bibinfo {author} {\bibfnamefont {E.}~\bibnamefont {{Armengaud}}}, \bibinfo {author} {\bibfnamefont {A.}~\bibnamefont {{Aviles}}}, \bibinfo {author} {\bibfnamefont {S.}~\bibnamefont {{Bailey}}}, \bibinfo {author} {\bibfnamefont {C.}~\bibnamefont {{Baltay}}}, \bibinfo {author} {\bibfnamefont {P.}~\bibnamefont {{Bansal}}}, \bibnamefont {and~others},\ }\href
  {https://doi.org/10.48550/arXiv.2503.14738} {\bibfield  {journal} {\bibinfo  {journal} {arXiv e-prints}\ ,\ \bibinfo {eid} {arXiv:2503.14738}} (\bibinfo {year} {2025}{\natexlab{b}})},\ \Eprint {https://arxiv.org/abs/2503.14738} {arXiv:2503.14738 [astro-ph.CO]} \BibitemShut {NoStop}%
\bibitem [{\citenamefont {{Casas}}\ \emph {et~al.}(2025)\citenamefont {{Casas}}, \citenamefont {{Herrera-Alcantar}}, \citenamefont {{Chaves-Montero}}, \citenamefont {{Cuceu}}, \citenamefont {{Font-Ribera}}, \citenamefont {{Lokken}}, \citenamefont {{Abdul-Karim}}, \citenamefont {{Ram{\'\i}rez-P{\'e}rez}}, \citenamefont {{Aguilar}}, \citenamefont {{Ahlen}}, \citenamefont {{Andrade}}, \citenamefont {{Armengaud}}, \citenamefont {{Aviles}}, \citenamefont {{Bailey}}, \citenamefont {{BenZvi}}, \citenamefont {{Bianchi}}, \citenamefont {{Brodzeller}}, \citenamefont {{Brooks}}, \citenamefont {{Canning}}, \citenamefont {{Carnero Rosell}}, \citenamefont {{Charles}}, \citenamefont {{Chaussidon}}, \citenamefont {{Claybaugh}}, \citenamefont {{Dawson}}, \citenamefont {{de la Macorra}}, \citenamefont {{de Mattia}}, \citenamefont {{Dey}}, \citenamefont {{Dey}}, \citenamefont {{Ding}}, \citenamefont {{Doel}}, \citenamefont {{Eisenstein}}, \citenamefont {{Elbers}}, \citenamefont {{Ferraro}}, \citenamefont {{Forero-Romero}},
  \citenamefont {{Garcia-Quintero}}, \citenamefont {{Garrison}}, \citenamefont {{Gazta{\~n}aga}}, \citenamefont {{Gil-Mar{\'\i}n}}, \citenamefont {{Gontcho}}, \citenamefont {{Gonzalez-Morales}}, \citenamefont {{Gordon}}, \citenamefont {{Gutierrez}}, \citenamefont {{Guy}}, \citenamefont {{Herbold}}, \citenamefont {{Honscheid}}, \citenamefont {{Howlett}}, \citenamefont {{Huterer}}, \citenamefont {{Ishak}}, \citenamefont {{Juneau}}, \citenamefont {{Kehoe}}, \citenamefont {{Kirkby}}, \citenamefont {{Kisner}}, \citenamefont {{Kremin}}, \citenamefont {{Lahav}}, \citenamefont {{Landriau}}, \citenamefont {{Le Goff}}, \citenamefont {{Le Guillou}}, \citenamefont {{Leauthaud}}, \citenamefont {{Levi}}, \citenamefont {{Li}}, \citenamefont {{Manera}}, \citenamefont {{Martini}}, \citenamefont {{Meisner}}, \citenamefont {{Mena-Fern{\'a}ndez}}, \citenamefont {{Miquel}}, \citenamefont {{Moustakas}}, \citenamefont {{Mu{\~n}oz Santos}}, \citenamefont {{Myers}}, \citenamefont {{Nadathur}}, \citenamefont {{Napolitano}},
  \citenamefont {{Niz}}, \citenamefont {{Noriega}}, \citenamefont {{Paillas}}, \citenamefont {{Palanque-Delabrouille}}, \citenamefont {{Percival}}, \citenamefont {{Pieri}}, \citenamefont {{Poppett}}, \citenamefont {{Prada}}, \citenamefont {{P{\'e}rez-R{\`a}fols}}, \citenamefont {{Ravoux}}, \citenamefont {{Rossi}}, \citenamefont {{Sanchez}}, \citenamefont {{Schlegel}}, \citenamefont {{Schubnell}}, \citenamefont {{Seo}}, \citenamefont {{Sinigaglia}}, \citenamefont {{Sprayberry}}, \citenamefont {{Tan}}, \citenamefont {{Tarl{\'e}}}, \citenamefont {{Taylor}}, \citenamefont {{Turner}}, \citenamefont {{Vargas-Maga{\~n}a}}, \citenamefont {{Walther}}, \citenamefont {{Weaver}}, \citenamefont {{Wolfson}}, \citenamefont {{Y{\`e}che}}, \citenamefont {{Zarrouk}},\ and\ \citenamefont {{Zhou}}}]{Y3.lya-s1.Casas.2025}%
  \BibitemOpen
  \bibfield  {author} {\bibinfo {author} {\bibfnamefont {L.}~\bibnamefont {{Casas}}}, \bibinfo {author} {\bibfnamefont {H.~K.}\ \bibnamefont {{Herrera-Alcantar}}}, \bibinfo {author} {\bibfnamefont {J.}~\bibnamefont {{Chaves-Montero}}}, \bibinfo {author} {\bibfnamefont {A.}~\bibnamefont {{Cuceu}}}, \bibinfo {author} {\bibfnamefont {A.}~\bibnamefont {{Font-Ribera}}}, \bibinfo {author} {\bibfnamefont {M.}~\bibnamefont {{Lokken}}}, \bibinfo {author} {\bibfnamefont {M.}~\bibnamefont {{Abdul-Karim}}}, \bibinfo {author} {\bibfnamefont {C.}~\bibnamefont {{Ram{\'\i}rez-P{\'e}rez}}}, \bibinfo {author} {\bibfnamefont {J.}~\bibnamefont {{Aguilar}}}, \bibinfo {author} {\bibfnamefont {S.}~\bibnamefont {{Ahlen}}}, \bibinfo {author} {\bibfnamefont {U.}~\bibnamefont {{Andrade}}}, \bibinfo {author} {\bibfnamefont {E.}~\bibnamefont {{Armengaud}}}, \bibinfo {author} {\bibfnamefont {A.}~\bibnamefont {{Aviles}}}, \bibinfo {author} {\bibfnamefont {S.}~\bibnamefont {{Bailey}}}, \bibinfo {author} {\bibfnamefont {S.}~\bibnamefont
  {{BenZvi}}}, \bibnamefont {and~others},\ }\href {https://doi.org/10.48550/arXiv.2503.14741} {\bibfield  {journal} {\bibinfo  {journal} {arXiv e-prints}\ ,\ \bibinfo {eid} {arXiv:2503.14741}} (\bibinfo {year} {2025})},\ \Eprint {https://arxiv.org/abs/2503.14741} {arXiv:2503.14741 [astro-ph.IM]} \BibitemShut {NoStop}%
\bibitem [{\citenamefont {{Brodzeller}}\ \emph {et~al.}(2025)\citenamefont {{Brodzeller}}, \citenamefont {{Wolfson}}, \citenamefont {{Santos}}, \citenamefont {{Ho}}, \citenamefont {{Tan}}, \citenamefont {{Pieri}}, \citenamefont {{Cuceu}}, \citenamefont {{Abdul-Karim}}, \citenamefont {{Aguilar}}, \citenamefont {{Ahlen}}, \citenamefont {{Anand}}, \citenamefont {{Andrade}}, \citenamefont {{Armengaud}}, \citenamefont {{Aviles}}, \citenamefont {{Bailey}}, \citenamefont {{Bault}}, \citenamefont {{Bianchi}}, \citenamefont {{Brooks}}, \citenamefont {{Canning}}, \citenamefont {{Casas}}, \citenamefont {{Charles}}, \citenamefont {{Chaussidon}}, \citenamefont {{Chaves-Montero}}, \citenamefont {{Chebat}}, \citenamefont {{Claybaugh}}, \citenamefont {{Dawson}}, \citenamefont {{de Belsunce}}, \citenamefont {{de la Macorra}}, \citenamefont {{de Mattia}}, \citenamefont {{Dey}}, \citenamefont {{Dey}}, \citenamefont {{Doel}}, \citenamefont {{Elbers}}, \citenamefont {{Ferraro}}, \citenamefont {{Font-Ribera}}, \citenamefont
  {{Forero-Romero}}, \citenamefont {{Garcia-Quintero}}, \citenamefont {{Garrison}}, \citenamefont {{Gazta{\~n}aga}}, \citenamefont {{Gontcho}}, \citenamefont {{Gonzalez-Morales}}, \citenamefont {{Green}}, \citenamefont {{Gutierrez}}, \citenamefont {{Guy}}, \citenamefont {{Hahn}}, \citenamefont {{Herbold}}, \citenamefont {{Herrera-Alcantar}}, \citenamefont {{Honscheid}}, \citenamefont {{Howlett}}, \citenamefont {{Huterer}}, \citenamefont {{Ishak}}, \citenamefont {{Juneau}}, \citenamefont {{Kehoe}}, \citenamefont {{Kisner}}, \citenamefont {{Kremin}}, \citenamefont {{Lahav}}, \citenamefont {{Lamman}}, \citenamefont {{Landriau}}, \citenamefont {{Le Goff}}, \citenamefont {{Le Guillou}}, \citenamefont {{Leauthaud}}, \citenamefont {{Levi}}, \citenamefont {{Li}}, \citenamefont {{Manera}}, \citenamefont {{Martini}}, \citenamefont {{Meisner}}, \citenamefont {{Mena-Fernandez}}, \citenamefont {{Miquel}}, \citenamefont {{Moustakas}}, \citenamefont {{Mu{\~n}oz-Guti{\'e}rrez}}, \citenamefont {{Myers}}, \citenamefont
  {{Nadathur}}, \citenamefont {{Napolitano}}, \citenamefont {{Noriega}}, \citenamefont {{Paillas}}, \citenamefont {{Palanque-Delabrouille}}, \citenamefont {{Percival}}, \citenamefont {{Poppett}}, \citenamefont {{Prada}}, \citenamefont {{P{\'e}rez-R{\`a}fols}}, \citenamefont {{Ram{\'\i}rez-P{\'e}rez}}, \citenamefont {{Ravoux}}, \citenamefont {{Rohlf}}, \citenamefont {{Rossi}}, \citenamefont {{Sanchez}}, \citenamefont {{Schlegel}}, \citenamefont {{Schubnell}}, \citenamefont {{Sinigaglia}}, \citenamefont {{Sprayberry}}, \citenamefont {{Tarl{\'e}}}, \citenamefont {{Taylor}}, \citenamefont {{Turner}}, \citenamefont {{Walther}}, \citenamefont {{Weaver}}, \citenamefont {{Y{\`e}che}}, \citenamefont {{Zhou}}, \citenamefont {{Zou}},\ and\ \citenamefont {{Zou}}}]{Y3.lya-s2.Brodzeller.2025}%
  \BibitemOpen
  \bibfield  {author} {\bibinfo {author} {\bibfnamefont {A.}~\bibnamefont {{Brodzeller}}}, \bibinfo {author} {\bibfnamefont {M.}~\bibnamefont {{Wolfson}}}, \bibinfo {author} {\bibfnamefont {D.~M.}\ \bibnamefont {{Santos}}}, \bibinfo {author} {\bibfnamefont {M.}~\bibnamefont {{Ho}}}, \bibinfo {author} {\bibfnamefont {T.}~\bibnamefont {{Tan}}}, \bibinfo {author} {\bibfnamefont {M.~M.}\ \bibnamefont {{Pieri}}}, \bibinfo {author} {\bibfnamefont {A.}~\bibnamefont {{Cuceu}}}, \bibinfo {author} {\bibfnamefont {M.}~\bibnamefont {{Abdul-Karim}}}, \bibinfo {author} {\bibfnamefont {J.}~\bibnamefont {{Aguilar}}}, \bibinfo {author} {\bibfnamefont {S.}~\bibnamefont {{Ahlen}}}, \bibinfo {author} {\bibfnamefont {A.}~\bibnamefont {{Anand}}}, \bibinfo {author} {\bibfnamefont {U.}~\bibnamefont {{Andrade}}}, \bibinfo {author} {\bibfnamefont {E.}~\bibnamefont {{Armengaud}}}, \bibinfo {author} {\bibfnamefont {A.}~\bibnamefont {{Aviles}}}, \bibinfo {author} {\bibfnamefont {S.}~\bibnamefont {{Bailey}}}, \bibnamefont {and~others},\
  }\href {https://doi.org/10.48550/arXiv.2503.14740} {\bibfield  {journal} {\bibinfo  {journal} {arXiv e-prints}\ ,\ \bibinfo {eid} {arXiv:2503.14740}} (\bibinfo {year} {2025})},\ \Eprint {https://arxiv.org/abs/2503.14740} {arXiv:2503.14740 [astro-ph.CO]} \BibitemShut {NoStop}%
\bibitem [{\citenamefont {{DESI Collaboration}}\ \emph {et~al.}(2025{\natexlab{c}})\citenamefont {{DESI Collaboration}}, \citenamefont {{Abdul-Karim}}, \citenamefont {{Aguilar}}, \citenamefont {{Ahlen}}, \citenamefont {{Allende Prieto}}, \citenamefont {{Alves}}, \citenamefont {{Anand}}, \citenamefont {{Andrade}}, \citenamefont {{Armengaud}}, \citenamefont {{Aviles}}, \citenamefont {{Bailey}}, \citenamefont {{Bault}}, \citenamefont {{BenZvi}}, \citenamefont {{Bianchi}}, \citenamefont {{Blake}}, \citenamefont {{Brodzeller}}, \citenamefont {{Brooks}}, \citenamefont {{Buckley-Geer}}, \citenamefont {{Burtin}}, \citenamefont {{Calderon}}, \citenamefont {{Canning}}, \citenamefont {{Carnero Rosell}}, \citenamefont {{Carrilho}}, \citenamefont {{Casas}}, \citenamefont {{Castander}}, \citenamefont {{Cereskaite}}, \citenamefont {{Charles}}, \citenamefont {{Chaussidon}}, \citenamefont {{Chaves-Montero}}, \citenamefont {{Chebat}}, \citenamefont {{Claybaugh}}, \citenamefont {{Cole}}, \citenamefont {{Cooper}}, \citenamefont
  {{Cuceu}}, \citenamefont {{Dawson}}, \citenamefont {{de Belsunce}}, \citenamefont {{de la Macorra}}, \citenamefont {{de Mattia}}, \citenamefont {{Deiosso}}, \citenamefont {{Della Costa}}, \citenamefont {{Dey}}, \citenamefont {{Dey}}, \citenamefont {{Ding}}, \citenamefont {{Doel}}, \citenamefont {{Edelstein}}, \citenamefont {{Eisenstein}}, \citenamefont {{Elbers}}, \citenamefont {{Fagrelius}}, \citenamefont {{Fanning}}, \citenamefont {{Ferraro}}, \citenamefont {{Font-Ribera}}, \citenamefont {{Forero-Romero}}, \citenamefont {{Garcia-Quintero}}, \citenamefont {{Garrison}}, \citenamefont {{Gazta{\~n}aga}}, \citenamefont {{Gil-Mar{\'\i}n}}, \citenamefont {{Gontcho}}, \citenamefont {{Gonzalez-Morales}}, \citenamefont {{Gordon}}, \citenamefont {{Green}}, \citenamefont {{Gutierrez}}, \citenamefont {{Guy}}, \citenamefont {{Hahn}}, \citenamefont {{Herbold}}, \citenamefont {{Herrera-Alcantar}}, \citenamefont {{Ho}}, \citenamefont {{Honscheid}}, \citenamefont {{Howlett}}, \citenamefont {{Huterer}}, \citenamefont
  {{Ishak}}, \citenamefont {{Juneau}}, \citenamefont {{Kara{\c{c}}ayl{\i}}}, \citenamefont {{Kehoe}}, \citenamefont {{Kent}}, \citenamefont {{Kirkby}}, \citenamefont {{Kisner}}, \citenamefont {{Kitaura}}, \citenamefont {{Koposov}}, \citenamefont {{Kremin}}, \citenamefont {{Lahav}}, \citenamefont {{Lamman}}, \citenamefont {{Landriau}}, \citenamefont {{Lang}}, \citenamefont {{Lasker}}, \citenamefont {{Le Goff}}, \citenamefont {{Le Guillou}}, \citenamefont {{Leauthaud}}, \citenamefont {{Levi}}, \citenamefont {{Li}}, \citenamefont {{Li}}, \citenamefont {{Lodha}}, \citenamefont {{Lokken}}, \citenamefont {{Magneville}}, \citenamefont {{Manera}}, \citenamefont {{Martini}}, \citenamefont {{Matthewson}}, \citenamefont {{McDonald}}, \citenamefont {{Meisner}}, \citenamefont {{Mena-Fern{\'a}ndez}}, \citenamefont {{Miquel}}, \citenamefont {{Moustakas}}, \citenamefont {{Mu{\~n}oz-Guti{\'e}rrez}}, \citenamefont {{Mu{\~n}oz-Santos}}, \citenamefont {{Myers}}, \citenamefont {{Newman}}, \citenamefont {{Niz}}, \citenamefont
  {{Noriega}}, \citenamefont {{Paillas}}, \citenamefont {{Palanque-Delabrouille}}, \citenamefont {{Pan}}, \citenamefont {{Percival}}, \citenamefont {{P{\'e}rez-R{\`a}fols}}, \citenamefont {{Pieri}}, \citenamefont {{Poppett}}, \citenamefont {{Prada}}, \citenamefont {{Rabinowitz}}, \citenamefont {{Raichoor}}, \citenamefont {{Ram{\'\i}rez-P{\'e}rez}}, \citenamefont {{Rashkovetskyi}}, \citenamefont {{Ravoux}}, \citenamefont {{Rich}}, \citenamefont {{Rockosi}}, \citenamefont {{Ross}}, \citenamefont {{Rossi}}, \citenamefont {{Ruhlmann-Kleider}}, \citenamefont {{Sanchez}}, \citenamefont {{Sanders}}, \citenamefont {{Satyavolu}}, \citenamefont {{Schlegel}}, \citenamefont {{Schubnell}}, \citenamefont {{Seo}}, \citenamefont {{Shafieloo}}, \citenamefont {{Sharples}}, \citenamefont {{Silber}}, \citenamefont {{Sinigaglia}}, \citenamefont {{Sprayberry}}, \citenamefont {{Tan}}, \citenamefont {{Tarl{\'e}}}, \citenamefont {{Taylor}}, \citenamefont {{Turner}}, \citenamefont {{Valdes}}, \citenamefont {{Vargas-Maga{\~n}a}},
  \citenamefont {{Walther}}, \citenamefont {{Weaver}}, \citenamefont {{Wolfson}}, \citenamefont {{Y{\`e}che}}, \citenamefont {{Zarrouk}}, \citenamefont {{Zhou}},\ and\ \citenamefont {{Zou}}}]{DESI.DR2.BAO.lya}%
  \BibitemOpen
  \bibfield  {author} {\bibinfo {author} {\bibnamefont {{DESI Collaboration}}}, \bibinfo {author} {\bibfnamefont {M.}~\bibnamefont {{Abdul-Karim}}}, \bibinfo {author} {\bibfnamefont {J.}~\bibnamefont {{Aguilar}}}, \bibinfo {author} {\bibfnamefont {S.}~\bibnamefont {{Ahlen}}}, \bibinfo {author} {\bibfnamefont {C.}~\bibnamefont {{Allende Prieto}}}, \bibinfo {author} {\bibfnamefont {O.}~\bibnamefont {{Alves}}}, \bibinfo {author} {\bibfnamefont {A.}~\bibnamefont {{Anand}}}, \bibinfo {author} {\bibfnamefont {U.}~\bibnamefont {{Andrade}}}, \bibinfo {author} {\bibfnamefont {E.}~\bibnamefont {{Armengaud}}}, \bibinfo {author} {\bibfnamefont {A.}~\bibnamefont {{Aviles}}}, \bibinfo {author} {\bibfnamefont {S.}~\bibnamefont {{Bailey}}}, \bibinfo {author} {\bibfnamefont {A.}~\bibnamefont {{Bault}}}, \bibinfo {author} {\bibfnamefont {S.}~\bibnamefont {{BenZvi}}}, \bibinfo {author} {\bibfnamefont {D.}~\bibnamefont {{Bianchi}}}, \bibinfo {author} {\bibfnamefont {C.}~\bibnamefont {{Blake}}}, \bibnamefont {and~others},\ }\href
  {https://doi.org/10.48550/arXiv.2503.14739} {\bibfield  {journal} {\bibinfo  {journal} {arXiv e-prints}\ ,\ \bibinfo {eid} {arXiv:2503.14739}} (\bibinfo {year} {2025}{\natexlab{c}})},\ \Eprint {https://arxiv.org/abs/2503.14739} {arXiv:2503.14739 [astro-ph.CO]} \BibitemShut {NoStop}%
\bibitem [{\citenamefont {{Elbers}}\ \emph {et~al.}(2025{\natexlab{a}})\citenamefont {{Elbers}}, \citenamefont {{Aviles}}, \citenamefont {{Noriega}}, \citenamefont {{Chebat}}, \citenamefont {{Menegas}}, \citenamefont {{Frenk}}, \citenamefont {{Garcia-Quintero}}, \citenamefont {{Gonzalez}}, \citenamefont {{Ishak}}, \citenamefont {{Lahav}}, \citenamefont {{Naidoo}}, \citenamefont {{Niz}}, \citenamefont {{Y{\`e}che}}, \citenamefont {{Abdul-Karim}}, \citenamefont {{Ahlen}}, \citenamefont {{Alves}}, \citenamefont {{Andrade}}, \citenamefont {{Armengaud}}, \citenamefont {{Bianchi}}, \citenamefont {{Brieden}}, \citenamefont {{Brodzeller}}, \citenamefont {{Brooks}}, \citenamefont {{Burtin}}, \citenamefont {{Calderon}}, \citenamefont {{Canning}}, \citenamefont {{Carnero Rosell}}, \citenamefont {{Casas}}, \citenamefont {{Castander}}, \citenamefont {{Charles}}, \citenamefont {{Chaussidon}}, \citenamefont {{Chaves-Montero}}, \citenamefont {{Claybaugh}}, \citenamefont {{Cole}}, \citenamefont {{Cuceu}}, \citenamefont
  {{Dawson}}, \citenamefont {{de la Macorra}}, \citenamefont {{de Mattia}}, \citenamefont {{Deiosso}}, \citenamefont {{Dey}}, \citenamefont {{Dey}}, \citenamefont {{Ding}}, \citenamefont {{Doel}}, \citenamefont {{Eisenstein}}, \citenamefont {{Ferraro}}, \citenamefont {{Font-Ribera}}, \citenamefont {{Forero-Romero}}, \citenamefont {{Garrison}}, \citenamefont {{Gazta{\~n}aga}}, \citenamefont {{Gil-Mar{\'\i}n}}, \citenamefont {{Gonzalez-Morales}}, \citenamefont {{Gutierrez}}, \citenamefont {{He}}, \citenamefont {{Herbold}}, \citenamefont {{Herrera-Alcantar}}, \citenamefont {{Howlett}}, \citenamefont {{Huterer}}, \citenamefont {{Kehoe}}, \citenamefont {{Kirkby}}, \citenamefont {{Kisner}}, \citenamefont {{Kremin}}, \citenamefont {{Lamman}}, \citenamefont {{Landriau}}, \citenamefont {{Le Guillou}}, \citenamefont {{Leauthaud}}, \citenamefont {{Levi}}, \citenamefont {{Li}}, \citenamefont {{Lodha}}, \citenamefont {{Magneville}}, \citenamefont {{Manera}}, \citenamefont {{Matthewson}}, \citenamefont {{Meisner}},
  \citenamefont {{Miquel}}, \citenamefont {{Moustakas}}, \citenamefont {{Nadathur}}, \citenamefont {{Newman}}, \citenamefont {{Paillas}}, \citenamefont {{Palanque-Delabrouille}}, \citenamefont {{Percival}}, \citenamefont {{Pieri}}, \citenamefont {{Prada}}, \citenamefont {{P{\'e}rez-R{\`a}fols}}, \citenamefont {{Rabinowitz}}, \citenamefont {{Ram{\'\i}rez-P{\'e}rez}}, \citenamefont {{Rashkovetskyi}}, \citenamefont {{Ravoux}}, \citenamefont {{Rivera}}, \citenamefont {{Rohlf}}, \citenamefont {{Rossi}}, \citenamefont {{Ruhlmann-Kleider}}, \citenamefont {{Samushia}}, \citenamefont {{Sanchez}}, \citenamefont {{Schlegel}}, \citenamefont {{Schubnell}}, \citenamefont {{Sinigaglia}}, \citenamefont {{Sprayberry}}, \citenamefont {{Tan}}, \citenamefont {{Tarl{\'e}}}, \citenamefont {{Taylor}}, \citenamefont {{Turner}}, \citenamefont {{Vargas-Maga{\~n}a}}, \citenamefont {{Verde}}, \citenamefont {{Walther}}, \citenamefont {{Weaver}}, \citenamefont {{Whitford}}, \citenamefont {{Wolfson}}, \citenamefont {{Zarrouk}},
  \citenamefont {{Zhao}},\ and\ \citenamefont {{Zou}}}]{Y3.cpe-s2.Elbers.2025}%
  \BibitemOpen
  \bibfield  {author} {\bibinfo {author} {\bibfnamefont {W.}~\bibnamefont {{Elbers}}}, \bibinfo {author} {\bibfnamefont {A.}~\bibnamefont {{Aviles}}}, \bibinfo {author} {\bibfnamefont {H.~E.}\ \bibnamefont {{Noriega}}}, \bibinfo {author} {\bibfnamefont {D.}~\bibnamefont {{Chebat}}}, \bibinfo {author} {\bibfnamefont {A.}~\bibnamefont {{Menegas}}}, \bibinfo {author} {\bibfnamefont {C.~S.}\ \bibnamefont {{Frenk}}}, \bibinfo {author} {\bibfnamefont {C.}~\bibnamefont {{Garcia-Quintero}}}, \bibinfo {author} {\bibfnamefont {D.}~\bibnamefont {{Gonzalez}}}, \bibinfo {author} {\bibfnamefont {M.}~\bibnamefont {{Ishak}}}, \bibinfo {author} {\bibfnamefont {O.}~\bibnamefont {{Lahav}}}, \bibinfo {author} {\bibfnamefont {K.}~\bibnamefont {{Naidoo}}}, \bibinfo {author} {\bibfnamefont {G.}~\bibnamefont {{Niz}}}, \bibinfo {author} {\bibfnamefont {C.}~\bibnamefont {{Y{\`e}che}}}, \bibinfo {author} {\bibfnamefont {M.}~\bibnamefont {{Abdul-Karim}}}, \bibinfo {author} {\bibfnamefont {S.}~\bibnamefont {{Ahlen}}}, \bibnamefont
  {and~others},\ }\href {https://doi.org/10.48550/arXiv.2503.14744} {\bibfield  {journal} {\bibinfo  {journal} {arXiv e-prints}\ ,\ \bibinfo {eid} {arXiv:2503.14744}} (\bibinfo {year} {2025}{\natexlab{a}})},\ \Eprint {https://arxiv.org/abs/2503.14744} {arXiv:2503.14744 [astro-ph.CO]} \BibitemShut {NoStop}%
\bibitem [{\citenamefont {{Qu}}\ \emph {et~al.}(2024)\citenamefont {{Qu}}, \citenamefont {{Sherwin}}, \citenamefont {{Madhavacheril}}, \citenamefont {{Han}}, \citenamefont {{Crowley}}, \citenamefont {{Abril-Cabezas}}, \citenamefont {{Ade}}, \citenamefont {{Aiola}}, \citenamefont {{Alford}}, \citenamefont {{Amiri}}, \citenamefont {{Amodeo}}, \citenamefont {{An}}, \citenamefont {{Atkins}}, \citenamefont {{Austermann}}, \citenamefont {{Battaglia}}, \citenamefont {{Battistelli}}, \citenamefont {{Beall}}, \citenamefont {{Bean}}, \citenamefont {{Beringue}}, \citenamefont {{Bhandarkar}}, \citenamefont {{Biermann}}, \citenamefont {{Bolliet}}, \citenamefont {{Bond}}, \citenamefont {{Cai}}, \citenamefont {{Calabrese}}, \citenamefont {{Calafut}}, \citenamefont {{Capalbo}}, \citenamefont {{Carrero}}, \citenamefont {{Carron}}, \citenamefont {{Challinor}}, \citenamefont {{Chesmore}}, \citenamefont {{Cho}}, \citenamefont {{Choi}}, \citenamefont {{Clark}}, \citenamefont {{C{\'o}rdova Rosado}}, \citenamefont {{Cothard}},
  \citenamefont {{Coughlin}}, \citenamefont {{Coulton}}, \citenamefont {{Dalal}}, \citenamefont {{Darwish}}, \citenamefont {{Devlin}}, \citenamefont {{Dicker}}, \citenamefont {{Doze}}, \citenamefont {{Duell}}, \citenamefont {{Duff}}, \citenamefont {{Duivenvoorden}}, \citenamefont {{Dunkley}}, \citenamefont {{D{\"u}nner}}, \citenamefont {{Fanfani}}, \citenamefont {{Fankhanel}}, \citenamefont {{Farren}}, \citenamefont {{Ferraro}}, \citenamefont {{Freundt}}, \citenamefont {{Fuzia}}, \citenamefont {{Gallardo}}, \citenamefont {{Garrido}}, \citenamefont {{Gluscevic}}, \citenamefont {{Golec}}, \citenamefont {{Guan}}, \citenamefont {{Halpern}}, \citenamefont {{Harrison}}, \citenamefont {{Hasselfield}}, \citenamefont {{Healy}}, \citenamefont {{Henderson}}, \citenamefont {{Hensley}}, \citenamefont {{Herv{\'\i}as-Caimapo}}, \citenamefont {{Hill}}, \citenamefont {{Hilton}}, \citenamefont {{Hilton}}, \citenamefont {{Hincks}}, \citenamefont {{Hlo{\v{z}}ek}}, \citenamefont {{Ho}}, \citenamefont {{Huber}}, \citenamefont
  {{Hubmayr}}, \citenamefont {{Huffenberger}}, \citenamefont {{Hughes}}, \citenamefont {{Irwin}}, \citenamefont {{Isopi}}, \citenamefont {{Jense}}, \citenamefont {{Keller}}, \citenamefont {{Kim}}, \citenamefont {{Knowles}}, \citenamefont {{Koopman}}, \citenamefont {{Kosowsky}}, \citenamefont {{Kramer}}, \citenamefont {{Kusiak}}, \citenamefont {{La Posta}}, \citenamefont {{Lague}}, \citenamefont {{Lakey}}, \citenamefont {{Lee}}, \citenamefont {{Li}}, \citenamefont {{Li}}, \citenamefont {{Limon}}, \citenamefont {{Lokken}}, \citenamefont {{Louis}}, \citenamefont {{Lungu}}, \citenamefont {{MacCrann}}, \citenamefont {{MacInnis}}, \citenamefont {{Maldonado}}, \citenamefont {{Maldonado}}, \citenamefont {{Mallaby-Kay}}, \citenamefont {{Marques}}, \citenamefont {{McMahon}}, \citenamefont {{Mehta}}, \citenamefont {{Menanteau}}, \citenamefont {{Moodley}}, \citenamefont {{Morris}}, \citenamefont {{Mroczkowski}}, \citenamefont {{Naess}}, \citenamefont {{Namikawa}}, \citenamefont {{Nati}}, \citenamefont {{Newburgh}},
  \citenamefont {{Nicola}}, \citenamefont {{Niemack}}, \citenamefont {{Nolta}}, \citenamefont {{Orlowski-Scherer}}, \citenamefont {{Page}}, \citenamefont {{Pandey}}, \citenamefont {{Partridge}}, \citenamefont {{Prince}}, \citenamefont {{Puddu}}, \citenamefont {{Radiconi}}, \citenamefont {{Robertson}}, \citenamefont {{Rojas}}, \citenamefont {{Sakuma}}, \citenamefont {{Salatino}}, \citenamefont {{Schaan}}, \citenamefont {{Schmitt}}, \citenamefont {{Sehgal}}, \citenamefont {{Shaikh}}, \citenamefont {{Sierra}}, \citenamefont {{Sievers}}, \citenamefont {{Sif{\'o}n}}, \citenamefont {{Simon}}, \citenamefont {{Sonka}}, \citenamefont {{Spergel}}, \citenamefont {{Staggs}}, \citenamefont {{Storer}}, \citenamefont {{Switzer}}, \citenamefont {{Tampier}}, \citenamefont {{Thornton}}, \citenamefont {{Trac}}, \citenamefont {{Treu}}, \citenamefont {{Tucker}}, \citenamefont {{Ullom}}, \citenamefont {{Vale}}, \citenamefont {{Van Engelen}}, \citenamefont {{Van Lanen}}, \citenamefont {{van Marrewijk}}, \citenamefont {{Vargas}},
  \citenamefont {{Vavagiakis}}, \citenamefont {{Wagoner}}, \citenamefont {{Wang}}, \citenamefont {{Wenzl}}, \citenamefont {{Wollack}}, \citenamefont {{Xu}}, \citenamefont {{Zago}},\ and\ \citenamefont {{Zheng}}}]{Qu:2023}%
  \BibitemOpen
  \bibfield  {author} {\bibinfo {author} {\bibfnamefont {F.~J.}\ \bibnamefont {{Qu}}}, \bibinfo {author} {\bibfnamefont {B.~D.}\ \bibnamefont {{Sherwin}}}, \bibinfo {author} {\bibfnamefont {M.~S.}\ \bibnamefont {{Madhavacheril}}}, \bibinfo {author} {\bibfnamefont {D.}~\bibnamefont {{Han}}}, \bibinfo {author} {\bibfnamefont {K.~T.}\ \bibnamefont {{Crowley}}}, \bibinfo {author} {\bibfnamefont {I.}~\bibnamefont {{Abril-Cabezas}}}, \bibinfo {author} {\bibfnamefont {P.~A.~R.}\ \bibnamefont {{Ade}}}, \bibinfo {author} {\bibfnamefont {S.}~\bibnamefont {{Aiola}}}, \bibinfo {author} {\bibfnamefont {T.}~\bibnamefont {{Alford}}}, \bibinfo {author} {\bibfnamefont {M.}~\bibnamefont {{Amiri}}}, \bibinfo {author} {\bibfnamefont {S.}~\bibnamefont {{Amodeo}}}, \bibinfo {author} {\bibfnamefont {R.}~\bibnamefont {{An}}}, \bibinfo {author} {\bibfnamefont {Z.}~\bibnamefont {{Atkins}}}, \bibinfo {author} {\bibfnamefont {J.~E.}\ \bibnamefont {{Austermann}}}, \bibinfo {author} {\bibfnamefont {N.}~\bibnamefont {{Battaglia}}},
  \bibnamefont {and~others},\ }\href {https://doi.org/10.3847/1538-4357/acfe06} {\bibfield  {journal} {\bibinfo  {journal} {\apj}\ }\textbf {\bibinfo {volume} {962}},\ \bibinfo {eid} {112} (\bibinfo {year} {2024})},\ \Eprint {https://arxiv.org/abs/2304.05202} {arXiv:2304.05202 [astro-ph.CO]} \BibitemShut {NoStop}%
\bibitem [{\citenamefont {{DESI Collaboration}}\ \emph {et~al.}(2024{\natexlab{c}})\citenamefont {{DESI Collaboration}}, \citenamefont {{Adame}}, \citenamefont {{Aguilar}}, \citenamefont {{Ahlen}}, \citenamefont {{Alam}}, \citenamefont {{Alexander}}, \citenamefont {{Alvarez}}, \citenamefont {{Alves}}, \citenamefont {{Anand}}, \citenamefont {{Andrade}}, \citenamefont {{Armengaud}}, \citenamefont {{Avila}}, \citenamefont {{Aviles}}, \citenamefont {{Awan}}, \citenamefont {{Bailey}}, \citenamefont {{Baltay}}, \citenamefont {{Bault}}, \citenamefont {{Behera}}, \citenamefont {{BenZvi}}, \citenamefont {{Beutler}}, \citenamefont {{Bianchi}}, \citenamefont {{Blake}}, \citenamefont {{Blum}}, \citenamefont {{Brieden}}, \citenamefont {{Brodzeller}}, \citenamefont {{Brooks}}, \citenamefont {{Brown}}, \citenamefont {{Buckley-Geer}}, \citenamefont {{Burtin}}, \citenamefont {{Calderon}}, \citenamefont {{Canning}}, \citenamefont {{Carnero Rosell}}, \citenamefont {{Cereskaite}}, \citenamefont {{Cervantes-Cota}}, \citenamefont
  {{Chabanier}}, \citenamefont {{Chaussidon}}, \citenamefont {{Chaves-Montero}}, \citenamefont {{Chen}}, \citenamefont {{Chen}}, \citenamefont {{Claybaugh}}, \citenamefont {{Cole}}, \citenamefont {{Cuceu}}, \citenamefont {{Davis}}, \citenamefont {{Dawson}}, \citenamefont {{de la Macorra}}, \citenamefont {{de Mattia}}, \citenamefont {{Deiosso}}, \citenamefont {{Demina}}, \citenamefont {{Dey}}, \citenamefont {{Dey}}, \citenamefont {{Ding}}, \citenamefont {{Doel}}, \citenamefont {{Edelstein}}, \citenamefont {{Eftekharzadeh}}, \citenamefont {{Eisenstein}}, \citenamefont {{Elliott}}, \citenamefont {{Fagrelius}}, \citenamefont {{Fanning}}, \citenamefont {{Ferraro}}, \citenamefont {{Ereza}}, \citenamefont {{Findlay}}, \citenamefont {{Flaugher}}, \citenamefont {{Font-Ribera}}, \citenamefont {{Forero-S{\'a}nchez}}, \citenamefont {{Forero-Romero}}, \citenamefont {{Frenk}}, \citenamefont {{Garcia-Quintero}}, \citenamefont {{Gazta{\~n}aga}}, \citenamefont {{Gil-Mar{\'\i}n}}, \citenamefont {{Gontcho}}, \citenamefont
  {{Gonzalez-Morales}}, \citenamefont {{Gonzalez-Perez}}, \citenamefont {{Gordon}}, \citenamefont {{Green}}, \citenamefont {{Gruen}}, \citenamefont {{Gsponer}}, \citenamefont {{Gutierrez}}, \citenamefont {{Guy}}, \citenamefont {{Hadzhiyska}}, \citenamefont {{Hahn}}, \citenamefont {{Hanif}}, \citenamefont {{Herrera-Alcantar}}, \citenamefont {{Honscheid}}, \citenamefont {{Hou}}, \citenamefont {{Howlett}}, \citenamefont {{Huterer}}, \citenamefont {{Ir{\v{s}}i{\v{c}}}}, \citenamefont {{Ishak}}, \citenamefont {{Juneau}}, \citenamefont {{Kara{\c{c}}ayl{\i}}}, \citenamefont {{Kehoe}}, \citenamefont {{Kent}}, \citenamefont {{Kirkby}}, \citenamefont {{Kitaura}}, \citenamefont {{Kong}}, \citenamefont {{Kremin}}, \citenamefont {{Krolewski}}, \citenamefont {{Lai}}, \citenamefont {{Lan}}, \citenamefont {{Landriau}}, \citenamefont {{Lang}}, \citenamefont {{Lasker}}, \citenamefont {{Le Goff}}, \citenamefont {{Le Guillou}}, \citenamefont {{Leauthaud}}, \citenamefont {{Levi}}, \citenamefont {{Li}}, \citenamefont {{Lodha}},
  \citenamefont {{Magneville}}, \citenamefont {{Manera}}, \citenamefont {{Margala}}, \citenamefont {{Martini}}, \citenamefont {{Maus}}, \citenamefont {{McDonald}}, \citenamefont {{Medina-Varela}}, \citenamefont {{Meisner}}, \citenamefont {{Mena-Fern{\'a}ndez}}, \citenamefont {{Miquel}}, \citenamefont {{Moon}}, \citenamefont {{Moore}}, \citenamefont {{Moustakas}}, \citenamefont {{Mudur}}, \citenamefont {{Mueller}}, \citenamefont {{Mu{\~n}oz-Guti{\'e}rrez}}, \citenamefont {{Myers}}, \citenamefont {{Nadathur}}, \citenamefont {{Napolitano}}, \citenamefont {{Neveux}}, \citenamefont {{Newman}}, \citenamefont {{Nguyen}}, \citenamefont {{Nie}}, \citenamefont {{Niz}}, \citenamefont {{Noriega}}, \citenamefont {{Padmanabhan}}, \citenamefont {{Paillas}}, \citenamefont {{Palanque-Delabrouille}}, \citenamefont {{Pan}}, \citenamefont {{Penmetsa}}, \citenamefont {{Percival}}, \citenamefont {{Pieri}}, \citenamefont {{Pinon}}, \citenamefont {{Poppett}}, \citenamefont {{Porredon}}, \citenamefont {{Prada}}, \citenamefont
  {{P{\'e}rez-Fern{\'a}ndez}}, \citenamefont {{P{\'e}rez-R{\`a}fols}}, \citenamefont {{Rabinowitz}}, \citenamefont {{Raichoor}}, \citenamefont {{Ram{\'\i}rez-P{\'e}rez}}, \citenamefont {{Ramirez-Solano}}, \citenamefont {{Rashkovetskyi}}, \citenamefont {{Ravoux}}, \citenamefont {{Rezaie}}, \citenamefont {{Rich}}, \citenamefont {{Rocher}}, \citenamefont {{Rockosi}}, \citenamefont {{Roe}}, \citenamefont {{Rosado-Marin}}, \citenamefont {{Ross}}, \citenamefont {{Rossi}}, \citenamefont {{Ruggeri}}, \citenamefont {{Ruhlmann-Kleider}}, \citenamefont {{Samushia}}, \citenamefont {{Sanchez}}, \citenamefont {{Saulder}}, \citenamefont {{Schlafly}}, \citenamefont {{Schlegel}}, \citenamefont {{Scholte}}, \citenamefont {{Schubnell}}, \citenamefont {{Seo}}, \citenamefont {{Sharples}}, \citenamefont {{Silber}}, \citenamefont {{Slosar}}, \citenamefont {{Smith}}, \citenamefont {{Sprayberry}}, \citenamefont {{Tan}}, \citenamefont {{Tarl{\'e}}}, \citenamefont {{Trusov}}, \citenamefont {{Vaisakh}}, \citenamefont {{Valcin}},
  \citenamefont {{Valdes}}, \citenamefont {{Vargas-Maga{\~n}a}}, \citenamefont {{Verde}}, \citenamefont {{Walther}}, \citenamefont {{Wang}}, \citenamefont {{Wang}}, \citenamefont {{Weaver}}, \citenamefont {{Weaverdyck}}, \citenamefont {{Wechsler}}, \citenamefont {{Weinberg}}, \citenamefont {{White}}, \citenamefont {{Wilson}}, \citenamefont {{Yu}}, \citenamefont {{Yu}}, \citenamefont {{Yuan}}, \citenamefont {{Y{\`e}che}}, \citenamefont {{Zaborowski}}, \citenamefont {{Zarrouk}}, \citenamefont {{Zhang}}, \citenamefont {{Zhao}}, \citenamefont {{Zhao}}, \citenamefont {{Zhou}},\ and\ \citenamefont {{Zou}}}]{DESI2024.II.KP3}%
  \BibitemOpen
  \bibfield  {author} {\bibinfo {author} {\bibnamefont {{DESI Collaboration}}}, \bibinfo {author} {\bibfnamefont {A.~G.}\ \bibnamefont {{Adame}}}, \bibinfo {author} {\bibfnamefont {J.}~\bibnamefont {{Aguilar}}}, \bibinfo {author} {\bibfnamefont {S.}~\bibnamefont {{Ahlen}}}, \bibinfo {author} {\bibfnamefont {S.}~\bibnamefont {{Alam}}}, \bibinfo {author} {\bibfnamefont {D.~M.}\ \bibnamefont {{Alexander}}}, \bibinfo {author} {\bibfnamefont {M.}~\bibnamefont {{Alvarez}}}, \bibinfo {author} {\bibfnamefont {O.}~\bibnamefont {{Alves}}}, \bibinfo {author} {\bibfnamefont {A.}~\bibnamefont {{Anand}}}, \bibinfo {author} {\bibfnamefont {U.}~\bibnamefont {{Andrade}}}, \bibinfo {author} {\bibfnamefont {E.}~\bibnamefont {{Armengaud}}}, \bibinfo {author} {\bibfnamefont {S.}~\bibnamefont {{Avila}}}, \bibinfo {author} {\bibfnamefont {A.}~\bibnamefont {{Aviles}}}, \bibinfo {author} {\bibfnamefont {H.}~\bibnamefont {{Awan}}}, \bibinfo {author} {\bibfnamefont {S.}~\bibnamefont {{Bailey}}}, \bibnamefont {and~others},\ }\href
  {https://doi.org/10.48550/arXiv.2411.12020} {\bibfield  {journal} {\bibinfo  {journal} {arXiv e-prints}\ ,\ \bibinfo {eid} {arXiv:2411.12020}} (\bibinfo {year} {2024}{\natexlab{c}})},\ \Eprint {https://arxiv.org/abs/2411.12020} {arXiv:2411.12020 [astro-ph.CO]} \BibitemShut {NoStop}%
\bibitem [{\citenamefont {{DESI Collaboration}}\ \emph {et~al.}(2024{\natexlab{d}})\citenamefont {{DESI Collaboration}}, \citenamefont {Adame}, \citenamefont {Aguilar}, \citenamefont {Ahlen}, \citenamefont {Alam}, \citenamefont {Alexander}, \citenamefont {Alvarez}, \citenamefont {Alves}, \citenamefont {Anand}, \citenamefont {Andrade}, \citenamefont {Armengaud}, \citenamefont {Avila}, \citenamefont {Aviles}, \citenamefont {Awan}, \citenamefont {Bailey}, \citenamefont {Baltay}, \citenamefont {Bault}, \citenamefont {Behera}, \citenamefont {BenZvi}, \citenamefont {Beutler}, \citenamefont {Bianchi}, \citenamefont {Blake}, \citenamefont {Blum}, \citenamefont {Brieden}, \citenamefont {Brodzeller}, \citenamefont {Brooks}, \citenamefont {Buckley-Geer}, \citenamefont {Burtin}, \citenamefont {Calderon}, \citenamefont {Canning}, \citenamefont {Rosell}, \citenamefont {Cereskaite}, \citenamefont {Cervantes-Cota}, \citenamefont {Chabanier}, \citenamefont {Chaussidon}, \citenamefont {Chaves-Montero}, \citenamefont {Chen},
  \citenamefont {Chen}, \citenamefont {Claybaugh}, \citenamefont {Cole}, \citenamefont {Cuceu}, \citenamefont {Davis}, \citenamefont {Dawson}, \citenamefont {de~la Macorra}, \citenamefont {de~Mattia}, \citenamefont {Deiosso}, \citenamefont {Dey}, \citenamefont {Dey}, \citenamefont {Ding}, \citenamefont {Doel}, \citenamefont {Edelstein}, \citenamefont {Eftekharzadeh}, \citenamefont {Eisenstein}, \citenamefont {Elliott}, \citenamefont {Fagrelius}, \citenamefont {Fanning}, \citenamefont {Ferraro}, \citenamefont {Ereza}, \citenamefont {Findlay}, \citenamefont {Flaugher}, \citenamefont {Font-Ribera}, \citenamefont {Forero-Sánchez}, \citenamefont {Forero-Romero}, \citenamefont {Garcia-Quintero}, \citenamefont {Gaztañaga}, \citenamefont {Gil-Marín}, \citenamefont {Gontcho}, \citenamefont {Gonzalez-Morales}, \citenamefont {Gonzalez-Perez}, \citenamefont {Gordon}, \citenamefont {Green}, \citenamefont {Gruen}, \citenamefont {Gsponer}, \citenamefont {Gutierrez}, \citenamefont {Guy}, \citenamefont {Hadzhiyska},
  \citenamefont {Hahn}, \citenamefont {Hanif}, \citenamefont {Herrera-Alcantar}, \citenamefont {Honscheid}, \citenamefont {Howlett}, \citenamefont {Huterer}, \citenamefont {Iršič}, \citenamefont {Ishak}, \citenamefont {Juneau}, \citenamefont {Karaçaylı}, \citenamefont {Kehoe}, \citenamefont {Kent}, \citenamefont {Kirkby}, \citenamefont {Kremin}, \citenamefont {Krolewski}, \citenamefont {Lai}, \citenamefont {Lan}, \citenamefont {Landriau}, \citenamefont {Lang}, \citenamefont {Lasker}, \citenamefont {Goff}, \citenamefont {Guillou}, \citenamefont {Leauthaud}, \citenamefont {Levi}, \citenamefont {Li}, \citenamefont {Linder}, \citenamefont {Lodha}, \citenamefont {Magneville}, \citenamefont {Manera}, \citenamefont {Margala}, \citenamefont {Martini}, \citenamefont {Maus}, \citenamefont {McDonald}, \citenamefont {Medina-Varela}, \citenamefont {Meisner}, \citenamefont {Mena-Fernández}, \citenamefont {Miquel}, \citenamefont {Moon}, \citenamefont {Moore}, \citenamefont {Moustakas}, \citenamefont {Mudur},
  \citenamefont {Mueller}, \citenamefont {Muñoz-Gutiérrez}, \citenamefont {Myers}, \citenamefont {Nadathur}, \citenamefont {Napolitano}, \citenamefont {Neveux}, \citenamefont {Newman}, \citenamefont {Nguyen}, \citenamefont {Nie}, \citenamefont {Niz}, \citenamefont {Noriega}, \citenamefont {Padmanabhan}, \citenamefont {Paillas}, \citenamefont {Palanque-Delabrouille}, \citenamefont {Pan}, \citenamefont {Penmetsa}, \citenamefont {Percival}, \citenamefont {Pieri}, \citenamefont {Pinon}, \citenamefont {Poppett}, \citenamefont {Porredon}, \citenamefont {Prada}, \citenamefont {Pérez-Fernández}, \citenamefont {Pérez-Ràfols}, \citenamefont {Rabinowitz}, \citenamefont {Raichoor}, \citenamefont {Ramírez-Pérez}, \citenamefont {Ramirez-Solano}, \citenamefont {Rashkovetskyi}, \citenamefont {Rezaie}, \citenamefont {Rich}, \citenamefont {Rocher}, \citenamefont {Rockosi}, \citenamefont {Roe}, \citenamefont {Rosado-Marin}, \citenamefont {Ross}, \citenamefont {Rossi}, \citenamefont {Ruggeri}, \citenamefont
  {Ruhlmann-Kleider}, \citenamefont {Samushia}, \citenamefont {Sanchez}, \citenamefont {Saulder}, \citenamefont {Schlafly}, \citenamefont {Schlegel}, \citenamefont {Schubnell}, \citenamefont {Seo}, \citenamefont {Sharples}, \citenamefont {Silber}, \citenamefont {Slosar}, \citenamefont {Smith}, \citenamefont {Sprayberry}, \citenamefont {Swanson}, \citenamefont {Tan}, \citenamefont {Tarlé}, \citenamefont {Trusov}, \citenamefont {Vaisakh}, \citenamefont {Valcin}, \citenamefont {Valdes}, \citenamefont {Vargas-Magaña}, \citenamefont {Verde}, \citenamefont {Walther}, \citenamefont {Wang}, \citenamefont {Wang}, \citenamefont {Weaver}, \citenamefont {Weaverdyck}, \citenamefont {Wechsler}, \citenamefont {Weinberg}, \citenamefont {White}, \citenamefont {Yu}, \citenamefont {Yu}, \citenamefont {Yuan}, \citenamefont {Yèche}, \citenamefont {Zaborowski}, \citenamefont {Zarrouk}, \citenamefont {Zhang}, \citenamefont {Zhao}, \citenamefont {Zhao}, \citenamefont {Zhou},\ and\ \citenamefont {Zou}}]{DESI2024.III.KP4}%
  \BibitemOpen
  \bibfield  {author} {\bibinfo {author} {\bibnamefont {{DESI Collaboration}}}, \bibinfo {author} {\bibfnamefont {A.~G.}\ \bibnamefont {Adame}}, \bibinfo {author} {\bibfnamefont {J.}~\bibnamefont {Aguilar}}, \bibinfo {author} {\bibfnamefont {S.}~\bibnamefont {Ahlen}}, \bibinfo {author} {\bibfnamefont {S.}~\bibnamefont {Alam}}, \bibinfo {author} {\bibfnamefont {D.~M.}\ \bibnamefont {Alexander}}, \bibinfo {author} {\bibfnamefont {M.}~\bibnamefont {Alvarez}}, \bibinfo {author} {\bibfnamefont {O.}~\bibnamefont {Alves}}, \bibinfo {author} {\bibfnamefont {A.}~\bibnamefont {Anand}}, \bibinfo {author} {\bibfnamefont {U.}~\bibnamefont {Andrade}}, \bibinfo {author} {\bibfnamefont {E.}~\bibnamefont {Armengaud}}, \bibinfo {author} {\bibfnamefont {S.}~\bibnamefont {Avila}}, \bibinfo {author} {\bibfnamefont {A.}~\bibnamefont {Aviles}}, \bibinfo {author} {\bibfnamefont {H.}~\bibnamefont {Awan}}, \bibinfo {author} {\bibfnamefont {S.}~\bibnamefont {Bailey}}, \bibnamefont {and~others},\ }\href
  {https://doi.org/10.48550/arXiv.2404.03000} {\bibfield  {journal} {\bibinfo  {journal} {arXiv e-prints}\ ,\ \bibinfo {pages} {arXiv:2404.03000}} (\bibinfo {year} {2024}{\natexlab{d}})},\ \Eprint {https://arxiv.org/abs/2404.03000} {arXiv:2404.03000 [astro-ph.CO]} \BibitemShut {NoStop}%
\bibitem [{\citenamefont {{DESI Collaboration}}\ \emph {et~al.}(2025{\natexlab{d}})\citenamefont {{DESI Collaboration}}, \citenamefont {{Adame}}, \citenamefont {{Aguilar}}, \citenamefont {{Ahlen}}, \citenamefont {{Alam}}, \citenamefont {{Alexander}}, \citenamefont {{Alvarez}}, \citenamefont {{Alves}}, \citenamefont {{Anand}}, \citenamefont {{Andrade}}, \citenamefont {{Armengaud}}, \citenamefont {{Avila}}, \citenamefont {{Aviles}}, \citenamefont {{Awan}}, \citenamefont {{Bailey}}, \citenamefont {{Baltay}}, \citenamefont {{Bault}}, \citenamefont {{Bautista}}, \citenamefont {{Behera}}, \citenamefont {{BenZvi}}, \citenamefont {{Beutler}}, \citenamefont {{Bianchi}}, \citenamefont {{Blake}}, \citenamefont {{Blum}}, \citenamefont {{Brieden}}, \citenamefont {{Brodzeller}}, \citenamefont {{Brooks}}, \citenamefont {{Buckley-Geer}}, \citenamefont {{Burtin}}, \citenamefont {{Calderon}}, \citenamefont {{Canning}}, \citenamefont {{Carnero Rosell}}, \citenamefont {{Cereskaite}}, \citenamefont {{Cervantes-Cota}},
  \citenamefont {{Chabanier}}, \citenamefont {{Chaussidon}}, \citenamefont {{Chaves-Montero}}, \citenamefont {{Chen}}, \citenamefont {{Chen}}, \citenamefont {{Claybaugh}}, \citenamefont {{Cole}}, \citenamefont {{Cuceu}}, \citenamefont {{Davis}}, \citenamefont {{Dawson}}, \citenamefont {{de la Cruz}}, \citenamefont {{de la Macorra}}, \citenamefont {{de Mattia}}, \citenamefont {{Deiosso}}, \citenamefont {{Dey}}, \citenamefont {{Dey}}, \citenamefont {{Ding}}, \citenamefont {{Ding}}, \citenamefont {{Doel}}, \citenamefont {{Edelstein}}, \citenamefont {{Eftekharzadeh}}, \citenamefont {{Eisenstein}}, \citenamefont {{Elliott}}, \citenamefont {{Fagrelius}}, \citenamefont {{Fanning}}, \citenamefont {{Ferraro}}, \citenamefont {{Ereza}}, \citenamefont {{Findlay}}, \citenamefont {{Flaugher}}, \citenamefont {{Font-Ribera}}, \citenamefont {{Forero-S{\'a}nchez}}, \citenamefont {{Forero-Romero}}, \citenamefont {{Garcia-Quintero}}, \citenamefont {{Gazta{\~n}aga}}, \citenamefont {{Gil-Mar{\'\i}n}}, \citenamefont {{Gontcho}},
  \citenamefont {{Gonzalez-Morales}}, \citenamefont {{Gonzalez-Perez}}, \citenamefont {{Gordon}}, \citenamefont {{Green}}, \citenamefont {{Gruen}}, \citenamefont {{Gsponer}}, \citenamefont {{Gutierrez}}, \citenamefont {{Guy}}, \citenamefont {{Hadzhiyska}}, \citenamefont {{Hahn}}, \citenamefont {{Hanif}}, \citenamefont {{Herrera-Alcantar}}, \citenamefont {{Honscheid}}, \citenamefont {{Howlett}}, \citenamefont {{Huterer}}, \citenamefont {{Ir{\v{s}}i{\v{c}}}}, \citenamefont {{Ishak}}, \citenamefont {{Juneau}}, \citenamefont {{Kara{\c{c}}ayl{\i}}}, \citenamefont {{Kehoe}}, \citenamefont {{Kent}}, \citenamefont {{Kirkby}}, \citenamefont {{Kremin}}, \citenamefont {{Krolewski}}, \citenamefont {{Lai}}, \citenamefont {{Lan}}, \citenamefont {{Landriau}}, \citenamefont {{Lang}}, \citenamefont {{Lasker}}, \citenamefont {{Le Goff}}, \citenamefont {{Le Guillou}}, \citenamefont {{Leauthaud}}, \citenamefont {{Levi}}, \citenamefont {{Li}}, \citenamefont {{Linder}}, \citenamefont {{Lodha}}, \citenamefont {{Magneville}},
  \citenamefont {{Manera}}, \citenamefont {{Margala}}, \citenamefont {{Martini}}, \citenamefont {{Maus}}, \citenamefont {{McDonald}}, \citenamefont {{Medina-Varela}}, \citenamefont {{Meisner}}, \citenamefont {{Mena-Fern{\'a}ndez}}, \citenamefont {{Miquel}}, \citenamefont {{Moon}}, \citenamefont {{Moore}}, \citenamefont {{Moustakas}}, \citenamefont {{Mueller}}, \citenamefont {{Mu{\~n}oz-Guti{\'e}rrez}}, \citenamefont {{Myers}}, \citenamefont {{Nadathur}}, \citenamefont {{Napolitano}}, \citenamefont {{Neveux}}, \citenamefont {{Newman}}, \citenamefont {{Nguyen}}, \citenamefont {{Nie}}, \citenamefont {{Niz}}, \citenamefont {{Noriega}}, \citenamefont {{Padmanabhan}}, \citenamefont {{Paillas}}, \citenamefont {{Palanque-Delabrouille}}, \citenamefont {{Pan}}, \citenamefont {{Penmetsa}}, \citenamefont {{Percival}}, \citenamefont {{Pieri}}, \citenamefont {{Pinon}}, \citenamefont {{Poppett}}, \citenamefont {{Porredon}}, \citenamefont {{Prada}}, \citenamefont {{P{\'e}rez-Fern{\'a}ndez}}, \citenamefont
  {{P{\'e}rez-R{\`a}fols}}, \citenamefont {{Rabinowitz}}, \citenamefont {{Raichoor}}, \citenamefont {{Ram{\'\i}rez-P{\'e}rez}}, \citenamefont {{Ramirez-Solano}}, \citenamefont {{Rashkovetskyi}}, \citenamefont {{Ravoux}}, \citenamefont {{Rezaie}}, \citenamefont {{Rich}}, \citenamefont {{Rocher}}, \citenamefont {{Rockosi}}, \citenamefont {{Roe}}, \citenamefont {{Rosado-Marin}}, \citenamefont {{Ross}}, \citenamefont {{Rossi}}, \citenamefont {{Ruggeri}}, \citenamefont {{Ruhlmann-Kleider}}, \citenamefont {{Samushia}}, \citenamefont {{Sanchez}}, \citenamefont {{Saulder}}, \citenamefont {{Schlafly}}, \citenamefont {{Schlegel}}, \citenamefont {{Schubnell}}, \citenamefont {{Seo}}, \citenamefont {{Sharples}}, \citenamefont {{Silber}}, \citenamefont {{Sinigaglia}}, \citenamefont {{Slosar}}, \citenamefont {{Smith}}, \citenamefont {{Sprayberry}}, \citenamefont {{Tan}}, \citenamefont {{Tarl{\'e}}}, \citenamefont {{Trusov}}, \citenamefont {{Vaisakh}}, \citenamefont {{Valcin}}, \citenamefont {{Valdes}}, \citenamefont
  {{Vargas-Maga{\~n}a}}, \citenamefont {{Verde}}, \citenamefont {{Walther}}, \citenamefont {{Wang}}, \citenamefont {{Wang}}, \citenamefont {{Weaver}}, \citenamefont {{Weaverdyck}}, \citenamefont {{Wechsler}}, \citenamefont {{Weinberg}}, \citenamefont {{White}}, \citenamefont {{Yu}}, \citenamefont {{Yu}}, \citenamefont {{Yuan}}, \citenamefont {{Y{\`e}che}}, \citenamefont {{Zaborowski}}, \citenamefont {{Zarrouk}}, \citenamefont {{Zhang}}, \citenamefont {{Zhao}}, \citenamefont {{Zhao}}, \citenamefont {{Zhou}}, \citenamefont {{Zou}},\ and\ \citenamefont {{DESI Collaboration}}}]{DESI2024.IV.KP6}%
  \BibitemOpen
  \bibfield  {author} {\bibinfo {author} {\bibnamefont {{DESI Collaboration}}}, \bibinfo {author} {\bibfnamefont {A.~G.}\ \bibnamefont {{Adame}}}, \bibinfo {author} {\bibfnamefont {J.}~\bibnamefont {{Aguilar}}}, \bibinfo {author} {\bibfnamefont {S.}~\bibnamefont {{Ahlen}}}, \bibinfo {author} {\bibfnamefont {S.}~\bibnamefont {{Alam}}}, \bibinfo {author} {\bibfnamefont {D.~M.}\ \bibnamefont {{Alexander}}}, \bibinfo {author} {\bibfnamefont {M.}~\bibnamefont {{Alvarez}}}, \bibinfo {author} {\bibfnamefont {O.}~\bibnamefont {{Alves}}}, \bibinfo {author} {\bibfnamefont {A.}~\bibnamefont {{Anand}}}, \bibinfo {author} {\bibfnamefont {U.}~\bibnamefont {{Andrade}}}, \bibinfo {author} {\bibfnamefont {E.}~\bibnamefont {{Armengaud}}}, \bibinfo {author} {\bibfnamefont {S.}~\bibnamefont {{Avila}}}, \bibinfo {author} {\bibfnamefont {A.}~\bibnamefont {{Aviles}}}, \bibinfo {author} {\bibfnamefont {H.}~\bibnamefont {{Awan}}}, \bibinfo {author} {\bibfnamefont {S.}~\bibnamefont {{Bailey}}}, \bibnamefont {and~others},\ }\href
  {https://doi.org/10.1088/1475-7516/2025/01/124} {\bibfield  {journal} {\bibinfo  {journal} {\jcap}\ }\textbf {\bibinfo {volume} {2025}},\ \bibinfo {eid} {124} (\bibinfo {year} {2025}{\natexlab{d}})},\ \Eprint {https://arxiv.org/abs/2404.03001} {arXiv:2404.03001 [astro-ph.CO]} \BibitemShut {NoStop}%
\bibitem [{\citenamefont {{DESI Collaboration}}\ \emph {et~al.}(2025{\natexlab{e}})\citenamefont {{DESI Collaboration}}, \citenamefont {{Adame}}, \citenamefont {{Aguilar}}, \citenamefont {{Ahlen}}, \citenamefont {{Alam}}, \citenamefont {{Alexander}}, \citenamefont {{Alvarez}}, \citenamefont {{Alves}}, \citenamefont {{Anand}}, \citenamefont {{Andrade}}, \citenamefont {{Armengaud}}, \citenamefont {{Avila}}, \citenamefont {{Aviles}}, \citenamefont {{Awan}}, \citenamefont {{Bahr-Kalus}}, \citenamefont {{Bailey}}, \citenamefont {{Baltay}}, \citenamefont {{Bault}}, \citenamefont {{Behera}}, \citenamefont {{BenZvi}}, \citenamefont {{Bera}}, \citenamefont {{Beutler}}, \citenamefont {{Bianchi}}, \citenamefont {{Blake}}, \citenamefont {{Blum}}, \citenamefont {{Brieden}}, \citenamefont {{Brodzeller}}, \citenamefont {{Brooks}}, \citenamefont {{Buckley-Geer}}, \citenamefont {{Burtin}}, \citenamefont {{Calderon}}, \citenamefont {{Canning}}, \citenamefont {{Carnero Rosell}}, \citenamefont {{Cereskaite}}, \citenamefont
  {{Cervantes-Cota}}, \citenamefont {{Chabanier}}, \citenamefont {{Chaussidon}}, \citenamefont {{Chaves-Montero}}, \citenamefont {{Chen}}, \citenamefont {{Chen}}, \citenamefont {{Claybaugh}}, \citenamefont {{Cole}}, \citenamefont {{Cuceu}}, \citenamefont {{Davis}}, \citenamefont {{Dawson}}, \citenamefont {{de la Macorra}}, \citenamefont {{de Mattia}}, \citenamefont {{Deiosso}}, \citenamefont {{Dey}}, \citenamefont {{Dey}}, \citenamefont {{Ding}}, \citenamefont {{Doel}}, \citenamefont {{Edelstein}}, \citenamefont {{Eftekharzadeh}}, \citenamefont {{Eisenstein}}, \citenamefont {{Elliott}}, \citenamefont {{Fagrelius}}, \citenamefont {{Fanning}}, \citenamefont {{Ferraro}}, \citenamefont {{Ereza}}, \citenamefont {{Findlay}}, \citenamefont {{Flaugher}}, \citenamefont {{Font-Ribera}}, \citenamefont {{Forero-S{\'a}nchez}}, \citenamefont {{Forero-Romero}}, \citenamefont {{Frenk}}, \citenamefont {{Garcia-Quintero}}, \citenamefont {{Gazta{\~n}aga}}, \citenamefont {{Gil-Mar{\'\i}n}}, \citenamefont {{Gontcho a Gontcho}},
  \citenamefont {{Gonzalez-Morales}}, \citenamefont {{Gonzalez-Perez}}, \citenamefont {{Gordon}}, \citenamefont {{Green}}, \citenamefont {{Gruen}}, \citenamefont {{Gsponer}}, \citenamefont {{Gutierrez}}, \citenamefont {{Guy}}, \citenamefont {{Hadzhiyska}}, \citenamefont {{Hahn}}, \citenamefont {{Hanif}}, \citenamefont {{Herrera-Alcantar}}, \citenamefont {{Honscheid}}, \citenamefont {{Howlett}}, \citenamefont {{Huterer}}, \citenamefont {{Ir{\v{s}}i{\v{c}}}}, \citenamefont {{Ishak}}, \citenamefont {{Juneau}}, \citenamefont {{Kara{\c{c}}ayl{\i}}}, \citenamefont {{Kehoe}}, \citenamefont {{Kent}}, \citenamefont {{Kirkby}}, \citenamefont {{Kremin}}, \citenamefont {{Krolewski}}, \citenamefont {{Lai}}, \citenamefont {{Lan}}, \citenamefont {{Landriau}}, \citenamefont {{Lang}}, \citenamefont {{Lasker}}, \citenamefont {{Le Goff}}, \citenamefont {{Le Guillou}}, \citenamefont {{Leauthaud}}, \citenamefont {{Levi}}, \citenamefont {{Li}}, \citenamefont {{Linder}}, \citenamefont {{Lodha}}, \citenamefont {{Magneville}},
  \citenamefont {{Manera}}, \citenamefont {{Margala}}, \citenamefont {{Martini}}, \citenamefont {{Maus}}, \citenamefont {{McDonald}}, \citenamefont {{Medina-Varela}}, \citenamefont {{Meisner}}, \citenamefont {{Mena-Fern{\'a}ndez}}, \citenamefont {{Miquel}}, \citenamefont {{Moon}}, \citenamefont {{Moore}}, \citenamefont {{Moustakas}}, \citenamefont {{Mueller}}, \citenamefont {{Mu{\~n}oz-Guti{\'e}rrez}}, \citenamefont {{Myers}}, \citenamefont {{Nadathur}}, \citenamefont {{Napolitano}}, \citenamefont {{Neveux}}, \citenamefont {{Newman}}, \citenamefont {{Nguyen}}, \citenamefont {{Nie}}, \citenamefont {{Niz}}, \citenamefont {{Noriega}}, \citenamefont {{Padmanabhan}}, \citenamefont {{Paillas}}, \citenamefont {{Palanque-Delabrouille}}, \citenamefont {{Pan}}, \citenamefont {{Penmetsa}}, \citenamefont {{Percival}}, \citenamefont {{Pieri}}, \citenamefont {{Pinon}}, \citenamefont {{Poppett}}, \citenamefont {{Porredon}}, \citenamefont {{Prada}}, \citenamefont {{P{\'e}rez-Fern{\'a}ndez}}, \citenamefont
  {{P{\'e}rez-R{\`a}fols}}, \citenamefont {{Rabinowitz}}, \citenamefont {{Raichoor}}, \citenamefont {{Ram{\'\i}rez-P{\'e}rez}}, \citenamefont {{Ramirez-Solano}}, \citenamefont {{Rashkovetskyi}}, \citenamefont {{Ravoux}}, \citenamefont {{Rezaie}}, \citenamefont {{Rich}}, \citenamefont {{Rocher}}, \citenamefont {{Rockosi}}, \citenamefont {{Roe}}, \citenamefont {{Rosado-Marin}}, \citenamefont {{Ross}}, \citenamefont {{Rossi}}, \citenamefont {{Ruggeri}}, \citenamefont {{Ruhlmann-Kleider}}, \citenamefont {{Samushia}}, \citenamefont {{Sanchez}}, \citenamefont {{Saulder}}, \citenamefont {{Schlafly}}, \citenamefont {{Schlegel}}, \citenamefont {{Schubnell}}, \citenamefont {{Seo}}, \citenamefont {{Shafieloo}}, \citenamefont {{Sharples}}, \citenamefont {{Silber}}, \citenamefont {{Slosar}}, \citenamefont {{Smith}}, \citenamefont {{Sprayberry}}, \citenamefont {{Tan}}, \citenamefont {{Tarl{\'e}}}, \citenamefont {{Taylor}}, \citenamefont {{Trusov}}, \citenamefont {{Ure{\~n}a-L{\'o}pez}}, \citenamefont {{Vaisakh}},
  \citenamefont {{Valcin}}, \citenamefont {{Valdes}}, \citenamefont {{Vargas-Maga{\~n}a}}, \citenamefont {{Verde}}, \citenamefont {{Walther}}, \citenamefont {{Wang}}, \citenamefont {{Wang}}, \citenamefont {{Weaver}}, \citenamefont {{Weaverdyck}}, \citenamefont {{Wechsler}}, \citenamefont {{Weinberg}}, \citenamefont {{White}}, \citenamefont {{Yu}}, \citenamefont {{Yu}}, \citenamefont {{Yuan}}, \citenamefont {{Y{\`e}che}}, \citenamefont {{Zaborowski}}, \citenamefont {{Zarrouk}}, \citenamefont {{Zhang}}, \citenamefont {{Zhao}}, \citenamefont {{Zhao}}, \citenamefont {{Zhou}},\ and\ \citenamefont {{Zhuang}}}]{DESI2024.VI.KP7A}%
  \BibitemOpen
  \bibfield  {author} {\bibinfo {author} {\bibnamefont {{DESI Collaboration}}}, \bibinfo {author} {\bibfnamefont {A.~G.}\ \bibnamefont {{Adame}}}, \bibinfo {author} {\bibfnamefont {J.}~\bibnamefont {{Aguilar}}}, \bibinfo {author} {\bibfnamefont {S.}~\bibnamefont {{Ahlen}}}, \bibinfo {author} {\bibfnamefont {S.}~\bibnamefont {{Alam}}}, \bibinfo {author} {\bibfnamefont {D.~M.}\ \bibnamefont {{Alexander}}}, \bibinfo {author} {\bibfnamefont {M.}~\bibnamefont {{Alvarez}}}, \bibinfo {author} {\bibfnamefont {O.}~\bibnamefont {{Alves}}}, \bibinfo {author} {\bibfnamefont {A.}~\bibnamefont {{Anand}}}, \bibinfo {author} {\bibfnamefont {U.}~\bibnamefont {{Andrade}}}, \bibinfo {author} {\bibfnamefont {E.}~\bibnamefont {{Armengaud}}}, \bibinfo {author} {\bibfnamefont {S.}~\bibnamefont {{Avila}}}, \bibinfo {author} {\bibfnamefont {A.}~\bibnamefont {{Aviles}}}, \bibinfo {author} {\bibfnamefont {H.}~\bibnamefont {{Awan}}}, \bibinfo {author} {\bibfnamefont {B.}~\bibnamefont {{Bahr-Kalus}}}, \bibnamefont {and~others},\ }\href
  {https://doi.org/10.1088/1475-7516/2025/02/021} {\bibfield  {journal} {\bibinfo  {journal} {\jcap}\ }\textbf {\bibinfo {volume} {2025}},\ \bibinfo {eid} {021} (\bibinfo {year} {2025}{\natexlab{e}})},\ \Eprint {https://arxiv.org/abs/2404.03002} {arXiv:2404.03002 [astro-ph.CO]} \BibitemShut {NoStop}%
\bibitem [{\citenamefont {{DESI Collaboration}}\ \emph {et~al.}(2024{\natexlab{e}})\citenamefont {{DESI Collaboration}}, \citenamefont {{Adame}}, \citenamefont {{Aguilar}}, \citenamefont {{Ahlen}}, \citenamefont {{Alam}}, \citenamefont {{Alexander}}, \citenamefont {{Allende Prieto}}, \citenamefont {{Alvarez}}, \citenamefont {{Alves}}, \citenamefont {{Anand}}, \citenamefont {{Andrade}}, \citenamefont {{Armengaud}}, \citenamefont {{Avila}}, \citenamefont {{Aviles}}, \citenamefont {{Awan}}, \citenamefont {{Bahr-Kalus}}, \citenamefont {{Bailey}}, \citenamefont {{Baltay}}, \citenamefont {{Bault}}, \citenamefont {{Behera}}, \citenamefont {{BenZvi}}, \citenamefont {{Beutler}}, \citenamefont {{Bianchi}}, \citenamefont {{Blake}}, \citenamefont {{Blum}}, \citenamefont {{Bonici}}, \citenamefont {{Brieden}}, \citenamefont {{Brodzeller}}, \citenamefont {{Brooks}}, \citenamefont {{Buckley-Geer}}, \citenamefont {{Burtin}}, \citenamefont {{Calderon}}, \citenamefont {{Canning}}, \citenamefont {{Carnero Rosell}}, \citenamefont
  {{Cereskaite}}, \citenamefont {{Cervantes-Cota}}, \citenamefont {{Chabanier}}, \citenamefont {{Chaussidon}}, \citenamefont {{Chaves-Montero}}, \citenamefont {{Chebat}}, \citenamefont {{Chen}}, \citenamefont {{Chen}}, \citenamefont {{Claybaugh}}, \citenamefont {{Cole}}, \citenamefont {{Cuceu}}, \citenamefont {{Davis}}, \citenamefont {{Dawson}}, \citenamefont {{de la Macorra}}, \citenamefont {{de Mattia}}, \citenamefont {{Deiosso}}, \citenamefont {{Dey}}, \citenamefont {{Dey}}, \citenamefont {{Ding}}, \citenamefont {{Doel}}, \citenamefont {{Edelstein}}, \citenamefont {{Eftekharzadeh}}, \citenamefont {{Eisenstein}}, \citenamefont {{Elbers}}, \citenamefont {{Elliott}}, \citenamefont {{Fagrelius}}, \citenamefont {{Fanning}}, \citenamefont {{Ferraro}}, \citenamefont {{Ereza}}, \citenamefont {{Findlay}}, \citenamefont {{Flaugher}}, \citenamefont {{Font-Ribera}}, \citenamefont {{Forero-S{\'a}nchez}}, \citenamefont {{Forero-Romero}}, \citenamefont {{Frenk}}, \citenamefont {{Garcia-Quintero}}, \citenamefont
  {{Garrison}}, \citenamefont {{Gazta{\~n}aga}}, \citenamefont {{Gil-Mar{\'\i}n}}, \citenamefont {{Gontcho}}, \citenamefont {{Gonzalez-Morales}}, \citenamefont {{Gonzalez-Perez}}, \citenamefont {{Gordon}}, \citenamefont {{Green}}, \citenamefont {{Gruen}}, \citenamefont {{Gsponer}}, \citenamefont {{Gutierrez}}, \citenamefont {{Guy}}, \citenamefont {{Hadzhiyska}}, \citenamefont {{Hahn}}, \citenamefont {{Hanif}}, \citenamefont {{Herrera-Alcantar}}, \citenamefont {{Honscheid}}, \citenamefont {{Howlett}}, \citenamefont {{Huterer}}, \citenamefont {{Ir{\v{s}}i{\v{c}}}}, \citenamefont {{Ishak}}, \citenamefont {{Joyce}}, \citenamefont {{Juneau}}, \citenamefont {{Kara{\c{c}}ayl{\i}}}, \citenamefont {{Kehoe}}, \citenamefont {{Kent}}, \citenamefont {{Kirkby}}, \citenamefont {{Kong}}, \citenamefont {{Koposov}}, \citenamefont {{Kremin}}, \citenamefont {{Krolewski}}, \citenamefont {{Lahav}}, \citenamefont {{Lai}}, \citenamefont {{Lan}}, \citenamefont {{Landriau}}, \citenamefont {{Lang}}, \citenamefont {{Lasker}},
  \citenamefont {{Le Goff}}, \citenamefont {{Le Guillou}}, \citenamefont {{Leauthaud}}, \citenamefont {{Levi}}, \citenamefont {{Li}}, \citenamefont {{Lodha}}, \citenamefont {{Magneville}}, \citenamefont {{Manera}}, \citenamefont {{Margala}}, \citenamefont {{Martini}}, \citenamefont {{Matthewson}}, \citenamefont {{Maus}}, \citenamefont {{McDonald}}, \citenamefont {{Medina-Varela}}, \citenamefont {{Meisner}}, \citenamefont {{Mena-Fern{\'a}ndez}}, \citenamefont {{Miquel}}, \citenamefont {{Moon}}, \citenamefont {{Moore}}, \citenamefont {{Moustakas}}, \citenamefont {{Mudur}}, \citenamefont {{Mueller}}, \citenamefont {{Mu{\~n}oz-Guti{\'e}rrez}}, \citenamefont {{Myers}}, \citenamefont {{Nadathur}}, \citenamefont {{Napolitano}}, \citenamefont {{Neveux}}, \citenamefont {{Newman}}, \citenamefont {{Nguyen}}, \citenamefont {{Nie}}, \citenamefont {{Niz}}, \citenamefont {{Noriega}}, \citenamefont {{Padmanabhan}}, \citenamefont {{Paillas}}, \citenamefont {{Palanque-Delabrouille}}, \citenamefont {{Pan}}, \citenamefont
  {{Penmetsa}}, \citenamefont {{Percival}}, \citenamefont {{Pieri}}, \citenamefont {{Pinon}}, \citenamefont {{Poppett}}, \citenamefont {{Porredon}}, \citenamefont {{Prada}}, \citenamefont {{P{\'e}rez-Fern{\'a}ndez}}, \citenamefont {{P{\'e}rez-R{\`a}fols}}, \citenamefont {{Rabinowitz}}, \citenamefont {{Raichoor}}, \citenamefont {{Ram{\'\i}rez-P{\'e}rez}}, \citenamefont {{Ramirez-Solano}}, \citenamefont {{Rashkovetskyi}}, \citenamefont {{Ravoux}}, \citenamefont {{Rezaie}}, \citenamefont {{Rich}}, \citenamefont {{Rocher}}, \citenamefont {{Rockosi}}, \citenamefont {{Roe}}, \citenamefont {{Rosado-Marin}}, \citenamefont {{Ross}}, \citenamefont {{Rossi}}, \citenamefont {{Ruggeri}}, \citenamefont {{Ruhlmann-Kleider}}, \citenamefont {{Samushia}}, \citenamefont {{Sanchez}}, \citenamefont {{Saulder}}, \citenamefont {{Schlafly}}, \citenamefont {{Schlegel}}, \citenamefont {{Schubnell}}, \citenamefont {{Seo}}, \citenamefont {{Shafieloo}}, \citenamefont {{Sharples}}, \citenamefont {{Silber}}, \citenamefont {{Slosar}},
  \citenamefont {{Smith}}, \citenamefont {{Sprayberry}}, \citenamefont {{Tan}}, \citenamefont {{Tarl{\'e}}}, \citenamefont {{Taylor}}, \citenamefont {{Trusov}}, \citenamefont {{Vaisakh}}, \citenamefont {{Valcin}}, \citenamefont {{Valdes}}, \citenamefont {{Valogiannis}}, \citenamefont {{Vargas-Maga{\~n}a}}, \citenamefont {{Verde}}, \citenamefont {{Walther}}, \citenamefont {{Wang}}, \citenamefont {{Wang}}, \citenamefont {{Weaver}}, \citenamefont {{Weaverdyck}}, \citenamefont {{Wechsler}}, \citenamefont {{Weinberg}}, \citenamefont {{White}}, \citenamefont {{Wilson}}, \citenamefont {{Yi}}, \citenamefont {{Yu}}, \citenamefont {{Yu}}, \citenamefont {{Yuan}}, \citenamefont {{Y{\`e}che}}, \citenamefont {{Zaborowski}}, \citenamefont {{Zarrouk}}, \citenamefont {{Zhang}}, \citenamefont {{Zhao}}, \citenamefont {{Zhao}}, \citenamefont {{Zhou}}, \citenamefont {{Zhuang}},\ and\ \citenamefont {{Zou}}}]{DESI2024.VII.KP7B}%
  \BibitemOpen
  \bibfield  {author} {\bibinfo {author} {\bibnamefont {{DESI Collaboration}}}, \bibinfo {author} {\bibfnamefont {A.~G.}\ \bibnamefont {{Adame}}}, \bibinfo {author} {\bibfnamefont {J.}~\bibnamefont {{Aguilar}}}, \bibinfo {author} {\bibfnamefont {S.}~\bibnamefont {{Ahlen}}}, \bibinfo {author} {\bibfnamefont {S.}~\bibnamefont {{Alam}}}, \bibinfo {author} {\bibfnamefont {D.~M.}\ \bibnamefont {{Alexander}}}, \bibinfo {author} {\bibfnamefont {C.}~\bibnamefont {{Allende Prieto}}}, \bibinfo {author} {\bibfnamefont {M.}~\bibnamefont {{Alvarez}}}, \bibinfo {author} {\bibfnamefont {O.}~\bibnamefont {{Alves}}}, \bibinfo {author} {\bibfnamefont {A.}~\bibnamefont {{Anand}}}, \bibinfo {author} {\bibfnamefont {U.}~\bibnamefont {{Andrade}}}, \bibinfo {author} {\bibfnamefont {E.}~\bibnamefont {{Armengaud}}}, \bibinfo {author} {\bibfnamefont {S.}~\bibnamefont {{Avila}}}, \bibinfo {author} {\bibfnamefont {A.}~\bibnamefont {{Aviles}}}, \bibinfo {author} {\bibfnamefont {H.}~\bibnamefont {{Awan}}}, \bibnamefont {and~others},\
  }\href {https://doi.org/10.48550/arXiv.2411.12022} {\bibfield  {journal} {\bibinfo  {journal} {arXiv e-prints}\ ,\ \bibinfo {eid} {arXiv:2411.12022}} (\bibinfo {year} {2024}{\natexlab{e}})},\ \Eprint {https://arxiv.org/abs/2411.12022} {arXiv:2411.12022 [astro-ph.CO]} \BibitemShut {NoStop}%
\bibitem [{\citenamefont {{Chevallier}}\ and\ \citenamefont {{Polarski}}(2001)}]{Chevallier:2001}%
  \BibitemOpen
  \bibfield  {author} {\bibinfo {author} {\bibfnamefont {M.}~\bibnamefont {{Chevallier}}}\ \bibnamefont {and}\ \bibinfo {author} {\bibfnamefont {D.}~\bibnamefont {{Polarski}}},\ }\href {https://doi.org/10.1142/S0218271801000822} {\bibfield  {journal} {\bibinfo  {journal} {International Journal of Modern Physics D}\ }\textbf {\bibinfo {volume} {10}},\ \bibinfo {pages} {213} (\bibinfo {year} {2001})},\ \Eprint {https://arxiv.org/abs/gr-qc/0009008} {arXiv:gr-qc/0009008 [gr-qc]} \BibitemShut {NoStop}%
\bibitem [{\citenamefont {{Linder}}(2003)}]{Linder2003}%
  \BibitemOpen
  \bibfield  {author} {\bibinfo {author} {\bibfnamefont {E.~V.}\ \bibnamefont {{Linder}}},\ }\href {https://doi.org/10.1103/PhysRevLett.90.091301} {\bibfield  {journal} {\bibinfo  {journal} {\prl}\ }\textbf {\bibinfo {volume} {90}},\ \bibinfo {eid} {091301} (\bibinfo {year} {2003})},\ \Eprint {https://arxiv.org/abs/astro-ph/0208512} {arXiv:astro-ph/0208512 [astro-ph]} \BibitemShut {NoStop}%
\bibitem [{\citenamefont {{Lodha}}\ \emph {et~al.}(2025)\citenamefont {{Lodha}}, \citenamefont {{Calderon}}, \citenamefont {{Matthewson}}, \citenamefont {{Shafieloo}}, \citenamefont {{Ishak}}, \citenamefont {{Pan}}, \citenamefont {{Garcia-Quintero}}, \citenamefont {{Huterer}}, \citenamefont {{Valogiannis}}, \citenamefont {{Ure{\~n}a-L{\'o}pez}}, \citenamefont {{Kamble}}, \citenamefont {{Parkinson}}, \citenamefont {{Kim}}, \citenamefont {{Zhao}}, \citenamefont {{Cervantes-Cota}}, \citenamefont {{Rohlf}}, \citenamefont {{Lozano-Rodr{\'\i}guez}}, \citenamefont {{Rom{\'a}n-Herrera}}, \citenamefont {{Abdul-Karim}}, \citenamefont {{Aguilar}}, \citenamefont {{Ahlen}}, \citenamefont {{Alves}}, \citenamefont {{Andrade}}, \citenamefont {{Armengaud}}, \citenamefont {{Aviles}}, \citenamefont {{BenZvi}}, \citenamefont {{Bianchi}}, \citenamefont {{Brodzeller}}, \citenamefont {{Brooks}}, \citenamefont {{Burtin}}, \citenamefont {{Canning}}, \citenamefont {{Carnero Rosell}}, \citenamefont {{Casas}}, \citenamefont
  {{Castander}}, \citenamefont {{Charles}}, \citenamefont {{Chaussidon}}, \citenamefont {{Chaves-Montero}}, \citenamefont {{Chebat}}, \citenamefont {{Claybaugh}}, \citenamefont {{Cole}}, \citenamefont {{Cuceu}}, \citenamefont {{Dawson}}, \citenamefont {{de la Macorra}}, \citenamefont {{de Mattia}}, \citenamefont {{Deiosso}}, \citenamefont {{Demina}}, \citenamefont {{Dey}}, \citenamefont {{Dey}}, \citenamefont {{Ding}}, \citenamefont {{Doel}}, \citenamefont {{Eisenstein}}, \citenamefont {{Elbers}}, \citenamefont {{Ferraro}}, \citenamefont {{Font-Ribera}}, \citenamefont {{Forero-Romero}}, \citenamefont {{Garrison}}, \citenamefont {{Gazta{\~n}aga}}, \citenamefont {{Gil-Mar{\'\i}n}}, \citenamefont {{Gontcho}}, \citenamefont {{Gonzalez-Morales}}, \citenamefont {{Gutierrez}}, \citenamefont {{Guy}}, \citenamefont {{Hahn}}, \citenamefont {{Herbold}}, \citenamefont {{Herrera-Alcantar}}, \citenamefont {{Honscheid}}, \citenamefont {{Howlett}}, \citenamefont {{Juneau}}, \citenamefont {{Kehoe}}, \citenamefont {{Kirkby}},
  \citenamefont {{Kisner}}, \citenamefont {{Kremin}}, \citenamefont {{Lahav}}, \citenamefont {{Lamman}}, \citenamefont {{Landriau}}, \citenamefont {{Le Guillou}}, \citenamefont {{Leauthaud}}, \citenamefont {{Levi}}, \citenamefont {{Li}}, \citenamefont {{Magneville}}, \citenamefont {{Manera}}, \citenamefont {{Martini}}, \citenamefont {{Meisner}}, \citenamefont {{Mena-Fern{\'a}ndez}}, \citenamefont {{Miquel}}, \citenamefont {{Moustakas}}, \citenamefont {{Mu{\~n}oz Santos}}, \citenamefont {{Mu{\~n}oz-Guti{\'e}rrez}}, \citenamefont {{Myers}}, \citenamefont {{Nadathur}}, \citenamefont {{Niz}}, \citenamefont {{Noriega}}, \citenamefont {{Paillas}}, \citenamefont {{Palanque-Delabrouille}}, \citenamefont {{Percival}}, \citenamefont {{Pieri}}, \citenamefont {{Poppett}}, \citenamefont {{Prada}}, \citenamefont {{P{\'e}rez-Fern{\'a}ndez}}, \citenamefont {{P{\'e}rez-R{\`a}fols}}, \citenamefont {{Ram{\'\i}rez-P{\'e}rez}}, \citenamefont {{Rashkovetskyi}}, \citenamefont {{Ravoux}}, \citenamefont {{Ross}}, \citenamefont
  {{Rossi}}, \citenamefont {{Ruhlmann-Kleider}}, \citenamefont {{Samushia}}, \citenamefont {{Sanchez}}, \citenamefont {{Schlegel}}, \citenamefont {{Schubnell}}, \citenamefont {{Seo}}, \citenamefont {{Sinigaglia}}, \citenamefont {{Sprayberry}}, \citenamefont {{Tan}}, \citenamefont {{Tarl{\'e}}}, \citenamefont {{Taylor}}, \citenamefont {{Turner}}, \citenamefont {{Vargas-Maga{\~n}a}}, \citenamefont {{Walther}}, \citenamefont {{Weaver}}, \citenamefont {{Wolfson}}, \citenamefont {{Y{\`e}che}}, \citenamefont {{Zarrouk}}, \citenamefont {{Zhou}},\ and\ \citenamefont {{Zou}}}]{Y3.cpe-s1.Lodha.2025}%
  \BibitemOpen
  \bibfield  {author} {\bibinfo {author} {\bibfnamefont {K.}~\bibnamefont {{Lodha}}}, \bibinfo {author} {\bibfnamefont {R.}~\bibnamefont {{Calderon}}}, \bibinfo {author} {\bibfnamefont {W.~L.}\ \bibnamefont {{Matthewson}}}, \bibinfo {author} {\bibfnamefont {A.}~\bibnamefont {{Shafieloo}}}, \bibinfo {author} {\bibfnamefont {M.}~\bibnamefont {{Ishak}}}, \bibinfo {author} {\bibfnamefont {J.}~\bibnamefont {{Pan}}}, \bibinfo {author} {\bibfnamefont {C.}~\bibnamefont {{Garcia-Quintero}}}, \bibinfo {author} {\bibfnamefont {D.}~\bibnamefont {{Huterer}}}, \bibinfo {author} {\bibfnamefont {G.}~\bibnamefont {{Valogiannis}}}, \bibinfo {author} {\bibfnamefont {L.~A.}\ \bibnamefont {{Ure{\~n}a-L{\'o}pez}}}, \bibinfo {author} {\bibfnamefont {N.~V.}\ \bibnamefont {{Kamble}}}, \bibinfo {author} {\bibfnamefont {D.}~\bibnamefont {{Parkinson}}}, \bibinfo {author} {\bibfnamefont {A.~G.}\ \bibnamefont {{Kim}}}, \bibinfo {author} {\bibfnamefont {G.~B.}\ \bibnamefont {{Zhao}}}, \bibinfo {author} {\bibfnamefont {J.~L.}\ \bibnamefont
  {{Cervantes-Cota}}}, \bibnamefont {and~others},\ }\href {https://doi.org/10.48550/arXiv.2503.14743} {\bibfield  {journal} {\bibinfo  {journal} {arXiv e-prints}\ ,\ \bibinfo {eid} {arXiv:2503.14743}} (\bibinfo {year} {2025})},\ \Eprint {https://arxiv.org/abs/2503.14743} {arXiv:2503.14743 [astro-ph.CO]} \BibitemShut {NoStop}%
\bibitem [{\citenamefont {{DES Collaboration}}\ \emph {et~al.}(2025)\citenamefont {{DES Collaboration}}, \citenamefont {{Abbott}}, \citenamefont {{Acevedo}}, \citenamefont {{Adamow}}, \citenamefont {{Aguena}}, \citenamefont {{Alarcon}}, \citenamefont {{Allam}}, \citenamefont {{Alves}}, \citenamefont {{Andrade-Oliveira}}, \citenamefont {{Annis}}, \citenamefont {{Armstrong}}, \citenamefont {{Avila}}, \citenamefont {{Bacon}}, \citenamefont {{Bechtol}}, \citenamefont {{Blazek}}, \citenamefont {{Bocquet}}, \citenamefont {{Brooks}}, \citenamefont {{Brout}}, \citenamefont {{Burke}}, \citenamefont {{Camacho}}, \citenamefont {{Camilleri}}, \citenamefont {{Campailla}}, \citenamefont {{Carnero Rosell}}, \citenamefont {{Carr}}, \citenamefont {{Carretero}}, \citenamefont {{Castander}}, \citenamefont {{Cawthon}}, \citenamefont {{Chan}}, \citenamefont {{Chang}}, \citenamefont {{Chen}}, \citenamefont {{Conselice}}, \citenamefont {{Costanzi}}, \citenamefont {{Crocce}}, \citenamefont {{da Costa}}, \citenamefont {{Pereira}},
  \citenamefont {{Davis}}, \citenamefont {{De Vicente}}, \citenamefont {{Deiosso}}, \citenamefont {{Desai}}, \citenamefont {{Diehl}}, \citenamefont {{Dodelson}}, \citenamefont {{Doux}}, \citenamefont {{Drlica-Wagner}}, \citenamefont {{Elvin-Poole}}, \citenamefont {{Everett}}, \citenamefont {{Ferrero}}, \citenamefont {{Fert{\'e}}}, \citenamefont {{Flaugher}}, \citenamefont {{Frieman}}, \citenamefont {{Galbany}}, \citenamefont {{Garc{\'\i}a-Bellido}}, \citenamefont {{Gatti}}, \citenamefont {{Gaztanaga}}, \citenamefont {{Giannini}}, \citenamefont {{Gruen}}, \citenamefont {{Gruendl}}, \citenamefont {{Gutierrez}}, \citenamefont {{Hartley}}, \citenamefont {{Herner}}, \citenamefont {{Hinton}}, \citenamefont {{Hollowood}}, \citenamefont {{Honscheid}}, \citenamefont {{Huterer}}, \citenamefont {{James}}, \citenamefont {{Jeffrey}}, \citenamefont {{Jeltema}}, \citenamefont {{Kessler}}, \citenamefont {{Lahav}}, \citenamefont {{Lee}}, \citenamefont {{Lee}}, \citenamefont {{Lidman}}, \citenamefont {{Lin}}, \citenamefont
  {{Lin}}, \citenamefont {{Marshall}}, \citenamefont {{Mena-Fern{\'a}ndez}}, \citenamefont {{Miquel}}, \citenamefont {{Muir}}, \citenamefont {{M{\"o}ller}}, \citenamefont {{Nichol}}, \citenamefont {{Palmese}}, \citenamefont {{Paterno}}, \citenamefont {{Percival}}, \citenamefont {{Pieres}}, \citenamefont {{Plazas Malag{\'o}n}}, \citenamefont {{Popovic}}, \citenamefont {{Porredon}}, \citenamefont {{Prat}}, \citenamefont {{Qu}}, \citenamefont {{Raveri}}, \citenamefont {{Rodriguez-Monroy}}, \citenamefont {{Romer}}, \citenamefont {{Rykoff}}, \citenamefont {{Sako}}, \citenamefont {{Samuroff}}, \citenamefont {{Sanchez}}, \citenamefont {{Sanchez Cid}}, \citenamefont {{Scolnic}}, \citenamefont {{Sevilla-Noarbe}}, \citenamefont {{Shah}}, \citenamefont {{Sheldon}}, \citenamefont {{Smith}}, \citenamefont {{Suchyta}}, \citenamefont {{Sullivan}}, \citenamefont {{Swanson}}, \citenamefont {{S{\'a}nchez}}, \citenamefont {{Tarle}}, \citenamefont {{Taylor}}, \citenamefont {{Thomas}}, \citenamefont {{To}}, \citenamefont
  {{Toribio San Cipriano}}, \citenamefont {{Toy}}, \citenamefont {{Troxel}}, \citenamefont {{Tucker}}, \citenamefont {{Vikram}}, \citenamefont {{Vincenzi}}, \citenamefont {{Walker}}, \citenamefont {{Weaverdyck}}, \citenamefont {{Weller}}, \citenamefont {{Wiseman}}, \citenamefont {{Yamamoto}},\ and\ \citenamefont {{Yanny}}}]{DES:2025bxy}%
  \BibitemOpen
  \bibfield  {author} {\bibinfo {author} {\bibnamefont {{DES Collaboration}}}, \bibinfo {author} {\bibfnamefont {T.~M.~C.}\ \bibnamefont {{Abbott}}}, \bibinfo {author} {\bibfnamefont {M.}~\bibnamefont {{Acevedo}}}, \bibinfo {author} {\bibfnamefont {M.}~\bibnamefont {{Adamow}}}, \bibinfo {author} {\bibfnamefont {M.}~\bibnamefont {{Aguena}}}, \bibinfo {author} {\bibfnamefont {A.}~\bibnamefont {{Alarcon}}}, \bibinfo {author} {\bibfnamefont {S.}~\bibnamefont {{Allam}}}, \bibinfo {author} {\bibfnamefont {O.}~\bibnamefont {{Alves}}}, \bibinfo {author} {\bibfnamefont {F.}~\bibnamefont {{Andrade-Oliveira}}}, \bibinfo {author} {\bibfnamefont {J.}~\bibnamefont {{Annis}}}, \bibinfo {author} {\bibfnamefont {P.}~\bibnamefont {{Armstrong}}}, \bibinfo {author} {\bibfnamefont {S.}~\bibnamefont {{Avila}}}, \bibinfo {author} {\bibfnamefont {D.}~\bibnamefont {{Bacon}}}, \bibinfo {author} {\bibfnamefont {K.}~\bibnamefont {{Bechtol}}}, \bibinfo {author} {\bibfnamefont {J.}~\bibnamefont {{Blazek}}}, \bibnamefont {and~others},\
  }\href {https://doi.org/10.48550/arXiv.2503.06712} {\bibfield  {journal} {\bibinfo  {journal} {arXiv e-prints}\ ,\ \bibinfo {eid} {arXiv:2503.06712}} (\bibinfo {year} {2025})},\ \Eprint {https://arxiv.org/abs/2503.06712} {arXiv:2503.06712 [astro-ph.CO]} \BibitemShut {NoStop}%
\bibitem [{\citenamefont {{Louis}}\ \emph {et~al.}(2025)\citenamefont {{Louis}}, \citenamefont {{La Posta}}, \citenamefont {{Atkins}}, \citenamefont {{Jense}}, \citenamefont {{Abril-Cabezas}}, \citenamefont {{Addison}}, \citenamefont {{Ade}}, \citenamefont {{Aiola}}, \citenamefont {{Alford}}, \citenamefont {{Alonso}}, \citenamefont {{Amiri}}, \citenamefont {{An}}, \citenamefont {{Austermann}}, \citenamefont {{Barbavara}}, \citenamefont {{Battaglia}}, \citenamefont {{Battistelli}}, \citenamefont {{Beall}}, \citenamefont {{Bean}}, \citenamefont {{Beheshti}}, \citenamefont {{Beringue}}, \citenamefont {{Bhandarkar}}, \citenamefont {{Biermann}}, \citenamefont {{Bolliet}}, \citenamefont {{Bond}}, \citenamefont {{Calabrese}}, \citenamefont {{Capalbo}}, \citenamefont {{Carrero}}, \citenamefont {{Chen}}, \citenamefont {{Chesmore}}, \citenamefont {{Cho}}, \citenamefont {{Choi}}, \citenamefont {{Clark}}, \citenamefont {{Cothard}}, \citenamefont {{Coughlin}}, \citenamefont {{Coulton}}, \citenamefont {{Crichton}},
  \citenamefont {{Crowley}}, \citenamefont {{Darwish}}, \citenamefont {{Devlin}}, \citenamefont {{Dicker}}, \citenamefont {{Duell}}, \citenamefont {{Duff}}, \citenamefont {{Duivenvoorden}}, \citenamefont {{Dunkley}}, \citenamefont {{Dunner}}, \citenamefont {{Embil Villagra}}, \citenamefont {{Fankhanel}}, \citenamefont {{Farren}}, \citenamefont {{Ferraro}}, \citenamefont {{Foster}}, \citenamefont {{Freundt}}, \citenamefont {{Fuzia}}, \citenamefont {{Gallardo}}, \citenamefont {{Garrido}}, \citenamefont {{Gerbino}}, \citenamefont {{Giardiello}}, \citenamefont {{Gill}}, \citenamefont {{Givans}}, \citenamefont {{Gluscevic}}, \citenamefont {{Goldstein}}, \citenamefont {{Golec}}, \citenamefont {{Gong}}, \citenamefont {{Guan}}, \citenamefont {{Halpern}}, \citenamefont {{Harrison}}, \citenamefont {{Hasselfield}}, \citenamefont {{Healy}}, \citenamefont {{Henderson}}, \citenamefont {{Hensley}}, \citenamefont {{Herv{\'\i}as-Caimapo}}, \citenamefont {{Hill}}, \citenamefont {{Hilton}}, \citenamefont {{Hilton}},
  \citenamefont {{Hincks}}, \citenamefont {{Hlo{\v{z}}ek}}, \citenamefont {{Ho}}, \citenamefont {{Hood}}, \citenamefont {{Hornecker}}, \citenamefont {{Huber}}, \citenamefont {{Hubmayr}}, \citenamefont {{Huffenberger}}, \citenamefont {{Hughes}}, \citenamefont {{Ikape}}, \citenamefont {{Irwin}}, \citenamefont {{Isopi}}, \citenamefont {{Joshi}}, \citenamefont {{Keller}}, \citenamefont {{Kim}}, \citenamefont {{Knowles}}, \citenamefont {{Koopman}}, \citenamefont {{Kosowsky}}, \citenamefont {{Kramer}}, \citenamefont {{Kusiak}}, \citenamefont {{Lague}}, \citenamefont {{Lakey}}, \citenamefont {{Lee}}, \citenamefont {{Li}}, \citenamefont {{Li}}, \citenamefont {{Limon}}, \citenamefont {{Lokken}}, \citenamefont {{Lungu}}, \citenamefont {{MacCrann}}, \citenamefont {{MacInnis}}, \citenamefont {{Madhavacheril}}, \citenamefont {{Maldonado}}, \citenamefont {{Maldonado}}, \citenamefont {{Mallaby-Kay}}, \citenamefont {{Marques}}, \citenamefont {{van Marrewijk}}, \citenamefont {{McCarthy}}, \citenamefont {{McMahon}},
  \citenamefont {{Mehta}}, \citenamefont {{Menanteau}}, \citenamefont {{Moodley}}, \citenamefont {{Morris}}, \citenamefont {{Mroczkowski}}, \citenamefont {{Naess}}, \citenamefont {{Namikawa}}, \citenamefont {{Nati}}, \citenamefont {{Nerval}}, \citenamefont {{Newburgh}}, \citenamefont {{Nicola}}, \citenamefont {{Niemack}}, \citenamefont {{Nolta}}, \citenamefont {{Orlowski-Scherer}}, \citenamefont {{Pagano}}, \citenamefont {{Page}}, \citenamefont {{Pandey}}, \citenamefont {{Partridge}}, \citenamefont {{Perez Sarmiento}}, \citenamefont {{Prince}}, \citenamefont {{Puddu}}, \citenamefont {{Qu}}, \citenamefont {{Ragavan}}, \citenamefont {{Ried Guachalla}}, \citenamefont {{Rogers}}, \citenamefont {{Rojas}}, \citenamefont {{Sakuma}}, \citenamefont {{Schaan}}, \citenamefont {{Schmitt}}, \citenamefont {{Sehgal}}, \citenamefont {{Shaikh}}, \citenamefont {{Sherwin}}, \citenamefont {{Sierra}}, \citenamefont {{Sievers}}, \citenamefont {{Sif{\'o}n}}, \citenamefont {{Simon}}, \citenamefont {{Sonka}}, \citenamefont
  {{Spergel}}, \citenamefont {{Staggs}}, \citenamefont {{Storer}}, \citenamefont {{Surrao}}, \citenamefont {{Switzer}}, \citenamefont {{Tampier}}, \citenamefont {{Thornton}}, \citenamefont {{Trac}}, \citenamefont {{Tucker}}, \citenamefont {{Ullom}}, \citenamefont {{Vale}}, \citenamefont {{Van Engelen}}, \citenamefont {{Van Lanen}}, \citenamefont {{Vargas}}, \citenamefont {{Vavagiakis}}, \citenamefont {{Wagoner}}, \citenamefont {{Wang}}, \citenamefont {{Wenzl}}, \citenamefont {{Wollack}},\ and\ \citenamefont {{Zheng}}}]{ACT:2025fju}%
  \BibitemOpen
  \bibfield  {author} {\bibinfo {author} {\bibfnamefont {T.}~\bibnamefont {{Louis}}}, \bibinfo {author} {\bibfnamefont {A.}~\bibnamefont {{La Posta}}}, \bibinfo {author} {\bibfnamefont {Z.}~\bibnamefont {{Atkins}}}, \bibinfo {author} {\bibfnamefont {H.~T.}\ \bibnamefont {{Jense}}}, \bibinfo {author} {\bibfnamefont {I.}~\bibnamefont {{Abril-Cabezas}}}, \bibinfo {author} {\bibfnamefont {G.~E.}\ \bibnamefont {{Addison}}}, \bibinfo {author} {\bibfnamefont {P.~A.~R.}\ \bibnamefont {{Ade}}}, \bibinfo {author} {\bibfnamefont {S.}~\bibnamefont {{Aiola}}}, \bibinfo {author} {\bibfnamefont {T.}~\bibnamefont {{Alford}}}, \bibinfo {author} {\bibfnamefont {D.}~\bibnamefont {{Alonso}}}, \bibinfo {author} {\bibfnamefont {M.}~\bibnamefont {{Amiri}}}, \bibinfo {author} {\bibfnamefont {R.}~\bibnamefont {{An}}}, \bibinfo {author} {\bibfnamefont {J.~E.}\ \bibnamefont {{Austermann}}}, \bibinfo {author} {\bibfnamefont {E.}~\bibnamefont {{Barbavara}}}, \bibinfo {author} {\bibfnamefont {N.}~\bibnamefont {{Battaglia}}}, \bibnamefont
  {and~others},\ }\href {https://doi.org/10.48550/arXiv.2503.14452} {\bibfield  {journal} {\bibinfo  {journal} {arXiv e-prints}\ ,\ \bibinfo {eid} {arXiv:2503.14452}} (\bibinfo {year} {2025})},\ \Eprint {https://arxiv.org/abs/2503.14452} {arXiv:2503.14452 [astro-ph.CO]} \BibitemShut {NoStop}%
\bibitem [{\citenamefont {{Naess}}\ \emph {et~al.}(2025)\citenamefont {{Naess}}, \citenamefont {{Guan}}, \citenamefont {{Duivenvoorden}}, \citenamefont {{Hasselfield}}, \citenamefont {{Wang}}, \citenamefont {{Abril-Cabezas}}, \citenamefont {{Addison}}, \citenamefont {{Ade}}, \citenamefont {{Aiola}}, \citenamefont {{Alford}}, \citenamefont {{Alonso}}, \citenamefont {{Amiri}}, \citenamefont {{An}}, \citenamefont {{Atkins}}, \citenamefont {{Austermann}}, \citenamefont {{Barbavara}}, \citenamefont {{Battaglia}}, \citenamefont {{Battistelli}}, \citenamefont {{Beall}}, \citenamefont {{Bean}}, \citenamefont {{Beheshti}}, \citenamefont {{Beringue}}, \citenamefont {{Bhandarkar}}, \citenamefont {{Biermann}}, \citenamefont {{Bolliet}}, \citenamefont {{Bond}}, \citenamefont {{Calabrese}}, \citenamefont {{Capalbo}}, \citenamefont {{Carrero}}, \citenamefont {{Chen}}, \citenamefont {{Chesmore}}, \citenamefont {{Cho}}, \citenamefont {{Choi}}, \citenamefont {{Clark}}, \citenamefont {{Cordova Rosado}}, \citenamefont
  {{Cothard}}, \citenamefont {{Coughlin}}, \citenamefont {{Coulton}}, \citenamefont {{Crichton}}, \citenamefont {{Crowley}}, \citenamefont {{Devlin}}, \citenamefont {{Dicker}}, \citenamefont {{Duell}}, \citenamefont {{Duff}}, \citenamefont {{Dunkley}}, \citenamefont {{Dunner}}, \citenamefont {{Embil Villagra}}, \citenamefont {{Fankhanel}}, \citenamefont {{Farren}}, \citenamefont {{Ferraro}}, \citenamefont {{Foster}}, \citenamefont {{Freundt}}, \citenamefont {{Fuzia}}, \citenamefont {{Gallardo}}, \citenamefont {{Garrido}}, \citenamefont {{Giardiello}}, \citenamefont {{Gill}}, \citenamefont {{Givans}}, \citenamefont {{Gluscevic}}, \citenamefont {{Golec}}, \citenamefont {{Gong}}, \citenamefont {{Halpern}}, \citenamefont {{Harrison}}, \citenamefont {{Healy}}, \citenamefont {{Henderson}}, \citenamefont {{Hensley}}, \citenamefont {{Herv{\'\i}as-Caimapo}}, \citenamefont {{Hill}}, \citenamefont {{Hilton}}, \citenamefont {{Hilton}}, \citenamefont {{Hincks}}, \citenamefont {{Hlo{\v{z}}ek}}, \citenamefont {{Ho}},
  \citenamefont {{Hood}}, \citenamefont {{Hornecker}}, \citenamefont {{Huber}}, \citenamefont {{Hubmayr}}, \citenamefont {{Huffenberger}}, \citenamefont {{Hughes}}, \citenamefont {{Ikape}}, \citenamefont {{Irwin}}, \citenamefont {{Isopi}}, \citenamefont {{Jense}}, \citenamefont {{Joshi}}, \citenamefont {{Keller}}, \citenamefont {{Kim}}, \citenamefont {{Knowles}}, \citenamefont {{Koopman}}, \citenamefont {{Kosowsky}}, \citenamefont {{Kramer}}, \citenamefont {{Kusiak}}, \citenamefont {{La Posta}}, \citenamefont {{Lagu{\"e}}}, \citenamefont {{Lakey}}, \citenamefont {{Lee}}, \citenamefont {{Li}}, \citenamefont {{Li}}, \citenamefont {{Limon}}, \citenamefont {{Lokken}}, \citenamefont {{Louis}}, \citenamefont {{Lungu}}, \citenamefont {{MacCrann}}, \citenamefont {{MacInnis}}, \citenamefont {{Madhavacheril}}, \citenamefont {{Maldonado}}, \citenamefont {{Maldonado}}, \citenamefont {{Mallaby-Kay}}, \citenamefont {{Marques}}, \citenamefont {{van Marrewijk}}, \citenamefont {{McCarthy}}, \citenamefont {{McMahon}},
  \citenamefont {{Mehta}}, \citenamefont {{Menanteau}}, \citenamefont {{Moodley}}, \citenamefont {{Morris}}, \citenamefont {{Mroczkowski}}, \citenamefont {{Namikawa}}, \citenamefont {{Nati}}, \citenamefont {{Nerval}}, \citenamefont {{Newburgh}}, \citenamefont {{Nicola}}, \citenamefont {{Niemack}}, \citenamefont {{Nolta}}, \citenamefont {{Orlowski-Scherer}}, \citenamefont {{Page}}, \citenamefont {{Pandey}}, \citenamefont {{Partridge}}, \citenamefont {{Perez Sarmiento}}, \citenamefont {{Prince}}, \citenamefont {{Puddu}}, \citenamefont {{Qu}}, \citenamefont {{Quiroga}}, \citenamefont {{Ragavan}}, \citenamefont {{Ried Guachalla}}, \citenamefont {{Rogers}}, \citenamefont {{Rojas}}, \citenamefont {{Sakuma}}, \citenamefont {{Schaan}}, \citenamefont {{Schmitt}}, \citenamefont {{Sehgal}}, \citenamefont {{Shaikh}}, \citenamefont {{Sherwin}}, \citenamefont {{Sierra}}, \citenamefont {{Sievers}}, \citenamefont {{Sif{\'o}n}}, \citenamefont {{Simon}}, \citenamefont {{Sonka}}, \citenamefont {{Spergel}}, \citenamefont
  {{Staggs}}, \citenamefont {{Storer}}, \citenamefont {{Surrao}}, \citenamefont {{Switzer}}, \citenamefont {{Tampier}}, \citenamefont {{Thornton}}, \citenamefont {{Trac}}, \citenamefont {{Tucker}}, \citenamefont {{Ullom}}, \citenamefont {{Vale}}, \citenamefont {{Van Engelen}}, \citenamefont {{Van Lanen}}, \citenamefont {{Vargas}}, \citenamefont {{Vavagiakis}}, \citenamefont {{Wagoner}}, \citenamefont {{Wenzl}}, \citenamefont {{Wollack}},\ and\ \citenamefont {{Zheng}}}]{2025arXiv250314451N}%
  \BibitemOpen
  \bibfield  {author} {\bibinfo {author} {\bibfnamefont {S.}~\bibnamefont {{Naess}}}, \bibinfo {author} {\bibfnamefont {Y.}~\bibnamefont {{Guan}}}, \bibinfo {author} {\bibfnamefont {A.~J.}\ \bibnamefont {{Duivenvoorden}}}, \bibinfo {author} {\bibfnamefont {M.}~\bibnamefont {{Hasselfield}}}, \bibinfo {author} {\bibfnamefont {Y.}~\bibnamefont {{Wang}}}, \bibinfo {author} {\bibfnamefont {I.}~\bibnamefont {{Abril-Cabezas}}}, \bibinfo {author} {\bibfnamefont {G.~E.}\ \bibnamefont {{Addison}}}, \bibinfo {author} {\bibfnamefont {P.~A.~R.}\ \bibnamefont {{Ade}}}, \bibinfo {author} {\bibfnamefont {S.}~\bibnamefont {{Aiola}}}, \bibinfo {author} {\bibfnamefont {T.}~\bibnamefont {{Alford}}}, \bibinfo {author} {\bibfnamefont {D.}~\bibnamefont {{Alonso}}}, \bibinfo {author} {\bibfnamefont {M.}~\bibnamefont {{Amiri}}}, \bibinfo {author} {\bibfnamefont {R.}~\bibnamefont {{An}}}, \bibinfo {author} {\bibfnamefont {Z.}~\bibnamefont {{Atkins}}}, \bibinfo {author} {\bibfnamefont {J.~E.}\ \bibnamefont {{Austermann}}}, \bibnamefont
  {and~others},\ }\href {https://doi.org/10.48550/arXiv.2503.14451} {\bibfield  {journal} {\bibinfo  {journal} {arXiv e-prints}\ ,\ \bibinfo {eid} {arXiv:2503.14451}} (\bibinfo {year} {2025})},\ \Eprint {https://arxiv.org/abs/2503.14451} {arXiv:2503.14451 [astro-ph.CO]} \BibitemShut {NoStop}%
\bibitem [{\citenamefont {Calabrese}\ \emph {et~al.}(2025)\citenamefont {Calabrese} \emph {et~al.}}]{ACT:2025tim}%
  \BibitemOpen
  \bibfield  {author} {\bibinfo {author} {\bibfnamefont {E.}~\bibnamefont {Calabrese}} \bibnamefont {and~others} (\bibinfo {collaboration} {ACT}),\ }\Eprint {https://arxiv.org/abs/2503.14454} {arXiv:2503.14454 [astro-ph.CO]}  (\bibinfo {year} {2025})\BibitemShut {NoStop}%
\bibitem [{\citenamefont {{Scolnic}}\ \emph {et~al.}(2022)\citenamefont {{Scolnic}}, \citenamefont {{Brout}}, \citenamefont {{Carr}}, \citenamefont {{Riess}}, \citenamefont {{Davis}}, \citenamefont {{Dwomoh}}, \citenamefont {{Jones}}, \citenamefont {{Ali}}, \citenamefont {{Charvu}}, \citenamefont {{Chen}}, \citenamefont {{Peterson}}, \citenamefont {{Popovic}}, \citenamefont {{Rose}}, \citenamefont {{Wood}}, \citenamefont {{Brown}}, \citenamefont {{Chambers}}, \citenamefont {{Coulter}}, \citenamefont {{Dettman}}, \citenamefont {{Dimitriadis}}, \citenamefont {{Filippenko}}, \citenamefont {{Foley}}, \citenamefont {{Jha}}, \citenamefont {{Kilpatrick}}, \citenamefont {{Kirshner}}, \citenamefont {{Pan}}, \citenamefont {{Rest}}, \citenamefont {{Rojas-Bravo}}, \citenamefont {{Siebert}}, \citenamefont {{Stahl}},\ and\ \citenamefont {{Zheng}}}]{Scolnic:2021amr}%
  \BibitemOpen
  \bibfield  {author} {\bibinfo {author} {\bibfnamefont {D.}~\bibnamefont {{Scolnic}}}, \bibinfo {author} {\bibfnamefont {D.}~\bibnamefont {{Brout}}}, \bibinfo {author} {\bibfnamefont {A.}~\bibnamefont {{Carr}}}, \bibinfo {author} {\bibfnamefont {A.~G.}\ \bibnamefont {{Riess}}}, \bibinfo {author} {\bibfnamefont {T.~M.}\ \bibnamefont {{Davis}}}, \bibinfo {author} {\bibfnamefont {A.}~\bibnamefont {{Dwomoh}}}, \bibinfo {author} {\bibfnamefont {D.~O.}\ \bibnamefont {{Jones}}}, \bibinfo {author} {\bibfnamefont {N.}~\bibnamefont {{Ali}}}, \bibinfo {author} {\bibfnamefont {P.}~\bibnamefont {{Charvu}}}, \bibinfo {author} {\bibfnamefont {R.}~\bibnamefont {{Chen}}}, \bibinfo {author} {\bibfnamefont {E.~R.}\ \bibnamefont {{Peterson}}}, \bibinfo {author} {\bibfnamefont {B.}~\bibnamefont {{Popovic}}}, \bibinfo {author} {\bibfnamefont {B.~M.}\ \bibnamefont {{Rose}}}, \bibinfo {author} {\bibfnamefont {C.~M.}\ \bibnamefont {{Wood}}}, \bibinfo {author} {\bibfnamefont {P.~J.}\ \bibnamefont {{Brown}}}, \bibnamefont
  {and~others},\ }\href {https://doi.org/10.3847/1538-4357/ac8b7a} {\bibfield  {journal} {\bibinfo  {journal} {\apj}\ }\textbf {\bibinfo {volume} {938}},\ \bibinfo {eid} {113} (\bibinfo {year} {2022})},\ \Eprint {https://arxiv.org/abs/2112.03863} {arXiv:2112.03863 [astro-ph.CO]} \BibitemShut {NoStop}%
\bibitem [{\citenamefont {{Rubin}}\ \emph {et~al.}(2023)\citenamefont {{Rubin}}, \citenamefont {{Aldering}}, \citenamefont {{Betoule}}, \citenamefont {{Fruchter}}, \citenamefont {{Huang}}, \citenamefont {{Kim}}, \citenamefont {{Lidman}}, \citenamefont {{Linder}}, \citenamefont {{Perlmutter}}, \citenamefont {{Ruiz-Lapuente}},\ and\ \citenamefont {{Suzuki}}}]{Rubin:2023}%
  \BibitemOpen
  \bibfield  {author} {\bibinfo {author} {\bibfnamefont {D.}~\bibnamefont {{Rubin}}}, \bibinfo {author} {\bibfnamefont {G.}~\bibnamefont {{Aldering}}}, \bibinfo {author} {\bibfnamefont {M.}~\bibnamefont {{Betoule}}}, \bibinfo {author} {\bibfnamefont {A.}~\bibnamefont {{Fruchter}}}, \bibinfo {author} {\bibfnamefont {X.}~\bibnamefont {{Huang}}}, \bibinfo {author} {\bibfnamefont {A.~G.}\ \bibnamefont {{Kim}}}, \bibinfo {author} {\bibfnamefont {C.}~\bibnamefont {{Lidman}}}, \bibinfo {author} {\bibfnamefont {E.}~\bibnamefont {{Linder}}}, \bibinfo {author} {\bibfnamefont {S.}~\bibnamefont {{Perlmutter}}}, \bibinfo {author} {\bibfnamefont {P.}~\bibnamefont {{Ruiz-Lapuente}}},\ \bibnamefont {and}\ \bibinfo {author} {\bibfnamefont {N.}~\bibnamefont {{Suzuki}}},\ }\href {https://doi.org/10.48550/arXiv.2311.12098} {\bibfield  {journal} {\bibinfo  {journal} {arXiv e-prints}\ ,\ \bibinfo {eid} {arXiv:2311.12098}} (\bibinfo {year} {2023})},\ \Eprint {https://arxiv.org/abs/2311.12098} {arXiv:2311.12098 [astro-ph.CO]}
  \BibitemShut {NoStop}%
\bibitem [{\citenamefont {{DES Collaboration}}\ \emph {et~al.}(2024)\citenamefont {{DES Collaboration}}, \citenamefont {{Abbott}}, \citenamefont {{Acevedo}}, \citenamefont {{Aguena}}, \citenamefont {{Alarcon}}, \citenamefont {{Allam}}, \citenamefont {{Alves}}, \citenamefont {{Amon}}, \citenamefont {{Andrade-Oliveira}}, \citenamefont {{Annis}}, \citenamefont {{Armstrong}}, \citenamefont {{Asorey}}, \citenamefont {{Avila}}, \citenamefont {{Bacon}}, \citenamefont {{Bassett}}, \citenamefont {{Bechtol}}, \citenamefont {{Bernardinelli}}, \citenamefont {{Bernstein}}, \citenamefont {{Bertin}}, \citenamefont {{Blazek}}, \citenamefont {{Bocquet}}, \citenamefont {{Brooks}}, \citenamefont {{Brout}}, \citenamefont {{Buckley-Geer}}, \citenamefont {{Burke}}, \citenamefont {{Camacho}}, \citenamefont {{Camilleri}}, \citenamefont {{Campos}}, \citenamefont {{Carnero Rosell}}, \citenamefont {{Carollo}}, \citenamefont {{Carr}}, \citenamefont {{Carretero}}, \citenamefont {{Castander}}, \citenamefont {{Cawthon}}, \citenamefont
  {{Chang}}, \citenamefont {{Chen}}, \citenamefont {{Choi}}, \citenamefont {{Conselice}}, \citenamefont {{Costanzi}}, \citenamefont {{da Costa}}, \citenamefont {{Crocce}}, \citenamefont {{Davis}}, \citenamefont {{DePoy}}, \citenamefont {{Desai}}, \citenamefont {{Diehl}}, \citenamefont {{Dixon}}, \citenamefont {{Dodelson}}, \citenamefont {{Doel}}, \citenamefont {{Doux}}, \citenamefont {{Drlica-Wagner}}, \citenamefont {{Elvin-Poole}}, \citenamefont {{Everett}}, \citenamefont {{Ferrero}}, \citenamefont {{Fert{\'e}}}, \citenamefont {{Flaugher}}, \citenamefont {{Foley}}, \citenamefont {{Fosalba}}, \citenamefont {{Friedel}}, \citenamefont {{Frieman}}, \citenamefont {{Frohmaier}}, \citenamefont {{Galbany}}, \citenamefont {{Garc{\'\i}a-Bellido}}, \citenamefont {{Gatti}}, \citenamefont {{Gaztanaga}}, \citenamefont {{Giannini}}, \citenamefont {{Glazebrook}}, \citenamefont {{Graur}}, \citenamefont {{Gruen}}, \citenamefont {{Gruendl}}, \citenamefont {{Gutierrez}}, \citenamefont {{Hartley}}, \citenamefont {{Herner}},
  \citenamefont {{Hinton}}, \citenamefont {{Hollowood}}, \citenamefont {{Honscheid}}, \citenamefont {{Huterer}}, \citenamefont {{Jain}}, \citenamefont {{James}}, \citenamefont {{Jeffrey}}, \citenamefont {{Kasai}}, \citenamefont {{Kelsey}}, \citenamefont {{Kent}}, \citenamefont {{Kessler}}, \citenamefont {{Kim}}, \citenamefont {{Kirshner}}, \citenamefont {{Kovacs}}, \citenamefont {{Kuehn}}, \citenamefont {{Lahav}}, \citenamefont {{Lee}}, \citenamefont {{Lee}}, \citenamefont {{Lewis}}, \citenamefont {{Li}}, \citenamefont {{Lidman}}, \citenamefont {{Lin}}, \citenamefont {{Malik}}, \citenamefont {{Marshall}}, \citenamefont {{Martini}}, \citenamefont {{Mena-Fern{\'a}ndez}}, \citenamefont {{Menanteau}}, \citenamefont {{Miquel}}, \citenamefont {{Mohr}}, \citenamefont {{Mould}}, \citenamefont {{Muir}}, \citenamefont {{M{\"o}ller}}, \citenamefont {{Neilsen}}, \citenamefont {{Nichol}}, \citenamefont {{Nugent}}, \citenamefont {{Ogando}}, \citenamefont {{Palmese}}, \citenamefont {{Pan}}, \citenamefont {{Paterno}},
  \citenamefont {{Percival}}, \citenamefont {{Pereira}}, \citenamefont {{Pieres}}, \citenamefont {{Malag{\'o}n}}, \citenamefont {{Popovic}}, \citenamefont {{Porredon}}, \citenamefont {{Prat}}, \citenamefont {{Qu}}, \citenamefont {{Raveri}}, \citenamefont {{Rodr{\'\i}guez-Monroy}}, \citenamefont {{Romer}}, \citenamefont {{Roodman}}, \citenamefont {{Rose}}, \citenamefont {{Sako}}, \citenamefont {{Sanchez}}, \citenamefont {{Sanchez Cid}}, \citenamefont {{Schubnell}}, \citenamefont {{Scolnic}}, \citenamefont {{Sevilla-Noarbe}}, \citenamefont {{Shah}}, \citenamefont {{Smith}}, \citenamefont {{Smith}}, \citenamefont {{Soares-Santos}}, \citenamefont {{Suchyta}}, \citenamefont {{Sullivan}}, \citenamefont {{Suntzeff}}, \citenamefont {{Swanson}}, \citenamefont {{S{\'a}nchez}}, \citenamefont {{Tarle}}, \citenamefont {{Taylor}}, \citenamefont {{Thomas}}, \citenamefont {{To}}, \citenamefont {{Toy}}, \citenamefont {{Troxel}}, \citenamefont {{Tucker}}, \citenamefont {{Tucker}}, \citenamefont {{Uddin}}, \citenamefont
  {{Vincenzi}}, \citenamefont {{Walker}}, \citenamefont {{Weaverdyck}}, \citenamefont {{Wechsler}}, \citenamefont {{Weller}}, \citenamefont {{Wester}}, \citenamefont {{Wiseman}}, \citenamefont {{Yamamoto}}, \citenamefont {{Yuan}}, \citenamefont {{Zhang}},\ and\ \citenamefont {{Zhang}}}]{DES:2024tys}%
  \BibitemOpen
  \bibfield  {author} {\bibinfo {author} {\bibnamefont {{DES Collaboration}}}, \bibinfo {author} {\bibfnamefont {T.~M.~C.}\ \bibnamefont {{Abbott}}}, \bibinfo {author} {\bibfnamefont {M.}~\bibnamefont {{Acevedo}}}, \bibinfo {author} {\bibfnamefont {M.}~\bibnamefont {{Aguena}}}, \bibinfo {author} {\bibfnamefont {A.}~\bibnamefont {{Alarcon}}}, \bibinfo {author} {\bibfnamefont {S.}~\bibnamefont {{Allam}}}, \bibinfo {author} {\bibfnamefont {O.}~\bibnamefont {{Alves}}}, \bibinfo {author} {\bibfnamefont {A.}~\bibnamefont {{Amon}}}, \bibinfo {author} {\bibfnamefont {F.}~\bibnamefont {{Andrade-Oliveira}}}, \bibinfo {author} {\bibfnamefont {J.}~\bibnamefont {{Annis}}}, \bibinfo {author} {\bibfnamefont {P.}~\bibnamefont {{Armstrong}}}, \bibinfo {author} {\bibfnamefont {J.}~\bibnamefont {{Asorey}}}, \bibinfo {author} {\bibfnamefont {S.}~\bibnamefont {{Avila}}}, \bibinfo {author} {\bibfnamefont {D.}~\bibnamefont {{Bacon}}}, \bibinfo {author} {\bibfnamefont {B.~A.}\ \bibnamefont {{Bassett}}}, \bibnamefont {and~others},\
  }\href {https://doi.org/10.3847/2041-8213/ad6f9f} {\bibfield  {journal} {\bibinfo  {journal} {\apjl}\ }\textbf {\bibinfo {volume} {973}},\ \bibinfo {eid} {L14} (\bibinfo {year} {2024})},\ \Eprint {https://arxiv.org/abs/2401.02929} {arXiv:2401.02929 [astro-ph.CO]} \BibitemShut {NoStop}%
\bibitem [{\citenamefont {{Planck Collaboration}}\ \emph {et~al.}(2020{\natexlab{a}})\citenamefont {{Planck Collaboration}}, \citenamefont {{Aghanim}}, \citenamefont {{Akrami}}, \citenamefont {{Ashdown}}, \citenamefont {{Aumont}}, \citenamefont {{Baccigalupi}}, \citenamefont {{Ballardini}}, \citenamefont {{Banday}}, \citenamefont {{Barreiro}}, \citenamefont {{Bartolo}}, \citenamefont {{Basak}}, \citenamefont {{Benabed}}, \citenamefont {{Bernard}}, \citenamefont {{Bersanelli}}, \citenamefont {{Bielewicz}}, \citenamefont {{Bock}}, \citenamefont {{Bond}}, \citenamefont {{Borrill}}, \citenamefont {{Bouchet}}, \citenamefont {{Boulanger}}, \citenamefont {{Bucher}}, \citenamefont {{Burigana}}, \citenamefont {{Butler}}, \citenamefont {{Calabrese}}, \citenamefont {{Cardoso}}, \citenamefont {{Carron}}, \citenamefont {{Casaponsa}}, \citenamefont {{Challinor}}, \citenamefont {{Chiang}}, \citenamefont {{Colombo}}, \citenamefont {{Combet}}, \citenamefont {{Crill}}, \citenamefont {{Cuttaia}}, \citenamefont {{de Bernardis}},
  \citenamefont {{de Rosa}}, \citenamefont {{de Zotti}}, \citenamefont {{Delabrouille}}, \citenamefont {{Delouis}}, \citenamefont {{Di Valentino}}, \citenamefont {{Diego}}, \citenamefont {{Dor{\'e}}}, \citenamefont {{Douspis}}, \citenamefont {{Ducout}}, \citenamefont {{Dupac}}, \citenamefont {{Dusini}}, \citenamefont {{Efstathiou}}, \citenamefont {{Elsner}}, \citenamefont {{En{\ss}lin}}, \citenamefont {{Eriksen}}, \citenamefont {{Fantaye}}, \citenamefont {{Fernandez-Cobos}}, \citenamefont {{Finelli}}, \citenamefont {{Frailis}}, \citenamefont {{Fraisse}}, \citenamefont {{Franceschi}}, \citenamefont {{Frolov}}, \citenamefont {{Galeotta}}, \citenamefont {{Galli}}, \citenamefont {{Ganga}}, \citenamefont {{G{\'e}nova-Santos}}, \citenamefont {{Gerbino}}, \citenamefont {{Ghosh}}, \citenamefont {{Giraud-H{\'e}raud}}, \citenamefont {{Gonz{\'a}lez-Nuevo}}, \citenamefont {{G{\'o}rski}}, \citenamefont {{Gratton}}, \citenamefont {{Gruppuso}}, \citenamefont {{Gudmundsson}}, \citenamefont {{Hamann}}, \citenamefont
  {{Handley}}, \citenamefont {{Hansen}}, \citenamefont {{Herranz}}, \citenamefont {{Hivon}}, \citenamefont {{Huang}}, \citenamefont {{Jaffe}}, \citenamefont {{Jones}}, \citenamefont {{Keih{\"a}nen}}, \citenamefont {{Keskitalo}}, \citenamefont {{Kiiveri}}, \citenamefont {{Kim}}, \citenamefont {{Kisner}}, \citenamefont {{Krachmalnicoff}}, \citenamefont {{Kunz}}, \citenamefont {{Kurki-Suonio}}, \citenamefont {{Lagache}}, \citenamefont {{Lamarre}}, \citenamefont {{Lasenby}}, \citenamefont {{Lattanzi}}, \citenamefont {{Lawrence}}, \citenamefont {{Le Jeune}}, \citenamefont {{Levrier}}, \citenamefont {{Lewis}}, \citenamefont {{Liguori}}, \citenamefont {{Lilje}}, \citenamefont {{Lilley}}, \citenamefont {{Lindholm}}, \citenamefont {{L{\'o}pez-Caniego}}, \citenamefont {{Lubin}}, \citenamefont {{Ma}}, \citenamefont {{Mac{\'\i}as-P{\'e}rez}}, \citenamefont {{Maggio}}, \citenamefont {{Maino}}, \citenamefont {{Mandolesi}}, \citenamefont {{Mangilli}}, \citenamefont {{Marcos-Caballero}}, \citenamefont {{Maris}},
  \citenamefont {{Martin}}, \citenamefont {{Mart{\'\i}nez-Gonz{\'a}lez}}, \citenamefont {{Matarrese}}, \citenamefont {{Mauri}}, \citenamefont {{McEwen}}, \citenamefont {{Meinhold}}, \citenamefont {{Melchiorri}}, \citenamefont {{Mennella}}, \citenamefont {{Migliaccio}}, \citenamefont {{Millea}}, \citenamefont {{Miville-Desch{\^e}nes}}, \citenamefont {{Molinari}}, \citenamefont {{Moneti}}, \citenamefont {{Montier}}, \citenamefont {{Morgante}}, \citenamefont {{Moss}}, \citenamefont {{Natoli}}, \citenamefont {{N{\o}rgaard-Nielsen}}, \citenamefont {{Pagano}}, \citenamefont {{Paoletti}}, \citenamefont {{Partridge}}, \citenamefont {{Patanchon}}, \citenamefont {{Peiris}}, \citenamefont {{Perrotta}}, \citenamefont {{Pettorino}}, \citenamefont {{Piacentini}}, \citenamefont {{Polenta}}, \citenamefont {{Puget}}, \citenamefont {{Rachen}}, \citenamefont {{Reinecke}}, \citenamefont {{Remazeilles}}, \citenamefont {{Renzi}}, \citenamefont {{Rocha}}, \citenamefont {{Rosset}}, \citenamefont {{Roudier}}, \citenamefont
  {{Rubi{\~n}o-Mart{\'\i}n}}, \citenamefont {{Ruiz-Granados}}, \citenamefont {{Salvati}}, \citenamefont {{Sandri}}, \citenamefont {{Savelainen}}, \citenamefont {{Scott}}, \citenamefont {{Shellard}}, \citenamefont {{Sirignano}}, \citenamefont {{Sirri}}, \citenamefont {{Spencer}}, \citenamefont {{Sunyaev}}, \citenamefont {{Suur-Uski}}, \citenamefont {{Tauber}}, \citenamefont {{Tavagnacco}}, \citenamefont {{Tenti}}, \citenamefont {{Toffolatti}}, \citenamefont {{Tomasi}}, \citenamefont {{Trombetti}}, \citenamefont {{Valiviita}}, \citenamefont {{Van Tent}}, \citenamefont {{Vielva}}, \citenamefont {{Villa}}, \citenamefont {{Vittorio}}, \citenamefont {{Wandelt}}, \citenamefont {{Wehus}}, \citenamefont {{Zacchei}},\ and\ \citenamefont {{Zonca}}}]{Planck-2018-likelihoods}%
  \BibitemOpen
  \bibfield  {author} {\bibinfo {author} {\bibnamefont {{Planck Collaboration}}}, \bibinfo {author} {\bibfnamefont {N.}~\bibnamefont {{Aghanim}}}, \bibinfo {author} {\bibfnamefont {Y.}~\bibnamefont {{Akrami}}}, \bibinfo {author} {\bibfnamefont {M.}~\bibnamefont {{Ashdown}}}, \bibinfo {author} {\bibfnamefont {J.}~\bibnamefont {{Aumont}}}, \bibinfo {author} {\bibfnamefont {C.}~\bibnamefont {{Baccigalupi}}}, \bibinfo {author} {\bibfnamefont {M.}~\bibnamefont {{Ballardini}}}, \bibinfo {author} {\bibfnamefont {A.~J.}\ \bibnamefont {{Banday}}}, \bibinfo {author} {\bibfnamefont {R.~B.}\ \bibnamefont {{Barreiro}}}, \bibinfo {author} {\bibfnamefont {N.}~\bibnamefont {{Bartolo}}}, \bibinfo {author} {\bibfnamefont {S.}~\bibnamefont {{Basak}}}, \bibinfo {author} {\bibfnamefont {K.}~\bibnamefont {{Benabed}}}, \bibinfo {author} {\bibfnamefont {J.~P.}\ \bibnamefont {{Bernard}}}, \bibinfo {author} {\bibfnamefont {M.}~\bibnamefont {{Bersanelli}}}, \bibinfo {author} {\bibfnamefont {P.}~\bibnamefont {{Bielewicz}}}, \bibnamefont
  {and~others},\ }\href {https://doi.org/10.1051/0004-6361/201936386} {\bibfield  {journal} {\bibinfo  {journal} {\aap}\ }\textbf {\bibinfo {volume} {641}},\ \bibinfo {eid} {A5} (\bibinfo {year} {2020}{\natexlab{a}})},\ \Eprint {https://arxiv.org/abs/1907.12875} {arXiv:1907.12875 [astro-ph.CO]} \BibitemShut {NoStop}%
\bibitem [{\citenamefont {{Planck Collaboration}}\ \emph {et~al.}(2020{\natexlab{b}})\citenamefont {{Planck Collaboration}}, \citenamefont {{Aghanim}}, \citenamefont {{Akrami}}, \citenamefont {{Ashdown}}, \citenamefont {{Aumont}}, \citenamefont {{Baccigalupi}}, \citenamefont {{Ballardini}}, \citenamefont {{Banday}}, \citenamefont {{Barreiro}}, \citenamefont {{Bartolo}}, \citenamefont {{Basak}}, \citenamefont {{Battye}}, \citenamefont {{Benabed}}, \citenamefont {{Bernard}}, \citenamefont {{Bersanelli}}, \citenamefont {{Bielewicz}}, \citenamefont {{Bock}}, \citenamefont {{Bond}}, \citenamefont {{Borrill}}, \citenamefont {{Bouchet}}, \citenamefont {{Boulanger}}, \citenamefont {{Bucher}}, \citenamefont {{Burigana}}, \citenamefont {{Butler}}, \citenamefont {{Calabrese}}, \citenamefont {{Cardoso}}, \citenamefont {{Carron}}, \citenamefont {{Challinor}}, \citenamefont {{Chiang}}, \citenamefont {{Chluba}}, \citenamefont {{Colombo}}, \citenamefont {{Combet}}, \citenamefont {{Contreras}}, \citenamefont {{Crill}},
  \citenamefont {{Cuttaia}}, \citenamefont {{de Bernardis}}, \citenamefont {{de Zotti}}, \citenamefont {{Delabrouille}}, \citenamefont {{Delouis}}, \citenamefont {{Di Valentino}}, \citenamefont {{Diego}}, \citenamefont {{Dor{\'e}}}, \citenamefont {{Douspis}}, \citenamefont {{Ducout}}, \citenamefont {{Dupac}}, \citenamefont {{Dusini}}, \citenamefont {{Efstathiou}}, \citenamefont {{Elsner}}, \citenamefont {{En{\ss}lin}}, \citenamefont {{Eriksen}}, \citenamefont {{Fantaye}}, \citenamefont {{Farhang}}, \citenamefont {{Fergusson}}, \citenamefont {{Fernandez-Cobos}}, \citenamefont {{Finelli}}, \citenamefont {{Forastieri}}, \citenamefont {{Frailis}}, \citenamefont {{Fraisse}}, \citenamefont {{Franceschi}}, \citenamefont {{Frolov}}, \citenamefont {{Galeotta}}, \citenamefont {{Galli}}, \citenamefont {{Ganga}}, \citenamefont {{G{\'e}nova-Santos}}, \citenamefont {{Gerbino}}, \citenamefont {{Ghosh}}, \citenamefont {{Gonz{\'a}lez-Nuevo}}, \citenamefont {{G{\'o}rski}}, \citenamefont {{Gratton}}, \citenamefont {{Gruppuso}},
  \citenamefont {{Gudmundsson}}, \citenamefont {{Hamann}}, \citenamefont {{Handley}}, \citenamefont {{Hansen}}, \citenamefont {{Herranz}}, \citenamefont {{Hildebrandt}}, \citenamefont {{Hivon}}, \citenamefont {{Huang}}, \citenamefont {{Jaffe}}, \citenamefont {{Jones}}, \citenamefont {{Karakci}}, \citenamefont {{Keih{\"a}nen}}, \citenamefont {{Keskitalo}}, \citenamefont {{Kiiveri}}, \citenamefont {{Kim}}, \citenamefont {{Kisner}}, \citenamefont {{Knox}}, \citenamefont {{Krachmalnicoff}}, \citenamefont {{Kunz}}, \citenamefont {{Kurki-Suonio}}, \citenamefont {{Lagache}}, \citenamefont {{Lamarre}}, \citenamefont {{Lasenby}}, \citenamefont {{Lattanzi}}, \citenamefont {{Lawrence}}, \citenamefont {{Le Jeune}}, \citenamefont {{Lemos}}, \citenamefont {{Lesgourgues}}, \citenamefont {{Levrier}}, \citenamefont {{Lewis}}, \citenamefont {{Liguori}}, \citenamefont {{Lilje}}, \citenamefont {{Lilley}}, \citenamefont {{Lindholm}}, \citenamefont {{L{\'o}pez-Caniego}}, \citenamefont {{Lubin}}, \citenamefont {{Ma}}, \citenamefont
  {{Mac{\'\i}as-P{\'e}rez}}, \citenamefont {{Maggio}}, \citenamefont {{Maino}}, \citenamefont {{Mandolesi}}, \citenamefont {{Mangilli}}, \citenamefont {{Marcos-Caballero}}, \citenamefont {{Maris}}, \citenamefont {{Martin}}, \citenamefont {{Martinelli}}, \citenamefont {{Mart{\'\i}nez-Gonz{\'a}lez}}, \citenamefont {{Matarrese}}, \citenamefont {{Mauri}}, \citenamefont {{McEwen}}, \citenamefont {{Meinhold}}, \citenamefont {{Melchiorri}}, \citenamefont {{Mennella}}, \citenamefont {{Migliaccio}}, \citenamefont {{Millea}}, \citenamefont {{Mitra}}, \citenamefont {{Miville-Desch{\^e}nes}}, \citenamefont {{Molinari}}, \citenamefont {{Montier}}, \citenamefont {{Morgante}}, \citenamefont {{Moss}}, \citenamefont {{Natoli}}, \citenamefont {{N{\o}rgaard-Nielsen}}, \citenamefont {{Pagano}}, \citenamefont {{Paoletti}}, \citenamefont {{Partridge}}, \citenamefont {{Patanchon}}, \citenamefont {{Peiris}}, \citenamefont {{Perrotta}}, \citenamefont {{Pettorino}}, \citenamefont {{Piacentini}}, \citenamefont {{Polastri}},
  \citenamefont {{Polenta}}, \citenamefont {{Puget}}, \citenamefont {{Rachen}}, \citenamefont {{Reinecke}}, \citenamefont {{Remazeilles}}, \citenamefont {{Renzi}}, \citenamefont {{Rocha}}, \citenamefont {{Rosset}}, \citenamefont {{Roudier}}, \citenamefont {{Rubi{\~n}o-Mart{\'\i}n}}, \citenamefont {{Ruiz-Granados}}, \citenamefont {{Salvati}}, \citenamefont {{Sandri}}, \citenamefont {{Savelainen}}, \citenamefont {{Scott}}, \citenamefont {{Shellard}}, \citenamefont {{Sirignano}}, \citenamefont {{Sirri}}, \citenamefont {{Spencer}}, \citenamefont {{Sunyaev}}, \citenamefont {{Suur-Uski}}, \citenamefont {{Tauber}}, \citenamefont {{Tavagnacco}}, \citenamefont {{Tenti}}, \citenamefont {{Toffolatti}}, \citenamefont {{Tomasi}}, \citenamefont {{Trombetti}}, \citenamefont {{Valenziano}}, \citenamefont {{Valiviita}}, \citenamefont {{Van Tent}}, \citenamefont {{Vibert}}, \citenamefont {{Vielva}}, \citenamefont {{Villa}}, \citenamefont {{Vittorio}}, \citenamefont {{Wandelt}}, \citenamefont {{Wehus}}, \citenamefont {{White}},
  \citenamefont {{White}}, \citenamefont {{Zacchei}},\ and\ \citenamefont {{Zonca}}}]{Planck-2018-cosmology}%
  \BibitemOpen
  \bibfield  {author} {\bibinfo {author} {\bibnamefont {{Planck Collaboration}}}, \bibinfo {author} {\bibfnamefont {N.}~\bibnamefont {{Aghanim}}}, \bibinfo {author} {\bibfnamefont {Y.}~\bibnamefont {{Akrami}}}, \bibinfo {author} {\bibfnamefont {M.}~\bibnamefont {{Ashdown}}}, \bibinfo {author} {\bibfnamefont {J.}~\bibnamefont {{Aumont}}}, \bibinfo {author} {\bibfnamefont {C.}~\bibnamefont {{Baccigalupi}}}, \bibinfo {author} {\bibfnamefont {M.}~\bibnamefont {{Ballardini}}}, \bibinfo {author} {\bibfnamefont {A.~J.}\ \bibnamefont {{Banday}}}, \bibinfo {author} {\bibfnamefont {R.~B.}\ \bibnamefont {{Barreiro}}}, \bibinfo {author} {\bibfnamefont {N.}~\bibnamefont {{Bartolo}}}, \bibinfo {author} {\bibfnamefont {S.}~\bibnamefont {{Basak}}}, \bibinfo {author} {\bibfnamefont {R.}~\bibnamefont {{Battye}}}, \bibinfo {author} {\bibfnamefont {K.}~\bibnamefont {{Benabed}}}, \bibinfo {author} {\bibfnamefont {J.~P.}\ \bibnamefont {{Bernard}}}, \bibinfo {author} {\bibfnamefont {M.}~\bibnamefont {{Bersanelli}}}, \bibnamefont
  {and~others},\ }\href {https://doi.org/10.1051/0004-6361/201833910} {\bibfield  {journal} {\bibinfo  {journal} {\aap}\ }\textbf {\bibinfo {volume} {641}},\ \bibinfo {eid} {A6} (\bibinfo {year} {2020}{\natexlab{b}})},\ \Eprint {https://arxiv.org/abs/1807.06209} {arXiv:1807.06209 [astro-ph.CO]} \BibitemShut {NoStop}%
\bibitem [{\citenamefont {{Pagano}}\ \emph {et~al.}(2020)\citenamefont {{Pagano}}, \citenamefont {{Delouis}}, \citenamefont {{Mottet}}, \citenamefont {{Puget}},\ and\ \citenamefont {{Vibert}}}]{2020_Sroll2_Pagano}%
  \BibitemOpen
  \bibfield  {author} {\bibinfo {author} {\bibfnamefont {L.}~\bibnamefont {{Pagano}}}, \bibinfo {author} {\bibfnamefont {J.~M.}\ \bibnamefont {{Delouis}}}, \bibinfo {author} {\bibfnamefont {S.}~\bibnamefont {{Mottet}}}, \bibinfo {author} {\bibfnamefont {J.~L.}\ \bibnamefont {{Puget}}},\ \bibnamefont {and}\ \bibinfo {author} {\bibfnamefont {L.}~\bibnamefont {{Vibert}}},\ }\href {https://doi.org/10.1051/0004-6361/201936630} {\bibfield  {journal} {\bibinfo  {journal} {\aap}\ }\textbf {\bibinfo {volume} {635}},\ \bibinfo {eid} {A99} (\bibinfo {year} {2020})},\ \Eprint {https://arxiv.org/abs/1908.09856} {arXiv:1908.09856 [astro-ph.CO]} \BibitemShut {NoStop}%
\bibitem [{\citenamefont {{Efstathiou}}\ and\ \citenamefont {{Gratton}}(2021)}]{Efstathiou:2021}%
  \BibitemOpen
  \bibfield  {author} {\bibinfo {author} {\bibfnamefont {G.}~\bibnamefont {{Efstathiou}}}\ \bibnamefont {and}\ \bibinfo {author} {\bibfnamefont {S.}~\bibnamefont {{Gratton}}},\ }\href {https://doi.org/10.21105/astro.1910.00483} {\bibfield  {journal} {\bibinfo  {journal} {The Open Journal of Astrophysics}\ }\textbf {\bibinfo {volume} {4}},\ \bibinfo {eid} {8} (\bibinfo {year} {2021})}\BibitemShut {NoStop}%
\bibitem [{\citenamefont {{Rosenberg}}\ \emph {et~al.}(2022)\citenamefont {{Rosenberg}}, \citenamefont {{Gratton}},\ and\ \citenamefont {{Efstathiou}}}]{Rosenberg:2022}%
  \BibitemOpen
  \bibfield  {author} {\bibinfo {author} {\bibfnamefont {E.}~\bibnamefont {{Rosenberg}}}, \bibinfo {author} {\bibfnamefont {S.}~\bibnamefont {{Gratton}}},\ \bibnamefont {and}\ \bibinfo {author} {\bibfnamefont {G.}~\bibnamefont {{Efstathiou}}},\ }\href {https://doi.org/10.1093/mnras/stac2744} {\bibfield  {journal} {\bibinfo  {journal} {\mnras}\ }\textbf {\bibinfo {volume} {517}},\ \bibinfo {pages} {4620} (\bibinfo {year} {2022})},\ \Eprint {https://arxiv.org/abs/2205.10869} {arXiv:2205.10869 [astro-ph.CO]} \BibitemShut {NoStop}%
\bibitem [{\citenamefont {{Carron}}\ \emph {et~al.}(2022)\citenamefont {{Carron}}, \citenamefont {{Mirmelstein}},\ and\ \citenamefont {{Lewis}}}]{Carron:2022}%
  \BibitemOpen
  \bibfield  {author} {\bibinfo {author} {\bibfnamefont {J.}~\bibnamefont {{Carron}}}, \bibinfo {author} {\bibfnamefont {M.}~\bibnamefont {{Mirmelstein}}},\ \bibnamefont {and}\ \bibinfo {author} {\bibfnamefont {A.}~\bibnamefont {{Lewis}}},\ }\href {https://doi.org/10.1088/1475-7516/2022/09/039} {\bibfield  {journal} {\bibinfo  {journal} {\jcap}\ }\textbf {\bibinfo {volume} {2022}},\ \bibinfo {eid} {039} (\bibinfo {year} {2022})},\ \Eprint {https://arxiv.org/abs/2206.07773} {arXiv:2206.07773 [astro-ph.CO]} \BibitemShut {NoStop}%
\bibitem [{\citenamefont {{Madhavacheril}}\ \emph {et~al.}(2024)\citenamefont {{Madhavacheril}}, \citenamefont {{Qu}}, \citenamefont {{Sherwin}}, \citenamefont {{MacCrann}}, \citenamefont {{Li}}, \citenamefont {{Abril-Cabezas}}, \citenamefont {{Ade}}, \citenamefont {{Aiola}}, \citenamefont {{Alford}}, \citenamefont {{Amiri}}, \citenamefont {{Amodeo}}, \citenamefont {{An}}, \citenamefont {{Atkins}}, \citenamefont {{Austermann}}, \citenamefont {{Battaglia}}, \citenamefont {{Battistelli}}, \citenamefont {{Beall}}, \citenamefont {{Bean}}, \citenamefont {{Beringue}}, \citenamefont {{Bhandarkar}}, \citenamefont {{Biermann}}, \citenamefont {{Bolliet}}, \citenamefont {{Bond}}, \citenamefont {{Cai}}, \citenamefont {{Calabrese}}, \citenamefont {{Calafut}}, \citenamefont {{Capalbo}}, \citenamefont {{Carrero}}, \citenamefont {{Challinor}}, \citenamefont {{Chesmore}}, \citenamefont {{Cho}}, \citenamefont {{Choi}}, \citenamefont {{Clark}}, \citenamefont {{C{\'o}rdova Rosado}}, \citenamefont {{Cothard}}, \citenamefont
  {{Coughlin}}, \citenamefont {{Coulton}}, \citenamefont {{Crowley}}, \citenamefont {{Dalal}}, \citenamefont {{Darwish}}, \citenamefont {{Devlin}}, \citenamefont {{Dicker}}, \citenamefont {{Doze}}, \citenamefont {{Duell}}, \citenamefont {{Duff}}, \citenamefont {{Duivenvoorden}}, \citenamefont {{Dunkley}}, \citenamefont {{D{\"u}nner}}, \citenamefont {{Fanfani}}, \citenamefont {{Fankhanel}}, \citenamefont {{Farren}}, \citenamefont {{Ferraro}}, \citenamefont {{Freundt}}, \citenamefont {{Fuzia}}, \citenamefont {{Gallardo}}, \citenamefont {{Garrido}}, \citenamefont {{Givans}}, \citenamefont {{Gluscevic}}, \citenamefont {{Golec}}, \citenamefont {{Guan}}, \citenamefont {{Hall}}, \citenamefont {{Halpern}}, \citenamefont {{Han}}, \citenamefont {{Harrison}}, \citenamefont {{Hasselfield}}, \citenamefont {{Healy}}, \citenamefont {{Henderson}}, \citenamefont {{Hensley}}, \citenamefont {{Herv{\'\i}as-Caimapo}}, \citenamefont {{Hill}}, \citenamefont {{Hilton}}, \citenamefont {{Hilton}}, \citenamefont {{Hincks}},
  \citenamefont {{Hlo{\v{z}}ek}}, \citenamefont {{Ho}}, \citenamefont {{Huber}}, \citenamefont {{Hubmayr}}, \citenamefont {{Huffenberger}}, \citenamefont {{Hughes}}, \citenamefont {{Irwin}}, \citenamefont {{Isopi}}, \citenamefont {{Jense}}, \citenamefont {{Keller}}, \citenamefont {{Kim}}, \citenamefont {{Knowles}}, \citenamefont {{Koopman}}, \citenamefont {{Kosowsky}}, \citenamefont {{Kramer}}, \citenamefont {{Kusiak}}, \citenamefont {{La Posta}}, \citenamefont {{Lague}}, \citenamefont {{Lakey}}, \citenamefont {{Lee}}, \citenamefont {{Li}}, \citenamefont {{Limon}}, \citenamefont {{Lokken}}, \citenamefont {{Louis}}, \citenamefont {{Lungu}}, \citenamefont {{MacInnis}}, \citenamefont {{Maldonado}}, \citenamefont {{Maldonado}}, \citenamefont {{Mallaby-Kay}}, \citenamefont {{Marques}}, \citenamefont {{McMahon}}, \citenamefont {{Mehta}}, \citenamefont {{Menanteau}}, \citenamefont {{Moodley}}, \citenamefont {{Morris}}, \citenamefont {{Mroczkowski}}, \citenamefont {{Naess}}, \citenamefont {{Namikawa}}, \citenamefont
  {{Nati}}, \citenamefont {{Newburgh}}, \citenamefont {{Nicola}}, \citenamefont {{Niemack}}, \citenamefont {{Nolta}}, \citenamefont {{Orlowski-Scherer}}, \citenamefont {{Page}}, \citenamefont {{Pandey}}, \citenamefont {{Partridge}}, \citenamefont {{Prince}}, \citenamefont {{Puddu}}, \citenamefont {{Radiconi}}, \citenamefont {{Robertson}}, \citenamefont {{Rojas}}, \citenamefont {{Sakuma}}, \citenamefont {{Salatino}}, \citenamefont {{Schaan}}, \citenamefont {{Schmitt}}, \citenamefont {{Sehgal}}, \citenamefont {{Shaikh}}, \citenamefont {{Sierra}}, \citenamefont {{Sievers}}, \citenamefont {{Sif{\'o}n}}, \citenamefont {{Simon}}, \citenamefont {{Sonka}}, \citenamefont {{Spergel}}, \citenamefont {{Staggs}}, \citenamefont {{Storer}}, \citenamefont {{Switzer}}, \citenamefont {{Tampier}}, \citenamefont {{Thornton}}, \citenamefont {{Trac}}, \citenamefont {{Treu}}, \citenamefont {{Tucker}}, \citenamefont {{Ullom}}, \citenamefont {{Vale}}, \citenamefont {{Van Engelen}}, \citenamefont {{Van Lanen}}, \citenamefont {{van
  Marrewijk}}, \citenamefont {{Vargas}}, \citenamefont {{Vavagiakis}}, \citenamefont {{Wagoner}}, \citenamefont {{Wang}}, \citenamefont {{Wenzl}}, \citenamefont {{Wollack}}, \citenamefont {{Xu}}, \citenamefont {{Zago}},\ and\ \citenamefont {{Zheng}}}]{Madhavacheril:ACT-DR6}%
  \BibitemOpen
  \bibfield  {author} {\bibinfo {author} {\bibfnamefont {M.~S.}\ \bibnamefont {{Madhavacheril}}}, \bibinfo {author} {\bibfnamefont {F.~J.}\ \bibnamefont {{Qu}}}, \bibinfo {author} {\bibfnamefont {B.~D.}\ \bibnamefont {{Sherwin}}}, \bibinfo {author} {\bibfnamefont {N.}~\bibnamefont {{MacCrann}}}, \bibinfo {author} {\bibfnamefont {Y.}~\bibnamefont {{Li}}}, \bibinfo {author} {\bibfnamefont {I.}~\bibnamefont {{Abril-Cabezas}}}, \bibinfo {author} {\bibfnamefont {P.~A.~R.}\ \bibnamefont {{Ade}}}, \bibinfo {author} {\bibfnamefont {S.}~\bibnamefont {{Aiola}}}, \bibinfo {author} {\bibfnamefont {T.}~\bibnamefont {{Alford}}}, \bibinfo {author} {\bibfnamefont {M.}~\bibnamefont {{Amiri}}}, \bibinfo {author} {\bibfnamefont {S.}~\bibnamefont {{Amodeo}}}, \bibinfo {author} {\bibfnamefont {R.}~\bibnamefont {{An}}}, \bibinfo {author} {\bibfnamefont {Z.}~\bibnamefont {{Atkins}}}, \bibinfo {author} {\bibfnamefont {J.~E.}\ \bibnamefont {{Austermann}}}, \bibinfo {author} {\bibfnamefont {N.}~\bibnamefont {{Battaglia}}},
  \bibnamefont {and~others},\ }\href {https://doi.org/10.3847/1538-4357/acff5f} {\bibfield  {journal} {\bibinfo  {journal} {\apj}\ }\textbf {\bibinfo {volume} {962}},\ \bibinfo {eid} {113} (\bibinfo {year} {2024})},\ \Eprint {https://arxiv.org/abs/2304.05203} {arXiv:2304.05203 [astro-ph.CO]} \BibitemShut {NoStop}%
\bibitem [{\citenamefont {{Planck Collaboration}}\ \emph {et~al.}(2020{\natexlab{c}})\citenamefont {{Planck Collaboration}}, \citenamefont {{Aghanim}}, \citenamefont {{Akrami}}, \citenamefont {{Arroja}}, \citenamefont {{Ashdown}}, \citenamefont {{Aumont}}, \citenamefont {{Baccigalupi}}, \citenamefont {{Ballardini}}, \citenamefont {{Banday}}, \citenamefont {{Barreiro}}, \citenamefont {{Bartolo}}, \citenamefont {{Basak}}, \citenamefont {{Battye}}, \citenamefont {{Benabed}}, \citenamefont {{Bernard}}, \citenamefont {{Bersanelli}}, \citenamefont {{Bielewicz}}, \citenamefont {{Bock}}, \citenamefont {{Bond}}, \citenamefont {{Borrill}}, \citenamefont {{Bouchet}}, \citenamefont {{Boulanger}}, \citenamefont {{Bucher}}, \citenamefont {{Burigana}}, \citenamefont {{Butler}}, \citenamefont {{Calabrese}}, \citenamefont {{Cardoso}}, \citenamefont {{Carron}}, \citenamefont {{Casaponsa}}, \citenamefont {{Challinor}}, \citenamefont {{Chiang}}, \citenamefont {{Colombo}}, \citenamefont {{Combet}}, \citenamefont {{Contreras}},
  \citenamefont {{Crill}}, \citenamefont {{Cuttaia}}, \citenamefont {{de Bernardis}}, \citenamefont {{de Zotti}}, \citenamefont {{Delabrouille}}, \citenamefont {{Delouis}}, \citenamefont {{D{\'e}sert}}, \citenamefont {{Di Valentino}}, \citenamefont {{Dickinson}}, \citenamefont {{Diego}}, \citenamefont {{Donzelli}}, \citenamefont {{Dor{\'e}}}, \citenamefont {{Douspis}}, \citenamefont {{Ducout}}, \citenamefont {{Dupac}}, \citenamefont {{Efstathiou}}, \citenamefont {{Elsner}}, \citenamefont {{En{\ss}lin}}, \citenamefont {{Eriksen}}, \citenamefont {{Falgarone}}, \citenamefont {{Fantaye}}, \citenamefont {{Fergusson}}, \citenamefont {{Fernandez-Cobos}}, \citenamefont {{Finelli}}, \citenamefont {{Forastieri}}, \citenamefont {{Frailis}}, \citenamefont {{Franceschi}}, \citenamefont {{Frolov}}, \citenamefont {{Galeotta}}, \citenamefont {{Galli}}, \citenamefont {{Ganga}}, \citenamefont {{G{\'e}nova-Santos}}, \citenamefont {{Gerbino}}, \citenamefont {{Ghosh}}, \citenamefont {{Gonz{\'a}lez-Nuevo}}, \citenamefont
  {{G{\'o}rski}}, \citenamefont {{Gratton}}, \citenamefont {{Gruppuso}}, \citenamefont {{Gudmundsson}}, \citenamefont {{Hamann}}, \citenamefont {{Handley}}, \citenamefont {{Hansen}}, \citenamefont {{Helou}}, \citenamefont {{Herranz}}, \citenamefont {{Hildebrandt}}, \citenamefont {{Hivon}}, \citenamefont {{Huang}}, \citenamefont {{Jaffe}}, \citenamefont {{Jones}}, \citenamefont {{Karakci}}, \citenamefont {{Keih{\"a}nen}}, \citenamefont {{Keskitalo}}, \citenamefont {{Kiiveri}}, \citenamefont {{Kim}}, \citenamefont {{Kisner}}, \citenamefont {{Knox}}, \citenamefont {{Krachmalnicoff}}, \citenamefont {{Kunz}}, \citenamefont {{Kurki-Suonio}}, \citenamefont {{Lagache}}, \citenamefont {{Lamarre}}, \citenamefont {{Langer}}, \citenamefont {{Lasenby}}, \citenamefont {{Lattanzi}}, \citenamefont {{Lawrence}}, \citenamefont {{Le Jeune}}, \citenamefont {{Leahy}}, \citenamefont {{Lesgourgues}}, \citenamefont {{Levrier}}, \citenamefont {{Lewis}}, \citenamefont {{Liguori}}, \citenamefont {{Lilje}}, \citenamefont {{Lilley}},
  \citenamefont {{Lindholm}}, \citenamefont {{L{\'o}pez-Caniego}}, \citenamefont {{Lubin}}, \citenamefont {{Ma}}, \citenamefont {{Mac{\'\i}as-P{\'e}rez}}, \citenamefont {{Maggio}}, \citenamefont {{Maino}}, \citenamefont {{Mandolesi}}, \citenamefont {{Mangilli}}, \citenamefont {{Marcos-Caballero}}, \citenamefont {{Maris}}, \citenamefont {{Martin}}, \citenamefont {{Martinelli}}, \citenamefont {{Mart{\'\i}nez-Gonz{\'a}lez}}, \citenamefont {{Matarrese}}, \citenamefont {{Mauri}}, \citenamefont {{McEwen}}, \citenamefont {{Meerburg}}, \citenamefont {{Meinhold}}, \citenamefont {{Melchiorri}}, \citenamefont {{Mennella}}, \citenamefont {{Migliaccio}}, \citenamefont {{Millea}}, \citenamefont {{Mitra}}, \citenamefont {{Miville-Desch{\^e}nes}}, \citenamefont {{Molinari}}, \citenamefont {{Moneti}}, \citenamefont {{Montier}}, \citenamefont {{Morgante}}, \citenamefont {{Moss}}, \citenamefont {{Mottet}}, \citenamefont {{M{\"u}nchmeyer}}, \citenamefont {{Natoli}}, \citenamefont {{N{\o}rgaard-Nielsen}}, \citenamefont
  {{Oxborrow}}, \citenamefont {{Pagano}}, \citenamefont {{Paoletti}}, \citenamefont {{Partridge}}, \citenamefont {{Patanchon}}, \citenamefont {{Pearson}}, \citenamefont {{Peel}}, \citenamefont {{Peiris}}, \citenamefont {{Perrotta}}, \citenamefont {{Pettorino}}, \citenamefont {{Piacentini}}, \citenamefont {{Polastri}}, \citenamefont {{Polenta}}, \citenamefont {{Puget}}, \citenamefont {{Rachen}}, \citenamefont {{Reinecke}}, \citenamefont {{Remazeilles}}, \citenamefont {{Renault}}, \citenamefont {{Renzi}}, \citenamefont {{Rocha}}, \citenamefont {{Rosset}}, \citenamefont {{Roudier}}, \citenamefont {{Rubi{\~n}o-Mart{\'\i}n}}, \citenamefont {{Ruiz-Granados}}, \citenamefont {{Salvati}}, \citenamefont {{Sandri}}, \citenamefont {{Savelainen}}, \citenamefont {{Scott}}, \citenamefont {{Shellard}}, \citenamefont {{Shiraishi}}, \citenamefont {{Sirignano}}, \citenamefont {{Sirri}}, \citenamefont {{Spencer}}, \citenamefont {{Sunyaev}}, \citenamefont {{Suur-Uski}}, \citenamefont {{Tauber}}, \citenamefont {{Tavagnacco}},
  \citenamefont {{Tenti}}, \citenamefont {{Terenzi}}, \citenamefont {{Toffolatti}}, \citenamefont {{Tomasi}}, \citenamefont {{Trombetti}}, \citenamefont {{Valiviita}}, \citenamefont {{Van Tent}}, \citenamefont {{Vibert}}, \citenamefont {{Vielva}}, \citenamefont {{Villa}}, \citenamefont {{Vittorio}}, \citenamefont {{Wandelt}}, \citenamefont {{Wehus}}, \citenamefont {{White}}, \citenamefont {{White}}, \citenamefont {{Zacchei}},\ and\ \citenamefont {{Zonca}}}]{Planck-2018-overview}%
  \BibitemOpen
  \bibfield  {author} {\bibinfo {author} {\bibnamefont {{Planck Collaboration}}}, \bibinfo {author} {\bibfnamefont {N.}~\bibnamefont {{Aghanim}}}, \bibinfo {author} {\bibfnamefont {Y.}~\bibnamefont {{Akrami}}}, \bibinfo {author} {\bibfnamefont {F.}~\bibnamefont {{Arroja}}}, \bibinfo {author} {\bibfnamefont {M.}~\bibnamefont {{Ashdown}}}, \bibinfo {author} {\bibfnamefont {J.}~\bibnamefont {{Aumont}}}, \bibinfo {author} {\bibfnamefont {C.}~\bibnamefont {{Baccigalupi}}}, \bibinfo {author} {\bibfnamefont {M.}~\bibnamefont {{Ballardini}}}, \bibinfo {author} {\bibfnamefont {A.~J.}\ \bibnamefont {{Banday}}}, \bibinfo {author} {\bibfnamefont {R.~B.}\ \bibnamefont {{Barreiro}}}, \bibinfo {author} {\bibfnamefont {N.}~\bibnamefont {{Bartolo}}}, \bibinfo {author} {\bibfnamefont {S.}~\bibnamefont {{Basak}}}, \bibinfo {author} {\bibfnamefont {R.}~\bibnamefont {{Battye}}}, \bibinfo {author} {\bibfnamefont {K.}~\bibnamefont {{Benabed}}}, \bibinfo {author} {\bibfnamefont {J.~P.}\ \bibnamefont {{Bernard}}}, \bibnamefont
  {and~others},\ }\href {https://doi.org/10.1051/0004-6361/201833880} {\bibfield  {journal} {\bibinfo  {journal} {\aap}\ }\textbf {\bibinfo {volume} {641}},\ \bibinfo {eid} {A1} (\bibinfo {year} {2020}{\natexlab{c}})},\ \Eprint {https://arxiv.org/abs/1807.06205} {arXiv:1807.06205 [astro-ph.CO]} \BibitemShut {NoStop}%
\bibitem [{\citenamefont {{Planck Collaboration}}\ \emph {et~al.}(2020{\natexlab{d}})\citenamefont {{Planck Collaboration}}, \citenamefont {{Aghanim}}, \citenamefont {{Akrami}}, \citenamefont {{Ashdown}}, \citenamefont {{Aumont}}, \citenamefont {{Baccigalupi}}, \citenamefont {{Ballardini}}, \citenamefont {{Banday}}, \citenamefont {{Barreiro}}, \citenamefont {{Bartolo}}, \citenamefont {{Basak}}, \citenamefont {{Benabed}}, \citenamefont {{Bernard}}, \citenamefont {{Bersanelli}}, \citenamefont {{Bielewicz}}, \citenamefont {{Bock}}, \citenamefont {{Bond}}, \citenamefont {{Borrill}}, \citenamefont {{Bouchet}}, \citenamefont {{Boulanger}}, \citenamefont {{Bucher}}, \citenamefont {{Burigana}}, \citenamefont {{Calabrese}}, \citenamefont {{Cardoso}}, \citenamefont {{Carron}}, \citenamefont {{Challinor}}, \citenamefont {{Chiang}}, \citenamefont {{Colombo}}, \citenamefont {{Combet}}, \citenamefont {{Crill}}, \citenamefont {{Cuttaia}}, \citenamefont {{de Bernardis}}, \citenamefont {{de Zotti}}, \citenamefont
  {{Delabrouille}}, \citenamefont {{Di Valentino}}, \citenamefont {{Diego}}, \citenamefont {{Dor{\'e}}}, \citenamefont {{Douspis}}, \citenamefont {{Ducout}}, \citenamefont {{Dupac}}, \citenamefont {{Efstathiou}}, \citenamefont {{Elsner}}, \citenamefont {{En{\ss}lin}}, \citenamefont {{Eriksen}}, \citenamefont {{Fantaye}}, \citenamefont {{Fernandez-Cobos}}, \citenamefont {{Finelli}}, \citenamefont {{Forastieri}}, \citenamefont {{Frailis}}, \citenamefont {{Fraisse}}, \citenamefont {{Franceschi}}, \citenamefont {{Frolov}}, \citenamefont {{Galeotta}}, \citenamefont {{Galli}}, \citenamefont {{Ganga}}, \citenamefont {{G{\'e}nova-Santos}}, \citenamefont {{Gerbino}}, \citenamefont {{Ghosh}}, \citenamefont {{Gonz{\'a}lez-Nuevo}}, \citenamefont {{G{\'o}rski}}, \citenamefont {{Gratton}}, \citenamefont {{Gruppuso}}, \citenamefont {{Gudmundsson}}, \citenamefont {{Hamann}}, \citenamefont {{Handley}}, \citenamefont {{Hansen}}, \citenamefont {{Herranz}}, \citenamefont {{Hivon}}, \citenamefont {{Huang}}, \citenamefont
  {{Jaffe}}, \citenamefont {{Jones}}, \citenamefont {{Karakci}}, \citenamefont {{Keih{\"a}nen}}, \citenamefont {{Keskitalo}}, \citenamefont {{Kiiveri}}, \citenamefont {{Kim}}, \citenamefont {{Knox}}, \citenamefont {{Krachmalnicoff}}, \citenamefont {{Kunz}}, \citenamefont {{Kurki-Suonio}}, \citenamefont {{Lagache}}, \citenamefont {{Lamarre}}, \citenamefont {{Lasenby}}, \citenamefont {{Lattanzi}}, \citenamefont {{Lawrence}}, \citenamefont {{Le Jeune}}, \citenamefont {{Levrier}}, \citenamefont {{Lewis}}, \citenamefont {{Liguori}}, \citenamefont {{Lilje}}, \citenamefont {{Lindholm}}, \citenamefont {{L{\'o}pez-Caniego}}, \citenamefont {{Lubin}}, \citenamefont {{Ma}}, \citenamefont {{Mac{\'\i}as-P{\'e}rez}}, \citenamefont {{Maggio}}, \citenamefont {{Maino}}, \citenamefont {{Mandolesi}}, \citenamefont {{Mangilli}}, \citenamefont {{Marcos-Caballero}}, \citenamefont {{Maris}}, \citenamefont {{Martin}}, \citenamefont {{Mart{\'\i}nez-Gonz{\'a}lez}}, \citenamefont {{Matarrese}}, \citenamefont {{Mauri}}, \citenamefont
  {{McEwen}}, \citenamefont {{Melchiorri}}, \citenamefont {{Mennella}}, \citenamefont {{Migliaccio}}, \citenamefont {{Miville-Desch{\^e}nes}}, \citenamefont {{Molinari}}, \citenamefont {{Moneti}}, \citenamefont {{Montier}}, \citenamefont {{Morgante}}, \citenamefont {{Moss}}, \citenamefont {{Natoli}}, \citenamefont {{Pagano}}, \citenamefont {{Paoletti}}, \citenamefont {{Partridge}}, \citenamefont {{Patanchon}}, \citenamefont {{Perrotta}}, \citenamefont {{Pettorino}}, \citenamefont {{Piacentini}}, \citenamefont {{Polastri}}, \citenamefont {{Polenta}}, \citenamefont {{Puget}}, \citenamefont {{Rachen}}, \citenamefont {{Reinecke}}, \citenamefont {{Remazeilles}}, \citenamefont {{Renzi}}, \citenamefont {{Rocha}}, \citenamefont {{Rosset}}, \citenamefont {{Roudier}}, \citenamefont {{Rubi{\~n}o-Mart{\'\i}n}}, \citenamefont {{Ruiz-Granados}}, \citenamefont {{Salvati}}, \citenamefont {{Sandri}}, \citenamefont {{Savelainen}}, \citenamefont {{Scott}}, \citenamefont {{Sirignano}}, \citenamefont {{Sunyaev}}, \citenamefont
  {{Suur-Uski}}, \citenamefont {{Tauber}}, \citenamefont {{Tavagnacco}}, \citenamefont {{Tenti}}, \citenamefont {{Toffolatti}}, \citenamefont {{Tomasi}}, \citenamefont {{Trombetti}}, \citenamefont {{Valiviita}}, \citenamefont {{Van Tent}}, \citenamefont {{Vielva}}, \citenamefont {{Villa}}, \citenamefont {{Vittorio}}, \citenamefont {{Wandelt}}, \citenamefont {{Wehus}}, \citenamefont {{White}}, \citenamefont {{White}}, \citenamefont {{Zacchei}},\ and\ \citenamefont {{Zonca}}}]{2020A&A...641A...8P}%
  \BibitemOpen
  \bibfield  {author} {\bibinfo {author} {\bibnamefont {{Planck Collaboration}}}, \bibinfo {author} {\bibfnamefont {N.}~\bibnamefont {{Aghanim}}}, \bibinfo {author} {\bibfnamefont {Y.}~\bibnamefont {{Akrami}}}, \bibinfo {author} {\bibfnamefont {M.}~\bibnamefont {{Ashdown}}}, \bibinfo {author} {\bibfnamefont {J.}~\bibnamefont {{Aumont}}}, \bibinfo {author} {\bibfnamefont {C.}~\bibnamefont {{Baccigalupi}}}, \bibinfo {author} {\bibfnamefont {M.}~\bibnamefont {{Ballardini}}}, \bibinfo {author} {\bibfnamefont {A.~J.}\ \bibnamefont {{Banday}}}, \bibinfo {author} {\bibfnamefont {R.~B.}\ \bibnamefont {{Barreiro}}}, \bibinfo {author} {\bibfnamefont {N.}~\bibnamefont {{Bartolo}}}, \bibinfo {author} {\bibfnamefont {S.}~\bibnamefont {{Basak}}}, \bibinfo {author} {\bibfnamefont {K.}~\bibnamefont {{Benabed}}}, \bibinfo {author} {\bibfnamefont {J.~P.}\ \bibnamefont {{Bernard}}}, \bibinfo {author} {\bibfnamefont {M.}~\bibnamefont {{Bersanelli}}}, \bibinfo {author} {\bibfnamefont {P.}~\bibnamefont {{Bielewicz}}}, \bibnamefont
  {and~others},\ }\href {https://doi.org/10.1051/0004-6361/201833886} {\bibfield  {journal} {\bibinfo  {journal} {\aap}\ }\textbf {\bibinfo {volume} {641}},\ \bibinfo {eid} {A8} (\bibinfo {year} {2020}{\natexlab{d}})},\ \Eprint {https://arxiv.org/abs/1807.06210} {arXiv:1807.06210 [astro-ph.CO]} \BibitemShut {NoStop}%
\bibitem [{\citenamefont {{Planck Collaboration}}\ \emph {et~al.}(2020{\natexlab{e}})\citenamefont {{Planck Collaboration}}, \citenamefont {{Akrami}}, \citenamefont {{Andersen}}, \citenamefont {{Ashdown}}, \citenamefont {{Baccigalupi}}, \citenamefont {{Ballardini}}, \citenamefont {{Banday}}, \citenamefont {{Barreiro}}, \citenamefont {{Bartolo}}, \citenamefont {{Basak}}, \citenamefont {{Benabed}}, \citenamefont {{Bernard}}, \citenamefont {{Bersanelli}}, \citenamefont {{Bielewicz}}, \citenamefont {{Bond}}, \citenamefont {{Borrill}}, \citenamefont {{Burigana}}, \citenamefont {{Butler}}, \citenamefont {{Calabrese}}, \citenamefont {{Casaponsa}}, \citenamefont {{Chiang}}, \citenamefont {{Colombo}}, \citenamefont {{Combet}}, \citenamefont {{Crill}}, \citenamefont {{Cuttaia}}, \citenamefont {{de Bernardis}}, \citenamefont {{de Rosa}}, \citenamefont {{de Zotti}}, \citenamefont {{Delabrouille}}, \citenamefont {{Di Valentino}}, \citenamefont {{Diego}}, \citenamefont {{Dor{\'e}}}, \citenamefont {{Douspis}}, \citenamefont
  {{Dupac}}, \citenamefont {{Eriksen}}, \citenamefont {{Fernandez-Cobos}}, \citenamefont {{Finelli}}, \citenamefont {{Frailis}}, \citenamefont {{Fraisse}}, \citenamefont {{Franceschi}}, \citenamefont {{Frolov}}, \citenamefont {{Galeotta}}, \citenamefont {{Galli}}, \citenamefont {{Ganga}}, \citenamefont {{Gerbino}}, \citenamefont {{Ghosh}}, \citenamefont {{Gonz{\'a}lez-Nuevo}}, \citenamefont {{G{\'o}rski}}, \citenamefont {{Gruppuso}}, \citenamefont {{Gudmundsson}}, \citenamefont {{Handley}}, \citenamefont {{Helou}}, \citenamefont {{Herranz}}, \citenamefont {{Hildebrandt}}, \citenamefont {{Hivon}}, \citenamefont {{Huang}}, \citenamefont {{Jaffe}}, \citenamefont {{Jones}}, \citenamefont {{Keih{\"a}nen}}, \citenamefont {{Keskitalo}}, \citenamefont {{Kiiveri}}, \citenamefont {{Kim}}, \citenamefont {{Kisner}}, \citenamefont {{Krachmalnicoff}}, \citenamefont {{Kunz}}, \citenamefont {{Kurki-Suonio}}, \citenamefont {{Lasenby}}, \citenamefont {{Lattanzi}}, \citenamefont {{Lawrence}}, \citenamefont {{Le Jeune}},
  \citenamefont {{Levrier}}, \citenamefont {{Liguori}}, \citenamefont {{Lilje}}, \citenamefont {{Lilley}}, \citenamefont {{Lindholm}}, \citenamefont {{L{\'o}pez-Caniego}}, \citenamefont {{Lubin}}, \citenamefont {{Mac{\'\i}as-P{\'e}rez}}, \citenamefont {{Maino}}, \citenamefont {{Mandolesi}}, \citenamefont {{Marcos-Caballero}}, \citenamefont {{Maris}}, \citenamefont {{Martin}}, \citenamefont {{Mart{\'\i}nez-Gonz{\'a}lez}}, \citenamefont {{Matarrese}}, \citenamefont {{Mauri}}, \citenamefont {{McEwen}}, \citenamefont {{Meinhold}}, \citenamefont {{Mennella}}, \citenamefont {{Migliaccio}}, \citenamefont {{Mitra}}, \citenamefont {{Molinari}}, \citenamefont {{Montier}}, \citenamefont {{Morgante}}, \citenamefont {{Moss}}, \citenamefont {{Natoli}}, \citenamefont {{Paoletti}}, \citenamefont {{Partridge}}, \citenamefont {{Patanchon}}, \citenamefont {{Pearson}}, \citenamefont {{Pearson}}, \citenamefont {{Perrotta}}, \citenamefont {{Piacentini}}, \citenamefont {{Polenta}}, \citenamefont {{Rachen}}, \citenamefont
  {{Reinecke}}, \citenamefont {{Remazeilles}}, \citenamefont {{Renzi}}, \citenamefont {{Rocha}}, \citenamefont {{Rosset}}, \citenamefont {{Roudier}}, \citenamefont {{Rubi{\~n}o-Mart{\'\i}n}}, \citenamefont {{Ruiz-Granados}}, \citenamefont {{Salvati}}, \citenamefont {{Savelainen}}, \citenamefont {{Scott}}, \citenamefont {{Sirignano}}, \citenamefont {{Sirri}}, \citenamefont {{Spencer}}, \citenamefont {{Suur-Uski}}, \citenamefont {{Svalheim}}, \citenamefont {{Tauber}}, \citenamefont {{Tavagnacco}}, \citenamefont {{Tenti}}, \citenamefont {{Terenzi}}, \citenamefont {{Thommesen}}, \citenamefont {{Toffolatti}}, \citenamefont {{Tomasi}}, \citenamefont {{Tristram}}, \citenamefont {{Trombetti}}, \citenamefont {{Valiviita}}, \citenamefont {{Van Tent}}, \citenamefont {{Vielva}}, \citenamefont {{Villa}}, \citenamefont {{Vittorio}}, \citenamefont {{Wandelt}}, \citenamefont {{Wehus}}, \citenamefont {{Zacchei}},\ and\ \citenamefont {{Zonca}}}]{Planck-2020-NPIPE}%
  \BibitemOpen
  \bibfield  {author} {\bibinfo {author} {\bibnamefont {{Planck Collaboration}}}, \bibinfo {author} {\bibfnamefont {Y.}~\bibnamefont {{Akrami}}}, \bibinfo {author} {\bibfnamefont {K.~J.}\ \bibnamefont {{Andersen}}}, \bibinfo {author} {\bibfnamefont {M.}~\bibnamefont {{Ashdown}}}, \bibinfo {author} {\bibfnamefont {C.}~\bibnamefont {{Baccigalupi}}}, \bibinfo {author} {\bibfnamefont {M.}~\bibnamefont {{Ballardini}}}, \bibinfo {author} {\bibfnamefont {A.~J.}\ \bibnamefont {{Banday}}}, \bibinfo {author} {\bibfnamefont {R.~B.}\ \bibnamefont {{Barreiro}}}, \bibinfo {author} {\bibfnamefont {N.}~\bibnamefont {{Bartolo}}}, \bibinfo {author} {\bibfnamefont {S.}~\bibnamefont {{Basak}}}, \bibinfo {author} {\bibfnamefont {K.}~\bibnamefont {{Benabed}}}, \bibinfo {author} {\bibfnamefont {J.~P.}\ \bibnamefont {{Bernard}}}, \bibinfo {author} {\bibfnamefont {M.}~\bibnamefont {{Bersanelli}}}, \bibinfo {author} {\bibfnamefont {P.}~\bibnamefont {{Bielewicz}}}, \bibinfo {author} {\bibfnamefont {J.~R.}\ \bibnamefont {{Bond}}},
  \bibnamefont {and~others},\ }\href {https://doi.org/10.1051/0004-6361/202038073} {\bibfield  {journal} {\bibinfo  {journal} {\aap}\ }\textbf {\bibinfo {volume} {643}},\ \bibinfo {eid} {A42} (\bibinfo {year} {2020}{\natexlab{e}})},\ \Eprint {https://arxiv.org/abs/2007.04997} {arXiv:2007.04997 [astro-ph.CO]} \BibitemShut {NoStop}%
\bibitem [{\citenamefont {Delouis}\ \emph {et~al.}(2019)\citenamefont {Delouis}, \citenamefont {Pagano}, \citenamefont {Mottet}, \citenamefont {Puget},\ and\ \citenamefont {Vibert}}]{Delouis:2019bub}%
  \BibitemOpen
  \bibfield  {author} {\bibinfo {author} {\bibfnamefont {J.~M.}\ \bibnamefont {Delouis}}, \bibinfo {author} {\bibfnamefont {L.}~\bibnamefont {Pagano}}, \bibinfo {author} {\bibfnamefont {S.}~\bibnamefont {Mottet}}, \bibinfo {author} {\bibfnamefont {J.~L.}\ \bibnamefont {Puget}},\ \bibnamefont {and}\ \bibinfo {author} {\bibfnamefont {L.}~\bibnamefont {Vibert}},\ }\href {https://doi.org/10.1051/0004-6361/201834882} {\bibfield  {journal} {\bibinfo  {journal} {Astron. Astrophys.}\ }\textbf {\bibinfo {volume} {629}},\ \bibinfo {pages} {A38} (\bibinfo {year} {2019})},\ \Eprint {https://arxiv.org/abs/1901.11386} {arXiv:1901.11386 [astro-ph.CO]} \BibitemShut {NoStop}%
\bibitem [{\citenamefont {de~Belsunce}\ \emph {et~al.}(2021)\citenamefont {de~Belsunce}, \citenamefont {Gratton}, \citenamefont {Coulton},\ and\ \citenamefont {Efstathiou}}]{deBelsunce:2021mec}%
  \BibitemOpen
  \bibfield  {author} {\bibinfo {author} {\bibfnamefont {R.}~\bibnamefont {de~Belsunce}}, \bibinfo {author} {\bibfnamefont {S.}~\bibnamefont {Gratton}}, \bibinfo {author} {\bibfnamefont {W.}~\bibnamefont {Coulton}},\ \bibnamefont {and}\ \bibinfo {author} {\bibfnamefont {G.}~\bibnamefont {Efstathiou}},\ }\href {https://doi.org/10.1093/mnras/stab2215} {\bibfield  {journal} {\bibinfo  {journal} {Mon. Not. Roy. Astron. Soc.}\ }\textbf {\bibinfo {volume} {507}},\ \bibinfo {pages} {1072} (\bibinfo {year} {2021})},\ \Eprint {https://arxiv.org/abs/2103.14378} {arXiv:2103.14378 [astro-ph.CO]} \BibitemShut {NoStop}%
\bibitem [{\citenamefont {{Brout}}\ \emph {et~al.}(2022)\citenamefont {{Brout}}, \citenamefont {{Scolnic}}, \citenamefont {{Popovic}}, \citenamefont {{Riess}}, \citenamefont {{Carr}}, \citenamefont {{Zuntz}}, \citenamefont {{Kessler}}, \citenamefont {{Davis}}, \citenamefont {{Hinton}}, \citenamefont {{Jones}}, \citenamefont {{Kenworthy}}, \citenamefont {{Peterson}}, \citenamefont {{Said}}, \citenamefont {{Taylor}}, \citenamefont {{Ali}}, \citenamefont {{Armstrong}}, \citenamefont {{Charvu}}, \citenamefont {{Dwomoh}}, \citenamefont {{Meldorf}}, \citenamefont {{Palmese}}, \citenamefont {{Qu}}, \citenamefont {{Rose}}, \citenamefont {{Sanchez}}, \citenamefont {{Stubbs}}, \citenamefont {{Vincenzi}}, \citenamefont {{Wood}}, \citenamefont {{Brown}}, \citenamefont {{Chen}}, \citenamefont {{Chambers}}, \citenamefont {{Coulter}}, \citenamefont {{Dai}}, \citenamefont {{Dimitriadis}}, \citenamefont {{Filippenko}}, \citenamefont {{Foley}}, \citenamefont {{Jha}}, \citenamefont {{Kelsey}}, \citenamefont {{Kirshner}},
  \citenamefont {{M{\"o}ller}}, \citenamefont {{Muir}}, \citenamefont {{Nadathur}}, \citenamefont {{Pan}}, \citenamefont {{Rest}}, \citenamefont {{Rojas-Bravo}}, \citenamefont {{Sako}}, \citenamefont {{Siebert}}, \citenamefont {{Smith}}, \citenamefont {{Stahl}},\ and\ \citenamefont {{Wiseman}}}]{Brout:2022}%
  \BibitemOpen
  \bibfield  {author} {\bibinfo {author} {\bibfnamefont {D.}~\bibnamefont {{Brout}}}, \bibinfo {author} {\bibfnamefont {D.}~\bibnamefont {{Scolnic}}}, \bibinfo {author} {\bibfnamefont {B.}~\bibnamefont {{Popovic}}}, \bibinfo {author} {\bibfnamefont {A.~G.}\ \bibnamefont {{Riess}}}, \bibinfo {author} {\bibfnamefont {A.}~\bibnamefont {{Carr}}}, \bibinfo {author} {\bibfnamefont {J.}~\bibnamefont {{Zuntz}}}, \bibinfo {author} {\bibfnamefont {R.}~\bibnamefont {{Kessler}}}, \bibinfo {author} {\bibfnamefont {T.~M.}\ \bibnamefont {{Davis}}}, \bibinfo {author} {\bibfnamefont {S.}~\bibnamefont {{Hinton}}}, \bibinfo {author} {\bibfnamefont {D.}~\bibnamefont {{Jones}}}, \bibinfo {author} {\bibfnamefont {W.~D.}\ \bibnamefont {{Kenworthy}}}, \bibinfo {author} {\bibfnamefont {E.~R.}\ \bibnamefont {{Peterson}}}, \bibinfo {author} {\bibfnamefont {K.}~\bibnamefont {{Said}}}, \bibinfo {author} {\bibfnamefont {G.}~\bibnamefont {{Taylor}}}, \bibinfo {author} {\bibfnamefont {N.}~\bibnamefont {{Ali}}}, \bibnamefont {and~others},\
  }\href {https://doi.org/10.3847/1538-4357/ac8e04} {\bibfield  {journal} {\bibinfo  {journal} {\apj}\ }\textbf {\bibinfo {volume} {938}},\ \bibinfo {eid} {110} (\bibinfo {year} {2022})},\ \Eprint {https://arxiv.org/abs/2202.04077} {arXiv:2202.04077 [astro-ph.CO]} \BibitemShut {NoStop}%
\bibitem [{\citenamefont {{Percival}}\ \emph {et~al.}(2002)\citenamefont {{Percival}}, \citenamefont {{Sutherland}}, \citenamefont {{Peacock}}, \citenamefont {{Baugh}}, \citenamefont {{Bland-Hawthorn}}, \citenamefont {{Bridges}}, \citenamefont {{Cannon}}, \citenamefont {{Cole}}, \citenamefont {{Colless}}, \citenamefont {{Collins}}, \citenamefont {{Couch}}, \citenamefont {{Dalton}}, \citenamefont {{De Propris}}, \citenamefont {{Driver}}, \citenamefont {{Efstathiou}}, \citenamefont {{Ellis}}, \citenamefont {{Frenk}}, \citenamefont {{Glazebrook}}, \citenamefont {{Jackson}}, \citenamefont {{Lahav}}, \citenamefont {{Lewis}}, \citenamefont {{Lumsden}}, \citenamefont {{Maddox}}, \citenamefont {{Moody}}, \citenamefont {{Norberg}}, \citenamefont {{Peterson}},\ and\ \citenamefont {{Taylor}}}]{Percival2002:astro-ph/0206256}%
  \BibitemOpen
  \bibfield  {author} {\bibinfo {author} {\bibfnamefont {W.~J.}\ \bibnamefont {{Percival}}}, \bibinfo {author} {\bibfnamefont {W.}~\bibnamefont {{Sutherland}}}, \bibinfo {author} {\bibfnamefont {J.~A.}\ \bibnamefont {{Peacock}}}, \bibinfo {author} {\bibfnamefont {C.~M.}\ \bibnamefont {{Baugh}}}, \bibinfo {author} {\bibfnamefont {J.}~\bibnamefont {{Bland-Hawthorn}}}, \bibinfo {author} {\bibfnamefont {T.}~\bibnamefont {{Bridges}}}, \bibinfo {author} {\bibfnamefont {R.}~\bibnamefont {{Cannon}}}, \bibinfo {author} {\bibfnamefont {S.}~\bibnamefont {{Cole}}}, \bibinfo {author} {\bibfnamefont {M.}~\bibnamefont {{Colless}}}, \bibinfo {author} {\bibfnamefont {C.}~\bibnamefont {{Collins}}}, \bibinfo {author} {\bibfnamefont {W.}~\bibnamefont {{Couch}}}, \bibinfo {author} {\bibfnamefont {G.}~\bibnamefont {{Dalton}}}, \bibinfo {author} {\bibfnamefont {R.}~\bibnamefont {{De Propris}}}, \bibinfo {author} {\bibfnamefont {S.~P.}\ \bibnamefont {{Driver}}}, \bibinfo {author} {\bibfnamefont {G.}~\bibnamefont {{Efstathiou}}},
  \bibnamefont {and~others},\ }\href {https://doi.org/10.1046/j.1365-8711.2002.06001.x} {\bibfield  {journal} {\bibinfo  {journal} {\mnras}\ }\textbf {\bibinfo {volume} {337}},\ \bibinfo {pages} {1068} (\bibinfo {year} {2002})},\ \Eprint {https://arxiv.org/abs/astro-ph/0206256} {arXiv:astro-ph/0206256 [astro-ph]} \BibitemShut {NoStop}%
\bibitem [{\citenamefont {{Calabrese}}\ \emph {et~al.}(2008)\citenamefont {{Calabrese}} \emph {et~al.}}]{Calabrese08}%
  \BibitemOpen
  \bibfield  {author} {\bibinfo {author} {\bibfnamefont {E.}~\bibnamefont {{Calabrese}}} \bibnamefont {and~others},\ }\href {https://doi.org/10.1103/PhysRevD.77.123531} {\bibfield  {journal} {\bibinfo  {journal} {PRD}\ }\textbf {\bibinfo {volume} {77}},\ \bibinfo {eid} {123531} (\bibinfo {year} {2008})},\ \Eprint {https://arxiv.org/abs/0803.2309} {arXiv:0803.2309 [astro-ph]} \BibitemShut {NoStop}%
\bibitem [{\citenamefont {{Hamidreza Mirpoorian}}\ \emph {et~al.}(2025)\citenamefont {{Hamidreza Mirpoorian}}, \citenamefont {{Jedamzik}},\ and\ \citenamefont {{Pogosian}}}]{Mirpoorian:2025rfp}%
  \BibitemOpen
  \bibfield  {author} {\bibinfo {author} {\bibfnamefont {S.}~\bibnamefont {{Hamidreza Mirpoorian}}}, \bibinfo {author} {\bibfnamefont {K.}~\bibnamefont {{Jedamzik}}},\ \bibnamefont {and}\ \bibinfo {author} {\bibfnamefont {L.}~\bibnamefont {{Pogosian}}},\ }\href {https://doi.org/10.48550/arXiv.2504.15274} {\bibfield  {journal} {\bibinfo  {journal} {arXiv e-prints}\ ,\ \bibinfo {eid} {arXiv:2504.15274}} (\bibinfo {year} {2025})},\ \Eprint {https://arxiv.org/abs/2504.15274} {arXiv:2504.15274 [astro-ph.CO]} \BibitemShut {NoStop}%
\bibitem [{\citenamefont {{Green}}\ and\ \citenamefont {{Meyers}}(2024)}]{Green24}%
  \BibitemOpen
  \bibfield  {author} {\bibinfo {author} {\bibfnamefont {D.}~\bibnamefont {{Green}}}\ \bibnamefont {and}\ \bibinfo {author} {\bibfnamefont {J.}~\bibnamefont {{Meyers}}},\ }\href {https://doi.org/10.48550/arXiv.2407.07878} {\bibfield  {journal} {\bibinfo  {journal} {arXiv e-prints}\ ,\ \bibinfo {eid} {arXiv:2407.07878}} (\bibinfo {year} {2024})},\ \Eprint {https://arxiv.org/abs/2407.07878} {arXiv:2407.07878 [astro-ph.CO]} \BibitemShut {NoStop}%
\bibitem [{\citenamefont {{Craig}}\ \emph {et~al.}(2024)\citenamefont {{Craig}}, \citenamefont {{Green}}, \citenamefont {{Meyers}},\ and\ \citenamefont {{Rajendran}}}]{Craig24}%
  \BibitemOpen
  \bibfield  {author} {\bibinfo {author} {\bibfnamefont {N.}~\bibnamefont {{Craig}}}, \bibinfo {author} {\bibfnamefont {D.}~\bibnamefont {{Green}}}, \bibinfo {author} {\bibfnamefont {J.}~\bibnamefont {{Meyers}}},\ \bibnamefont {and}\ \bibinfo {author} {\bibfnamefont {S.}~\bibnamefont {{Rajendran}}},\ }\href {https://doi.org/10.1007/JHEP09(2024)097} {\bibfield  {journal} {\bibinfo  {journal} {Journal of High Energy Physics}\ }\textbf {\bibinfo {volume} {2024}},\ \bibinfo {eid} {97} (\bibinfo {year} {2024})},\ \Eprint {https://arxiv.org/abs/2405.00836} {arXiv:2405.00836 [astro-ph.CO]} \BibitemShut {NoStop}%
\bibitem [{\citenamefont {{Elbers}}\ \emph {et~al.}(2025{\natexlab{b}})\citenamefont {{Elbers}}, \citenamefont {{Frenk}}, \citenamefont {{Jenkins}}, \citenamefont {{Li}},\ and\ \citenamefont {{Pascoli}}}]{Elbers24}%
  \BibitemOpen
  \bibfield  {author} {\bibinfo {author} {\bibfnamefont {W.}~\bibnamefont {{Elbers}}}, \bibinfo {author} {\bibfnamefont {C.~S.}\ \bibnamefont {{Frenk}}}, \bibinfo {author} {\bibfnamefont {A.}~\bibnamefont {{Jenkins}}}, \bibinfo {author} {\bibfnamefont {B.}~\bibnamefont {{Li}}},\ \bibnamefont {and}\ \bibinfo {author} {\bibfnamefont {S.}~\bibnamefont {{Pascoli}}},\ }\href {https://doi.org/10.1103/PhysRevD.111.063534} {\bibfield  {journal} {\bibinfo  {journal} {\prd}\ }\textbf {\bibinfo {volume} {111}},\ \bibinfo {eid} {063534} (\bibinfo {year} {2025}{\natexlab{b}})},\ \Eprint {https://arxiv.org/abs/2407.10965} {arXiv:2407.10965 [astro-ph.CO]} \BibitemShut {NoStop}%
\bibitem [{\citenamefont {{Aiola}}\ \emph {et~al.}(2020)\citenamefont {{Aiola}}, \citenamefont {{Calabrese}}, \citenamefont {{Maurin}}, \citenamefont {{Naess}}, \citenamefont {{Schmitt}}, \citenamefont {{Abitbol}}, \citenamefont {{Addison}}, \citenamefont {{Ade}}, \citenamefont {{Alonso}}, \citenamefont {{Amiri}}, \citenamefont {{Amodeo}}, \citenamefont {{Angile}}, \citenamefont {{Austermann}}, \citenamefont {{Baildon}}, \citenamefont {{Battaglia}}, \citenamefont {{Beall}}, \citenamefont {{Bean}}, \citenamefont {{Becker}}, \citenamefont {{Bond}}, \citenamefont {{Bruno}}, \citenamefont {{Calafut}}, \citenamefont {{Campusano}}, \citenamefont {{Carrero}}, \citenamefont {{Chesmore}}, \citenamefont {{Cho}}, \citenamefont {{Choi}}, \citenamefont {{Clark}}, \citenamefont {{Cothard}}, \citenamefont {{Crichton}}, \citenamefont {{Crowley}}, \citenamefont {{Darwish}}, \citenamefont {{Datta}}, \citenamefont {{Denison}}, \citenamefont {{Devlin}}, \citenamefont {{Duell}}, \citenamefont {{Duff}}, \citenamefont
  {{Duivenvoorden}}, \citenamefont {{Dunkley}}, \citenamefont {{D{\"u}nner}}, \citenamefont {{Essinger-Hileman}}, \citenamefont {{Fankhanel}}, \citenamefont {{Ferraro}}, \citenamefont {{Fox}}, \citenamefont {{Fuzia}}, \citenamefont {{Gallardo}}, \citenamefont {{Gluscevic}}, \citenamefont {{Golec}}, \citenamefont {{Grace}}, \citenamefont {{Gralla}}, \citenamefont {{Guan}}, \citenamefont {{Hall}}, \citenamefont {{Halpern}}, \citenamefont {{Han}}, \citenamefont {{Hargrave}}, \citenamefont {{Hasselfield}}, \citenamefont {{Helton}}, \citenamefont {{Henderson}}, \citenamefont {{Hensley}}, \citenamefont {{Hill}}, \citenamefont {{Hilton}}, \citenamefont {{Hilton}}, \citenamefont {{Hincks}}, \citenamefont {{Hlo{\v{z}}ek}}, \citenamefont {{Ho}}, \citenamefont {{Hubmayr}}, \citenamefont {{Huffenberger}}, \citenamefont {{Hughes}}, \citenamefont {{Infante}}, \citenamefont {{Irwin}}, \citenamefont {{Jackson}}, \citenamefont {{Klein}}, \citenamefont {{Knowles}}, \citenamefont {{Koopman}}, \citenamefont {{Kosowsky}},
  \citenamefont {{Lakey}}, \citenamefont {{Li}}, \citenamefont {{Li}}, \citenamefont {{Li}}, \citenamefont {{Lokken}}, \citenamefont {{Louis}}, \citenamefont {{Lungu}}, \citenamefont {{MacInnis}}, \citenamefont {{Madhavacheril}}, \citenamefont {{Maldonado}}, \citenamefont {{Mallaby-Kay}}, \citenamefont {{Marsden}}, \citenamefont {{McMahon}}, \citenamefont {{Menanteau}}, \citenamefont {{Moodley}}, \citenamefont {{Morton}}, \citenamefont {{Namikawa}}, \citenamefont {{Nati}}, \citenamefont {{Newburgh}}, \citenamefont {{Nibarger}}, \citenamefont {{Nicola}}, \citenamefont {{Niemack}}, \citenamefont {{Nolta}}, \citenamefont {{Orlowski-Sherer}}, \citenamefont {{Page}}, \citenamefont {{Pappas}}, \citenamefont {{Partridge}}, \citenamefont {{Phakathi}}, \citenamefont {{Pisano}}, \citenamefont {{Prince}}, \citenamefont {{Puddu}}, \citenamefont {{Qu}}, \citenamefont {{Rivera}}, \citenamefont {{Robertson}}, \citenamefont {{Rojas}}, \citenamefont {{Salatino}}, \citenamefont {{Schaan}}, \citenamefont {{Schillaci}},
  \citenamefont {{Sehgal}}, \citenamefont {{Sherwin}}, \citenamefont {{Sierra}}, \citenamefont {{Sievers}}, \citenamefont {{Sifon}}, \citenamefont {{Sikhosana}}, \citenamefont {{Simon}}, \citenamefont {{Spergel}}, \citenamefont {{Staggs}}, \citenamefont {{Stevens}}, \citenamefont {{Storer}}, \citenamefont {{Sunder}}, \citenamefont {{Switzer}}, \citenamefont {{Thorne}}, \citenamefont {{Thornton}}, \citenamefont {{Trac}}, \citenamefont {{Treu}}, \citenamefont {{Tucker}}, \citenamefont {{Vale}}, \citenamefont {{Van Engelen}}, \citenamefont {{Van Lanen}}, \citenamefont {{Vavagiakis}}, \citenamefont {{Wagoner}}, \citenamefont {{Wang}}, \citenamefont {{Ward}}, \citenamefont {{Wollack}}, \citenamefont {{Xu}}, \citenamefont {{Zago}},\ and\ \citenamefont {{Zhu}}}]{Aiola:2020}%
  \BibitemOpen
  \bibfield  {author} {\bibinfo {author} {\bibfnamefont {S.}~\bibnamefont {{Aiola}}}, \bibinfo {author} {\bibfnamefont {E.}~\bibnamefont {{Calabrese}}}, \bibinfo {author} {\bibfnamefont {L.}~\bibnamefont {{Maurin}}}, \bibinfo {author} {\bibfnamefont {S.}~\bibnamefont {{Naess}}}, \bibinfo {author} {\bibfnamefont {B.~L.}\ \bibnamefont {{Schmitt}}}, \bibinfo {author} {\bibfnamefont {M.~H.}\ \bibnamefont {{Abitbol}}}, \bibinfo {author} {\bibfnamefont {G.~E.}\ \bibnamefont {{Addison}}}, \bibinfo {author} {\bibfnamefont {P.~A.~R.}\ \bibnamefont {{Ade}}}, \bibinfo {author} {\bibfnamefont {D.}~\bibnamefont {{Alonso}}}, \bibinfo {author} {\bibfnamefont {M.}~\bibnamefont {{Amiri}}}, \bibinfo {author} {\bibfnamefont {S.}~\bibnamefont {{Amodeo}}}, \bibinfo {author} {\bibfnamefont {E.}~\bibnamefont {{Angile}}}, \bibinfo {author} {\bibfnamefont {J.~E.}\ \bibnamefont {{Austermann}}}, \bibinfo {author} {\bibfnamefont {T.}~\bibnamefont {{Baildon}}}, \bibinfo {author} {\bibfnamefont {N.}~\bibnamefont {{Battaglia}}},
  \bibnamefont {and~others},\ }\href {https://doi.org/10.1088/1475-7516/2020/12/047} {\bibfield  {journal} {\bibinfo  {journal} {\jcap}\ }\textbf {\bibinfo {volume} {2020}},\ \bibinfo {eid} {047} (\bibinfo {year} {2020})},\ \Eprint {https://arxiv.org/abs/2007.07288} {arXiv:2007.07288 [astro-ph.CO]} \BibitemShut {NoStop}%
\bibitem [{\citenamefont {Feldman}\ and\ \citenamefont {Cousins}(1998)}]{Feldman:1997qc}%
  \BibitemOpen
  \bibfield  {author} {\bibinfo {author} {\bibfnamefont {G.~J.}\ \bibnamefont {Feldman}}\ \bibnamefont {and}\ \bibinfo {author} {\bibfnamefont {R.~D.}\ \bibnamefont {Cousins}},\ }\href {https://doi.org/10.1103/PhysRevD.57.3873} {\bibfield  {journal} {\bibinfo  {journal} {Phys. Rev. D}\ }\textbf {\bibinfo {volume} {57}},\ \bibinfo {pages} {3873} (\bibinfo {year} {1998})},\ \Eprint {https://arxiv.org/abs/physics/9711021} {arXiv:physics/9711021} \BibitemShut {NoStop}%
\bibitem [{\citenamefont {James}\ and\ \citenamefont {Roos}(1975)}]{JamesMinuit1975}%
  \BibitemOpen
  \bibfield  {author} {\bibinfo {author} {\bibfnamefont {F.}~\bibnamefont {James}}\ \bibnamefont {and}\ \bibinfo {author} {\bibfnamefont {M.}~\bibnamefont {Roos}},\ }\href {https://doi.org/10.1016/0010-4655(75)90039-9} {\bibfield  {journal} {\bibinfo  {journal} {Comput. Phys. Commun.}\ }\textbf {\bibinfo {volume} {10}},\ \bibinfo {pages} {343} (\bibinfo {year} {1975})}\BibitemShut {NoStop}%
\bibitem [{\citenamefont {Dembinski}\ and\ \citenamefont {et~al.}(2020)}]{iminuit}%
  \BibitemOpen
  \bibfield  {author} {\bibinfo {author} {\bibfnamefont {H.}~\bibnamefont {Dembinski}}\ \bibnamefont {and}\ \bibinfo {author} {\bibfnamefont {P.~O.}\ \bibnamefont {et~al.}}\ }\href {https://doi.org/10.5281/zenodo.3949207} {10.5281/zenodo.3949207} (\bibinfo {year} {2020})\BibitemShut {NoStop}%
\end{thebibliography}%


\appendix 

\section{Combining \textbf{\emph{Planck}} and ACT} \label{appendix}

\begin{figure*}
    \centering
    \includegraphics[width=\textwidth]{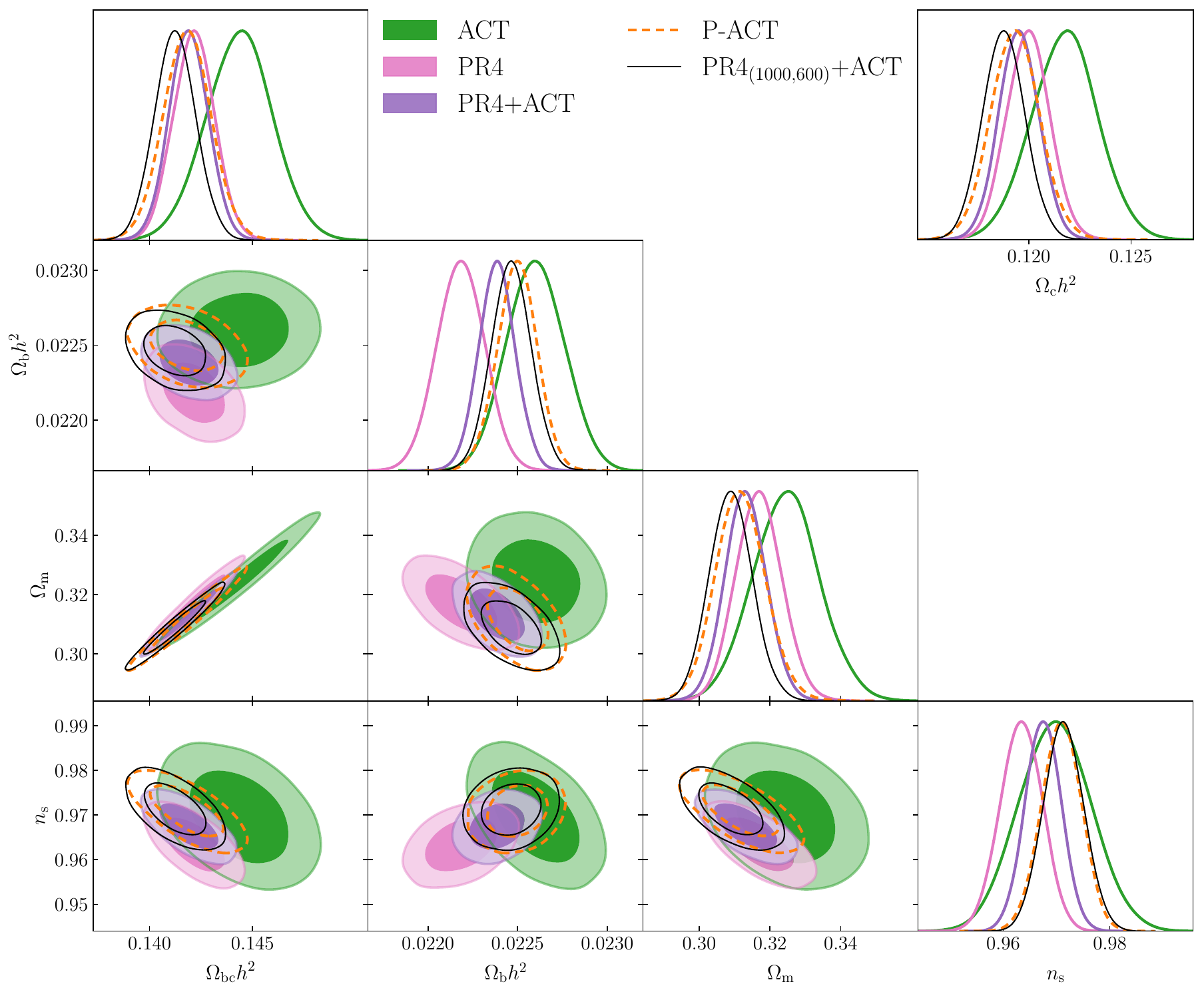}
    \caption{The 68\% and 95\% confidence contours for the parameters $\Omega_\mathrm{bc} h^2$, $\Ob h^2$, $\Om$, and $n_{\rm s}$ using various CMB datasets. The green contours show the ACT data described in \cite{ACT:2025fju} with CMB lensing, while the pink contours correspond to the baseline CMB dataset used in \cite{DESI.DR2.BAO.cosmo}. The baseline CMB dataset of this work is shown in purple, and the P-ACT and PR4$_{(1000,600)}$+ACT results are shown in orange and black unfilled contours, respectively. The $\Ob h^2$-$n_{\rm s}$ panel shows the positive correlation between these parameters as measured by PR4, opposite to the negative correlation measured by ACT (see \cite{Aiola:2020}). The corresponding 1D posterior distributions on $\Ocdm h^2$ are shown in the upper right corner.
    }
    \label{fig:cmb_cuts}
\end{figure*}

We describe how the CMB variations used in this work affect the cosmological parameters within $\Lambda$CDM, under the different multipole cuts used to combine {\it Planck} and ACT DR6. \cref{fig:cmb_cuts} shows the four parameters $\Omega_\mathrm{bc} h^2$, $\Ob h^2$, $\Om$, and $n_{\rm s}$, under ACT, PR4 and several CMB combinations. Here, the parameter $\Omega_\mathrm{bc} h^2$ controls several elements in early physics such as the epoch of matter-radiation equality and the acoustic scale of the CMB peaks, while $\Ob h^2$ affects the baryon-to-photon density ratio. Also, the low-redshift data from background probes is affected by $\Omega_m$, and $n_{\rm s}$ characterizes the scale dependence of the primordial power spectrum of scalar perturbations. Additionally, \cref{fig:cmb_cuts} also shows the 1D posterior for $\Ocdm h^2$ which is a key parameter for the deviations from $(w_0,w_a)=(1,0)$ in $w_0w_a$CDM when combining DESI with CMB.

The ACT data favor higher $\Omega_\mathrm{bc} h^2$ and $\Ob h^2$ values compared to PR4, leading to a small $\rd$. The higher value of $\Omega_\mathrm{c} h^2$ measured from ACT with respect to PR4 also leads to ACT alone being less consistent with DESI. An interesting feature is the correlation observed between $\Ob h^2$ and $n_{\rm s}$ when measured with ACT and PR4 \cite{Aiola:2020}. {\it Planck}, constraining large scales, measures a positive correlation, while ACT, on small scales, measures a negative correlation, making the two surveys highly complementary. This difference in the correlation of $\Ob h^2$ and $n_{\rm s}$  between the two CMB experiments, along with the fact that the measured values of $n_{\rm s}$ and particularly $\Ob h^2$ from ACT are higher than the {\it Planck} prediction, control the behavior of the joint constraints in the higher parameter space.

As defined in \cref{sec:method_data}, the PR4+ACT combination combines PR4 and ACT such that we do not have any overlapping multipole regions and that we can get more constraining power by using each survey in the regime where its uncertainties are lower. \cref{fig:cmb_cuts} shows how PR4+ACT provides a tight constraint that falls between ACT and PR4 following the degeneracy direction of PR4. This constraint means higher $\Ob h^2$ and $n_{\rm s}$ values compared to PR4, with the central values of these parameters landing between the PR4 and ACT constraints. The effect on the rest of the parameters can be understood from the correlations with these parameters. For example, we observe that for PR4  high values of $\Ob h^2$ and $n_{\rm s}$ correspond to low values of $\Omega_\mathrm{bc} h^2$, as observed in \cref{fig:cmb_cuts}. A similar effect occurs for $\Om$ and $\Ocdm h^2$, leading to joint constraints in PR4+ACT to measure lower values of $\Om$ and $\Ocdm h^2$ compared to PR4. Furthermore, the fact that {\it Planck} and ACT individually prefer a higher value of $\Om$ compared to any of their joint combinations (either P-ACT or PR4+ACT) becomes relevant for the tensions shown in \cref{fig:lcdm_constraints}, leading to better consistency with DESI when combining {\it Planck} and ACT. Thus, the inferred values on these parameters play a key role in the consistency with the DESI data and therefore can affect the evidence for evolving dark energy.

\cref{fig:cmb_cuts} also shows the comparison between PR4+ACT and the results from P-ACT and PR4$_{(1000,600)}$+ACT. We observe that the datasets are consistent with each other, with the combinations using PR4 showing slightly tighter constraints. The precision improvement on the $\Omega_\mathrm{c} h^2$ measurement of PR4+ACT is 19\% with respect to P-ACT, coming primarily from adopting PR4 instead of PR3 and secondarily due to our multipole cuts. In fact, the PR4+ACT is 5\% more precise in $\Ocdm h^2$ than PR4$_{(1000,600)}$+ACT, coming exclusively from our choice of the $\ell$ cuts. Finally, the combinations P-ACT and PR4$_{(1000,600)}$+ACT contain more information from ACT than in the case of PR4+ACT. This implies that both P-ACT and PR4$_{(1000,600)}$+ACT measure higher values of $\Ob h^2$ with respect to PR4+ACT, therefore leading to lower values of $\Omega_\mathrm{bc} h^2$, $\Om$ and $\Ocdm h^2$.

\section{Frequentist analysis for neutrino mass constraints} \label{appendix_B}

When constraining the neutrino mass in \cref{sec:neutrinos}, we have seen that the posteriors tend to peak at $\sumnu=0$, which is an indication that the physical prior $\sumnu>0$ impacts our constraints. To further investigate this behavior, we perform a profile likelihood analysis for the combination of DESI DR2 BAO and ACT data\footnote{For computational efficiency, we use the ACT-lite likelihood, a compressed version of the multi-frequency likelihood (\texttt{MFLike}) provided by the ACT team~\cite{ACT:2025fju}. Differences with \texttt{MFLike} are expected to be negligible, given that the inferred cosmological parameters agree to within $0.1\sigma$, as shown in Appendix F of~\cite{ACT:2025fju}.} in the $\Lambda$CDM and $w_0w_a$CDM models, and explore how these constraints are affected by the inclusion of SNe data.

\begin{figure}
    \centering
        \includegraphics[width=\columnwidth]{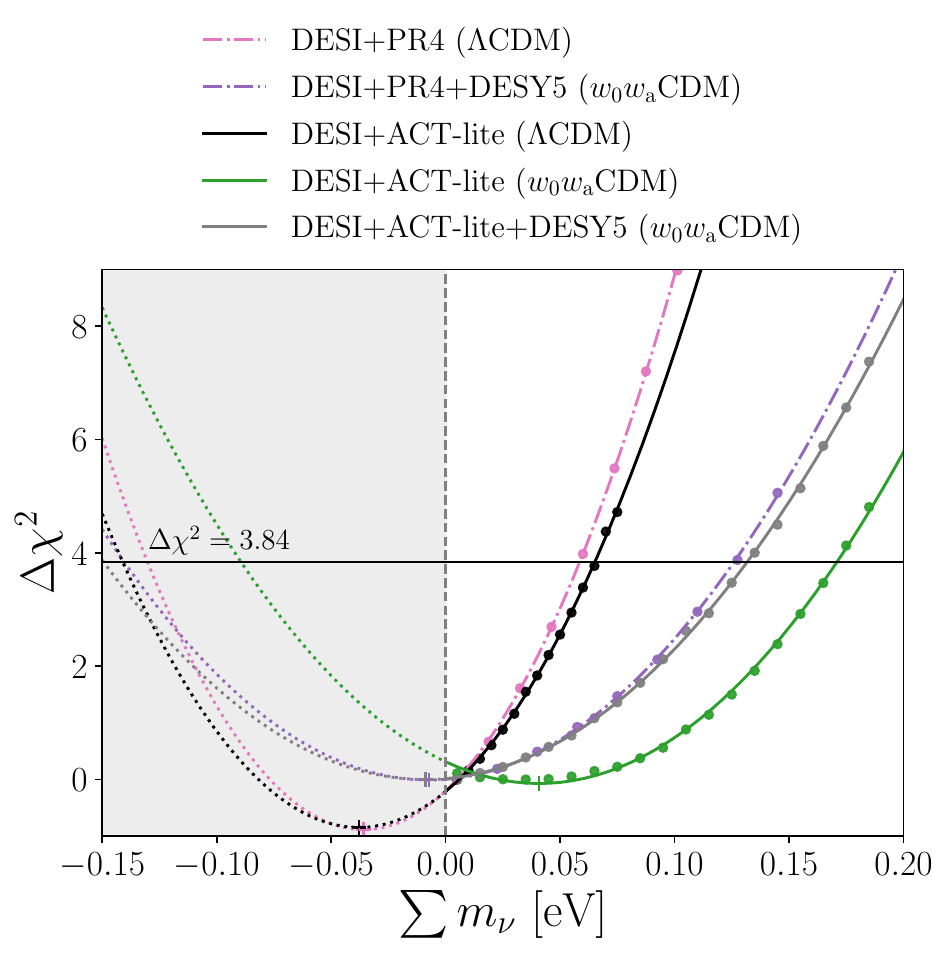}
    \caption{Profile likelihoods for $\sumnu$ from different combinations of datasets and cosmological models. The green curve corresponds to the combination of DESI DR2 BAO and ACT-lite in the $w_0w_a$CDM model, while the gray curve additionally includes SNe data from DESY5. 
     For completeness, the combination DESI DR2 BAO+PR4+DESY5 is shown in purple. 
     The inclusion of SNe information shifts the profile back into the negative $\sumnu$ region, in agreement with the Bayesian results presented in~\cref{sec:neutrinos}. 
     For comparison, we also show $\Lambda$CDM profiles using ACT-lite (black) and PR4 (pink), both with minima lying in the unphysical (negative) region. 
    }
    \label{fig:profile}
\end{figure}

\begin{table}
  \centering
  \resizebox{\columnwidth}{!}{
  \begin{tabular}{l c c c}
    \toprule
    \toprule
    Model/Dataset &   $\mu_0$ [eV] &   $\sigma$ [eV] & 95\% c.l [eV] \\
    \toprule
     DESI+PR4 (\lcdm) &  $-0.036$ &  $0.043$ & $<0.053$ \\
     DESI+ACT-lite (\lcdm) & $-0.038$  &  $0.048$  &$<0.060$ \\
     DESI+ACT-lite (\wowacdm)  &    $ 0.041$ &   $0.066$     &$<0.170$ \\
     DESI+PR4+DESY5  (\wowacdm) &  $-0.007$ &  $0.068$ & $<0.126$ \\
     DESI+ACT-lite+DESY5  (\wowacdm) &    $-0.009$ & $0.072$ & $<0.132$ \\
     \bottomrule
     \bottomrule
  \end{tabular}}
  \caption{Profile likelihood parameters for various dataset combinations and cosmological models. We report the minimum $\mu_0$, the scale $\sigma$, and the 95\% c.l. computed using the Feldman-Cousins prescription \cite{Feldman:1997qc}.
  }
  \label{tab:profile_stats}
\end{table}

We follow the same methodology as in~\cite{Y3.cpe-s2.Elbers.2025}, where the profile likelihood is evaluated for several fixed values of $\sumnu$, while maximizing the likelihood $\mathcal{L}$ (or equivalently minimizing $\chi^2 = -2 \log \mathcal{L}$) with respect to all other cosmological and nuisance parameters. As in that work, we perform a numerical minimization of the log-likelihood using the \texttt{Minuit} algorithm \cite{JamesMinuit1975}, via its Python interface, \texttt{iminuit} \cite{iminuit}. The resulting profiles, shown in \cref{fig:profile}, closely follow a parabolic fit parameterized by the minimum $\mu_0$ and its scale $\sigma$. The latter can be interpreted as the constraining power of the corresponding data combination. The minima $\mu_0$, the scale $\sigma$, and the 95\% c.l. are summarized in \cref{tab:profile_stats}.

The profile likelihood results are in good agreement with the Bayesian findings. In both approaches, the combination of DESI DR2 BAO and CMB data in the $w_0w_a$CDM model yields a peak in the positive neutrino mass region—either in the profile likelihood or in the posterior.
The inclusion of SNe data primarily shifts the profile toward lower values, reaching the boundary of the physical region, while the scale $\sigma$ is only moderately affected. Consequently, the tightening of the estimated upper limit is mainly driven by this shift rather than by a substantial change in the uncertainty scale.

\cref{fig:profile} shows that for the $\Lambda$CDM model, combining DESI DR2 BAO with either ACT-lite or PR4 leads to a profile likelihood minimum that lies outside the physical region, consistent with the Bayesian results presented in \cref{sec:neutrinos}. The minima from both CMB combinations remain almost unaffected, with the only difference being that ACT-lite exhibits weaker constraining power, as also observed in the Bayesian analysis. This latter behavior is also seen in the $w_0w_a$CDM model.

Finally, one can compare the upper bounds on $\sumnu$ derived from both Bayesian and frequentist analyses. In \lcdm{}, the profile likelihood leads to a smaller bound than the Bayesian method by up to $0.013\,\text{eV}$, while in \wowacdm{}, they become very consistent, differing by up to $0.001\,\text{eV}$ at most. This difference can be attributed to the position of the profiles and posteriors with regard to the $\sumnu > 0$ eV bound. In the \wowacdm{} model, the profiles have minima close to or greater than zero, which brings the situation closer to a regular, uninterrupted Gaussian distribution where Bayesian and frequentist frameworks are expected to coincide.

\end{document}